\documentclass[10pt,oneside,psfig]{book}
\usepackage{amsfonts}
\usepackage{latexsym}
\usepackage{epsfig}

\newcommand{\lapp}{\mathrel{\vcenter{\hbox{\tiny \ooalign{\raise 3.25pt
        \hbox{$<$}\crcr $\sim$}}}}}
\newcommand{\gapp}{\mathrel{\vcenter{\hbox{\tiny \ooalign{\raise 3.25pt
        \hbox{$>$}\crcr $\sim$}}}}}
\newcommand{\eqdef}{\!\!\mathrel{\vcenter{\hbox{ \ooalign{\raise 4.75pt
        \hbox{${\textsf{\tiny{\,def}}}$}\crcr $=$}}}}}

\newcommand{\bi}{\begin{itemize}}
\newcommand{\ei}{\end{itemize}}

\newcommand{\forget}[1]{\iffalse#1\fi}
\newcommand{\forgetmenot}[1]{\iftrue#1\fi}

\newcommand{\be}{\begin{equation}}
\newcommand{\ee}{\end{equation}}

\newcommand{\emp}[1]{{\sffamily{\bfseries{#1}}}}

\newcommand{\sfrac}[2]{{\textstyle\frac{#1}{#2}}}
\renewcommand{\:}[2]{{\textstyle\frac{#1}{#2}}}
\renewcommand{\;}[2]{{\frac{#1}{#2}}}

\newcommand{\FRW}{FLRW}
\newcommand{\T}{T}
\newcommand{\ti}{t}

\newcommand{\ba}{\begin{eqnarray}}
\newcommand{\ea}{\end{eqnarray}}
\newcommand{\del}{\nabla}
\newcommand{\pha}{\phantom{h}}
\newcommand{\phan}[1]{\phantom{#1}}
\newcommand{\pd}{\partial}

\newcommand{\li}{&\!\!\!\!}
\newcommand{\lan}{\left\langle}
\newcommand{\ran}{\right\rangle}
\newcommand{\<}{\langle}
\renewcommand{\>}{\rangle}

\newcommand{\etal}{\emph{et~al.}}
\newcommand{\mofz}{magnitude-redshift}

\newcommand{\hu}{km~s$^{-1}$~Mpc$^{-1}$}
\newcommand{\cmb}{_{\scriptscriptstyle{CMB}}}
\newcommand{\dab}{D\c{a}browski}

\newcommand{\udot}{\dot{u}}

\newcommand{\sdel}{\widetilde{\del}}
\newcommand{\ghat}{\hat{g}}
\newcommand{\uhat}{\hat{u}}
\newcommand{\thhat}{\hat{\theta}}
\newcommand{\rothat}{\hat{\omega}}
\newcommand{\ahat}{\hat{\dot{u}}}
\newcommand{\delhat}{\widehat{\del}}

\newcommand{\Ghat}{\hat{G}}
\newcommand{\qdot}{\dot{Q}}

\newcommand{\obs}{^{^{\tiny{\textsf{obs}}}}}

\newtheorem{thm}{Theorem}

\newtheorem{cor}{Corollary}

\newcommand{\astroncite}[2]{\relax}

\begin{document}


\bibliographystyle{astron}

\begin{titlepage}
\begin{center}
{ {\textsc{\large On the Observational Characteristics of Inhomogeneous
Cosmologies:
\protect\\ 
Undermining the Cosmological Principle\\ or\\
 Have cosmologists put all their EGS in one basket?}}

\vskip 3cm

{\textsf{\large A Ph.D. Thesis submitted to\\ The University of Glasgow.}}

\vskip 3cm
{\textsc{\large  Christopher A. Clarkson B.Sc.}}
\vskip 2cm
\large{\textsc{December 1999}}
}
\end{center}

\vskip 3cm
\raggedright{
\textsf{\normalsize Astronomy and Astrophysics Group,\\
 Department of Physics and Astronomy,\\
 University of Glasgow,\\
 Glasgow, G12 8QQ.}
}
\vskip 3cm
\raggedleft{\copyright~C. A. Clarkson, 1999}

\end{titlepage}
\pagenumbering{roman}

\forgetmenot{
\vskip 8cm
\centerline{\psfig{file={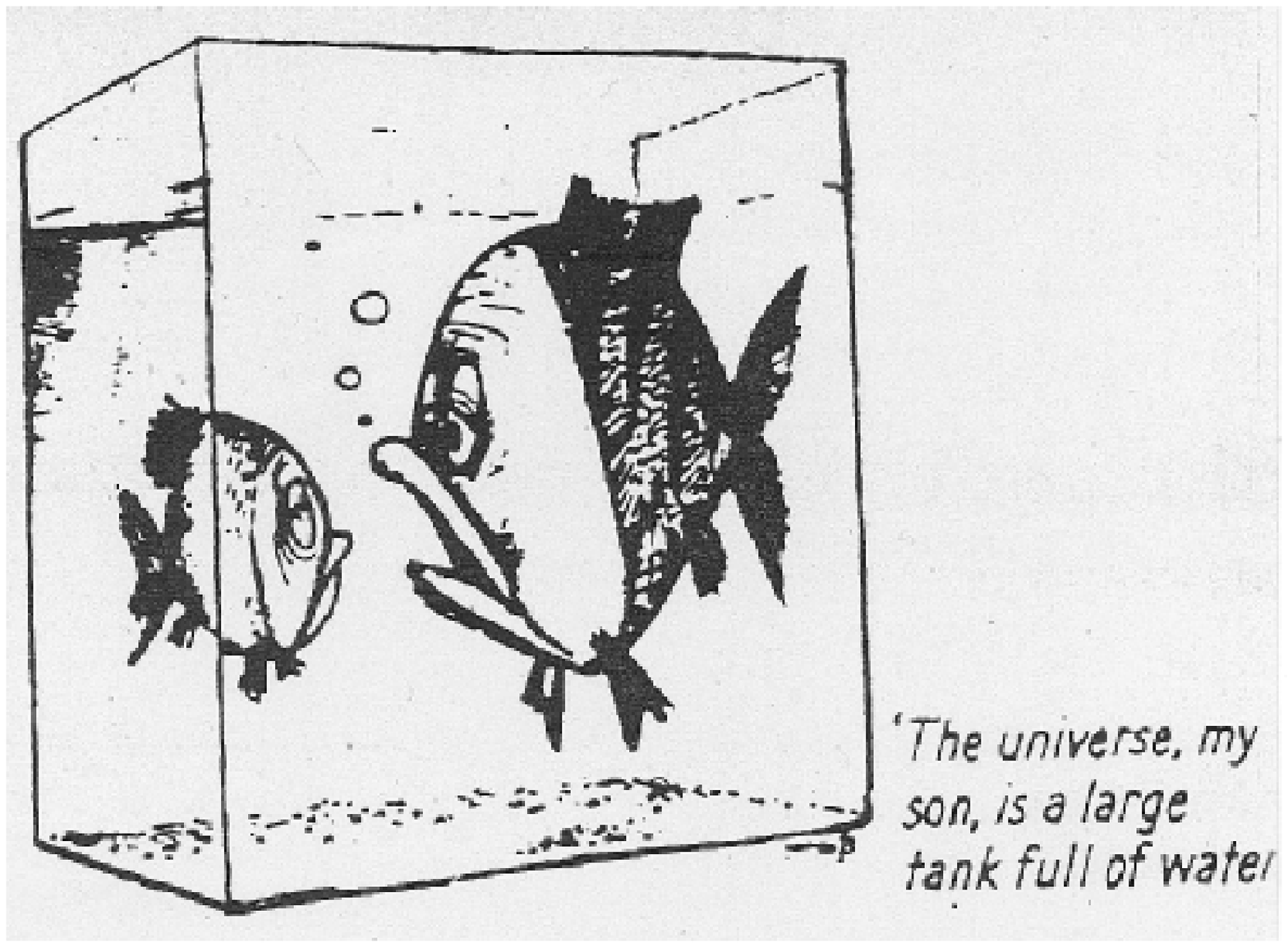},width=14cm,angle=0}}
\begin{center}
{\Large\textsf{The Cosmological Principle}}
\end{center}

\clearpage
}

\forgetmenot{
\forgetmenot{
\chapter*{Acknowledgements}


I think I have had a pretty easy time doing my PhD. In all aspects people have
been willing to help me along, and I have no idea why. Friends with higher
income have been willing to keep me as intoxicated as I can possibly manage to
be; while at university there has been an enthusiasm on the part of some to do
my PhD for me. Perhaps being small and ginger is a good thing after all: it
would seem to create enough sorrow and pity in people to enable me to meander
along without worry. I'm sure it won't last.

I have never had to go without a pint due to the care and cash of Dan, Rois,
and Mike, who have helped my potential as an alcoholic no end. Any time I had
spent all my money on Gold beer at the Garage (approximately one month after
the arrival of each grant cheque) I have always been welcome for a meal and a
drink, and more recently, accommodation (although I'm not quite sure about the
welcome bit; they may be too polite to tell me to leave). Thanks also to Roy
for keeping me up-to-date on his bird/harrison situation, making me feel
better, and for sharing my passion for free jazz.

I would like to thank my sister, Catherine, and her newly acquired husband,
Bill, for getting into enough well timed scrapes across the globe and causing
my parents enough worry to keep off my back. Despite their time spent in panic,
they have still been able to visit and buy me a meal, and give a lecture. My
brother, Nicky, has always been around at the right time for us to poke some
armless fun at. I should like to thank my sister, Jo, for trying with the
utmost love and dedication the difficult project of making the entire family
give up smoking.

All the astro-types have been very pleasant. Thanks to the various
office-mates; Aidan, Iain, and Stephane, who have had to put up with me and my
clothes lying about. Thanks to Suzanne for providing a cheery light in the
first year; and to Guillian for always bitching about stuff, and Noelle for her
Irish patriotism ($33\:13$). More recently, Helen for taking care of me in
various womanly ways, and cooking lovely tuna pie. Special thanks to Martin,
who has taken on the role of official supervisor, and has been of great help in
general, especially in knowing $H_0$. Daphne has of course managed to care for
me within her busy schedule of taking care of everyone.
Thanks to Graeme, the super-computer
manager, who has made all the computers work~-- the first one! Thanks also to
Norman who knows everything about C and \LaTeX, and to Windy who would be happy
to help when Norman wasn't about.

On the outside Glasgow front (a place many Glasweigans have never heard of), I
would like to thank Bruce Bassett for being loud and taking care of me in South
Africa. I would also like to thank Roy Maartens for his recent help on the EGS
stuff, and for encouragement in general. Thanks too to Alan Coley for putting
up cash to offer me a way out. Also thanks to Mariusz \dab\ who initiated the
study of Stephani models. Thanks to Mr Boag for starting off my interest.

For what it's worth, Richard Barrett is the person who deserves the most
gratitude. Since I first came here he has stuck his oar in and made me work.
When my first supervisor, John Simmons, retired and the exhausting, but very
enjoyable, mental task of discussing physics with him stopped, Richard, seeing
a new mind to warp and control, stepped in to assume the unpaid supervisory
role. It if wasn't for him I would now unhappily be expert in analysing stuff
called data, and possibly even employable. As it happened, I have been able to
devote time to a far more interesting area of physics. Richard and I have
worked closely throughout my PhD, and this thesis is as much his work as mine.
His input, encouragement, and general `world advice' have been much
appreciated, if generally unheeded. (I have been made to point out, because he
is anally retentive, that he is not responsible for the mistakes, which are all
my own wok).

On the boring side I have been funded by PPARC, and the Bradly-Watts
foundation. I would like to thank the department for helping fund trips to
Corsica and South Africa. Most of the plots were created using Maple V, except
for figures \ref{bestfit-a-q}~-- \ref{bestfit-a-T}, which were created with IDL
and Guillian. The picture on the second page is from Stephani~(1990).

}
\forgetmenot{
\forgetmenot{
\sf
\chapter*{Summary}
\label{summary chapter-----------------------------}

This thesis concerns the compatibility of inhomogeneous cosmologies with our
present understanding of the universe. It is a problem of some interest to find
the class of all relativistic cosmological models which are capable of
providing a reasonable `fit' to the universe. One can imagine building up an
(infinite-dimensional) parameter space containing all cosmological models. At
any time the understanding of the universe would represent a blob in parameter
space in which, presumably, the real universe would sit. This thesis, in some
respects, is part of this process. We consider Stephani models, which are a
generalisation of the standard Friedmann-Lema\^{\i}tre-Robertson-Walker (FLRW)
models, which can be thought of as FLRW models with curvature which changes
over time. This changing curvature reflects the existence of spatial pressure
gradients which leads to an acceleration of the fundamental observers. Thus
these models generalise the `dust' assumption of standard cosmology.

Models normally considered in the `classification scheme' approach are usually
homogeneous dust or barotropic perfect fluid models. They are anisotropic
however, and thus generalise the FLRW models. Most importantly, because these
models are homogeneous, they satisfy the Copernican principle. The crucial
aspect of this work is the retention of the Copernican principle~-- an
assumption regarded by many as crucial to cosmology. It states that we are not
at a special location in the universe.  This is a vital aspect of the original
work in this thesis: consideration of an inhomogeneous model, while retaining
the Copernican principle has, as far as the author is aware, not been
considered in detail before.

One may formulate the Copernican principle in many ways, from assuming we are
not at a special location, to assuming that all (or most) locations are
equivalent, which more or less forces homogeneity. Because the models
considered here \emph{are} inhomogeneous, they cannot satisfy the stronger
version of the Copernican principle entirely~-- all locations will not be
equivalent. However, we may demand that they are \emph{observationally}
indistinguishable. This is the tactic we use here. \emph{En route} to this goal
we must therefore calculate all observable quantities at any location in the
spacetime. Certain properties of the Stephani models we consider allow us to do
this exactly; consequently, many results of this thesis present, for the first
time, observational relations for a class of inhomogeneous cosmological models
which are exact, and valid for any observer position in the spacetime.

It may reasonably be claimed that the standard model is perfectly acceptable.
However, a number of the properties of the models considered here do make them
rather appealing. For example it is shown in \S\ref{horizon} these models do
not suffer from the horizon problem which is prevalent in the standard model.
Also, the current conflict between cosmologists measuring a small but non-zero
cosmological constant, and particle physics requiring it to be either zero or
one hundred and twenty orders of magnitude higher, may be motivation in itself
for considering cosmologies with a somewhat non-standard matter content.

In chapter~\ref{introduction chapter------------------------} a brief review of
the Copernican and cosmological principles in in homogeneous cosmologies is
presented. A discussion is given of the present understanding of the important
parameters of the standard model, some methods used to find these parameters,
and some of the problems encountered.

In chapter~\ref{relativistic cosmol chapter-----------------} an overview of
relativistic cosmology is presented. This is necessarily incomplete, with an
emphasis on deriving observable relations, and on the FLRW models. The 1+3
formalism is presented. Some new results are presented concerning conformally
related spacetimes (which turn out to be useful but not necessary for deriving
the observational quantities in the Stephani models). FLRW models are then
reviewed. We give a discussion of the generalised definitions of the Hubble
constant and deceleration parameter in non-standard cosmologies.

In chapter~\ref{EGS chapter---------------------------------} we discuss a
theorem of fundamental importance to cosmology -- the Ehlers-Geren-Sachs
theorem~(1968). This condition singles out spacetimes which will allow an
isotropic cosmic microwave background (CMB). When applied to a dust universe it
says that the existence of an isotropic CMB for every observer in the spacetime
implies that the universe is FLRW. We extend this theorem to include the case
of non-geodesic observers (in a perfect fluid model), which singles out a
subclass of the Stephani models with symmetry. These models form the basis of
the rest of the thesis. This chapter is a slightly extended version of Clarkson
and Barrett~(1999).

Chapter~\ref{step chapter obs derivations----------------} takes the models
which were singled out in chapter~\ref{EGS
chapter---------------------------------} and derives the observational
relations for these models. Consideration is given to non-central observers,
and the observational relations are then derived for any location in the
spacetime.

The following two chapters examine the models in some detail from every
observer position. The `worst case'~-- ie, the observer position most
restrictive on the parameter space~-- is singled out for each constraint, and
it is shown that there is a large area of parameter space which is allowed by
the tests we consider. In addition it is shown that some of the allowed models
are distinctly inhomogeneous. Chapter~\ref{step chapter
1------------------------------} is the most thorough, and deals exclusively
with the spherically symmetric subclass of the models derived in
chapter~\ref{EGS chapter---------------------------------}.

}
\rm

\tableofcontents
\listoffigures

\forgetmenot{
\chapter{Introduction and Review}
\label{introduction chapter------------------------}
\pagenumbering{arabic}
\forget{
Ever since the first application of the scientific method there have been
periods with seemingly insurmountable problems to be overcome, leading to brief
periods with comparatively few and supposedly minor problems; at this stage the
belief is that a particular area of science is more or less solved, up to
details. Cosmology is by no means exempt from this, and with the present
complacency and conviction in the standard model being pervasive in the
scientific and popular literature, it would appear that we are entering a
golden age. It is believe that the origin, fate, and global structure of the
universe is understood, while the various parameters which supposedly
characterise the universe only need to be measured in more detail. This is
probably true; however, cosmology as a science is unique in that we \emph{must}
make broad assumptions. For example, we must, in some sense, extrapolate the
known laws of physics across the universe, since we can't go there, we can only
look. This is the very nature of physics. The purpose of this thesis is to
examine the consequences of one of these assumptions; it is preferable to
determine as much as we can about the global structure of the universe on the
basis of what we can \emph{know} rather than what we can make up; religious
people have been doing that since time began. }

Since Hubble discovered the expansion of the universe and the homogeneous and
isotropic expanding models of Friedman, Lema\^\i tre, Robertson, and Walker
(FLRW) were accepted as the correct model of the universe, there has been
relatively little consideration of alternative cosmological models. It is
natural that `mainstream' cosmology focuses on the understanding of the
simplest acceptable models. However, it is also important to consider other
possibilities; seventy years concentrating on one class of models is likely to
lead to undue conviction in these highly special solutions. It is important to
examine the assumptions on which cosmology is based, in the hope of improving
our understanding of the universe. Indeed, it is essential that the assumptions
can be tested wherever possible. This thesis is an attempt to do just that.

The assumption I will investigate in this thesis is that we are geodesic, or
freely falling, observers. There are a number of reasons this assumption has
been used: firstly it simplifies things enormously; secondly, if one `imagines'
galaxies floating about in space, then it seems `obvious' that they must be
freely falling; one assumes that galaxies are like particles of dust (the
classic billiard ball approach to a physical system) which, in itself, implies
geodesic observers.

As convincing as the geodesic assumption is, a standard FLRW dust universe
cannot, on its own, satisfy the latest supernovae Ia (SNIa) results (Riess
\etal,~1998\nocite{riess-98}, Schmidt \etal,~1998\nocite{schmidt-98} and
Perlmutter \emph{et al.},~1999), which imply that the rate of expansion of the
universe is \emph{increasing}; galaxies are moving apart faster and faster as
time goes on. The simplest generalisation to dust, which solves this problem is
the cosmological constant, or vacuum energy density: a concept which has been
invoked and rejected as each new crisis is faced by cosmologists
(figure~\ref{cosmological_constant_over_time}).
\begin{figure}[t]
\centerline{\psfig{file={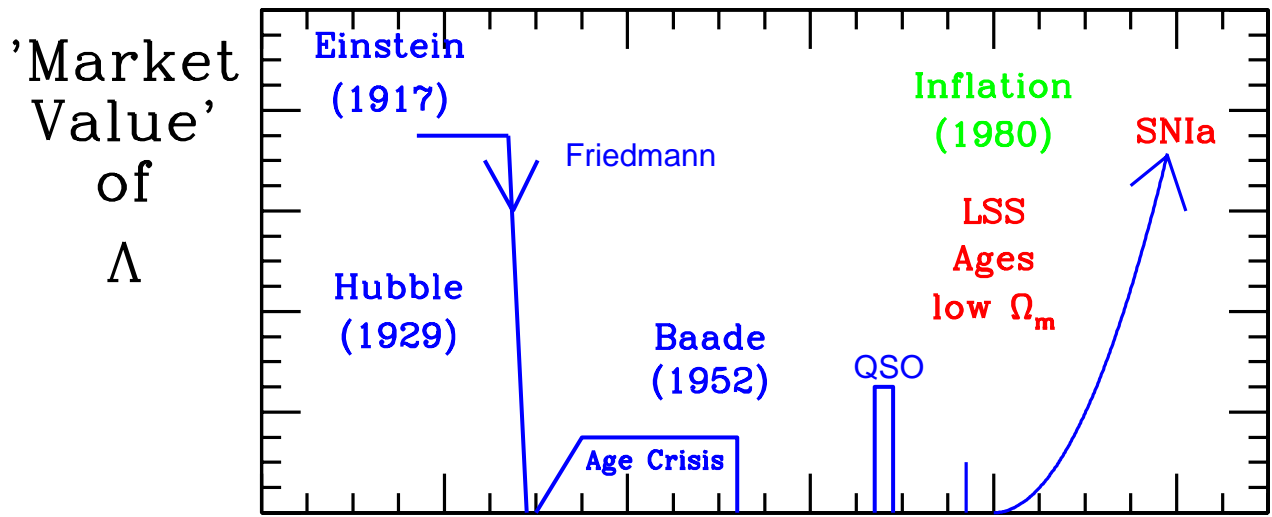}
,width=\textwidth,angle=0}}
\caption{\small The variation in the accepted value of the cosmological constant over time, from Freedman (1999).
\label{cosmological_constant_over_time} }
\end{figure}
\nocite{free99} It was originally invoked by Einstein because of the belief at
the time that the universe was static. It was only when Hubble discovered that
the universe is expanding and the FLRW models were generally accepted that an
alternative route was available. This led to the standard big bang cosmology.

It is possible to achieve an accelerated expansion of the universe without
invoking the cosmological constant, but this requires a large negative
pressure: note that gravity effectively becomes repulsive whenever the pressure
is large and negative enough, that is when
\be
\mu+3p<0;
\ee
(where $\mu,~p$ are the energy density and pressure respectively) which is
precisely when the \emp{strong energy condition} fails (see
\S\ref{energy_cond} and \S\ref{energy_cond_hyp}). This happens because
$\mu+3p$ is the \emp{effective gravitational mass}. This is most elegantly
shown in the \emp{Raychaudhuri equation}~(\ref{raychaudhuri}), which is the
fundamental equation of gravitational attraction. It says that the expansion
rate $\theta$ changes with time according to
\be
\dot\theta\sim-\:12(\mu+3p)+\Lambda
\ee
where $\Lambda$ is the cosmological constant. In fact, it is easy to see
(\S\ref{Perfect Fluids and Dust Models}) that a cosmological constant is
indistinguishable from a constant pressure; we may acknowledge this and
identify the effective active gravitational mass with
\be
\mu+3p-2\Lambda
\ee
which shows why a cosmological constant may lead to an increasing expansion
rate.

\forget{
Specifically we consider a subclass of the Stephani models (Stephani, 1967a,b).
These are the class of perfect fluid spacetimes which are conformally flat
expanding perfect fluids which have zero shear and rotation, but have
non-geodesic observers. The subclass we consider from an observational aspect
are those derived in chapter~\ref{EGS chapter---------------------------------}
which are shown to allow an isotropic cosmic microwave background (CMB), and
are thus relevant from a cosmological point of view.}

Central to this thesis is an assumption we shall keep: the \emp{Copernican
principle}. There is no precise definition of the Copernican principle. In its
weakest form it states that we are not at the center of the universe. This is
not particularly useful for this thesis, so we will take a slightly stronger
version: we are in a `typical' location as observers in the universe. (This is
sometimes known as the weak cosmological principle; Ellis, 1975.) Stronger
still would be to say that \emph{all} observers are equivalent, which would
then force homogeneity. This is the \emp{cosmological principle} (CP), and
cannot be satisfied for inhomogeneous models. The CP follows from (either
version of) the Copernican principle if we assume perfect isotropy about
ourselves. With these assumptions we are lead to the standard homogeneous and
isotropic model of the universe. We aim to show that the homogeneity of the
universe \emph{does not follow} from the Copernican principle, given that
observations about ourselves are not perfectly isotropic. This allows the CP,
and thus the standard model to be questioned.

\forget{
These are usually dropped when considering non-standard models, such as the
spherically symmetric LTB model. In homogeneous models it is easy to retain the
Copernican principle as all points are equivalent by definition.

 In fact, because the models are
inhomogeneous the Copernican principle cannot be completely satisfied; so we
adopt a slightly weaker version: that all locations are observationally
indistinguishable to within the current precision of cosmology today. We do
this by considering observational evidence at any location in the spacetime,
and showing, not that all locations are indistinguishable as in the FLRW
models, but that all locations are \emph{observationally} indistinguishable,
within current observational constraints. Thus it is demonstrated that
homogeneity of the universe \emph{does not follow} from the Copernican
principle, when combined with \emph{observational} data, and follows, in fact,
only if the local isotropy about ourselves is perfect. We are a long way,
observationally speaking, from demonstrating this, although things are pointing
in that direction. This puts doubt in the cosmological principle, and thus the
standard model.}

Although it essentially a philosophical assumption, the Copernican principle
must be taken seriously (Ellis~1975\nocite{ellis75}): in order to reject an
inhomogeneous cosmological model on the basis of its conflicting with the
observed isotropy of the universe it is necessary to consider \emph{all}
observer positions and to show that \emph{for most observers in that spacetime}
the anisotropy observed is too large to be compatible with observations. We
could adopt this view of the Copernican principle in this thesis, but, for
simplicity, we will require consistency with observations for \emph{all}
observers~-- our results will thus be rather stronger than is strictly required
by the Copernican Principle. In fact, it is only really possible to consider
the weaker version of the Copernican principle when the spatial sections of the
universe have finite volume (or, at least, when the number of observers, as
measured by the integral over the spatial sections of the number density of
particles~-- \emph{cf}.~\S\ref{obs-constr}~-- is finite), otherwise the
expression `most observers' has very little meaning. For infinite universes the
stronger version of the Copernican principle must be adopted. With our stronger
definition we are thus prepared for all eventualities (although it will turn
out that the models we consider are all `finite').

Non-central locations are rarely considered in the literature in the analysis
of inhomogeneous cosmologies however, owing to the mathematical difficulties
that this usually entails. Having said that, Humphreys \etal~(1997) made a
study of Tolman-Bondi models from a non-central location and applied their
results to a `Great Attractor' model. Most other non-central analyses, however,
look only at perturbations of standard \FRW\ models.

Having adopted (a strong version of) the Copernican principle, the question
then arises as to whether the observed isotropy of the universe, when required
to hold at every point, forces homogeneity, thus validating the CP. Well, the
nearby universe is distinctly lumpy, so it would be difficult to claim there is
isotropy on that basis. However, the CMB is isotropic to one part in 10$^{5}$.
Together with the EGS theorem (Ehlers, Geren and
Sachs~(1968)\nocite{ehlers-68}), or rather, the almost EGS theorem of Stoeger,
Maartens, and Ellis~(1995)\nocite{stoeg-95} this allows us to say that within
our past lightcone the universe is almost \FRW\ (ie,~almost homogeneous and
isotropic), \emph{provided} the fundamental observers in the universe follow
geodesics (that is, as long as the fundamental fluid is dust). What the almost
EGS theorem achieves is to provide support for the CP without the need to
blithely assume the (near) isotropy of \emph{every} observable: isotropy of the
CMB alone is enough to ensure the validity of the CP (for geodesic observers).

The EGS theorem is crucial to this thesis. It states that if all observers in
an expanding dust universe see an isotropic radiation field, then that universe
is FLRW. As the observed high isotropy of the CMB so well established, we can
combine it with the Copernican principle as the starting point of this work. We
generalise the EGS theorem to the case of an irrotational perfect fluid (ie,
allowing for acceleration), to find a class of models which generalise the FLRW
model. These models are a subclass of the inhomogeneous Stephani spacetimes
with symmetry. We are left with a class of spacetimes which allow an isotropic
CMB \emph{for all observers}. We then study these models from all observer
locations to show that these inhomogeneous models are acceptable given present
observational constraints. They thus satisfy the Copernican principle while
being inhomogeneous.

\forget{Since, however, the Stephani models do not satisfy the conditions of the almost
EGS theorem (the acceleration is non-zero), its conclusions do not apply, we
need not be restricted \emph{a priori} by it. In fact we show in
chapter~\ref{EGS chapter---------------------------------} that a subclass of
the Stephani models actually satisfy the EGS theorem, and thus allow an
isotropic CMB. It is the subclass we derive here which are discussed from other
observational considerations.}

A number of inhomogeneous or anisotropic cosmological models have been studied
in relation to the CP. The homogeneous but anisotropic Bianchi and
Kantowki-Sachs models (Kantowski and Sachs~1966\nocite{kant-sachs66}; see
Ellis~1998,\nocite{ell98}~\S6 and references therein) have been investigated
with regard to the time evolution of the anisotropy. It can be shown, for
example, that there exist Bianchi models for which a significant phase of their
evolution is spent in a near-\FRW\ state, even though at early and late times
they may be highly anisotropic (again, see~\S6 of Ellis~1998 and references
therein). Of the inhomogeneous models that arise in cosmological applications,
by far the most common are the (Lema\^\i tre-)Tolman-Bondi dust spacetimes
(Tolman~1934; Bondi~1947\nocite{tolm34}\nocite{bond47}). These are used both as
global inhomogeneous cosmologies~-- probably the most important papers being
Hellaby and Lake (1984,~1985\nocite{hel-lak-I84}\nocite{hel-lak85}), studying
geometrical aspects, and Rindler and Suson (1989)\nocite{rindler-sus89},
Goicoechea and Martin-Mirones (1987)\nocite{goi-mm97}, Schneider and
C\'el\'erier~(1999)\nocite{sch-cel99}, C\'el\'erier~(1999)\nocite{cel99}, and
Maartens \etal~(1996)\nocite{maar-95} investigating observational aspects~--
and also as models of local, nonlinear perturbations (over- or under-densities)
in an \FRW\ background (Tomita~1995, 1996\nocite{tom96}\nocite{tomita95};
Moffat and Tatarski~1995\nocite{mof-tat95}; Krasi\'{n}ski~1998; Nakao
\etal~1995\nocite{nakao-95}). See also Krasi\'{n}ski (1998)\nocite{kras98} for
a review.

There has been some consideration of Stephani solutions (Stephani
1967a,b\nocite{step67a}\nocite{step67b}; see also Kramer
\etal~1980\nocite{kram-80} and Krasi\'{n}ski
1983,~1997\nocite{Kras83}\nocite{kras97}). These are the most general
conformally flat perfect fluid solutions~-- and obviously therefore contain the
\FRW\ models. They differ from \FRW\ models in general because they have
inhomogeneous pressure, which leads to acceleration of the fundamental
observers. D\c{a}browski and Hendry~(1998) fitted a certain subclass of these
models to the first SNIa data of Perlmutter \etal~(1997) using a low-order
series expansion of the magnitude-redshift relation for central observers
derived in D\c{a}browski~(1995), and found that they were significantly older
than the \FRW\ models that fit that data. \forget{In Barrett and Clarkson
(1999)\nocite{bar-cl99} (Paper~I) we extended those results, obtaining
\emph{exact} distance-redshift and number count-redshift relations, valid for
all~$z$, and showing that it would always be possible to find a more than
acceptable fit to any data that could also be fit by an \FRW\ model (for
plausible ranges of the \FRW\ parameters $H_0$, $\Omega_0$
and~$\Omega_\Lambda$). The best-fit models were again consistently $1-4$~Gyr
older than their \FRW\ counterparts.}

One unusual feature of the Stephani models is their matter content. The usual
perfect-fluid interpretation precludes the existence of a barotropic equation
of state in general (because the density is homogeneous but the pressure is
not), although they can be provided with a strict thermodynamic scheme (Bona
and Coll~1988\nocite{bon-col88})~-- it has recently been shown explicitly that
they may be given a physically reasonable interpretation
(Sussman~1999)\nocite{sus99}. Moreover, even individual fluid elements can
behave in a rather exotic manner, having negative pressure, for example
(\emph{cf}.~\S\ref{energy_cond}). For these reasons, amongst others,
Lorenz-Petzold~(1986)\nocite{lorpet86} has claimed that Stephani models are not
a viable description of the universe, but Krasi\'nski~(1997, p.170) argues
rather vigorously that this conclusion is incorrect, as do we. Other
cosmologies have sometimes been ruled out \emph{a priori} because the
stress-energy tensor does not behave in the correct manner, or for other
reasons such as the lack of an \FRW\ limit~-- see Krasi\'{n}ski
(1997)\nocite{kras97} for a complete review. However, it has become
increasingly difficult to avoid the conclusion that the expansion of the
universe is accelerating, with the type~Ia supernovae data being the latest and
strongest evidence for this. It may be taken as evidence that there is some
kind of `negative pressure' driving the expansion of the universe. In standard
\FRW\ models, this must correspond to an inflationary scenario, with
$\Lambda>0$, or a matter content of the universe which is mostly scalar field
(`quintessence'~-- see Frampton 1999\nocite{framp99}; Liddle
1999\nocite{lid99}; Coble, Dodelson and Frieman 1997\nocite{coble-97}; Liddle
and Scherrer 1998\nocite{lid-sch98}; see also Goliath and Ellis
1998\nocite{gol-ellis98} for a discussion of the dynamical effects associated
with~$\Lambda$). Either way, the real universe is behaving in a manner that is
at odds with `everyday' physics, so it is not appropriate to rule out Stephani
models for exhibiting similar behaviour.

We now briefly review FLRW models and then discuss the observational
constraints primarily used in this thesis.

\section{The FLRW Models}

The standard model of modern cosmology are homogeneous and isotropic expanding
FLRW models. They are parameterised by three independent functions of time;
$\{H_0, \Omega_0, \Omega_\Lambda\}$. The present day expansion rate is given by
the Hubble constant, $H_0$; the density of matter today is given by $\Omega_0$,
which is normalised with $H_0$; and $\Omega_\Lambda$ represents a possible
cosmological constant.

These models expand from a big bang\footnote{Some models may `bounce'
indefinitely depending on the model parameters.} into the universe's present
state. The models may recollapse after a finite time or expand indefinitely
into the future: they are called `closed' or `open' respectively depending on
this future fate, while the limiting case between the two possibilities is
called `flat'. The different possibilities depend on whether the matter density
and cosmological constant (in the combination $\Omega_0+ \Omega_\Lambda$) are
large enough to halt the expansion of the universe. The time since the big bang
depends crucially on $H_0$. A significant problem facing cosmologists is an
accurate measurement of these parameters.

\section{The Hubble constant}

The \emp{Hubble constant}, $H_0$, represents the present day expansion rate of
the universe. It is estimated, not surprisingly, by measuring the recession
velocities of nearby galaxies, which move according to the \emp{Hubble law};
\be
v\approx H_0 d,
\ee
where $v$ is the recession velocity and $d$ is the distance to nearby galaxies.
(This relation holds only in the `local' universe.)

Measuring Hubble's constant has proved to be extremely difficult in practice;
despite 70 years searching it still would be difficult to claim anything better
than 10\% accuracy. In fact it is only within the past few years that the
`factor of 2' uncertainty has been resolved. One of the biggest problems lies
in calculating distance accurately: this becomes more difficult the further
away galaxies lie. Distant galaxies are preferable because their random or
peculiar velocities are substantially smaller than the Hubble expansion. The
most common method of distance measurement is to use the observed magnitude. To
infer distance from this one must obviously know the actual luminosity~-- and
one must therefore use objects which have a narrow range of intrinsic
luminosity which is independent of distance. A good example of such an object
are Cepheid variables which pulsate regularly~-- the period of which is known
to correlate to their luminosity. The most promising example of distant
standard candles are the supernovae Ia (SNIa), which can outshine whole
galaxies~-- and are thus observable to great distances. The maximum brightness,
and subsequent light curve shape is know to correlate to luminosity. Other
examples include spiral galaxies whose rotation velocity correlates with
luminosity (the Tully-Fisher relation).

The main error in using these standard candles\footnote{Actually, a standard
candle is an object with a very narrow range of brightness, but there is no
need to labour this point here.} arises because they must be combined together
to obtain $H_0$ via the `cosmic distance ladder'. One of the main sources of
error in $H_0$ lies in finding the period-luminosity relation  for Cepheids,
which is a consequence of errors in the distance to the Large Magellanic Cloud.

These methods are finally reaching conclusion, albeit with different groups
reaching slightly different mean values; finally, however, all the results seem
to be consistent. At the 95\% confidence level
Freedman~(1999b)\nocite{freed99b} quotes $57<H_0<85$~\hu. Despite this many
would argue that $H_0=65\pm5$~\hu\ is a safe bet; see Trimble and
Aschwanden~(1999)\nocite{tri-ash99} for an entertaining discussion of this.

There are promising new methods for measuring $H_0$ from observations of
objects at large distances. These include time delay in gravitational lensing
events, and use of the Sunyaev-Zel'dovich effect from the X-ray emitting gas in
clusters affecting the cosmic microwave background (CMB) spectrum.

Measurements of $H_0$ from gravitational lensing make use of distant variable
sources which are lensed, producing multiple images. The different paths the
light travels will result in a time delay effect, which is measurable.
Measurements of the angular separation of the two images then allow $H_0$ to be
determined. The problem with this is that the mass distribution of the lensing
galaxy will not be known independently. This, combined with the difficulty of
finding a suitable system makes gravitational lensing a test for the future
(Freedman~1999a\nocite{free99}).

The Sunyaev-Zel'dovich effect is the scattering of the CMB photons off
electrons in the X-ray emitting gas of rich clusters
(Sunyaev-Zel'dovich~1969\nocite{sun-zel69}). This results in a change in the
CMB spectrum. The X-ray flux is distance dependent, but the SZ temperature
fluctuations are not, therefore $H_0$ may be determined. As with the
gravitational lensing measurements, it is still early days, with large
uncertainties, and only a few clusters.

\forget{
 and a whole host of problems arise due to
the \emp{evolution} of the galaxies in question; that is, because the galaxies
are far away, they are also seen in the distant past, and it becomes difficult
to measure or infer their intrinsic size or luminosity. There are ways round
this problem by using \emp{standard candles}, such as cepheids or RR-Lyraes, of
know intrinsic luminosity from local studies, to infer the luminositys of
distant galaxies. The most promising of these are the SNIa used recently --
although they are used to specify the density parameters discussed below, once
a value for $H_0$ is settled upon.

In terms of \emp{luminosity distance}, $r_L$ (cf,~\S\ref{Luminosity-Distance
Relations}), the apparent magnitude $m$ of a galaxy of intrinsic magnitude $M$
is given by the \emp{Pogson equation};
\be
m-M-25=5\log_{10}r_L+K
\ee
where $K$ is the \emp{$K$-correction}. The $K$-correction is dependent on
colour, and is not always known. $M$ will have evolution effects in it making
the uncertainties in this equation quite large in general. $H_0$ is obtained
once a cosmological model is specified as the lowest order term in the function
$r_L$. All the uncertainties in this relation, especially $M$, make the
uncertainties in finding $H_0$ quite large. Combined with this, one has to
measure distant objects in the hope that their peculiar velocities are much
smaller than their recession velocities.

The latest results from the HST key project indicate
\be
H_0=69\pm10 \hbox{\hu}.
\ee
See Ferrarese \etal~(1999)\nocite{fer-99} for details, and
Freedman~(1999b)\nocite{freed99b} and references therein for a thorough
discussion. Most people agree that $H_0=65\pm5$~\hu\ is a safe bet; see Trimble
and Aschwanden~(1999)\nocite{tri-ash99} for an entertaining discussion of this.
However, one can only say at the 95\% confidence level that the range is
$57<H_0<85$ \hu. }

\section{The Density Parameters $\Omega_0$ and $\Omega_\Lambda$, and the Curvature}

In the standard model the energy density, curvature, and cosmological constant
are all related by the normalised (with respect to $H_0$) parameters,
$\{\Omega_0,\Omega_k,\Omega_\Lambda\}$ by
\be
\Omega_0+\Omega_\Lambda+\Omega_k=1.
\ee
We see that determining any two of the three will give the third. Determining
these will determine the origin and fate of the universe.
\forget{
or
\be
\:32\Omega_0+\Omega_k=1+q_0
\ee
where $q_0=\:12\Omega_0-\Omega_\Lambda$ is the \emp{deceleration parameter}
(cf, \S\ref{FLRW Models}). Unfortunately the measurable parameters are $q_0$
from `close' measurements ($z\sim1$) and $\Omega_0+\Omega_\Lambda$ from the CMB
multipoles. See
\S\ref{Observable Quantities in FLRW Models} for details.}

The density and curvature of the universe remain in question, and this
uncertainty has led to the \emp{dark matter problem}. The main problems here
lie in the fact that the luminous matter density of galaxies nearby give a
value of $\Omega_0\sim0.01$ (\emp{Big Bang Nucleosynthesis} also gives similar,
but slightly higher, results~-- see Wainwright and Ellis~1997, or
Olive,~1999\nocite{oli99}, for details), while dynamical studies of the same
galaxies suggest that $\Omega_0\gapp0.3$~-- ie, they are gravitationally bound
by a far greater mass than we see. This means that most of the matter in the
galaxies is `dark'. There have been plenty of suggestions as to what it may be,
see Carr~(1994\nocite{car94}) for further details. In addition to this the
latest SNIa measurements suggest that the deceleration parameter
$q_0=\:12\Omega_0-\Omega_\Lambda<0$ which necessarily implies that
$\Omega_\Lambda>0$~-- see figure~\ref{perlmutter exclusion plot}. If the SNIa
results are correct, then there is no way, in the standard model, to avoid
invoking a cosmological constant or \emp{scalar field}\footnote{Also known as a
\emp{quintessence} model~-- see later.} model.

Measurements of these parameters from `local' ($z\sim1$) observations (eg,
SNIa) , and from observations of the CMB ($z\sim1000$) complement each other
(Tegmark, Eisenetein and Hu, 1998a\nocite{teg-98b}; Tegmark
\etal,~1998b\nocite{teg-98b}; Eisenstein, Hu and Tegmark,
1998a,b\nocite{eis-98a}\nocite{eis-98b}; see also Tegmark, 1999\nocite{teg99};
White, 1998\nocite{whi98}; Efstathiou \etal,~1998\nocite{efs-98}; and
\S\ref{Observable Quantities in FLRW Models}). The SNIa determine the quantity $\Omega_0-\Omega_\Lambda$, while CMB
results give $\Omega_0+\Omega_\Lambda$ (ie, the curvature). See
\S\ref{Observable Quantities in FLRW Models} for details why. This means that
the two data sets will give all of $\{\Omega_0,\Omega_k,\Omega_\Lambda\}$.

\paragraph{Supernovae Results}

Riess \etal~(1998)\nocite{riess-98}, Schmidt \etal~(1998)\nocite{schmidt-98}
and Perlmutter \emph{et al.}~(1999). These all suggest that $\Lambda$ is
distinctly non-zero and positive. Accurate measurements of the SNIa constrain
the deceleration parameter because they measure the change in the expansion
rate with distance. They are observed at roughly half the age of the universe.

The primary errors in using SNIa for the determination of the density
parameters are similar as for $H_0$. In addition, the SNIa may evolve (as has
been suggested by Drell, Loredo, and Wasserman~1999\nocite{dre-99}) ie,
intrinsic brightness changes with distance. This simply amounts to not knowing
precisely enough how intrinsically bright the objects in question are at the
time of emission. The SNIa have caused such a stir initially, because they are
thought not have this problem.

Perlmutter \etal~(1999) quote their best fit as
$\Omega_0=0.73,~\Omega_\Lambda=1.32$; while their best fit results assuming a
flat universe is $\Omega_0=0.28,~\Omega_\Lambda=0.72$ with errors of $\pm0.14$.
Regardless of the type of fit its claimed that $\Lambda>0$ at the 99\%
confidence level. The other group achieve similar results.

\paragraph{CMB Measurements}
\label{CMB Measurements}

The CMB was discovered in 1965 by Penzias and Wilson\nocite{pen-wil65}, and
interpreted as a `relic' of the big bang by Dicke~\etal~(1965)\nocite{dic-65}.

The CMB is observed today to be a blackbody at a temperature of
$T_0=2.734\pm0.01$K, with a dipole moment of $T_1=3.343\pm0.016\times10^{-3}$K
and quadrupole moment as big as $T_2=2.8\times10^{-5}$K (see Mather
\etal,~1994\nocite{mat-94}; Partridge~1997\nocite{part97} for details and
references). It was emitted at a time when the radiation was no longer hot
enough to keep Hydrogen ionised, causing it to decouple from matter, which
happens at $T_{_\mathsf{dec}}\sim 3000$K. At this time, the small perturbations
(created by inflation) which were present in the universe left their mark on
the CMB surface. The baryons (electrons and protons) fell into the potential
wells created by the small perturbations. Because the baryons were still
coupled to the photons, the photon pressure acted as a restoring force against
the motion of the baryons into the potential wells. This led to acoustic
oscillations. These oscillations may be decomposed into their Fourier modes,
with the first acoustic peak at $l\sim200/\sqrt{\Omega_0+\Omega_\Lambda}$~--
see figure~\ref{cmb_data}.
\begin{figure}[t!]
\centerline{\psfig{file={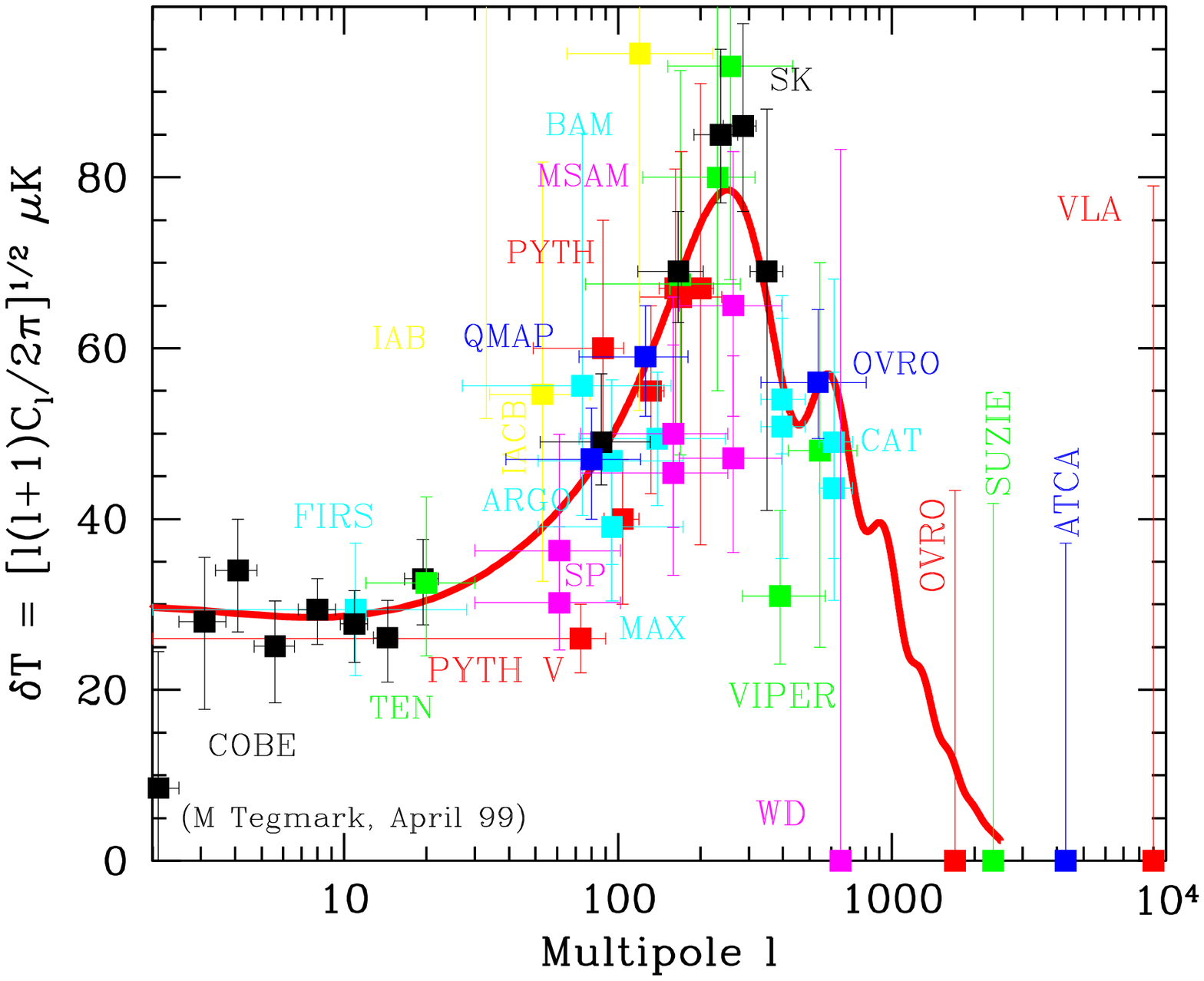}
,width=\textwidth,angle=0}}
\caption{\small The CMB multipoles as a function of temperature.\newline From Max
Tegmark's home page: {\texttt{http://www.sns.ias.edu/\char126
max/foregrounds.html}}.
\label{cmb_data} }
\end{figure}

Because the position of the first Doppler peak depends essentially only on the
curvature, the CMB is a good test of $\Omega_0+\Omega_\Lambda$. From
figure~\ref{cmb_data}, we can see that the curve favours a flat universe~-- the
first peak is around $l\sim200$. The estimated curvature is
$\Omega_0+\Omega_\Lambda=1\pm0.2$. See Turner~(1999)\nocite{tur99},
Straumann~(1999)\nocite{str99}, or Rocha~(1999)\nocite{roc99}.

This is still a preliminary result, but the issue should be settled soon with
some new satellites being launched in the next year or so (eg, MAP, Plank,
etc.). These results are not particularly relevant for the work contained in
this thesis.

The SNIa and CMB measurements may be combined to give more accurate information
on the density parameters~-- see \S\ref{Observable Quantities in FLRW Models}.
They can also be used to find the `equation of state' of the universe; see
Perlmutter \etal~(1999), or Efstathiou~(1999)\nocite{efs99}.

\section{Age}

There are two ways to determine the age of the universe. Determinations of the
ages of `old' objects such as globular clusters in the local universe provide a
lower bound to the age of the universe. Alternatively, given a model of the
universe, a measurement of $H_0$ provides an estimate of the age of the
universe. Both are plagued with uncertainty and there has been considerable
disagreement until very recently, with  globular clusters being significantly
older than the age as implied by the expansion rate.

From $H_0$ we can obtain a range of ages from 8~Gyr for an Einstein-de Sitter
model, all the way up to 16-17~Gyr for a low density flat model, both depending
on the value of $H_0$. Given a set of parameters $\{H_0, \Omega_0,
\Omega_\Lambda\}$ the age can be calculated simply from
\be
T=\int_0^T dt=\int_0^\infty\;{dz}{H(z)(1+z)},
\ee
where $H(z)$ is given by equation~(\ref{H(z)}).

Measuring the ages of globular clusters is quite a complicated process; it is
based on a number of complementary methods. It combines estimates from stellar
models, with separate evaluations of the stars in the clusters turning off from
the main sequence. As stellar models are not perfect uncertainties of about 7\%
arise. The largest uncertainties in the cluster ages arise from estimating
their distance~-- and hence their intrinsic luminosities which are required to
calibrate the stellar models. Distance estimates are made from parallax
measurements, and from distance indicators such as RR Lyraes. As in estimating
$H_0$, calibration of these stars relies on knowing the distance to the Large
Magellanic Cloud, which again introduces considerable uncertainties.

The ages of the globular clusters have traditionally been quite high~--
certainly too high for a flat $\Lambda=0$ model, and led to the `age problem'
as it was known. However recent recalibration of RR-Lyraes has led to a
considerable reduction in the ages of these to about 11-14 Gyrs (Chaboyer
\etal,~1998; Krauss,~1999\nocite{kra99}); also recent parallax measurements
using the Hipparcos satellite confirm these results (Reid, 1997\nocite{rei97};
Gratton \etal,~1997\nocite{gra-97}). For a high value of $H_0$ or a high
density model, the margin of error between cosmological and globular cluster
ages is still very small. See Freedman~(1999b) for discussion and references.

The age of the universe is important for this thesis because the Stephani
models we consider here were initially put forward for their `naturally' high
age, and as a solution to the age crisis (\dab\ and
Hendry,~1998\nocite{dab-hen98}). We constrain the Stephani models later using
the age.

\forget{

We extend the work of a previous paper (Paper~I) on the viability of
inhomogeneous cosmological models to give a more detailed analysis of a larger
class of spherically symmetric Stephani models. In particular, we derive exact
observational relations for all observer positions, not only for the central
observers considered before.

In contrast to Paper~I, we do not consider in detail the `local' constraints
derived from galaxy surveys, supernovae and other data at relatively low
redshift ($z\lapp 5$), but concentrate instead on broader physical properties:
age, the physical acceptability of the matter (ie,~energy conditions) and, most
importantly of all, the CMB anisotropy. However, we do choose the epoch of
observation to give a value of~$H_0$ compatible with observations, and we
demand that the models exhibit no unfortunate features at low redshift that are
obviously ruled out by observations (namely, zeros in the distance-redshift
relation characteristic of the closed spatial sections in these spacetimes). We
also demonstrate that there is \emph{no horizon problem} for any of the models
we consider here.

Stephani models are distinguished from \FRW\ models by the presence of
inhomogeneous pressure, which results in an acceleration of the fluid. We
briefly consider the influence of this non-geodesic flow on low-$z$
observations.

A crucial aspect of this work is the application of the \emph{Copernican
principle}: for a specific model to be acceptable we demand that it must be
consistent with presently available observational constraints (especially
anisotropy constraints) for most (or all) observer positions on some suitably
chosen spacelike hypersurface.

The most important results of the paper are presented as an exclusion plot in
the 2-D parameter space of the models. We show that there is a region of
parameter space not ruled out by the constraints we consider and containing
models which are significantly inhomogeneous.

As far as the authors are aware, this work represents the first exact analysis
of the observational properties of an inhomogeneous cosmological model for all
observer positions.



\section{Introduction.}

The work presented here is part of an ongoing project to shed light on some of
the assumptions underlying standard cosmology by considering the constraints
imposed by observations on inhomogeneous cosmological models. In particular, we
consider whether the \FRW\ models are the only viable candidates for a
cosmological model by examining the observational and physical properties of
Stephani cosmologies. A pivotal part of this analysis is the cosmological
principle (CP) and its relation to the Copernican principle and the observed
isotropy of the universe about us: does it really follow from the high degree
of isotropy of the cosmic microwave background radiation (CMB) that the
universe is homogeneous and isotropic?

A number of inhomogeneous or anisotropic cosmological models have been studied
in relation to the CP. The homogeneous but anisotropic Bianchi and
Kantowki-Sachs models (Kantowski and Sachs~1966\nocite{kant-sachs66}; see
Ellis~1998,\nocite{ell98}~\S6 and references therein) have been investigated
with regard to the time evolution of the anisotropy. It can be shown, for
example, that there exist Bianchi models for which a significant phase of their
evolution is spent in a near-\FRW\ state, even though at early and late times
they may be highly anisotropic (again, see~\S6 of Ellis~1998 and references
therein). Of the inhomogeneous models that arise in cosmological applications,
by far the most common are the Tolman-Bondi dust spacetimes (Tolman~1934;
Bondi~1947\nocite{tolm34}\nocite{bond47}). These are used both as global
inhomogeneous cosmologies~-- probably the most important papers being Hellaby
and Lake (1984,~1985\nocite{hel-lak-I84}\nocite{hel-lak85}), studying
geometrical aspects, and Rindler and Suson (1989)\nocite{rindler-sus89},
Goicoechea and Martin-Mirones (1987)\nocite{goi-mm97}, and Maartens
\etal~(1995)\nocite{maar-95} investigating observational aspects~-- and also as
models of local, nonlinear perturbations (over- or under-densities) in an \FRW\
background (Tomita~1995, 1996\nocite{tom96}\nocite{tomita95}; Moffat and
Tatarski~1995\nocite{mof-tat95}; Krasi\'{n}ski~1998; Nakao
\etal~1995\nocite{nakao-95}). See also Krasi\'{n}ski (1998)\nocite{kras98} for
a review.

However, there has been some consideration of Stephani solutions (Stephani
1967a,b\nocite{step67a}\nocite{step67b}; see also Kramer
\etal~1980\nocite{kram-80} and Krasi\'{n}ski
1983,~1997\nocite{Kras83}\nocite{kras97}). These are the most general
conformally flat perfect fluid solutions~-- and obviously therefore contain the
\FRW\ models. They differ from \FRW\ models in general because they have
inhomogeneous pressure, which leads to acceleration of the fundamental
observers. D\c{a}browski and Hendry~(1998) fitted a certain subclass of these
models to the first SNIa data of Perlmutter \etal~(1997) using a low-order
series expansion of the magnitude-redshift relation for central observers
derived in D\c{a}browski~(1995), and found that they were significantly older
than the \FRW\ models that fit that data. In Barrett and Clarkson
(1999)\nocite{bar-cl99} (Paper~I) we extended those results, obtaining
\emph{exact} distance-redshift and number count-redshift relations, valid for
all~$z$, and showing that it would always be possible to find a more than
acceptable fit to any data that could also be fit by an \FRW\ model (for
plausible ranges of the \FRW\ parameters $H_0$, $\Omega_0$
and~$\Omega_\Lambda$). The best-fit models were again consistently $1-4$~Gyr
older than their \FRW\ counterparts.

One alarming feature of the Stephani models is their matter content. The usual
perfect-fluid interpretation precludes the existence of a barotropic equation
of state in general (because the density is homogeneous but the pressure is
not), although they can be provided with a strict thermodynamic scheme (Bona
and Coll~1988\nocite{bon-col88}). Moreover, even individual fluid elements can
behave in a rather exotic manner, having negative pressure, for example
(\emph{cf}.~\S\ref{energy_cond}). For these reasons, amongst others,
Lorenz-Petzold~(1986)\nocite{lorpet86} has claimed that Stephani models are not
a viable description of the universe, but Krasi\'nski~(1997, p.170) argues
rather vigorously that this conclusion is incorrect, as do we. Other
cosmologies have sometimes been ruled out \emph{a priori} because the
stress-energy tensor does not behave in the correct manner, or for other
reasons such as the lack of an \FRW\ limit~-- see Krasi\'{n}ski
(1997)\nocite{kras97} for a complete review. However, it has become
increasingly difficult to avoid the conclusion that the expansion of the
universe is accelerating, with the type~Ia supernovae data of Perlmutter
\etal~(1998b) being the latest and strongest evidence for this (see also
Perlmutter \etal~1997a,b,
1998a,b;\nocite{Perl-97}\nocite{perl-97b}\nocite{perl-98a}\nocite{perlm-98}
Krauss 1998\nocite{kraus98}; Peebles 1998\nocite{peeb98}). This is clear
evidence that there is some kind of `negative pressure' driving the expansion
of the universe. In standard \FRW\ models, this must correspond to an
inflationary scenario, with $\Lambda>0$, or a matter content of the universe
which is mostly scalar field (`quintessence'~-- see Frampton
1999\nocite{framp99}; Liddle 1999\nocite{lid99}; Coble, Dodelson and Frieman
1997\nocite{coble-97}; Liddle and Scherrer 1998\nocite{lid-sch98}; see also
Goliath and Ellis 1998\nocite{gol-ellis98} for a discussion of the dynamical
effects associated with~$\Lambda$). Either way, the real universe is behaving
in a manner that is at odds with `everyday' physics, so it is not justifiable
to rule out Stephani models for exhibiting similar behaviour.

The aim of this paper is to extend the results of Paper~I to discuss the new
effects that arise when the observer is at a non-central location. Principally,
we are interested in the anisotropy that this will introduce in observed
quantities such as redshift, and we wish to compare these anisotropies with
presently available observational constraints to see whether it is possible to
rule out the Stephani models we consider on the basis of their anisotropy. Of
course, we can always make the anisotropy as small as we like by only
permitting observations from very close to the centre. But this is a very
special position, so that such a resolution to the anisotropy problem would be
in conflict with the Copernican principle (although it is may be possible to
circumvent this by invoking anthropic arguments, see Ellis \etal~1978).

Although it is an assumption based purely on philosophy, the Copernican
principle must be taken seriously (Ellis~1975\nocite{ellis75}): in order to
reject an inhomogeneous cosmological model on the basis of its conflicting with
the observed isotropy of the universe it is necessary to consider \emph{all}
observer positions and to show that \emph{for most observers in that spacetime}
the anisotropy observed is too large to be compatible with observations. We
could adopt this view of the Copernican principle in this paper, but, for
simplicity, we will require consistency with observations for \emph{all}
observers~-- our results will thus be rather stronger than is strictly required
by the Copernican Principle. In fact, it is only really possible to consider
the weaker version of the Copernican principle when the spatial sections of the
universe have finite volume (or, at least, when the number of observers, as
measured by the integral over the spatial sections of the number density of
particles~-- \emph{cf}.~\S\ref{obs-constr}~-- is finite), otherwise the
expression `most observers' has very little meaning. For infinite universes the
stronger version of the Copernican principle must be adopted. With our stronger
definition we are thus prepared for all eventualities (although it will turn
out that the models we consider are all `finite'). Non-central locations are
rarely considered in the literature in the analysis of inhomogeneous
cosmologies however, owing to the mathematical difficulties that this usually
entails. Having said that, Humphreys \etal~(1997) made a study of Tolman-Bondi
models from a non-central location and applied their results to a `Great
Attractor' model. Most other non-central analyses, though, look only at
perturbations of standard \FRW\ models.

Having adopted (a strong version of) the Copernican principle, the question
then arises as to whether the observed isotropy of the universe, when required
to hold at every point, forces homogeneity, thus validating the CP and
rendering our work futile. Well, the nearby universe is distinctly lumpy, so it
would be difficult to claim there is isotropy on that basis. However, the CMB
is isotropic to one part in 10$^{3}$. Together with the EGS theorem (Ehlers,
Geren and Sachs~(1968)\nocite{ehlers-68}), or rather, the almost EGS theorem of
Stoeger, Maartens, and Ellis~(1995)\nocite{stoeg-95} this allows us to say that
within our past lightcone the universe is almost \FRW\ (ie,~almost homogeneous
and isotropic), \emph{provided} the fundamental observers in the universe
follow geodesics (that is, as long as the fundamental fluid is dust). What the
almost EGS theorem achieves is to provide support for the CP without the need
to blithely assume the (near) isotropy of \emph{every} observable: isotropy of
the CMB alone is enough to ensure the validity of the CP (for geodesic
observers). Since, however, the Stephani models do not satisfy the conditions
of the almost EGS theorem (the acceleration is non-zero), its conclusions do
not apply, we need not be restricted \emph{a priori} by it. It will turn out,
actually, and we will show in detail in Paper~III, that the models we consider
provide a solid counterexample to the spirit of the EGS and almost~EGS
theorems, being inhomogeneous but appearing highly isotropic.

In the following section we describe the spherically symmetric Stephani models
and discuss their physical and geometrical properties, and we present in some
detail the particular two-parameter subclass of the Stephani models that we
will be studying. It will be shown that these models do not have particle
horizons, in contrast to standard \FRW\ models. The energy conditions will be
used to constrain the model parameters, leaving a manageable parameter set to
be investigated further. Then, in~\S\ref{noncent}, the transformation to
coordinates centred on any observer will be derived. In~\S\ref{obs-constr} the
various distance-redshift relations are presented for the models we consider,
and observations are applied to constrain the values of the model parameters.
The constraints we address are: the value of Hubble's constant, age, `size'
(the meaning of which will be explained in~\S\ref{size}) and, most importantly,
the anisotropy of the CMB. We show that after applying these constraints there
remains a region of parameter space containing models consistent with all of
the constraints. Furthermore, we demonstrate in~\S\ref{inhomog} that many of
the models not excluded by the constraints of~\S\ref{obs-constr} are distinctly
inhomogeneous. We devote some space, in~\S\ref{local-dipole}, to a qualitative
discussion of the constraints on the acceleration and inhomogeneity from
`local' (ie,~$z\lapp 5$) observations. The significance of these results is
discussed in~\S\ref{conclusions}. In two appendices we present derivations of
important results that would otherwise disturb the flow of the discussion.

BITS }

}

}
\forgetmenot{
\forgetmenot{
\chapter{\textsc{Relativistic Cosmology}}
\label{relativistic cosmol chapter-----------------}

The goal of cosmology is to find a model that best describes the universe on
any particular scale, preferably on all scales. The framework which is usually
used, and will be used throughout this thesis, is that of General Relativity
(GR) as discussed in eg, Wald (1984)\nocite{wald};
Stephani~(1990\nocite{step90}); Misner, Thorne and Wheeler~(1971\nocite{mtw});
Schutz~(1990\nocite{schutz-rel}); Hawking and Ellis~(1973\nocite{hawk-ell});
that is to say spacetime is a manifold, which has a geometry described by a
Lorentz metric $g_{ab}$, and associated connection~$\Gamma^a_{~bc}$, containing
matter whose physical properties are described by the energy-momentum tensor,
$T_{ab}$. The curvature of the spacetime is felt via the Riemann tensor,
$R_{abcd}$, which gives the lack of commutivity of derivatives when parallel
transporting vectors in a curved manifold (Schutz~1980\nocite{sch80});
\be
\left(\del_a\del_b-\del_b\del_a\right)\xi_c=R_{abcd}\xi^d,
\label{Ricci identities}
\ee
for any vector field $\xi^a$; these are the Ricci identities. The curvature of
spacetime and the matter content interact via Einstein's field equations,
\be
G_{ab}\eqdef R_{ab}-\sfrac{1}{2}Rg_{ab}=T_{ab}-\Lambda g_{ab},\label{EFE}
\ee
where $G_{ab}$ is the Einstein tensor, $R_{ab}$ and $R$ are the Ricci tensor
and scalar, and $\Lambda$ is the cosmological constant.\footnote{In this
chapter units where $8\pi G=c=1$ will be used; in later chapters, where
observational quantities are important, they will be put into the equations.}
These equations give local conservation of energy from the geometry of the
spacetime. The vector formed by taking the 4-divergence of $T_{ab}$,
$\del_bT^{ab}$ represents the local creation or loss of energy. The Bianchi
identities,
\be
\del_{[a}R_{bc]}^{\pha\pha de}=0,\label{bianchi identities}
\ee
when twice contracted (over the $ac$ and $ed$ indices) imply $\del_bG^{ab}=0$,
from~(\ref{EFE}) so that,
\be
\del_bT^{ab}=0,\label{conservation of EFE}
\ee
so that energy is locally conserved.

For any model of the universe, we must bear in mind that we have to extract
\emp{observational} predictions from it;~(\ref{EFE}) are not particularly
intuitive. Rather than just start from a metric and see what we get, it is
desirable to break~(\ref{EFE}) down into a simpler (or at least more intuitive)
set of equations while retaining their covariant character, and at the same
time introducing quantities that are directly measurable.

\section{The Covariant 1+3 Formulation of Fluids in GR}

In this thesis, we don't make much use of the 1+3 formalism (Ehlers 1993), but
it is necessary for the results of
\S\ref{conformal trans section}, and~\S\ref{QCDM}, and also for a proper
understanding of modern relativistic cosmology.

For a complete cosmological model one must specify not only a metric defined on
a manifold, but also a family of \emp{fundamental observers}. These worldlines
may be used to represent galaxies at late times (where the galaxies sit on the
worldlines) or radiation at early times. It is usually assumed however that at
any particular epoch there is, on average, one dominant or fundamental
congruence of worldlines, eg, that defined by the CMB.  The combinations of
such a set of fluids will be described briefly later. These worldlines will
have a velocity
\be
u^a\eqdef\frac{dx^a}{d\tau},
\ee
where $\tau$ is the proper time measured along the worldlines, such that
$u_au^a=-1$; ie, it is timelike. This velocity field is very important and its
properties (together with a matter description) can be used to give invariant
definitions of many cosmological models.

This velocity field can be used to look at tensors \emp{along} these
worldlines, and \emp{orthogonal} to them. This is the `1+3' splitting of the
spacetime into `temporal' and `spatial'  parts; it is also covariant because
$u^a$ can be defined uniquely and without any coordinates which would be
required for splitting the spacetime into space+time.

Given $u^a$ we define a \emp{projection tensor} by
\be
h_{ab}\eqdef g_{ab}+u_au_b,
\ee
which defines a `3-metric' orthogonal to the congruence, and satisfies
\be
h_a^{\pha a}=3, \pha h_{ab}u^b=0;
\ee
ie, it is orthogonal to $u^a$, it therefore projects things into the
instantaneous rest space of the congruence.

We also make use of the projected alternating tensor (a generalisation of the
Levi-Civita symbol)
\be
\eta_{abc}\eqdef\eta_{abcd}u^d,
\ee
where
\be
\eta_{abcd}\eqdef-\sqrt{|g|}
\delta^0{}_{[a}\delta^1{}_b\delta^2{}_c\delta^3{}_{d]}
\ee
is the spacetime alternating tensor (the 4D Levi-Civita symbol).


We use angle brackets to denote the projected, symmetric, trace-free  part of a
2nd rank tensor,
\be
F_{\langle ab\rangle}\eqdef \left(h_{(a}^{\pha c}h_{b)}^{\pha
d}-\sfrac{1}{3}h_{ab}h^{cd}\right)F_{cd},
\ee
(and for a 1-form or vector, $K_{\<a\>}\eqdef h_a^{\pha b}K_b$) so that any
projected 2nd rank tensor has the irreducible covariant decomposition
\be
F_{ab}=\:{1}{3}Fh_{ab}+\eta_{abc}F^c+F_{\< ab\>},
\ee
where $F=F_{cd}h^{cd}$ is the spatial trace, and $F_a=\:12\eta_{abc}F^{bc}$ is
the spatial dual vector of the antisymmetric part of $F_{ab}$. In the 1+3
covariant formalism, all irreducible quantities are either scalars, projected
vectors or projected, symmetric, trace-free tensors.

\forget{We can also define a generalised vector or cross product of vectors
\be
K\times L_a\eqdef(K\times L)_a\eqdef\eta_{abc}K^aL^b;
\ee
and 2nd rank tensors
\be
S\times T_a\eqdef(S\times T)_a\eqdef\eta_{abc}S^{bd}T^{c}_{\pha d}.
\ee}

With this projection tensor and velocity field, we can define two derivatives;
firstly, differentiation along the fluid flow -- a time derivative, denoted by
a dot,
\be
\dot{F}_{ab}\eqdef u^c\del_c F_{ab};
\ee
and also the derivative projected orthogonal to the flow lines -- a spatial
derivative,
\be
\sdel_c F_{ab}\eqdef h^d_{\pha c} h^e_{\pha a} h^f_{\pha b} \del_d
F_{ef}.
\ee
This new tensor is entirely orthogonal to $u^a$; contraction of any index with
$u^a$ is zero. $\sdel_a$ is no longer a proper 3-dimensional covariant
derivative however, because in general,
\be
\sdel_{[a}\sdel_{b]}Q\neq0;\label{sdel-doesn't-commute}
\ee
which means that the congruence is no longer orthogonal to the spacelike
surfaces.

\forget{We can also generalise the divergence and curl of normal vector calculus to the
1+3 covariant spatial counterparts;
\be
\sdel\cdot K\eqdef\sdel_aK^a,~~~\sdel\times K_a\eqdef \eta_{abc}\sdel^bK^c,
\ee
for vectors; and, for 2nd rank tensors
\be
\sdel\cdot F_a\eqdef\sdel^bF_{ab},~~~
\sdel\times F_{ab}\eqdef \eta_{cd(a}\sdel^cF_{b)}^{~~d}.
\ee}

The covariant derivative of any scalar can be split into orthogonal and
parallel parts;
\be
\del_aQ=\sdel_aQ-\dot{Q}u_a;
\ee
while derivatives of vectors and tensors can be split into their irreducible
covariant bits, see~Maartens, Gebbie and Ellis~(1999)\nocite{maar-99} for the
equations.

\subsection{Kinematics of the Fundamental Congruence}

The derivative of the fundamental congruence $\del_au_b$ can be split into
irreducible quantities. These are:

{\bi {\item The \emp{expansion},
\be
\theta\eqdef \del_au^a=\sdel_au^a,\label{expansion def}
\ee
which is the trace of the derivative of $u^a$ and also the \emp{3-divergence}
of the congruence. It represents the volume rate of expansion of the fluid
elements. In general, one can associate a fundamental or average length scale
by
\be
\frac13\theta\equiv\frac{\dot{\ell}}{\ell}\label{average length scale-def}
\ee
which describes the volume change of the fluid.} {\item The \emp{acceleration},
\be
\dot{u}^a\eqdef u^b\del_bu^a,
\ee
is the time rate of change of $u^a$; it describes motion of the flow moving
under forces other than gravity alone. This comment can be understood by
thinking about a Schwarzschild black hole: if someone sits at a constant proper
distance from the black hole, then the force of gravity will pull them into the
center. In order to stay at a constant proper distance from it, then there must
be some non-gravitational force pushing outwards. Acceleration is zero if and
only if the flow is geodesic (freely falling observers).} {\item The
\emp{shear},
\be
\sigma_{ab}\eqdef \sdel_{\langle a}u_{b\rangle},
\ee
is the rate of shearing of the congruence; ie, the trace-free symmetric part,
which describes how the congruence will distort in time. While it doesn't
affect the volume change of the congruence, the relative distances of objects
will change because of the presence of shear. Its magnitude is defined by
$\sigma^2\eqdef\sfrac{1}{2}\sigma_{ab}\sigma^{ab}$, and $\sigma=0$
$\Leftrightarrow$ $\sigma_{ab}=0$.} {\item The \emp{rotation},
\be
\omega_{ab}\eqdef \sdel_{[a}u_{b]}\label{rotation def}
\ee
describes the rate of rotation of the congruence; it is the orthogonally
projected anti symmetric part of the derivative of the flow. Objects will
change their position in the sky because of rotation. It also has magnitude
$\omega^2\eqdef
\sfrac{1}{2}\omega_{ab}\omega^{ab}$, with
$\omega=0$~$\Leftrightarrow$~$\omega_{ab}=0$. We also define a rotation vector
by $\omega^a\eqdef\sfrac{1}{2}\eta^{abc}\omega_{bc}$.}
\ei}

The covariant derivative can now be decomposed as
\be
\del_au_b=-u_a\dot{u}_b+\sfrac{1}{3}\theta h_{ab}+\sigma_{ab}+\omega_{ab},
\ee
giving a covariant representation of the changing congruence in terms of
invariantly defined quantities -- which also give a physical breakdown of
different aspects to the kinematics of the congruence. This equation is of
fundamental importance to cosmology -- and relativity in general -- and was
first given by Ehlers~(1961), translated recently as Ehlers~(1993). See also
Ellis~(1971)\nocite{ellis71}.

We are now in a position to understand and expand~(\ref{sdel-doesn't-commute}),
and a derivation is probably useful:
\ba
\sdel_a\sdel_bQ\li=\li h_{a}^{\pha d}h_{b}^{\pha e}\del_d h_{e}^{\pha c}
\del_c Q\nonumber\\
\li=\li \dot{Q}\sdel_bu_a+h_{a}^{\pha c}h_{b}^{\pha d}\del_c \del_d Q,\nonumber
\ea
which implies, using~(\ref{rotation def}),
\be
\sdel_{[a}\sdel_{b]}Q=-\omega_{ab}\dot{Q}\label{sdel comute}
\ee
for any scalar field $Q$. This result was first presented in Bruni, Dunsby, and
Ellis~(1992)\nocite{bruni-92}. It shows explicitly that the fundamental
congruence will define a spatial `3-metric' orthogonal to the flow line if and
only if the rotation is zero, because that is when $\sdel$ defines a proper
covariant derivative.

\subsection{The Energy-Momentum Tensor}

Regardless of the particular matter present, \emp{any} energy-momentum tensor
can be decomposed with respect to the chosen fundamental congruence in the
following manner:
\be
T_{ab}\eqdef \mu
u_au_b+ph_{ab}+2q_{(a}u_{b)}+\pi_{ab},\label{E-M_Tensor_general}
\ee
with each of the quantities having a physical interpretation:
\bi
\item The \emp{relativistic energy density}
\be\label{energy density def}
\mu\eqdef u^au^bT_{ab}=u^au^bG_{ab}-\Lambda;
\ee
\item The \emp{isotropic pressure}
\be\label{pressure def}
p\eqdef \sfrac{1}{3}h^{ab}T_{ab}=\sfrac{1}{3}h^{ab}G_{ab}+\Lambda;
\ee
\item The \emp{energy flux}, or relativistic momentum density
\be\label{energy flux def}
q_a\eqdef T_{\< a\> b}u^b,
\ee
which can be interpreted as the heat flow relative to $u^a$; it satisfies
$q_au^a=0$;
\item The \emp{anisotropic pressure}
\be\label{anisotropic pressure def}
\pi_{ab}\eqdef T_{\langle ab\rangle},
\ee
which is trace-free, $\pi_{a}^{\pha a}=0$, and $u^a\pi_{ab}=0$.
\ei

\subsection{The Splitting of the Weyl Tensor}

The trace-free part of the Riemann tensor is the Weyl tensor\footnote{Also
known as the conformal tensor, due to its invariant nature under conformal
transformations; see \S\ref{conformal trans section}.}, $C_{abcd}$, which
describes the free gravitational field -- tidal forces and gravitational waves
-- and to some extent describes the null structure of the spacetime. It is
defined by
\be
R_{abcd}=C_{abcd}+g_{a[c}R_{d]b}-\:13Rg_{a[c}g_{d]b}.\label{weyl def}
\ee

In the 1+3 splitting of the spacetime, the Weyl tensor can be split too, into
`electric' and `magnetic' parts:
\ba
&E_{ab}&\eqdef C_{abcd}u^cu^d\\
&H_{ab}&\eqdef\sfrac{1}{2}\eta_{acde}C^{cd}_{\pha\pha bf}u^eu^f.
\ea
Both of them are symmetric, trace-free, and orthogonal to $u^a$.

\section{Splitting Einstein's Equations}

If we take Einstein's equations~(\ref{EFE}), the twice-contracted Bianchi
identities, the Bianchi identities~(\ref{bianchi identities}), or
using~(\ref{weyl def})
\be
\del^dC_{abcd}=\del_{[a}\left(-R_{b]c}+\sfrac{1}{6}Rg_{b]c}\right)
\ee
and the Ricci identities~(\ref{Ricci identities}) for the congruence $u^a$, and
separate out all the independent parts, we arrive at a set of \emp{evolution}
and \emp{constraint} equations describing the structure of spacetime. Not all
of these are used in this thesis, but they are included for completeness.
\newline\newline\textsc{Evolution}:
\bi
\item The \emp{energy conservation equation}
\be
\dot{\mu}+\theta(\mu+p)+\sdel_aq^a=-2\udot_aq^a-\sigma_{ab}\pi^{ab};
\label{energy_cons_gen}
\ee
shows that, for a perfect fluid, the expansion correlates directly to the
change in energy density along the flow lines;
\item The \emp{Raychaudhuri equation} (expansion evolution)
\be
\dot{\theta}+\sfrac{1}{3}\theta^2=\sdel_a\udot^a+\udot_a\udot^a-2\sigma^2+2
\omega^2-
\sfrac{1}{2}(\mu+3p)+\Lambda,\label{raychaudhuri}
\ee
which is the equation of gravitational attraction. This shows that a positive
cosmological constant, or rotation, or acceleration will make the expansion
rate increase -- which is the accepted scenario at present, see later --
whereas shear will slow the expansion rate. In a sense this is obvious; one
would expect the `rotation of the universe' to increase the expansion rate, and
similarly a large \emp{negative pressure} (or a positive cosmological constant
-- see~(\ref{cosmological const is pressure})) will do the same;
\item The \emp{momentum conservation equation}
\be
\dot{q}^{\< a\>}+\sfrac{4}{3}\theta q^a+\sdel^ap+\sdel_b\pi^{ab}=
-\sigma^{ab}q_b-(\mu+p)\udot^a-\udot_b\pi^{ab}-\eta^{abc}\omega_bq_c,
\label{moment_cons}
\ee
which shows that a perfect fluid can have acceleration only if there are
spatial pressure gradients present;
\item The \emp{shear and rotation evolution equations}
\ba
\dot{\omega}^{\< a\>}+\sfrac{2}{3}\theta\omega^a\li=
\li \sfrac{1}{2}\eta^{abc}\sdel_b\udot_c+\sigma^{ab}\omega_b,
\label{rotation evolution}\\
\dot{\sigma}^{\< ab\>}+\sfrac{2}{3}\theta\sigma^{ab}\li=\li\sdel^{\< a}
\udot^{b\>}+\udot^{\< a}\udot^{b\>}-\sigma^{\< a}_{\pha\pha c}
\sigma^{b\> c}-
\omega^{\< a}\omega^{b\>}-E^{ab}+\sfrac{1}{2}\pi^{ab};\label{shear evolution}
\ea
the second equation shows that a fluid with $\omega_a=\sigma_{ab}=\udot_a=0$
will have the electric part of the Weyl tensor proportional to the anisotropic
pressure.
\item The evolution of $E_{ab}$ and $H_{ab}$,
\ba
\dot{E}^{\< ab\>}+\theta E^{ab}-\eta^{cd\< a}\sdel_cH^{b\>}_{\pha d}+
\sfrac{1}{2}
\dot{\pi}^{\< ab\>}=-\sfrac{1}{2}\li\li[\sdel^{\< a}q^{b\>}+(\mu+p)
\sigma^{ab}]\nonumber
+3\sigma^{\< a}_{\pha\pha c}\left(E^{b\> c}-
\sfrac{1}{6}\pi^{b\> c}\right)\\-\udot^{\< a}q^{b\>}-
\sfrac{1}{6}\theta\pi^{ab}\li+\li
\eta^{cd\< a}\left(2\udot_cH^{b\>}_{\pha d}+\omega_c[E^{b\>}_{\pha d}+
\sfrac{1}{2}\pi^{b\>}_{\pha d}]\right),\\
\dot{H}^{\< ab\>}+\theta H^{ab}+\eta^{cd\< a}\sdel_cE^{b\>}_{\pha d}-
\sfrac{1}{2}\eta^{cd\< a}\sdel_c\pi^{b\>}_{\pha d}\li=\li3\sigma^{\<
a}_{\pha\pha c}H^{b\> c}+\sfrac{3}{2}\omega^{\< a}q^{b\>}\nonumber\\\li -\li
\eta^{cd\< a}\left(2\udot_cE^{b\>}_{\pha d}-\omega_cH^{b\>}_{\pha d}-
\sfrac{1}{2}\sigma^{b\>}_{\pha c}q_d\right).\label{H dot eqn}
\ea

\ei
\textsc{Constraint:}
\bi
\item The \emp{shear and vorticity divergence equations}
\ba
\sdel_a\omega^a\li=\li\udot_a\omega^a\\
\sdel_b\sigma^{ab}\li=\li\sfrac{2}{3}\sdel^a\theta-\eta^{abc}
\left(\sdel_b\omega_c+2\udot_b\omega_c\right) -q^a.\label{div shear}
\ea
The second equation is crucial in \S\ref{QCDM}, where it is used to set energy
flux equal to zero in a QCDM model with an isotropic radiation field.
\item The \emp{`curl $\sigma$' equation}
\be
\eta^{cd\< a}\sdel_c\sigma^{b\>}_{\pha d}-H^{ab}=2\udot^{\< a}
\omega^{b\>}+
\sdel^{\< a}\omega^{b\>}
\ee
\item The \emp{$E_{ab}$ and $H_{ab}$ divergence equations}
\ba
\sdel_bE^{ab}-3\omega_bH^{ab}\li=\li\sfrac{1}{3}[\sdel^a\mu-\theta q^a]+
\sfrac{1}{2}[
\sigma^{ab}q_b-\sdel_b\pi^{ab}]\nonumber\\
\li\li~~~~~~~+\eta^{abc}\left(\sigma_{bd}H^d_{\pha c}-
\sfrac{3}{2}\omega_b q_c\right),\\
\sdel_bH^{ab}+3\omega_bE^{ab}\li=\li\sfrac{1}{2}\omega_b\pi^{ab}-(\mu+p)
\omega^a\nonumber\\
&&~~~~~-
\eta^{abc}\left(\sfrac{1}{2}\sdel_bq_c+\sigma_{bd}[E^d_{\pha c}+
\sfrac{1}{2}\pi^d_{\pha c}]\right).\label{div H}
\ea

\ei

\section{1+3 Splitting Under a Conformal Transformation}
\label{conformal trans section}

A \emp{conformal transformation} is an angle preserving transformation that
changes lengths and volumes. The importance of these types of transformations
lies in the fact that, under a conformal transformation, \emph{the null
structure of the spacetime is preserved:} indeed, it trivially follows that the
causal structure is preserved. We also have the important property that the the
Weyl tensor, $C_{abc}^{\phantom{abc}d}$, is invariant (note that one index must
be raised) so that a conformal transformation will introduce no tidal forces or
gravitational waves; that is, a conformal transformation will only introduce
`non-gravitational' forces and matter into the new spacetime (by changing
$R_{ab}$ and thus the matter tensor $T_{ab}$ via Einstein's equations).

We perform the conformal transformation
\be
g_{ab}=e^{2Q}\ghat_{ab},\pha u^a=e^{-Q}\uhat^a,\pha u_a=e^Q\uhat_a;
\ee
where $Q>0$ is an arbitrary function, $u^a$ is a velocity vector with respect
to $g_{ab}$: $g_{ab}u^au^b=u_au^a=-1$; and $\uhat^a$ is the conformally related
(parallel) velocity vector, and is normalised with respect to $\ghat_{ab}$:
$\ghat_{ab}\uhat^a\uhat^b=-1$.\footnote{When performing conformal
transformations, confusion can arise over which metric to use when performing
contractions; this will be avoided here by always using $g_{ab}$: ie,
$v^av_a\eqdef g_{ab}v^av^b$.} The covariant derivative of any one-form field
$v_a$ transforms as
\be
\del_av_b=\delhat_av_b-2Q_{(a}v_{b)}+g_{ab}Q^cv_c,
\ee
where $Q_a\equiv Q_{,a}=\sdel_aQ-\dot{Q}u_a$. The expansion
($\theta=\del_au^a$), acceleration ($\udot^a=u^b\del_bu^a$), rotation
($\omega_{ab}=\sdel_{[a}u_{b]}$), and shear ($\sigma_{ab}=\sdel_{\<
a}u_{b\rangle}$) of the two velocity congruences are related by:
\ba
\thhat\li=\li e^{Q}(\theta-3\dot{Q})\nonumber\\
\ahat_a\li=\li\dot{u}_a-\sdel_a Q\nonumber\\
\rothat_{ab}\li=\li e^{-Q}\omega_{ab}\nonumber\\
\hat{\sigma}_{ab}\li=\li e^{-Q}\sigma_{ab}.\label{kinematic_trans}
\ea
The equation for the acceleration corrects equation~(6.14) of Kramer
\etal~(1980). These show that a conformal transformation mat induce
acceleration and expansion into the new spacetime, by not shear or rotation: in
particular, a conformally flat model must have shear and rotation vanishing (as
in eg, the Stephani models considered later). \emph{With respect to $g_{ab}$},
a dot denotes differentiation along the fluid flow -- a time derivative,
$\dot{F}_{ab}=u^c\del_c F_{ab}$; and $\sdel_a$ is the derivative projected
orthogonal to the flow lines -- a spatial derivative, $\sdel_c F_{ab}=h^d_{\pha
c} h^e_{\pha a} h^f_{\pha b} \del_d F_{ef}$, where $h_{ab}=g_{ab}+u_au_b$ is
the usual projection tensor.

The Einstein tensor transforms as (see Wald~1982)
\be
G_{ab}=\Ghat_{ab}-2\del_aQ_b-2Q_aQ_b+g_{ab}\left[2\del_cQ^c-Q^2\right],
\label{G_transform}
\ee
where $\Ghat_{ab}$ is the Einstein tensor of $\ghat_{ab}$, and $Q^2=Q_aQ^a$.
For clarity with the above, we decompose derivatives of $Q$ into time and space
derivatives:
\ba
&Q_a=\sdel_a Q-\qdot u_a,\nonumber\\
\del_bQ_a=&\!\!\!\sdel_a\sdel_bQ+u_au_b\left(\ddot{Q}-\udot^c\sdel_cQ\right)-
\qdot\left(
\frac{1}{3}\theta h_{ab}+\sigma_{ab}-\omega_{ab}\right)\nonumber\\
&\!\!+2u_{(a}\left[-\sdel_{b)}\qdot+
\frac{1}{3}\theta\sdel_{b)}Q+\left(\sigma_{b)}^{\pha c}+\omega_{b)}^{\pha c}
\right)
\sdel_cQ\right].
\label{Q_derivatives}
\ea
We also write $\hat T_{ab}=\Ghat_{ab}$, and $T_{ab}=G_{ab}$ as general fluids,
\emph{both with respect to $u^a$}:
\ba
&\hat{G}_{ab}=\hat\mu\hat u_a\hat u_b +\hat{p}\hat h_{ab} +2\hat{q}_{(a}\hat
u_{b)}+\hat{\pi}_{ab},\label{Gbar}
\\ &{G}_{ab}=\mu u_a u_b +p h_{ab}+2{q}_{(a}u_{b)}+{\pi}_{ab};\label{G_fluid}
\ea
where $\{\hat\mu,\hat{p},\hat{q}_a, \hat{\pi}_{ab}\}$, and
$\{\mu,p,q_a,\pi_{ab}\}$ are the energy density, isotropic pressure, heat flux,
and anisotropic pressure of $\Ghat_{ab}$ and $G_{ab}$ respectively.

We can decompose $G_{ab}$ given by~(\ref{G_transform}) into the fluid variables
in~(\ref{G_fluid}) by using~(\ref{Q_derivatives}) in the following covariant
manner:
\ba
\mu\li=\li u^au^bG_{ab}=e^{2Q}\hat\mu-3\qdot\left(\qdot-\frac{2}{3}\theta\right)-
2\sdel_a\sdel^aQ+\sdel_aQ\sdel^aQ,\\
p\li=\li\frac{1}{3}h^{ab}G_{ab}=e^{2Q}\hat{p}+\left(\qdot-\frac{4}{3}\theta\right)
\qdot-2\ddot{Q}+
\frac{4}{3}\sdel_a\sdel^aQ-\frac{5}{3}\sdel_aQ\sdel^aQ\\
&&~~~~~~~~~~~~~~~~~~~~~~~~~~~~~~~~~~~~~~~~~~~~~~~~~~~~~~~~~+2\udot_c\sdel^cQ,\nonumber
\\ q_a\li=\li-u^bG_{\langle
a\rangle b}=e^{Q}\hat{q}_a-2\sdel_a\qdot+2\left(\frac{1}{3}\theta-\qdot\right)
\sdel_aQ+2\left(\sigma_a^{\pha b}+\omega_a^{\pha b}\right)\sdel_bQ\\
\pi_{ab}\li=\li G_{\langle ab\rangle}=\hat{\pi}_{ab}+2\qdot\sigma_{ab}-2
\sdel_{\< a}\sdel_{b\>}Q-2\sdel_{\< a}Q\sdel_{b\>}Q.
\ea

\forget{

\subsection{Interpreting the matter: Examples}

Under the conformal transformation
\be
g_{ab}=e^{2Q}\ghat_{ab}
\ee
where $Q$ is defined by
\be
Q_{,a}=\dot{u}_a-\frac{1}{3}\theta u_a,
\ee
the Einstein tensor becomes:
\ba
G_{ab}=\tilde{G}_{ab}&-&2(\del_a\dot{u}_b+\dot{u}_a\dot{u}_b)+
\frac{2}{3}\theta(\omega_{ab}-u_a\dot{u}_b) +\frac{2}{3}\theta_{,a}u_b\nonumber\\
&+&g_{ab}\left(2\del_c\dot{u}^c-\dot{u}^2
-\frac{1}{3}\theta^2-\frac{2}{3}\dot{\theta}\right),
\ea
where $\tilde{G}_{ab}$ is the Einstein tensor of the metric $\ghat_{ab}$.
Without loss of generality we can write this as a general fluid with respect to
$u^a$;
\be
\tilde{G}_{ab}=\rho u_a u_b +P h_ab +2\tilde{q}_{(a}u_{b)}+\tilde{\pi}_{ab}.
\ee
These quantities may or may not be interpreted in the usual fluid fashion; it
has no significance at this stage. What must have some interpretation are the
fluid quantities associated with $G_{ab}$. These are decomposed as follows:
\begin{eqnarray}
\mu&=u^au^bG_{ab}&=\rho+\frac{1}{3}\theta^2+3\dot{u}^2-2\del_c\dot{u}^c\\
p&=\frac{1}{3}h^{ab}G_{ab}&=P-\frac{2}{3}\dot{\theta}-\frac{1}{3}\theta^2-
\dot{u}^2+
\frac{4}{3}\del_c\dot{u}^c\\
q_a&=-h_a^{\pha b}u^c G_{bc}&
=\tilde{q}_a+\frac{2}{3}\sdel_a\theta+2\udot^c\omega_{ca}\\
\pi_{ab}&=
G_{\langle ab\rangle}&
=\tilde{\pi}_{ab}+\frac{2}{3}\theta\omega_{ab}-2(\sdel_a\udot_b+\udot_a\udot_b)+
\frac{2}{3}\del_c\udot^c h_{ab}
\end{eqnarray}

}

\section{Relativistic Thermodynamics}

Any realistic cosmological model must include some sort of \emp{thermodynamic
scheme}. This means that we expect the laws of thermodynamics to hold
throughout the evolution of the universe.

Together with the conservation equations~(\ref{conservation of EFE}) the
energy-momentum tensor~(\ref{E-M_Tensor_general}), which lead to the
equations~(\ref{energy_cons_gen}) and~(\ref{moment_cons}), there are a number
of other conservation equations which one may or may not impose upon the fluid.
See Ehlers~(1971)\nocite{ehl71} for derivations and a discussion of kinetic
theory in GR.

Firstly, we normally assume that the comoving \emp{number density} is
conserved; ie,
\be
\del_a(N^a)\equiv0,\label{conservation-number
density}
\ee
where $N^a\eqdef nu^a$ and $n$ is the number density of particles -- see
Krasinski~(1997)\nocite{kras97}, or Maartens~(1996)\nocite{maar96}.
Now,~(\ref{conservation-number density}) is equivalent to the condition
\be
\dot{n}+\theta n=0\pha\Leftrightarrow\pha n\ell^3={\mathrm{const.}},
\label{cons-number density-int}
\ee
where~(\ref{average length scale-def}) was used to demonstrate that the number
of particles in a comoving volume is constant.

However, in a more realistic model this may not be true at some stages of the
evolution, see for example Gunzig \etal~(1997)\nocite{gunz-97} where a particle
creation rate is introduced to model creation of radiation due to inflation.
Other generalisations include adding a viscous pressure term; eg, Coley, van
den Hoogen, and Maartens~(1996)\nocite{col-96}, or
Maartens~(1995)\nocite{maar95}.

\subsubsection{\emp{The First Law of Thermodynamics}}

The first law is the Gibbs equation which applies in equilibrium:
\be
TdS\equiv d\left(\frac{\mu}{n}\right)+pd\left(\frac{1}{n}\right),\label{First
law thermodynamics-Gibbs}
\ee
where $S$ is the \emp{entropy density}, and $T$ is the temperature of the
fluid; or
\be
T\del_aSdx^a=\del_a\left[\frac{\mu}{n}\right]dx^a+p\del_a\left[\frac{1}{n}
\right]dx^a;
\ee
which becomes, upon dividing by an increment of proper time along the
congruence, and substituting from~(\ref{cons-number density-int}),
\be
\dot{S}=\frac{1}{Tn}\left(\dot{\mu}+\theta(\mu+p)\right)\label{entropy change}
\ee
which shows that $\dot{S}=0$ for a perfect fluid~(cf,~(\ref{energy_cons_gen})):
ie, entropy per particle along a particular flow line remains constant.

\subsubsection{\emp{The Second Law}}

Entropy flux is generally defined as a vector;
\be
S^a=Snu^a+\frac{R^a}{T}
\ee
where $R^a$ represents all dissipative processes, and $S$ and $T$ are related
by ((\ref{First law thermodynamics-Gibbs})~-- ie, they are scalars defined in
local equilibrium). In the simplest cases $R^a$ is assumed zero, or equated
with the energy flux. The second law states $\del_aS^a\geq0$, where equality
applies in equilibrium (where $R^a=0$).

Equations~(\ref{conservation-number density}), and~(\ref{First law
thermodynamics-Gibbs}) comprise a \emp{thermodynamic scheme} which implies that
the right hand side of~(\ref{First law thermodynamics-Gibbs}) has an
integrating factor~-- see Krasi\'nski, Quevedo and
Sussman~(1997)\nocite{kras-97}. These are additional constraints on
cosmological models. (This is used in
\S\ref{Symmetry and Thermodynamic Schemes}.)

\section{Fluids in Cosmology}

Equation~(\ref{E-M_Tensor_general}) is a general equation describing any fluid.
Different types of fluid have different energy-momentum tensors, which we will
now discuss in order of importance for cosmological models.

The most important fluids used in cosmology are \emp{dust} solutions, often
with a cosmological constant. The simplest of these are the FLRW models, which
are generally believed to represent the universe on a suitably large scale;
perturbations of these models give a more realistic representation of the
universe. However, simple does not mean correct and therefore other models have
been studied to give a different view of the universe; among these are the dust
Lemaitre-Tolman (LT) models and Bianchi models -- the latter often used because
they are `close to FLRW' for some suitable length of time, and therefore give a
better understanding of FLRW models themselves.

Generalisation of a simple dust solution to a \emp{perfect fluid} is a
necessary, but not always simple, task if one wants a model of the universe
before recombination; most solutions of Einstein's equations which are used as
cosmological models are perfect fliuds (in particular, a scalar field can
always be written as a perfect fluid).

\forget{
A \emp{scalar field} model can be used to model the early universe (inflation),
but as a cosmological model, in classical GR, will not necessarily differ from
the perfect fluid models; the physics will change however\footnote{However, one
need not take the gradient of the scalar field to be parallel to the
fundamental congruence; this will effectively result in the matter being
described by~(\ref{E-M_Tensor_general}).}. }

Here it will be useful to note that a cosmological constant can always be
absorbed into the energy-momentum tensor by redefining the density and
pressure.
\ba
G_{ab}\li=\li T_{ab}+\Lambda g_{ab}\nonumber\\
\li=\li\mu u_au_b +\Lambda h_{ab} - \Lambda u_au_b\nonumber\\
\li=\li(\mu-\Lambda)u_au_b+\Lambda h_{ab};\label{cosmological const is pressure}
\ea
ie, a perfect fluid with constant pressure. A cosmological constant does not
need to be studied separately in the other perfect fluid cases.

\subsection{Perfect Fluids and Dust Models}
\label{Perfect Fluids and Dust Models}

A perfect fluid occurs when $\pi_{ab}=q^a=0$; in the case of dust\footnote{Also
known as an incoherent fluid.} (`CDM') we also have $p=0$, which necessarily
implies (by~(\ref{moment_cons})) that the fundamental congruence is geodesic;
$\udot^a\equiv0$. These models are implicitly in \emp{equilibrium}; that is,
the dynamics are completely reversible and no heat is generated from friction
or anything else ($\dot{S}=0$).

From the energy conservation equation~(\ref{energy_cons_gen}), we get the
simple relation for the expansion,
\be
\theta=-\frac{\dot{\mu}}{(\mu+p)};\label{energy_cons_PF}
\ee
and similarly for the acceleration~(\ref{moment_cons}),
\be
\udot_a=-\frac{\sdel_ap}{\mu+p}.\label{mom_cons_PF}
\ee

However, the Raychaudhuri equation remains unchanged -- emphasising that it is
an equation relating the important kinematic quantities of the flow, rather
than the more complex anisotropic pressures, energy flux or Weyl curvature. We
can substitute~(\ref{energy_cons_PF}) and~(\ref{mom_cons_PF}) into the
Raychaudhuri equation~(\ref{raychaudhuri}) to get a second order differential
equation for the matter quantities.

From the Gibbs relation~(\ref{First law thermodynamics-Gibbs}) (which
\emph{defines} the temperature) the entropy of a perfect fluid cannot change
along the congruence~(cf,~\ref{entropy change});
\be
\dot{S}=0.
\ee

The adiabatic speed of sound is given by
\be
c_s=\sqrt{\frac{\dot{p}}{\dot{\mu}}}
\ee
which must be less than the speed of light.

Perfect fluids can be split into different types, according to their dependence
of~$\mu$ on~$p$.

\subsubsection{\emp{A Barotropic Equation of State}}

Generally a useful simplifying assumption to make upon the fluid is that of a
\emp{barotropic} equation of state (EOS). It is a fairly \emph{ad hoc}
assumption which simply requires that the pressure depends only on the density:
$p\equiv p(\mu)$. It is adopted usually for mathematical simplicity rather than
for any physical reasons\footnote{J. Ehlers, private communication.}. This
functional dependence can in principle take any form. There are a few results
which follow from this simple EOS.

From~(\ref{mom_cons_PF}) we get
\be
\sdel_{[a}\udot_{b]}=0\pha \Leftrightarrow\pha \eta^{abc}\sdel_b\udot_c=0,
\ee
so that rotation is conserved,
\be
\dot{\omega}^{\< a\>}+\sfrac{2}{3}\theta\omega^a= \sigma^{ab}\omega_b,
\ee
cf~(\ref{rotation evolution}).

For these types of cosmological models the entropy is a global constant;
(from~\ref{First law thermodynamics-Gibbs}),
\be
\sdel_aS=0;
\ee
which implies $\del_aS=0$. This is an \emp{isentropic} fluid -- an unrealistic
situation. \forget{If we take temperature to be barotropic also ($T=T(\mu)$)
then the same conclusion is reached; in addition we have
\be
T\propto\exp\int\frac{dp}{\mu(p)+p}.
\ee
}

\subsubsection{\emp{Dust}: $p\equiv0$.}

This is the most common fluid form used for a cosmological model. Its
simplicity lies in making the motion geodesic, and thus eliminating many terms
in Einstein's equations.

In this case the energy density scales inversely with volume. In general, one
defines the \emp{characteristic length scale} of a model by
\be
\frac13 \theta=\frac{\dot{\ell}}{\ell},\label{average_length_scale}
\ee
hence, by~(\ref{energy_cons_PF}) we have
\be\label{energy densit dust prop ell^-3}
\frac{3\dot{\ell}}{\ell}=-\frac{\dot{\mu}}{\mu}\Rightarrow \mu\sim {\ell}^{-3}.
\ee

Dust is a particularly simple cosmological model to deal with. It models the
late universe of galaxies as `particles' of dust moving under gravity alone. It
requires that the random velocities of the galaxies be negligible -- ie, the
temperature is very low, so may not be used in the radiation dominated era.

\forget{
In the case of the FLRW models -- and indeed all irrotational dust models -- a
cosmological constant must be introduced if the expansion rate is to be able to
increase~(\ref{raychaudhuri}). As the latest Supernovae Ia results trickle in,
a negative \emp{deceleration parameter}~($q_0$) \emph{is} being measured: this
poses real problems for these types of alternative models. If one has to
introduce a cosmological constant in these then one may as well stick to the
FLRW models??Horizon problematic. }

\subsubsection{\emp{$\gamma$-law equation of state}}

This is a classic simple type of perfect fluid, which can describe a number of
situations. We have an equation of state of the form
\be
p=(\gamma-1)\mu:\pha1\leq\gamma\leq2,
\ee
where $\gamma$ is constant. $\gamma=1$ corresponds to dust, described above.
Again, from~(\ref{energy_cons_PF}), we have $\mu\sim {\ell}^{-3\gamma}$. The
other important case is that of radiation, $\gamma=4/3$.

If $\gamma=2$, then we have a `stiff fluid', in which the speed of sound equals
that of light. This is, in a sense, the equation of state of the `ether'. It is
usually discounted as a physically reasonable form of matter as it fails the
energy conditions, but some scalar field models have an effective equation of
state of this form. A cosmological constant may be accounted for by $\gamma=0,
\mu=\Lambda$.

In the standard model (FLRW), we have a \emp{multi-fluid} description: we have
a mixture of (non-interacting) matter and radiation, with the radiation
dominating at early times. If we have the characteristic length scale $\ell$ in
an expanding universe, then radiation density falls off as $\ell^{-4}$, but the
matter density of the dust decreases as $\ell^{-3}$: hence, if we proceed
forward in time from the initial singularity ($\ell=0$), then radiation will
dominate early on, but matter will take on the dominant role later.

\subsection{Scalar Fields}

Scalar fields are fashionable at the moment for their use in describing
inflationary scenarios in the early universe, but they may be present today
(quintessence). There is no need to go into any physical detail here, but the
basics are required for
\S\ref{QCDM}. A \emp{scalar field} has an energy-momentum tensor of the form
\be
T_{ab}\eqdef
\del_a\phi\del_b\phi-g_{ab}\left(\:12\del_c\phi\del^c\phi+V(\phi)\right),
\ee
where $V$ must satisfy the Klein-Gordon equation;
\be
V'(\phi)=\del^a\del_a\phi.
\ee
If $\del_a\phi$ is timelike then we may define the velocity field
\be
u^a=\;{\del^a\phi}{\sqrt{-\del_b\phi\del^b\phi}}
\ee
so that $T_{ab}$ has the form of a perfect fluid with effective energy density
and pressure
\be
\mu=-\:12\del_a\phi\del^a\phi+V(\phi),~~~p=-\:12\del_a\phi\del^a\phi-V(\phi).
\ee

\subsection{Velocity Fields}

In general there is no canonical definition of the `correct' fundamental
velocity field to use and the energy-momentum tensor will have a different
interpretation depending on the chosen congruence. That is, it is not obvious
whether fundamental observers should be identified with the natural congruence
in, say, a perfect fluid, or with some other velocity field (Coley and Tupper,
1983, 1984, 1985\nocite{col-tup83, col-tup84, col-tup85}). This is fundamental
to the 1+3 formalism: the decomposition above is velocity field dependent.

If we take a perfect fluid
\be
T_{ab}=\hat\mu\hat u_a\hat u_b+\hat p\hat h_{ab}
\ee
and consider this with respect to another congruence $u^a$ defined such that
\be
\hat u^a\eqdef\Gamma(u^a+v^a)
\ee
with $v^au_a=0$ (ie, $v^i$ is the 3-velocity relative to $\hat u^a$) and
$\Gamma=1/\sqrt{1-v^2}$ -- basically a Lorentz boost -- then $T_{ab}$ will have
the general form (\ref{E-M_Tensor_general}), with the relative energy densities
etc., related by
\ba
\mu\li=\li \hat\mu+\Gamma^2v^2(\hat\mu+\hat p)\nonumber\\
p\li=\li\hat p+\:13\Gamma^2v^2(\hat\mu+\hat p)\nonumber\\
 q_a\li=\li\Gamma^2(\hat\mu+\hat p)v_a\nonumber\\
 \pi_{ab}\li=\li \Gamma^2(\hat\mu+\hat p)\left\{v_av_b-\:13v^2h_{ab}\right\}.
\ea
(See Wainwright and Ellis, 1997.) We can see that for a perfect fluid a
relative energy flux will always be introduced by a change of velocity field.
This is required for \S\ref{QCDM}.

In the case of a general fluid with respect to another velocity field, the
transformations are given in Maartens, Gebbie and Ellis~(1999).

\subsection{More General Fluids}

The perfect fluids considered up to now are clearly unphysical: the entropy is
constant and there is no frictional heating. While this is not a problem if all
we want is a rough and ready model to work with (eg, the standard model), we
require something a little more involved for a decent model of the universe:
thermodynamic processes in the real universe are (probably) not reversible.
Unfortunately, there doesn't seem to be a well formulated theory of
relativistic dissipative fluid mechanics, and there only seem to be a couple of
cases occasionally used, such as bulk viscosity to describe inflation using the
truncated Israel-Stewart theory of irreversible thermodynamics (Israel and
Stewart, 1979, \nocite{isr-ste79} 1980\nocite{isr-ste80}); see Maartens~(1996)
for a review. See also Gariel and Le Denmat~(1994)\nocite{gar-led94}.

\section{The CMB Anisotropies}
\label{CMB anisotropies section}

In \S\ref{CMB Measurements} we discussed the observed characteristics of the
anisotropies in the CMB radiation.  Relativistic cosmology must explain the
effects which cause these anisotropies \emph{before} decoupling (ie,
perturbations of the spacetime structure, density fluctuations etc.) and
effects that occur \emph{after} the last scattering surface which affect the
size of the anisotropies as the photons travel to us (eg, lensing by structure,
the Rees-Sciama effect~-- the changing redshift of the CMB photons as they pass
through varying gravitational potentials; Rees and Sciama
1968\nocite{ree-sci68}~-- interstellar dust~-- Finkbeiner and
Schlegel~1999\nocite{fin-sch99})

The first attempt at a relativistic treatment was by Sachs and
Wolfe~(1967\nocite{sac-wol67}) by integrating the null geodesics in a
perturbation of a flat $\Lambda=0$ FLRW model to examine the redshift function
along the null rays. This has been done in a gauge-invariant and covariant 1+3
way by Challinor and Lasenby~(1998, 1999\nocite{cha-las98}\nocite{cha-las99})
for scalar perturbations. These approaches have been generalised by Maartens,
Gebbie and Ellis~(1999)\nocite{maar-99} to include some more non-linear
effects~-- see also Maartens~(1999)\nocite{maar99},
Challinor~(1999)\nocite{chal99}, Gebbie and Ellis~(1999)\nocite{geb-ell99}.
The essential steps, after choosing a suitable velocity field, are to decompose
the photon distribution function, the collision term in the Boltzmann equation
and the temperature fluctuation into spherical harmonics to get the multipole
moments for an inhomogeneous spacetime.

The covariant and gauge invariant results are presented in the above
references, and are a bit of a mess. In this thesis we will only require a
small part of the calculations which give additional evolution equations for
radiation other than those given by~(\ref{energy_cons_gen})-(\ref{H dot eqn}).

For a radiation field, the energy-momentum tensor is given by
\be
T_{\!_R}^{ab}(x^i)=\mu_{\!_R}u^au^b+\:13\mu_{\!_R}h^{ab} + 2q_{\!_R}^{(a}u^{b)}
+
\pi_{\!_R}^{ab}=
\int p^ap^b f(x^i,p^c)d^3p,\label{energy-mom-for-rad}
\ee
where
\be
p^a\eqdef E(u^a+e^a):~~~E=-p^au_a,
\ee
is the photon momentum, with $e^a$ a unit spacelike vector, and
$d^3p=EdEd\Omega$ is the volume element on the future null cone at $x^i$.
$f(x^i,p^c)$ is the \emp{photon distribution function}, which gives the number
density of photons at $x^i$ with momenta $p^a$, and must satisfy the
\emp{Boltzmann equation},
\be
\;{df}{dv}={\cal C}(f),
\ee
with ${\cal C}$ being change of $f$ along the geodesics (parameterised by $v$)
from all types of collisions, absorptions, etc., and which is effectively zero
after decoupling. The terms in~(\ref{energy-mom-for-rad}) define the first
three \emp{multipole moments}: $\mu_{\!_R}$ is the monopole, and represents the
average temperature over the whole sky via the Stefan-Boltzmann law;
\be
\lan T^4\ran_{_{_{\mathsf{all~sky}}}}\propto\mu_{\!_R},
\ee
so if the distribution function is a Planck distribution, then the temperature
is that of a black body. Any fluctuations in brightness across the sky are
given by the higher order multipoles,
$q_{\!_R}^a,~\pi_{\!_R}^{ab},~\Pi^{a_1a_2\cdots a_\ell}$. Now, if the
observers' frame ($v^a$) is moving non-relativistically relative to the CMB
frame ($u^a$) then these evolve according to
\ba
\dot{\mu}_{\!_R} + \:43\theta\mu_{\!_R}
+ \sdel_{a}q_{\!_R}^a + 2\udot_aq_{\!_R}^a + \sigma_{ab}\pi_{\!_R}^{ab} \li=
\li
n_{_E}\sigma_{_T}\left(\:43\mu_{\!_R}v^2 - q_{\!_R}^av_{a}\,\right) + {\cal
O}[3]\label{monopole evolution-CMB} \\
\dot{q}_{\!_R}^{\< a\>} + \:43\theta q_{\!_R}^a
+ \:43\mu_{\!_R}\udot^a + \:13\sdel^a\mu_{\!_R}\li +\li
\sdel_{b}\pi_{\!_R}^{ab}
 +  \sigma^a{}_bq_{\!_R}^b +\nonumber\\
\eta^{abc}\omega_{b}q_{\!_Rc} + \udot_b\pi_{\!_R}^{ab} \li = \li
n_{_E}\sigma_{_T}\left(\:43\mu_{\!_R}v^a - q_{\!_R}^a +
\pi_{\!_R}^{ab}v_{b}\right) + {\cal{O}}[3]\label{dipole evolution-CMB}
\ea
which are the new forms of~(\ref{energy_cons_gen}), and~(\ref{moment_cons}) for
the case when the radiation field is not quite aligned with the (baryonic)
observer. There are evolution equations for all the higher order multipoles as
well, which are not given in a simple fluid description: the \emp{quadrupole}
evolves according to
\ba
\dot{\pi}_{\!_R}^{\< ab\>}\li +\li \:43\theta
\pi_{\!_R}^{ab} + \:{8}{15}\mu_{\!_R}\sigma^{ab}
+ \:25\sdel^{\< a}q_{\!_R}^{b\>} + {8\pi\over35}\sdel_{c}\Pi^{abc}
\nonumber\\ & +\li  2\udot^{\< a}q_{\!_R}^{b\>} +
2\omega^c\eta_{cd}{}^{\< a}\pi_{\!_R}^{b\> d} + \:27\sigma_c{}^{\<
a}\pi_{\!_R}^{b\> c} - {32\pi\over315}\sigma_{cd}\Pi^{abcd}
\nonumber\\
{}&&{} = -n_{\!_E}\sigma_{\!_T}\left(\:{9}{10}\pi_{\!_R}^{ab} -
\:15q_{\!_R}^{\< a}v^{b\>} -
{8\pi\over35}\Pi^{abc}v_{c} - \:25\mu_{\!_R}v^{\< a} v^{b\>}\right) + {\cal
O}[3]\label{quadrupole evolution-CMB}
\ea
while the higher multipoles ($\ell> 3$) evolve according to
\ba
\dot{\Pi}^{\< A_\ell\>}\li+\li\:43 \theta \Pi^{A_\ell} + \sdel^{\<
a_\ell}\Pi^{A_{\ell-1}\>} + {(\ell+1)\over(2\ell+3)} \sdel_b\Pi^{bA_\ell}
\nonumber\\
&&{} - \ {(\ell+1)(\ell-2)\over(2\ell+3)} \udot_b \Pi^{bA_\ell} + (\ell+3)
\udot^{\< a_\ell} \Pi^{A_{\ell-1}\>} +
\ell \omega^b \eta_{bc}{}^{\< a_\ell} \Pi^{A_{\ell-1}\> c}
\nonumber\\
&&{} - \ {(\ell-1)(\ell+1)(\ell+2)\over(2\ell+3)(2\ell+5)}
\sigma_{bc} \Pi^{bcA_\ell} + {5\ell\over(2\ell+3)}
\sigma_b{}^{\< a_\ell} \Pi^{A_{\ell-1}\> b}
- (\ell+2) \sigma^{\< a_{\ell}a_{\ell-1}} \Pi^{A_{\ell-2}\>}
\nonumber\\
&&{} = - n_{\!_E}\sigma_{\!_T}\left[\ \Pi^{A_\ell} - \Pi^{\< A_{\ell-1}}
v^{a_\ell\>} - \left({\ell+1\over 2\ell+3}\right)
\Pi^{A_\ell a} v_{a}\ \right] + {\cal O}[3].\label{higher-order multipoles}
\ea
For $\ell=3$, the $\Pi^{\langle A_{\ell-1}}v^{{a_\ell}\rangle}$ term on the
right-hand side of equation (\ref{higher-order multipoles}) must be multiplied
by ${3\over2}$. $n_{\!_E}$ is the \emp{free electron number density} and
$\sigma_{\!_T}$ is the \emp{Thompson scattering cross section}. The series
expansion terminates at ${\cal O}[3]={\cal O}(\epsilon v^2,v^3)$ where
$\epsilon$ is a small parameter to represent how close the radiation and
baryonic frames are: there is no neglect of physical and geometric quantities.
All that is assumed is that the matter moves non-relativisticaly.

These results are required for the EGS theorem later: without them (especially
without (\ref{quadrupole evolution-CMB})) Theorem~\ref{isoradthm} cannot be
proved in the 1+3 formalism.

\section{Observations in the 1+3 Formalism}

Observations in cosmology are made by observing light\footnote{Or gravitational
waves, or neutrinos.} which has travelled on our past light cone. They are made
essentially from one spacetime point -- `here and now'. All we can hope to know
directly from this is where the light is coming from and how it travelled here.
The point of cosmology is to infer as much information about the universe as
possible, hopefully with as few assumptions as possible. As a generic problem,
the absolute limit to the amount we can know has been shown by Ellis
\etal~(1985)\nocite{ellis-85} to be that part of the universe enclosed in our
past lightcone, even assuming infinite precision in the observations (and
assuming Einstein's equations~(\ref{EFE})).

The part relativistic cosmology has to play is to determine the cosmological
model (or indeed the \emph{entire class} of models) which fits the observations
best: that is, what is the metric~$g_{ab}$ and matter content~$T_{ab}\equiv
G_{ab}$ (and necessarily the fundamental congruence,~$u^a$) of the universe?
This is distinct from other areas of cosmology insofar as it need not, in the
first approximation, describe the details of structure, or how the structure
formed. It must, however, be able to describe gross features like the CMB; for
example, the anisotropy of the CMB has been used to limit some global
properties of~$u^a$ such as the shear -- see Ellis, Treciokas and
Matravers~(1983a,b)\nocite{ellis-83-I}\nocite{ellis-83-II}; Stoeger, Maartens
and Ellis~(1995); Maartens, Ellis and Stoeger~(1995); and
Theorem~\ref{isoradthm}.

We must be able to relate the light that is observed to the metric~$g_{ab}$ and
velocity field~$u^a$. The most important methods are also the most direct:
light observed from discrete sources, and light from the CMB.

\subsection{Observable Quantities From Discrete Sources}

The foundations of this subject come from the now `famous in the right circles'
paper by Kristian and Sachs~(1966)\nocite{kris-sac66}, who first treated the
subject in a covariant manner (see also MacCallum and
Ellis~(1970)\nocite{Mac-Ellis-II}). Their method was to identify the various
\emp{distance measures} available in cosmology and relate them to
\emp{redshift} using the intrinsic and observed brightnesses (or equivalently,
magnitudes) of the sources.

\subsubsection{Luminosity-Distance Relations}
\label{Luminosity-Distance Relations}

If we observe a star to have some flux $\cal F$ that has some intrinsic
luminosity $\cal L$ then its \emp{luminosity distance} is defined to be
\be\label{luniosity distance def}
r_L\eqdef\sqrt{\frac{\cal L}{4\pi \cal F}}.
\ee
This is a simple extension of the variation of the brightness of objects in
non-relativistic physics because the flux that we observe not only decreases
because of the decreased number-density of photons, but also because each
photon is losing energy due to the expansion of the universe\footnote{For ease
of discussion I will assume throughout this thesis that the universe at some
time (ie, now) is expanding -- although models can be constructed which are
static that give similar observed redshifts -- but the analysis applies in any
relativistic model.}. This loss of energy of each photon is given by its
\emp{redshift}, $z$;
\be
1+z\eqdef\frac{\lambda_{{\mathrm{observed}}}}{\lambda_{{\mathrm{
emitted}}}}\equiv\frac{\nu_{{\mathrm{emitted}}}}{\nu_{{\mathrm{
observed}}}}=\frac{\delta t_o}{\delta t_e},
\ee
which shows redshift to be a time dilation effect. Although this last equality
is true for photons,\footnote{This is true in the \emp{geometrical optics
approximation} which basically says that photons move on null geodesics.} it is
actually true for any null-connected points in any (smooth) spacetime.
Redshift, therefore, reflects the stretching effect of expansion of spacetime;
since it is a directly measurable quantity, it is of fundamental importance. It
is therefore of some importance to write down other quantities in terms of
redshift. There is, however, a problem, and one which is quite difficult in
practice. If a galaxy is observed to have a redshift $z$ then determining how
much of this is cosmological, $z_c$, and how much is due to the relative
motions of source ($z_e$) and observer ($z_o$) requires a careful study of the
galaxy movement and a knowledge of our relative motion. The redshifts are
related by
\be
1+z=(1+z_c)(1+z_o)(1+z_e),
\ee
although the $z_o$ term can be inferred from a knowledge of the CMB dipole --
the CMB frame.

\subsubsection{Redshift}

Redshift can be treated in a more rigorous way. By solving Maxwell's equations
on a pseudo-Riemannian manifold in a charge and current free region, and
assuming that the wavelength of light is small compared to the spacetime
curvature (the \emp{geometrical optics approximation}), we find that light is
tangent to null surfaces of constant phase $\phi$, and therefore travels on
null geodesics. So if a photon's velocity is given by a null (geodesic) vector
$k^a\eqdef\del^a\phi$ ($\Leftrightarrow$ $\del_{[a}k_{b]}=0$):
\be
0=\del_a\phi\del^a\phi=k^ak_a;\phan{aaaa}  k^a\del_ak^b=0,
\ee
(the first of these is the \emp{eikonal equation}~-- see Ehlers and Newman
1999\nocite{ehl-new99}) then the angle between this vector and some velocity
field $u^a$ is the relative (angular) frequency of the photon as measured by an
observer traveling on $u^a$,
\be
\omega=-u^ak_a.
\ee
This frequency is clearly observer dependent; an observer moving in a different
manner would measure a different frequency (doppler shift). If a photon travels
between two points then the \emp{redshift} is the relative change of frequency
between the two points;
\be
1+z\eqdef\frac{\left.u_ak^a\right|_{\mathrm{e}}}{\left.u_bk^b\right|_{\mathrm{o}}}.
\ee
We can see this more explicitly by calculating the rate of change of frequency
along the photon path;
\be\label{relative frequency change}
k^a\del_a\omega=k^a\del_a(-u_bk^b)=-k^ak^b\del_au_b=
\left(-\sfrac13\theta h_{ab}+\udot_au_b-\sigma_{ab}\right)k^ak^b.
\ee
$k^a$ is a null vector, and so can be written as a linear combination of a
component parallel and orthogonal to $u^a$, viz;
\be\label{k^a parallel and perpendicular}
k^a=-u_bk^b(u^a+e^a).
\ee
where $e^a$ is a normalised spacelike vector orthogonal to $u^a$: $e^ae_a=1$;
$u_ae^a=0$; it simply defines the direction of the photon relative to $u^a$.
Substituting this into~(\ref{relative frequency change}) we find
\begin{equation}\label{relative frequency change2}
k^a\del_a\omega=-\omega^2\left(\sfrac13\theta+e^a\udot_a+
e^ae^b\sigma_{ab}\right).
\end{equation}
It is now clear that the relative direction of the photon is important when
considering frequency change; important effects also come from the kinematics
of $u^a$. For example expansion will decrease the photon frequency and thus
increase the wavelength, making it redshifted. However, the change in frequency
caused by shear and acceleration is direction dependent: acceleration can
either increase or decrease the frequency, depending on the direction of the
incoming photon relative to the direction of acceleration; similarly shear will
decrease the frequency by varying amounts depending on the direction of the
incoming photon and the \emp{principle directions} of the shear (ie, its
eigenvectors).

\subsubsection {The Redshift Structure of Conformally Related Spacetimes.}
\label{conformal redshift section}

If we have two conformally related metrics $g_{ab}=e^{2Q}\hat{g}_{ab}$:~$Q>0$,
then null geodesics, $g_{ab}k^ak^b=0$, are conformally invariant:
\be
k^b\nabla_b k^a =0\Rightarrow k^b\widehat{\nabla}_b k^a \propto k^a,
\ee
where $\nabla_a$ and $\widehat{\nabla}_a$ are the covariant derivatives
associated with $g_{ab}$ and $\hat{g}_{ab}$ respectively. Although the null
geodesics associated with $\widehat{\del}_a$ are non-affinely parameterised,
the structure of the lightcones is identical for both spacetimes (see, for
example Wald 1984\nocite{wald}). The affine parameters $\lambda$,
$\hat{\lambda}$ associated with the $\del_a$-geodesics and
$\widehat{\del}_a$-geodesics are related by
\be
\frac{d\lambda}{d\hat{\lambda}}=q e^{2Q}.
\ee
with $q$ constant. Now we associate a different null vector with each metric:
$k^a$ is tangent vector to a geodesic with affine parameter $\lambda$, and
$\hat{k}^a$ is tangent to a geodesic with affine parameter $\hat{\lambda}$. In
some basis, $x^{\alpha}$
\be
\frac{d\pha}{d\hat{\lambda}}=\hat{k}^{\alpha}\frac{\partial \pha}{\partial
x^{\alpha}}= qe^{2Q}\frac{d\pha}{d\lambda}= q e^{2Q} k^{\alpha}\frac{\partial
\pha}{\partial x^{\alpha}};
\ee
since the basis vectors are linearly independent, we now have established the
general result
\be
\hat{k}^{a}=qe^{2Q} k^a.
\ee
Hence, if the geodesics with respect to $\widehat{\nabla}_a$ are known
explicitly then we can find them easily for $\del_a$.

Suppose we have a metric $g_{ab}$ in comoving coordinates,
$u^{\alpha}=-|g_{00}|^{-1/2}\delta^{\alpha}_0$, which is conformally related to
another metric $\hat{g}_{ab}$ and there exists a coordinate transformation
$x^{\alpha'}(x^{\alpha})$ : $g_{\alpha\beta} \longrightarrow g_{\alpha'\beta'}$
where $g_{\alpha'\beta'}=\exp{[2Q(x^{\gamma'})]}\hat{g}_{\alpha'\beta'}$
explicitly in the primed coordinate system. This means that the frequency of a
photon travelling along a null geodesic becomes
\be
u_{\alpha}k^{\alpha}=u_{\alpha}\frac{\pd x^{\alpha}}{\pd
x^{\alpha'}}k^{\alpha'} =-|g_{00}|^{-1/2}\delta^{\alpha}_0\frac{\pd t}{\pd
x^{\alpha'}}k^{\alpha'}=
-\frac{|g_{00}|^{-1/2}\delta^{\alpha}_0}{qe^{2Q}}\frac{\pd t} {\pd
x^{\alpha'}}\hat{k}^{\alpha'}
\ee

It is then straight forward to show that if $g_{\alpha'\beta'}$ is in comoving
coordinates the redshift becomes
\be
1+z=\frac {\left. u_{a}k^{a}\right|_{_{\mathsf
Galaxy}}}{\left.u_{b}k^{b}\right|_{_{\mathsf Observer}}}=
\frac {\left. u_{0}k^{0}\right|_{_{G}}}{\left.u_{0}k^{0}\right|_{_{O}}}=
e^{{Q|_{_{O}}}-{Q|_{_{G}}}}(1+\hat{z}). \label{redshift for conf rel g}
\ee
where $\hat{z}$ is the redshift associated with $\hat{g}_{ab}$.

\subsubsection{Area and Luminosity Distance}
\label{area dist section}

In addition to~(\ref{luniosity distance def}) as a distance measure, we can
also measure the \emp{angular size} of an object, $d\Omega$. If this object has
an intrinsic proper area $dS$, then the \emp{area distance} is defined by the
ratio of these;
\begin{equation}\label{area distance def}
r_A^2\eqdef\frac{dS}{d\Omega}.
\end{equation}
This is related to the luminosity distance~(\ref{luniosity distance def}) by
the \emp{reciprocity theorem}, which states
\begin{equation}\label{reciprocity theorem}
r_L=r_A(1+z)^2,
\end{equation}
as first proved by Etherington~(1933)\nocite{eth33}. It is essentially a
geometrical result, but can also be viewed as a time dilation effect, relating
geodesics traveling up and down the null cone. See figure~\ref{galaxy-obs-FIG}.
\begin{figure}[t!]
\centerline{\psfig{file={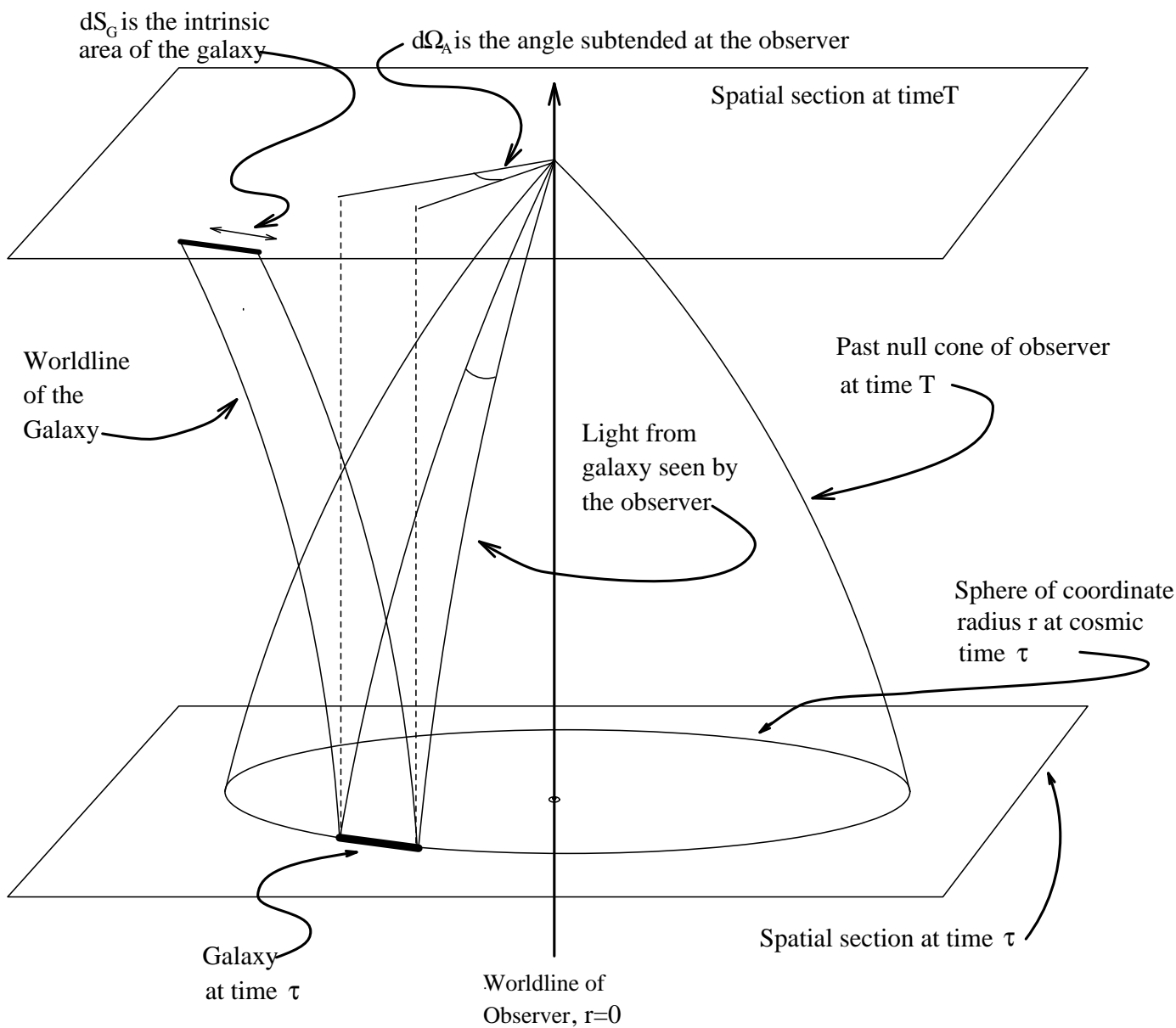},width=\textwidth,angle=0}}
\caption{\small An observer measuring the area distance of a galaxy.
\label{galaxy-obs-FIG} }
\end{figure}%

Area distance is measurable, and can be found by integrating the geodesic
deviation equation~-- see MacCallum and Ellis~(1970). Using~(\ref{reciprocity
theorem}), and~(\ref{luniosity distance def}) we can relate the metric to
directly measurable quantities: redshift, luminosity and area
distance;\footnote{Relating area distance to the metric and redshift is not
simple in general, and results are model dependent.}
\begin{equation}\label{luniosity-redshift-area distance}
{\cal{F}}=\frac{{\cal{L}}}{4\pi r_A^2(1+z)^4}.
\end{equation}

\subsubsection{The Magnitude-Redshift Relation}
\label{magnitude-redshift general expansion}

If we take the logarithm of~(\ref{luniosity-redshift-area distance}) we get the
\emp{magnitude-redshift} relation;
\begin{equation}\label{magnitude-redshift-area distance}
m-M-25=5\log_{10}r_L.
\end{equation}

As it stands though,~(\ref{magnitude-redshift-area distance}) is virtually
useless until a cosmological model is given (or, at the very least, a metric).
However, the cunning Kristian and Sachs~(1966) managed to expand $r_A$ in a
power series in $z$, and thus making a generalised \emp{magnitude-redshift}
relation. The result of that expansion is, in the notation of MacCallum and
Ellis~(1970)\nocite{Mac-Ellis-II},
\ba\label{m-z-general}
m\li-\li M-25=5\log_{10}z-5\log_{10}
\left.K^aK^b\del_au_b\right|_o+
\sfrac52\log_{10}e\left\{
\left[\left.4-\frac{K^aK^bK^c\del_a\del_bu_c}{(K^dK^e\del_du_e)^2}
\right|_O\right]z\right.
\nonumber\\
 \!\!\!\!\li-\li\left[2+\left.\left.\frac{R_{ab}K^aK^b}{6(K^cK^d\del_cu_d)^2}-
\frac{3(K^aK^bK^c\del_a\del_bu_c)^2}{4(K^dK^e\del_du_e)^4}
+
\frac{K^aK^bK^cK^d\del_a\del_b\del_cu_d}{3(K^eK^f\del_eu_f)^3}\right]
\!\!\right|_Oz^2
\right\}\nonumber\\\li\li
~~~~~~~~~~~~~~~~~~~~~~~~~~~~~~~~~~~~~~~~~~~~~~~~~~~~~~~~~~~~~~~~~~~~~~~+{\cal
O}(z^3),
\ea
where
\be
K^a\eqdef\frac{k^a}{\left.u^bk_b\right|_O}\phan{xxxx}\hbox{ie,}\phan{xxxx}
\left.K^a\right|_O=\left.-(u^a+e^a)\right|_O.
\ee

Obviously,~(\ref{m-z-general}) is extremely complicated for a general
cosmological model -- even for the lowest order terms. It is also questionable
how accurate this series expansion will be when truncated at low order in
redshift. Certainly it will be useless for $z>1$ unless the exact function
$m(z)$ has a very special form, or all terms $>{\cal O}(z)$ die off very
rapidly~-- see figures~\ref{flat FLRW m(z) vs series} and~\ref{stephani m(z) vs
series}.

\subsubsection{Number Count-Distance Relations}

A potentially interesting area of observational cosmology comes from
\emp{number counts} as a function of either redshift or magnitude. It works by
simply counting galaxies at a certain redshift in a solid angle of the sky
$d\Omega$ up to some limiting magnitude.  If we know the (mean) density of
galaxies in the spacetime then by counting how many we see in this volume will
give us information about the spacetime geometry.

We do not use number-counts in this thesis (although the relevant formulae may
be derived) because there is large uncertainty in the source evolution function
(Mustapha, Hellaby and Ellis,~1998\nocite{mus-hel98}~-- although it is unclear
if their result holds if multi-colour observations are taken into
account\footnote{B. Bassett, private communication.}).


\section{FLRW Models}
\label{FLRW Models}

The \emp{standard model} of modern cosmology is based on a hypothesis: the
\emp{Cosmological Principle} (CP; Ellis 1975). One of the key issues in this
thesis is the validity of the CP. The cosmological principle assumes perfect
isotropy about our location and extrapolates this to every other location using
the \emp{Copernican principle} which necessarily implies homogeneity of the
universe.\footnote{Any universe which is isotropic about three points will be
homogeneous.} Once the CP is in place then one must conclude that the universe
has an FLRW form. The proof of this is intuitive, and can be found eg, in
Wald~(1984).

The metric of FLRW models has the form (in isotropic coordinates)
\be\label{FLRW-metric}
ds^2=-dt^2+\frac{a(t)^2}{(1+\sfrac14 kr^2)^2}\left(dr^2+r^2d\Omega^2\right)
\ee
where
\be
d\Omega^2\eqdef d\vartheta^2+\sin^2\vartheta d\varphi^2;
\ee
$k$ is a constant which characterises the spatial hypersurfaces of the model:
$k<,=,>,0$ corresponds to hyperbolic, flat, or spherical
geometry\footnote{These are often called open, flat, or closed geometries on
the assumption that only the spherically symmetric case has closed spatial
sections. This is not true in general and an open model may not be infinite --
a non-trivial topology may be imposed to give eg, an open model with closed
spatial sections (eg, Luminet and Roukema~1999\nocite{lum-rou99}; Cornish and
Spergel,~1999,\nocite{cor-spe99} show that such a model is \emph{favoured} over
an infinite open model, on the basis of COBE data.}. Almost all the properties
of the FLRW models follow from~(\ref{FLRW-metric}). The fundamental velocity
field is given by
\be
u^0=1;\phan{xxx}u^i=0,
\ee
which has expansion
\be
\theta(t)=3\frac{\dot a(t)}{a(t)},
\ee
with the shear, rotation, and acceleration vanishing. The energy-momentum
tensor automatically has the form of a perfect fluid spacetime with respect to
this congruence. In fact, the FLRW models can be invariantly classed as perfect
fluid spacetimes with a fundamental congruence that has
$\sigma_{ab}=\omega_{ab}=\udot^a=0$, as can be proven from the field
equations~(\ref{energy_cons_gen}-\ref{div H})~-- see Krasi\'nski~(1997).

The energy density and pressure are given by
\ba
\mu\li=\li3\frac{k}{a^2}+3\left(\frac{\dot a}{a}\right)^2-\Lambda,
\label{Friedmann equation}\\
p\li=\li-2\frac{\ddot a}{a}-\frac{k}{a^2}-\left(\frac{\dot
a}{a}\right)^2+\Lambda\label{pressure for FLRW},
\ea
as can be found using~(\ref{energy density def}) and~(\ref{pressure def}).
Eq.~(\ref{Friedmann equation})~is the Friedmann equation. Combining these gives
the Raychaudhuri equation (cf,~(\ref{raychaudhuri}))
\be
\frac{\ddot a}{a}=-\:16(\mu+3p)+\:13\Lambda.\label{raychaudhuri for FLRW}
\ee
The Friedmann equation~(\ref{Friedmann equation}) and~(\ref{raychaudhuri for
FLRW}) gives equation~(\ref{energy_cons_gen}).

The function $a(t)$ is free, but will have a specific form once the
thermodynamics is settled upon; this usually takes the form of a barotropic
equation of state, or multiple (non-)interacting fluids. As far as the standard
model is concerned, the matter is assumed to be radiation until decoupling and
dust thereafter. Once the matter specified~(\ref{Friedmann equation})
and~(\ref{raychaudhuri for FLRW}) give a differential equation for $a(t)$,
which can be solved in principle -- see eg, Ellis~(1998).

We can define various scalars to get an idea about the dynamics of any
particular model. We define the \emp{Hubble scalar} as the expansion rate
\be
H\eqdef\sfrac13\theta=\frac{\dot\ell}{\ell}=\frac{\dot a}{a};\label{hubble
const def}
\ee
the \emp{deceleration parameter}
\be
q\eqdef-\frac{\ddot\ell\ell}{\dot\ell^2}=-\frac{\ddot aa}{\dot a^2};\label{q0
def}
\ee
and the \emp{density parameter}
\be\label{omega_0 def}
\Omega\eqdef\frac{\mu}{3H^2}=\frac{k}{H^2a^2}-\frac{\Lambda}{3H^2}+1.
\ee
where the last equality (from~(\ref{Friedmann equation})) shows that the
density parameter may be used to characterise the spatial surfaces in a
$\Lambda=0$ model. We can also define
\be
\Omega_{\Lambda}\eqdef\frac{\Lambda}{3H^2}
\ee
and
\be
\Omega_k\eqdef-\frac{k}{a^2H^2},
\ee
which implies that the deceleration parameter becomes, using~(\ref{raychaudhuri
for FLRW}),
\be
q=\sfrac12\Omega\left(1+\frac{p}{\mu}\right)-\Omega_{\Lambda};
\ee
for dust, we have the simpler, more familiar, form
\be
q=\sfrac12\Omega-\Omega_{\Lambda}.
\ee
From~(\ref{omega_0 def}) we have the normalisation condition
\be
\Omega+\Omega_{\Lambda}+\Omega_k=1,
\ee
which means, even though all the parameters are functions of time above, that
we can always characterise the curvature of the spatial surfaces in the
following way:

\begin{table}[h!]
\begin{center}{
\textsf{
\begin{tabular}{|c|c|c|c|}
  \hline
  $\Omega+\Omega_{\Lambda}>1$ & closed spatial surfaces  & $k>0$ & $\Omega_k<0$ \\
  $\Omega+\Omega_{\Lambda}=1$ & flat spatial surfaces & $k=0$ & $\Omega_k=0$ \\
  $\Omega+\Omega_{\Lambda}<1$ & open spatial surfaces & $k<0$ & $\Omega_k>0$ \\
  \hline
\end{tabular}}\label{FLRW table}
}
\end{center}
\end{table}

\subsection{The Magnitude-Redshift Relation}
\label{sec m-z-FLRW}

The \mofz\ relation is quite easy to derive due to the symmetry of the
solution, as can be found in, eg, Peebles~(1993)\nocite{peebles}. The key steps
are outlined below. (Factors of $c$, the speed of light, are included in this
section.) We are aiming to find the function $m(z)$ -- a relationship between
\emph{measurable} quantities. The information we have about the spacetime is
the metric~(\ref{FLRW-metric}), the Friedmann equation~(\ref{Friedmann
equation}), and the Raychaudhuri equation~(\ref{raychaudhuri for FLRW}). What
we don't know is the function $a(t)$ which is going to be needed if we want to
compare the models to data. The usual route is to assume that the universe is
dust after decoupling, and solve~(\ref{pressure for FLRW}) for $a(t)$.

The \mofz\ relation in the case of $\Lambda=0$ was first solved by
Mattig~(1958)\nocite{mat58}. The general formula for vanishing pressure was
first found by Kaufman~(1971)\nocite{kauf71} after the initial generalization
by Kaufman and Schucking~(1971)\nocite{kauf-sch71} to include $\Lambda>0$ in a
closed model.

We will derive it explicitly here to facilitate comparison with later
derivations for the Stephani models.

\paragraph{\emp{Redshift}:} For a general spherically symmetric spacetime it
is necessary to solve the geodesic equation for the null ray connecting the
galaxy to the observer in order to determine the galaxy's redshift. More
specifically, for a null ray from the galaxy, $G$, at~$(r,\tau)$ to the
observer, $O$, at~$(0,\tau_0)$, the tangent vector along the ray $k^a$ is
obtained as a solution of the geodesic equation. Once we have $k^a$ along the
ray, though, the redshift is obtained immediately from (cf, Ellis 1998)
\begin{equation}
    1+z = {\nu_G\over \nu_O}
      \equiv {u_a k^a|_G \over u_a k^a|_O},
\label{zdef}
\end{equation}
where $u^a$ is the four-velocity of the perfect fluid, ie,~of the galaxy~(G) or
the observer~(O). Since $u^a$ is a four-velocity it is normalised by
\begin{equation}
           g_{ab}u^a u^b = -1,
\label{guu1}
\end{equation}
and in comoving coordinates only $u^0$ is nonzero, so that~(\ref{guu1})
completely fixes~$u^a$. However, the high symmetry of \FRW\ models means that
the redshift can be obtained directly without having to integrate the geodesic
equation -- see eg, Wald~(1984) for a derivation. However we can use a novel
approach outlined in \S\ref{conformal redshift section}, which can be
implemented if we write the metric in the conformally static form
\be\label{FLRW metric conformal time}
ds^2=a(t(\eta))^2\left[d\eta^2+(1+\:14kr^2)^{-2}(dr^2+r^2d\Omega)\right]
\ee
where $\eta$ is the \emp{conformal time coordinate};
\be
\eta\eqdef\int\frac{dt}{a(t)}.
\ee
Using~(\ref{redshift for conf rel g}), we immediately have (because the part in
square brackets in~(\ref{FLRW metric conformal time}) is static and there are
no gravitational redshifts)
\begin{equation}
              1+z = {a_0\over a(t)};
\label{redshift FLRW}
\end{equation}
where $a_0\equiv a(T)$ is the scale factor today.

\paragraph{\emp{Lookback time}:} Consider radial null ray
(ie,~$d\vartheta=d\varphi=ds^2=0$) from a source at $(r,\tau)$ reaching the
observer ($r=0$) at time $\tau_0$. The relationship between $\tau_0$, $\tau$
and~$r$ can be obtained directly from the metric symmetry guaranteeing that the
raypath is purely radial so it is parameterised by a function~$\tau(r)$ by
integrating
\be\label{roft FLRW}
        {dr\over (1+\:14kr^2)} = c {dt\over a(t)}.
\ee
between $r$ and $r=0$. Now we have, on using~(\ref{redshift FLRW}),
\be
\frac{dt}{a(t)}=\frac{da}{a^2}\frac{1}{H}=-\frac{dz}{a_0H},
\ee
where $H$ is given by~(\ref{hubble const def}). From~(\ref{energy densit dust
prop ell^-3}) we have
\be
\mu=\mu_0\frac{a(t)^3}{a_0^3}
\ee
so that~(\ref{Friedmann equation}) becomes, using~(\ref{redshift FLRW}) once
again,
\be
H(z)^2=H_0^2\left[\Omega_0(1+z)^3+\Omega_{k_0}(1+z)^2+\Omega_{{\Lambda}_0}\right];
\label{H(z)}
\ee
where a subscript `$0$' means the present day value; viz;
\be
\Omega_0=\frac{\mu_0}{3H_0^2},\pha\Omega_{k_{0}}=\frac{k}{3H_0^2},
\pha\Omega_{\Lambda_0}=\frac{\Lambda}{3H_0^2}.
\ee
Together with~(\ref{roft FLRW}) this gives the function $r(z)$:
\be
\left.
      \begin{array}{c|c}
      k>0&\frac{2}{\sqrt{k}}\tan^{-1}\left(\:12\sqrt{k}r\right)\\
      k=0&r\\
      k<0&\frac{2}{\sqrt{|k|}}\tanh^{-1}\left(\:12\sqrt{|k|}r\right)
      \end{array}
      \right\}
=\;{1}{a_0H_0}\int_0^z\;{dz'}{\sqrt{\Omega_0(1+z)^3+\Omega_{k_0}(1+z)^2+
      \Omega_{{\Lambda}_0}}}.
\ee
We now have the coordinate distance as a function of redshift. We must now
relate this to some measurable distance quantity.

\paragraph{\emp{Angular diameter distance}:} Angular size distance, $r_A$,
is the ratio of an objects physical diameter to its apparent angular diameter.
As is well known, in spherically symmetric spacetimes the angular size distance
of an object as seen from the centre is given by (the square root of) the
coefficient in front of the angular components of the metric ($d\Omega^2$)
evaluated at the time the light was emitted.

For any inward radial null rays, in the plane~$\theta=\pi/2$, say, the angle
coordinate $\phi$ obviously does not change along the path. This means that the
apparent diameter,~$\delta_G$, of a galaxy is just determined by the difference
between the $\phi$ coordinates at either edge of the galaxy, at the time the
light we observe was emitted (ie,~at $t(r)$): $\delta_G = \delta\phi$. The
\emph{physical} diameter,~$D$ of the galaxy is given from the metric (with
$dt=dr=d\vartheta=0$):
\[
D = \frac{a(t(r))r}{1+\:14kr^2}\delta\phi = \;{a(t(r))r}{1+\:14kr^2} \delta_G
\]
($a(t(r))r/({1+\:14kr^2})$ just being the coefficient in the angular part of
the metric). The angular diameter distance,~$r_A$, is simply defined to satisfy
$D=r_A\delta_G$, so that
\begin{equation}
r_A(z) = \;{a(t)r}{1+\:14kr^2} = \;{a_0}{1+\:14kr(z)^2}{r(z)\over 1+z}
\label{rAF}
\end{equation}
for \FRW\ models, where we have used~(\ref{roft FLRW}) in the second equality.

The area distance can behave in quite an odd way: in non-relativistic
situations when things move further away they get smaller. This is not
necessarily what happens in a cosmological model: as something moves away it
may get smaller at first but then start to get bigger again. This is shown in
Fig~\ref{area dist vs redshift FLRW} where we can see that in a closed model
this \emp{refocusing} occurs at some redshift.
\begin{figure}[here!]
\centerline{\psfig{file=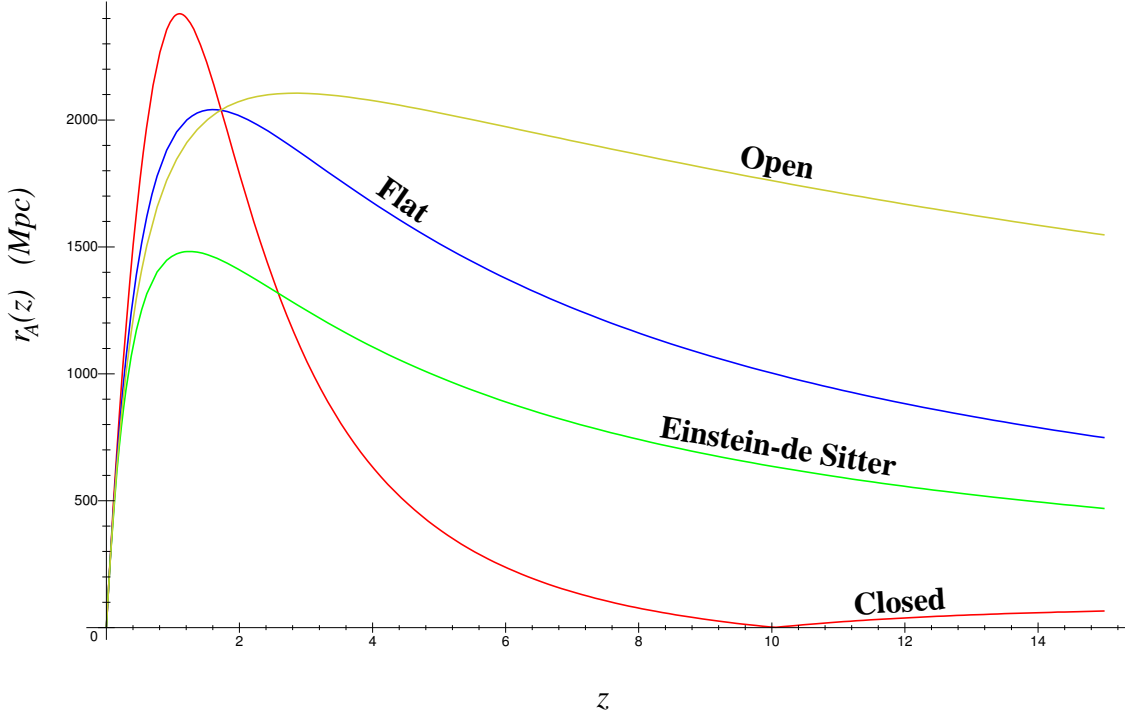,width=10cm,angle=90}}
\caption{\small Area distance as a function of redshift for a variety of
`standard' models. The closed model has
$\{\Omega_0,\Omega_{\Lambda_0}\}$=$\{0.3,1.5\}$, Einstein-de Sitter (1932) has
$\{1,0\}$, flat: $\{0.3,0.7\}$ and the open model has $\{0.1,0\}$. All the
models have $H_0=60$~\hu. Note that all the curves match up for low enough
redshift.}\label{area dist vs redshift FLRW}
\end{figure}\nocite{eis-des32}
This is because the light rays are effectively traversing an expanding sphere,
so a galaxy situated at the antipode to a particular observer will be spread
all over the observers sky. In (dust) FLRW models, this refocusing only occurs
once as $z\rightarrow\infty$ because the expansion is always just fast enough
that the light rays never `catch it up.'

\paragraph{{Luminosity distance}:} The key to deriving the luminosity
distance,~$r_L$, is the \emph{reciprocity theorem} (see Ellis and MacCallum,
1970, and references therein), which allows us to write
\begin{equation}
                        r_L = (1+z)^2 r_A.
\label{recipro}
\end{equation}
This equation is exact, and completely general, applying in \emph{any}
spacetime. Combining this with~(\ref{rAF}) gives
\begin{equation}
              r_L(z) = \;{(1+z)a_0r(z)}{1+\:14kr(z)^2}
\label{rLF}
\end{equation}
for \FRW\ models. Note that we can write this in the more familiar form
\be
r_L(z)= (1+z)a_0\tilde r(z):\left\{
      \begin{array}{c|c}
      \pha\tilde r(z)=\;{1}{\sqrt{k}}\sin2\tan^{-1}\:12\sqrt{k}r(z)& k>0,\\
      \tilde{r}(z)=r(z) & k=0,\\
      \pha\tilde r(z)=\;{1}{\sqrt{|k|}}\sinh2\tanh^{-1}\:12\sqrt{|k|}r(z)& k<0.
      \end{array}
      \right.
\ee
The function~$\tilde r(z)$ is plotted in Fig.~\ref{metric dist vs redshift
FLRW}. Hence we have the result
\be\label{luminosity redshift FLRW}
r_L(z)=(1+z)a_0\left\{\left.
      \begin{array}{c}
      \sin[\sqrt{k}\\
      1\\
      \sinh[\sqrt{|k|}
      \end{array}\right\}
\;{1}{a_0H_0}\int_0^z\;{dz'}{\sqrt{\Omega_0(1+z)^3+\Omega_{k_0}(1+z)^2+
      \Omega_{{\Lambda}_0}}}\right]
      \begin{array}{|c}
      k>0\\
      k=0\\
      k<0.
      \end{array}
\ee
\begin{figure}[here!]
\centerline{\psfig{file=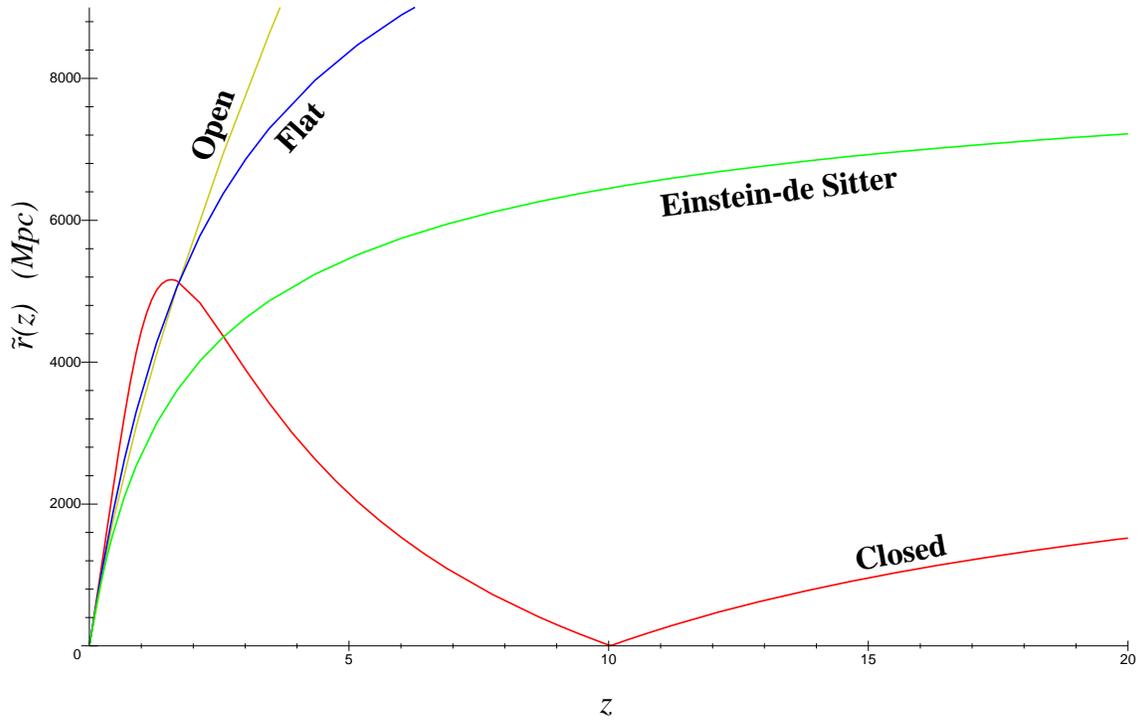,width=10cm,angle=90}}
\caption{\small Coordinate distance as a function of redshift the
models of figure~\ref{area dist vs redshift FLRW}. }\label{metric dist vs
redshift FLRW}
\end{figure}

\paragraph{\emp{Magnitude-redshift relation}:}
The magnitude-redshift relation for an object of absolute magnitude~$M$ then
follows from~(\ref{luminosity redshift FLRW}) and
\ba
m(z)-M\li-\li25 = 5\log_{10}\left.(1+z)a_0\times\left.\right.\right.\\&&
\left\{\left.
      \begin{array}{c}
      \sin[\sqrt{k}\\
      1\\
      \sinh[\sqrt{|k|}
      \end{array}\right\}
\;{1}{a_0H_0}\int_0^z\;{dz'}{\sqrt{\Omega_0(1+z)^3+\Omega_{k_0}(1+z)^2+
      \Omega_{{\Lambda}_0}}}\right]
      \begin{array}{c}
      k>0\\
      k=0\\
      k<0.
      \end{array}\nonumber
\label{pogson-FLRW}
\ea
In Fig~\ref{magnitude vs redshift FLRW} is shown the $m-z$ relation for the
same models as Fig.~\ref{area dist vs redshift FLRW}.
\begin{figure}[here!]
\centerline{\psfig{file=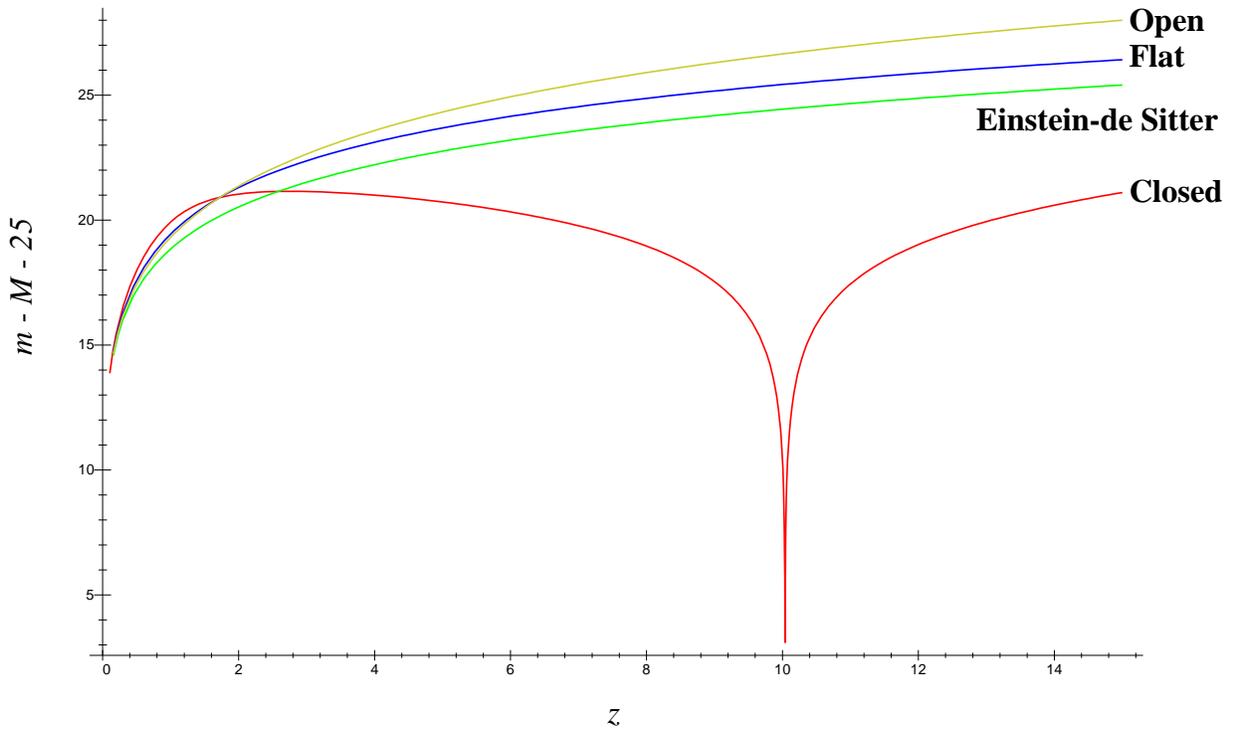,width=10cm,angle=90}}
\caption{\small Apparent magnitude as a function of redshift for the models of figure~\ref{area dist vs redshift FLRW}}\label{magnitude vs redshift FLRW}
\end{figure}
The closed model has the refocusing issue as before; the apparent magnitude
becomes singular at that point.

As the integral in~(\ref{pogson-FLRW}) is elliptic, it cannot be evaluated
explicitly. However, a series relation can be found relatively easily;
\ba\label{magnitude-redshift FLRW series}
m\li-\li M-25=5\log_{10}\;{z}{H_0}\nonumber\\ \li+\li\;52\log_{10}e\left\{(1-
q_0)z +\left[\;14(3q_0+1)(q_0-1) - 2\Omega_{\Lambda_0}\right]z^2 + {\cal
O}(z^3)\right\}.
\ea
We can now consider quantitatively the comments made in
\S\ref{magnitude-redshift general expansion} on the validity of the low order
expansion. Consider Fig~\ref{flat FLRW m(z) vs series}: the plot shows the
difference between the exact $m(z)$ function and the series
form~(\ref{magnitude-redshift FLRW series}) with an increasing power of
redshift~$z$ up to order~$9$.
\begin{figure}[t!]
\centerline{\psfig{file=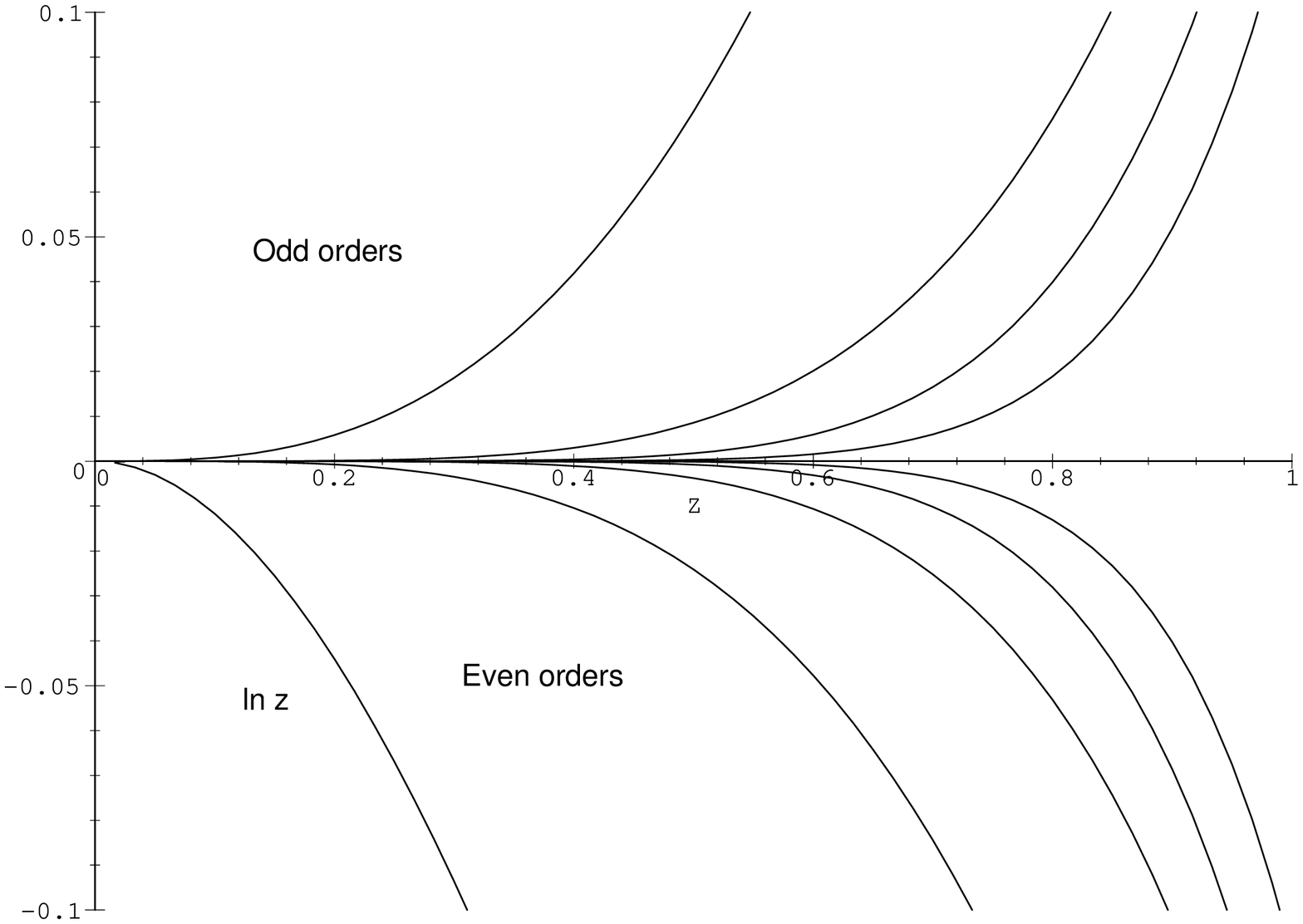,width=10cm,angle=-90}}
\caption{\small The exact \mofz\ relation minus the same function as a power
series relation to various orders.  $|m_{_{\mathrm exact}}-m_{_{\mathrm
series}}|>0.1$ would give a significant detectable error. }\label{flat FLRW
m(z) vs series}
\end{figure}
What we can see from figure~\ref{flat FLRW m(z) vs series} is that all the
series expansions diverge from the true function to more than 0.1mag (which is
roughly the errors in the magnitudes of the SNIa) at some redshift less than
unity.
\forget{For example, if we consider the lowest order curve, then at
$z\approx0.3$ and $\Delta m=|m_{_{\mathrm exact}}-m_{_{\mathrm
series}}|\approx0.1$, so an error?? in $H_0$ will occur $\sim 1$~\hu. Similarly
for the second order curve an error in $q_0$ of $\sim0.2$ will be found.} While
not massively important for the FLRW models at low redshift, for supernovae at
high redshift $z\sim 1$ it will be increasingly important to use the exact
relation, as even the ninth order expansion diverges from the true result at
these redshift, by a measurable amount.

If we jump ahead somewhat and consider the issue for a non-FLRW model (the
Stephani model considered in substantial detail later) then we can see from
Fig~(\ref{stephani m(z) vs series}) that the situation is far worse: the series
expansion can diverge very dramatically even at low redshift.
\begin{figure}[here!]
\centerline{\psfig{file=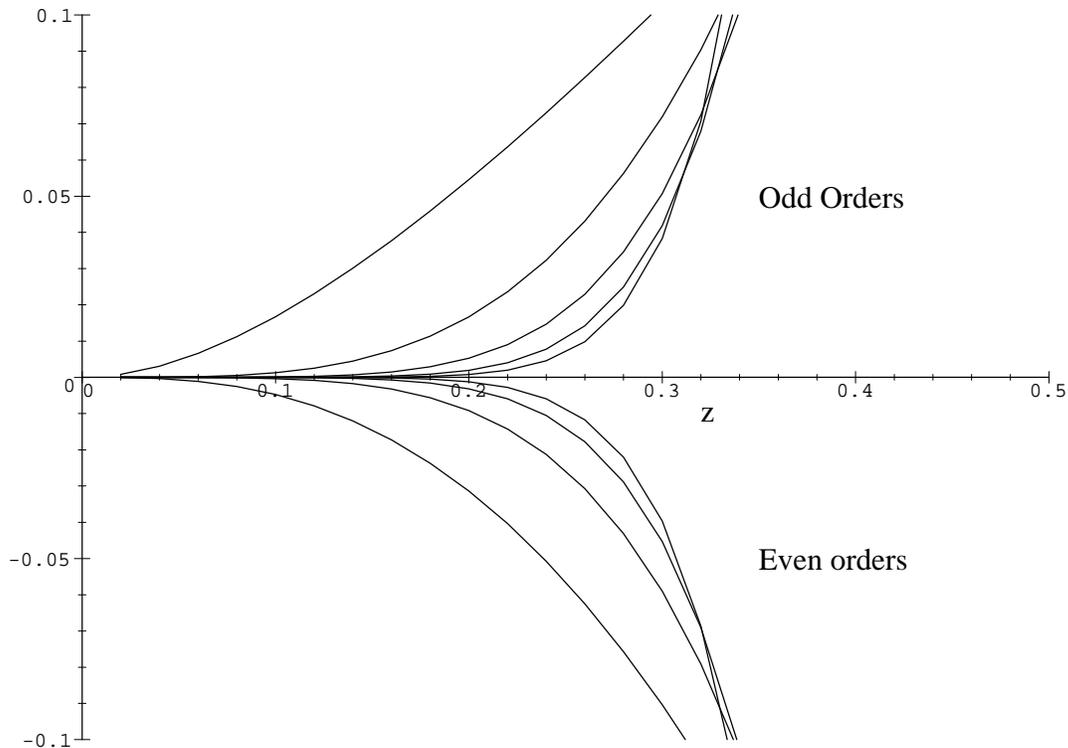,width=10cm,angle=-90}}
\caption{\small As in figure~\ref{flat FLRW
m(z) vs series}, but for the Stephani models considered later.}\label{stephani
m(z) vs series}
\end{figure}

We shall come back to the series form of the \mofz\ relation later in relation
to general cosmological models.


\subsection{Observable Quantities in FLRW Models}
\label{Observable Quantities in FLRW Models}

For a dust FLRW model there are only 3 independent parameters to fit from data:
$\{H_0, \Omega_0, \Omega_{\Lambda_0}\}$. In practice, these have boiled down to
the equivalent set $\{H_0, q_0, \Omega_{0}\}$. This arises from the fact that
most measurements are made at $z\lapp1$, or $z\sim1000$ at the CMB. Although
the measurement of all three parameters may be made in principle from any
observations, it is essentially impossible to determine any more than two of
these from observations in a narrow range of redshift. This has been
demonstrated  explicitly in a series of papers by Tegmark, Eisenstein and Hu
(Tegmark, Eisenetein and Hu, 1998a\nocite{teg-98b}; Tegmark
\etal,~1998b\nocite{teg-98b}; Eisenstein, Hu and Tegmark,
1998a,b\nocite{eis-98a}\nocite{eis-98b}; see also Tegmark, 1999\nocite{teg99};
White, 1998\nocite{whi98}; Efstathiou \etal,~1998\nocite{efs-98}; see Ehlers
and Rindler,~1989\nocite{ehl-rin89}, who first introduced the `phase plane'
arguments used here). Essentially they have shown that the set of possible
$\Omega_0,\Omega_{\Lambda_0}$ found from low-redshift observations will
complement the set of values preferred by high redshift experiments (ie, CMB
results). (It should be noted that they stress the need to fit all the data
sets at the same time, instead of taking their conclusions separately.)

For the low-redshift experiments ($z\lapp1$), we can see from the series
expansion of $m(z)$~(\ref{magnitude-redshift FLRW series}) that they will be
able to determine $q_0$ to some accuracy, but the extra detail required for a
measurement of $ \Omega_0$ and $\Omega_{\Lambda_0}$ separately would need
knowledge of the ${\cal O}(z^2)$ term. For example, the latest supernovae Ia
results of Perlmutter \etal\ (Perlmutter\etal,1999) give the results of their
data as an exclusion plot in the $ \Omega_0 -
\Omega_{\Lambda_0}$ plane,~Fig.~\ref{perlmutter exclusion plot}.
\begin{figure}[p!]
\centerline{\psfig{file=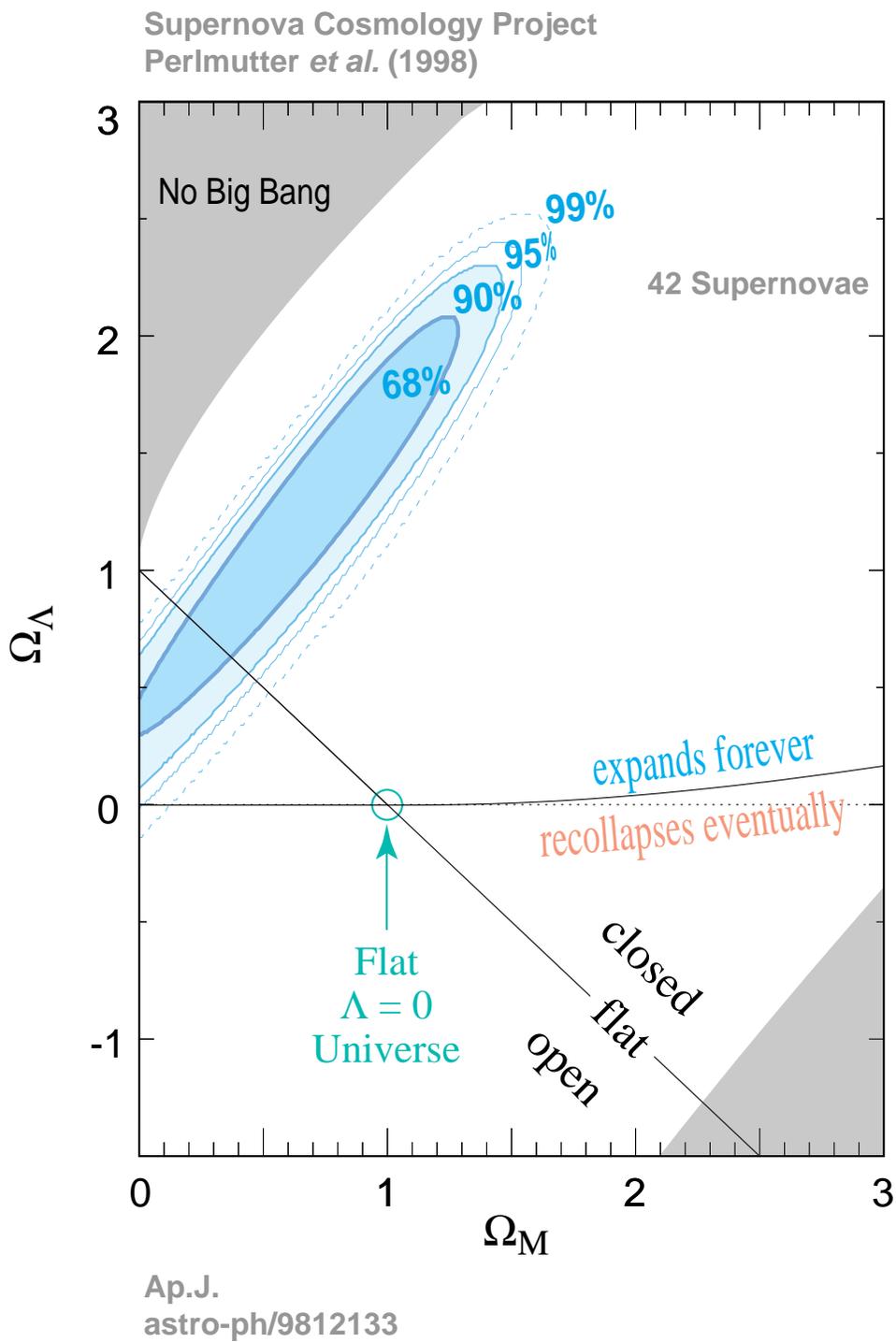,width=13cm,angle=0}}
\caption{\small The best fit to the $m(z)$ function for the supernovae
data of Perlmutter \etal,~(1998b). the confidence regions do nothing for
determining $ \Omega_0$ and $\Omega_{\Lambda_0}$. }\label{perlmutter exclusion
plot}
\end{figure}
We can see in this figure that their best fit regions do not provide any
substantial constraint on $\Omega_0$ and $\Omega_{\Lambda_0}$, but they do
provide quite good constraints in $q_0$.

At high redshift a very similar thing happens, but in a less obvious way. The
series expansion of the $m(z)$ curve is clearly useless for any intuitive idea
of what's going on. However, if we consider the CMB to have an intrinsic and
apparent magnitude, then we can examine the exact $m-z$ relation at $z=1000$
and plot the contours of constant $m$ on the $\Omega_0-\Omega_{\Lambda_0}$
phase plane. See Fig.~\ref{phase om vs ol z3}, where $m-M-25$ is plotted in the
$\Omega_0-\Omega_{\Lambda_0}$ plane, with contours of constant $m-M-25$ marked
on it. In practice we can calculate $m-M-25$ for any object from observations
(giving $m$) and assumptions about the objects intrinsic brightness ($M$). Thus
we can select just one contour in~(\ref{phase om vs ol z3}), and get some
limitations on $\Omega_0$ and $\Omega_{\Lambda_0}$. Just as in the SNIa
results, the constraints will be weak, with any particular contour selecting a
large range of allowed values for $\Omega_0$ and $\Omega_{\Lambda_0}$ (unless
$\Omega_0$ is small). It should be noted that this is a very simple analysis,
because the degeneracy of the CMB in the $\Omega_0-\Omega_{\Lambda_0}$ plane
comes from the position of the first Doppler peak (cf, figure~\ref{cmb_data})
rather than simply the $m-z$ relation, but the principle is the same.

Tegmark \etal\ have noticed that when the results from the CMB are taken
\emph{together} with the SNIa results, then it is possible to get very precise
predictions of $\Omega_0$ and $\Omega_{\Lambda_0}$ \emph{separately}. If we
have a look at~Fig.~\ref{phase om vs ol z3} we can see why.
\begin{figure}[p!]
\centerline{\psfig{file=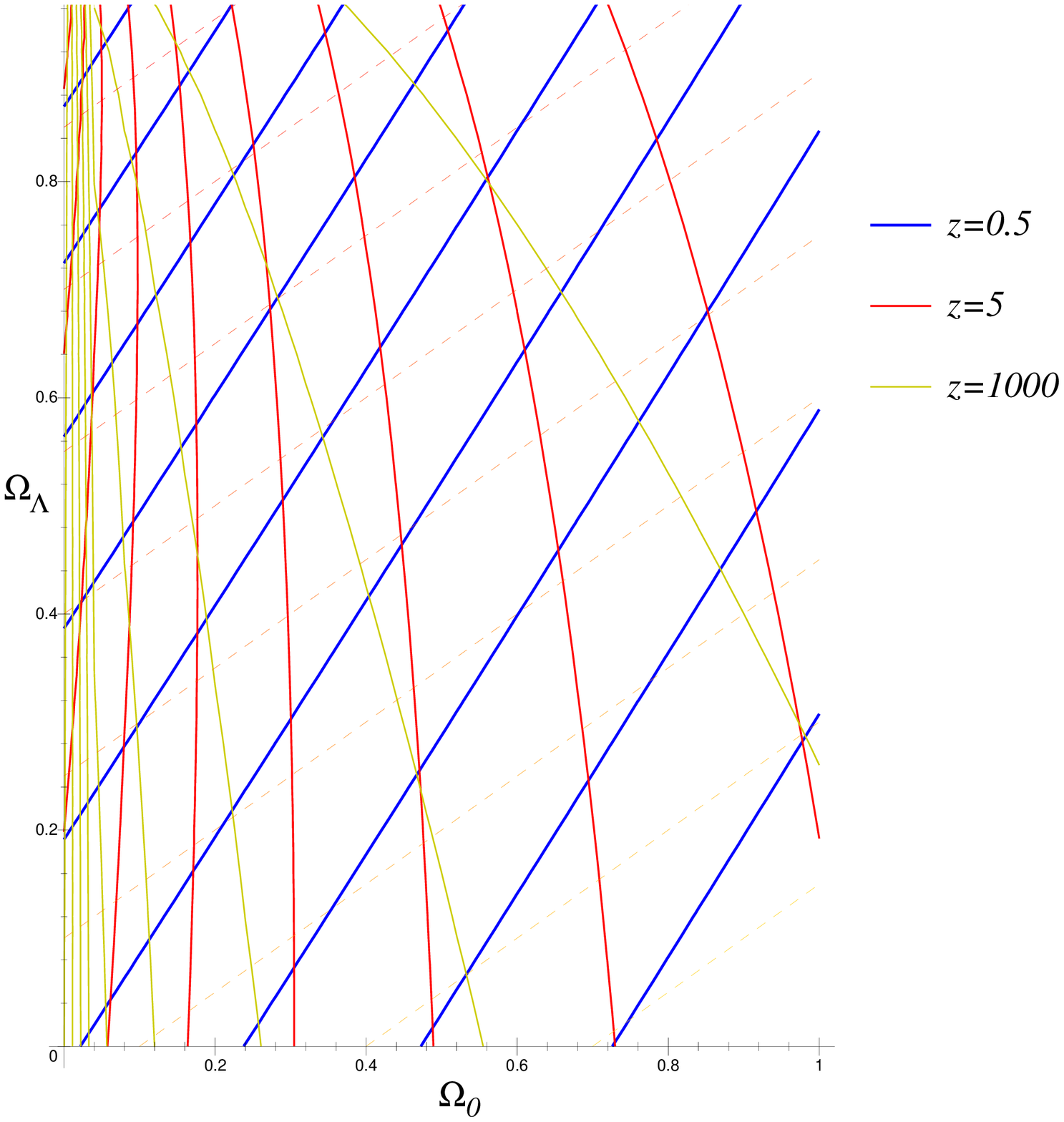,
width=\textwidth,angle=0}}
\caption{\small The phase diagram for the $m-z$ relation for $z=0.5,5,1000$
presented in the $ \Omega_0-\Omega_{\Lambda_0}$ plane. We can see that the
contours move round the diagram with increasing redshift. The faint dashed
lines are constant $q_0$, which shows that low redshift observations will give
good results for $q_0$. }\label{phase om vs ol z3}
\end{figure}
We have plotted contours, as described above, for redshifts of 0.5, 5, and
1000. Typically, a set of observations at low redshift will determine one of
the blue contours (obviously this means a set of contours in a small
neighborhood of some mean or `best fit' contour); similarly, the CMB
measurements will determine a similar set of the yellow contours. Because these
contours intersect at a fairly large angle, it is only the overlap of these
curves where the real universe will lie (provided the universe is FLRW of
course).

For completeness, we show in Fig,~\ref{complimentarity1} the results of
Tegmark, Eisenstein and Hu~(1998a) who computed this overlap for \emph{future}
SNIa and CMB observations.
\begin{figure}[here!]
\centerline{\psfig{file=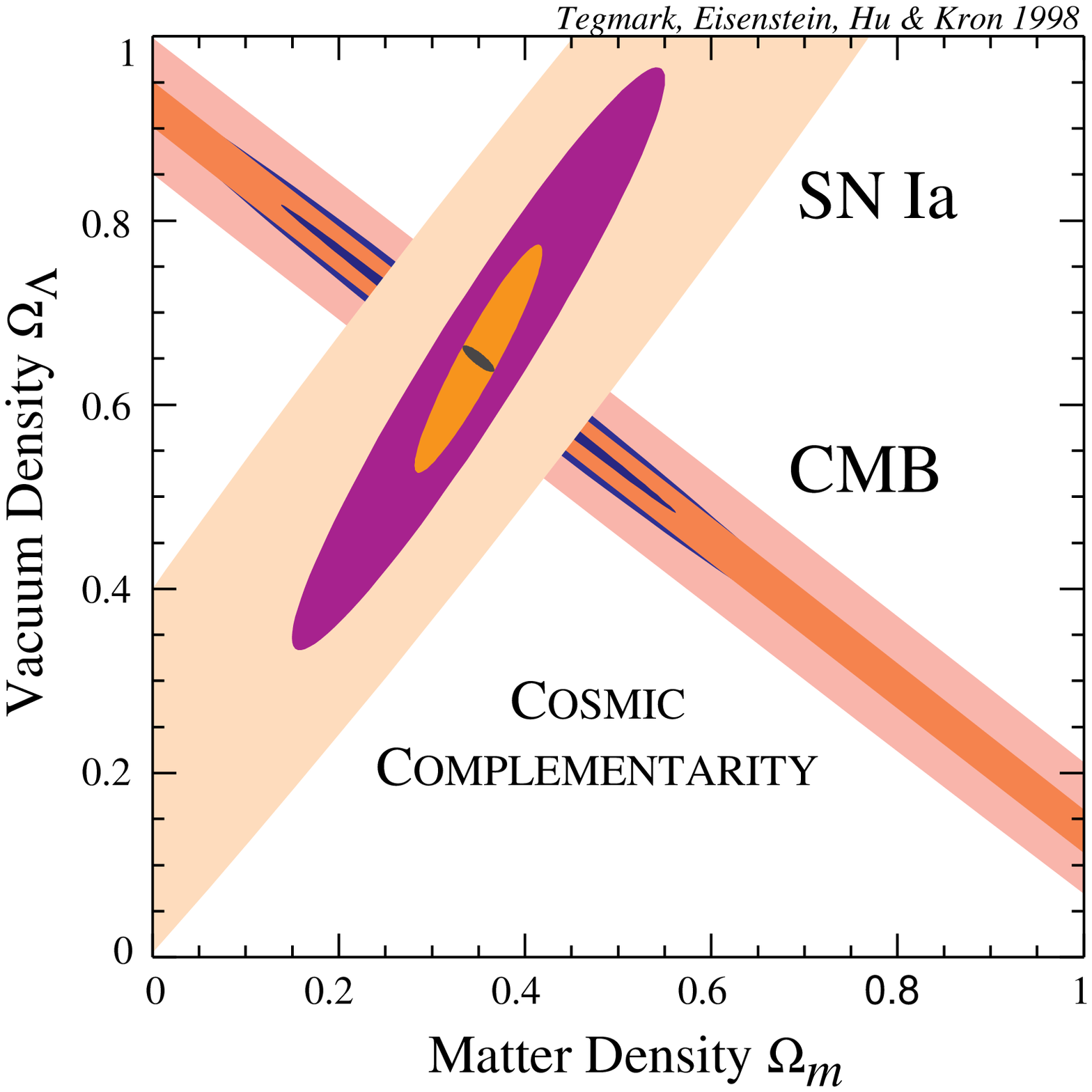,width=11cm,angle=0}}
\caption{\small  A futureistic diagram of what might be expected from
SNIa and CMB complementing each other in the future. The errors on $ \Omega_0$
and $\Omega_{\Lambda_0}$ (small black ellipse) are extremely small (by
cosmology standards) in comparison with the independent results. Plagiarised
from Tegmark, Eisenstein and Hu~(1998a).}\label{complimentarity1}
\end{figure}

We also show another stolen picture, Fig,~\ref{complimentarity2}, which gives
the same sort of results as Fig,~\ref{phase om vs ol z3}, but for
number-counts, age and growth as well.
\begin{figure}[here!]
\centerline{\psfig{file=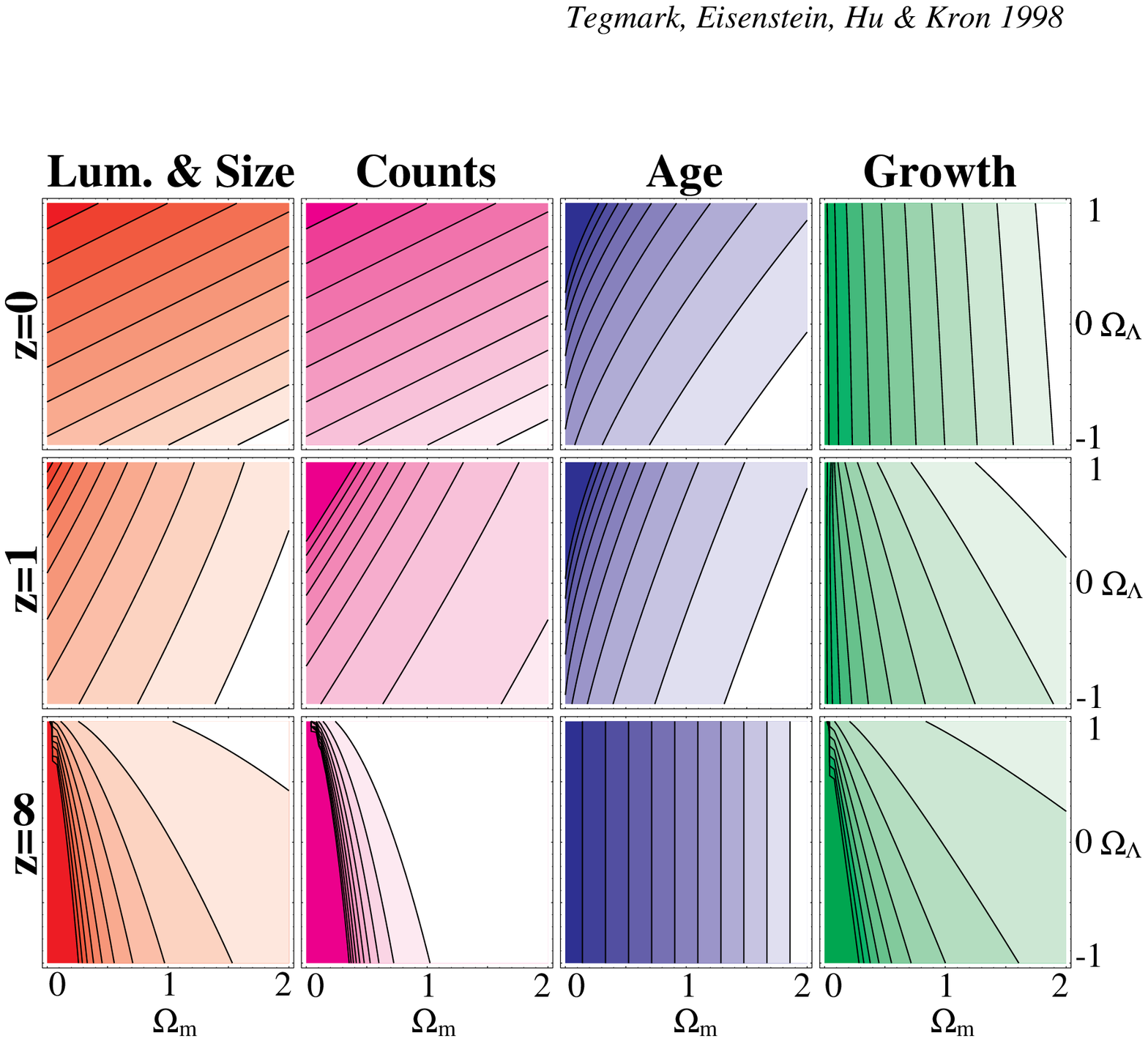,width=11cm,angle=0}}
\caption{\small Extra results similar to~(\ref{phase om vs ol z3}) for other
number-counts, age and growth as well. From Tegmark~\etal,~(1998b). }
\label{complimentarity2}
\end{figure}

What is also interesting about~figure~\ref{phase om vs ol z3} is something that
none of the papers above have mentioned; a quick look at the figure shows the
contours chang angle from $\sim45^\circ$ to $>90^\circ$. This means that there
will be some redshift ($z\sim3$) for which the lines will be vertical; ie,
$\Omega_0$ will be determined to very high accuracy from just one set of
experiments independent of $\Omega_\Lambda$. In fact it may be worth
specifically looking for eg, quasars or SNIa at this redshift to give a very
accurate determination of $\Omega_0$ (and hence $\Omega_{\Lambda_0}$ from other
redshift surveys).


\section{Observable Quantities in a General Spacetime}
\label{Observable Quantities in a General Spacetime}

In the FLRW models, there is a simple relation between the fundamental length
scale $a$ and the observable parameters $H_0, q_0$ etc. It is interesting to
determine the correct covariant definitions of $H_0$ and $q_0$ for a general
cosmological model and to ask whether the FLRW $\Omega_\Lambda$ generalises to
other cosmological models. Almost invariably these parameters are measured
fitting distance-redshift data to an FLRW model (essentially
eq.~(\ref{magnitude vs redshift FLRW})). What can this say about other
inhomogeneous models; and what could be causing a negative $q_0$ other than a
cosmological constant?

The standard generalisations of $H_0$ and $q_0$ are based on the
\emp{fundamental length scale} defined by~(\ref{average length scale-def})
simply by generalising the definitions for FLRW models,~(\ref{hubble const
def}) and~(\ref{q0 def}), viz;
\ba
H\li=\li\;{\dot\ell}{\ell}\equiv\;13\theta;\label{hubble scalar genera def}\\
q\li=\li-\;{\ddot\ell\ell}{\dot\ell^2}.\label{q def general}
\ea
Using the Raychaudhuri equation~(\ref{raychaudhuri}) we find simply that
\be
q=\;12\Omega\left(1+\;p\mu\right)-\Omega_{\Lambda}+\;1H\left(2(\sigma^2
-\omega^2)-\;13\left[\sdel_a\udot^a+\udot^2\right]\right).\label{q-from
raychaudhuri}
\ee
(See eq. (52) in Ellis,~1998. Note that his statement about the acceleration
terms being small from CMB anisotropies is false~-- see Chapter~\ref{EGS
chapter---------------------------------}.)

The definition of the expansion~$\theta$ is the volume change along $u^a$:
defining~$H$ via~(\ref{hubble scalar genera def}) is misleading. It gives the
\emph{average} distance change to neighbouring particles along $u^a$; whereas
the \emph{actual} distance change to a particle in direction $e^a$ is given by
\be
\;{(\delta\ell)^\cdot}{\delta\ell}=\:13\theta+\sigma_{ab}e^ae^b=H;
\ee
as one would expect, the shear will affect the distance to neighbouring
particles, as some may be moving towards or away from you relative to the
overall expansion~-- see Ehlers~(1993). Now this is a good definition of the
Hubble scalar, and has been used before. Often, the \emp{expansion tensor} is
defined as (eg, Wainwright and Ellis~1997\nocite{wain-ell97})
\be
\Theta_{ab}\eqdef\sigma_{ab}+\:13\theta h_{ab}
\ee
so we can now define
\be
H\eqdef \Theta_{ab}e^ae^b\label{hubble def proper}
\ee
to give a covariant, direction-dependent definition of the Hubble scalar. This
has been used in Humphreys \etal~(1997)\nocite{humph-97}.

If we want to compare a non-standard cosmological model with the real universe
(or with an FLRW model, with values of $H_0$ etc. derived from them) then we
need a generalisation of $H_0$ and $q_0$ suitable for a decent comparison with
data. It has been suggested that equating the various order terms in the
generalised $m-z$ series relation with those of the FLRW series would give the
proper generalised $H_0$ and $q_0$ (Humphreys \etal~1997; Paczy\'{n}ski and
Piran~1990\nocite{pac-pir90};
Dom\'inguez-Tenreiro~1981a,b\nocite{dom81a,dom81b}; also see
Palle~1999\nocite{pal99}). So, comparing~(\ref{m-z-general})
with~(\ref{magnitude-redshift FLRW series}) we can define
\ba
\left.H\obs\right|_{0}\li\eqdef\li\left.K^aK^b\del_au_b\right|_0,
\label{hubble def from series}\\
\left.q\obs\right|_0\li\eqdef\li\left.\;{K^aK^bK^c\del_a\del_bu_c}
{(K^dK^e\del_du_e)^2}\right|_0-3.\label{q0 from series}
\ea
Expanding $H\obs$ gives
\be
H\obs_0=\underbrace{\:13\theta}_{\mathsf{monopole}}+
\underbrace{\udot_ae^a}_{\mathsf{dipole}}+
\underbrace{\sigma_{ab}e^ae^b}_{\mathsf{quadrupole}};
\label{hubble from series}
\ee
where the quantities on the right hand side are understood to be evaluated at
the present time. This definition does not have the physical significance
of~(\ref{hubble def proper}); rather it is the zero point of the linear Hubble
relation. To compare a non-standard model to an FLRW model \emph{at low
redshift}, then~(\ref{hubble from series}) is the definition which must be
used. Eq,~(\ref{hubble from series}) shows that a general inhomogeneous
cosmological model has the effect of inducing a dipole and quadrupole in the
$m-z$ relation as $z\rightarrow0$. However, this conclusion only concerns
measurements made at very small redshift, say $z\lapp0.1$, because, as we have
seen in \S\ref{sec m-z-FLRW}, we cannot be sure that the low-order terms in the
series expansion will give a function even approximating the real thing for
larger redshift (say, $z\gapp0.1$). In other words, at higher redshifts (eg,
for SNIa) higher order terms may modify these moments to some degree.

\forget{??What can we say about arbitrary cosmologies from~(\ref{hubble from series})
on the basis of existing results? If we consider measurements of $H_0$ made
from say lensing, only two systems have been found to date and give a value
$H_0=72^{+11.85}_{-9.75}=82.8\pm10.8$ \hu\ (Tonry and
Franx,~1998\nocite{ton-fra98}). Similarly, constraints on $H_0$ from the HST
Key Project reveal $H_0=72\pm12$ \hu\ (eg, Madore \etal,~1998\nocite{mad-98}).
However, Hendry and Rauzy~(1999)\nocite{??}?? Willick??}

All of these low redshift surveys essentially use the \emp{linear Hubble Law}
which, for a general cosmological model reads
\be
z=\left.K^aK^b\del_au_b\right|_0r_A+{\cal{O}}(r_A^2)=H\obs_0r_A+{\cal{O}}(r_A^2).
\ee
If we write this in the more suggestive form
\be
z\simeq
H_0r\left(1+\;{\udot_ae^a}{H_0}+\;{\sigma_{ab}e^ae^b}{H_0}\right),\label{zlocal}
\ee
where $H_0$ is the expansion rate~(\ref{hubble scalar genera def}) as in the
FLRW models (basically the all sky average expansion rate) Equation
(\ref{zlocal}) reflect the effects of local small-scale inhomogeneities, ie,
perturbations of the smooth background model. If we and add in these local
effects, (this is often done in terms of spherical harmonics, eg, Reg\"os and
Szalay~1989;\nocite{reg-sza89} or Branchini \etal~(1999)\nocite{bra-99}; Tadros
\etal~1999\nocite{tad-99}~-- but we shall just use $\cos\vartheta$ expansion)
represented by constants $b_1, b_2, b_3\cdots$, we have
\be
z\simeq H_0r\left(1+\;{\udot_ae^a}{H_0}+\;{\sigma_{\<ab\>}e^ae^b} {H_0}\right)
+ b_1\cos\vartheta+b_2r\cos\vartheta+b_3r\cos^2\vartheta+\cdots,\label{hubble
law gen}
\ee
where $r_A$ has been replaced by $r$ because all definitions of $r$ agree at
very small redshift. $b_1$~corresponds to the \emp{bulk flow} of the local
group moving relative to the CMB frame, and $b_3$~often corresponds to a
shearing effect from our infall to the Great Attractor and Virgo cluster -- see
eg,~Lilje, Yahil and Jones~(1986)\nocite{lil-86} who first reported the effect.
However, there is no \emph{physical} local effect that will produce $b_2\neq0$.
It is certainly not clear at all that $b_2=0$ on \emph{observational} grounds;
this leaves \emph{acceleration as the only effect that will produce a dipole in
the Hubble law which grows linearly with distance.}

If we try to determine from equation~(\ref{hubble law gen}) from galaxy surveys
then we must be able to determine how much of the observed dipole is bulk flow
and how much may be attributed to any possible acceleration term; ie, whether
any of the dipole grows with distance. In principle we may be able to limit the
acceleration quite strongly in this way. However, preliminary results using the
IRAS catalogue only constrain
\be
\;{\lan\udot_ae^a\ran_{_\mathsf{half~sky}}}{H_0}<0.2.\label{acceleration constraint}
\ee
(Clarkson, Rauzy and Barrett, in preparation.) This is not strict at all, but
it is likely that a more complete survey such as the PSCz catalogue may provide
stronger constraints.

The generalised $q\obs_0$ is not nearly so simple; expanding~(\ref{q0 from
series}), and substituting for~$\dot\theta$, $\dot\sigma_{\<ab\>}$~and
$\sdel_b\sigma^{ab}$ from~(\ref{raychaudhuri}),~(\ref{shear evolution})
and~(\ref{div shear}) respectively we find
\be
\begin{array}{ll|c}
q\obs_0\li={H\obs_0}^{-2}\left\{\:16\mu+\:12p-\:13\Lambda
 -\:23\sdel_a\udot^a
 +\:65\sigma^2-\:23\omega^2\right.&\mathsf{monopole}\nonumber\\
& \left.+e^a\left[-\:23\theta \udot_a-\:{3}{5}\sdel_a\theta-\ddot
 u_a+\udot^b\sigma_{ab}+\:25q_a
 \right.\right.&\nonumber\\
 &\hfill\left.\left.
 +\:35\eta_a^{\pha bc}\left(\sdel_b\omega_c
 +2\udot_b\omega_c\right)\right]\right.&\mathsf{dipole}\nonumber\\
& \left.+e^ae^b\left[-2\sdel_{\<a}\udot_{b\>}+\udot_{\<a}\udot_{b\>}
 +E_{\<ab\>}-\:12\pi_{\<ab\>}+
\:97\sigma^{\vphantom{c}}_{c\<a}\sigma_{b\>}^{\phantom{b\>}c}
 \right.\right.&\nonumber\\
 &\hfill\left.\left.
 +\omega_{\<a}\omega_{b\>}-
\:23\sigma^{\vphantom{c}}_{c\<a}\omega_{b\>}^{\phantom{b\>}c}
 \right]\right.&\mathsf{quadrupole}\nonumber\\
&\hfill \left.+
e^ae^be^c\left[-5\udot_{\<a}\sigma_{bc\>}-\sdel_{\<a}\sigma_{bc\>}\right]
 -3e^ae^be^ce^d\sigma_{\<ab}\sigma_{cd\>}\right\}
 .&\mathsf{higher\pha multipoles}\label{q_0-observable}
\end{array}
\ee
This is somewhat complicated for an observable quantity. Measurements of $q_0$
are nowhere near accurate enough to detect any multipole moments; in fact it is
only recently that the \emph{sign} of the deceleration parameter has been
decided upon, let alone a variation in the sky. However, if we consider the
mean value of~(\ref{q_0-observable}) around the sky, then we
have\footnote{Obviously the multipoles of odd order are zero when averaged over
the sky, and the even order multipoles have an average value of zero because
they are contractions of traceless tensors.}
\be
\left\<{H\obs}^2 q\obs\right\>_{_\mathsf{all~sky}}=\:16\mu+\:12p-\:13\Lambda
 -\:23\sdel_a\udot^a
 +\:65\sigma^2-\:23\omega^2,
\ee
which, we can compare to the definition given by the length scale~(\ref{q-from
raychaudhuri}):
\be
\left\<H^2
q\right\>_{_\mathsf{all~sky}}
=\:16\mu+\:12p-\:13\Lambda-\:13\sdel_a\udot^a-\:13\udot_a\udot^a
+2(\sigma^2-\omega^2).
\ee
We find that the shear and rotation become less important in the
observationally derived relation, while the acceleration terms become
relatively \emph{more} important.
More interestingly though is the contribution from the
acceleration vector $\udot^2$, has \emph{disappeared}: it will actually
contribute nothing to the value of the deceleration.

}
\forgetmenot{
\chapter{The Ehlers-Geren-Sachs Theorem and Some Generalisations}
\label{EGS chapter---------------------------------}
\markright{{\large CHAPTER \ref{EGS chapter---------------------------------}.
SOME GENERALISED EGS THEOREMS}}

The high isotropy of the CMB is usually taken as strong evidence that the
universe is homogeneous and isotropic, ie,~is well described by an FLRW model.
The principle justification for this is an important theorem of Ehlers, Geren
and Sachs~(1968)\nocite{ehlers-68} (based on earlier work by Tauber and
Weinberg~1961\nocite{tau-wei61}), which states that if all observers in an
expanding, dust universe measure an isotropic CMB then the universe is FLRW and
the cosmological principle is valid. The importance of this theorem lies in the
fact that it permits the homogeneity and isotropy of the universe to be deduced
not from measurements of the actual isotropy of the universe about us, but from
only measurements of the CMB, combined with the Copernican principle (that is,
the assumption that all observers in the universe see the same degree of
isotropy). The Copernican principle is often regarded as a powerful but
untestable assumption in cosmology, although there are suggestions that it may
be testable using the Sunyaev-Zeldovich effect, for example
(Goodman~1995\nocite{goo95}). Here we simply assume that the Copernican
principle is valid and study the consequences of applying it to the observed
high degree of isotropy of the CMB. That is, we examine spacetimes with an
isotropic CMB for \emph{all} observers.

The EGS theorem has been generalised by Treciokas and
Ellis~(1971)\nocite{trec-ell71} to include an isotropic collision term.
Ferrando, Morales, and Portilla~(1992)\nocite{ferran-92} find the general form
of the energy-momentum tensor and Einstein's equations for spacetimes with an
isotropic radiation field, and consider some special cases with anisotropic
pressure. It has also been shown by Stoeger, Maartens and
Ellis~(1995)\nocite{sto-maa95} that the EGS theorem almost holds when applied
to an almost isotropic radiation field.

There are counterexamples to the spirit of the EGS theorem (that is, when some
of the assumptions are relaxed the result fails to hold). In particular, Ellis,
Maartens and Nel~(1978)\nocite{ellis-78} show that the result does not hold if
the expansion is zero (which is obviously not relevant to cosmology), and
Ferrando \etal~(1992) emphasise that homogeneity does not follow if there is
anisotropic pressure in the energy-momentum tensor. \forget{Most importantly
for the work presented here, though, is the result of Barrett and
Clarkson~(1999), which shows that when the assumption of geodesic observers is
relaxed there exist inhomogeneous perfect fluid (or scalar field) cosmologies
with an isotropic CMB. (In fact, we show in this paper that the Stephani models
considered in Barrett and Clarkson~(1999) are representatives of the only
perfect fluid spacetimes that admit an isotropic CMB for all fundamental
observers.)} Nilsson \etal~(1999)\nocite{nils-99} provide a counterexample to
the almost EGS result when the Weyl curvature is not negligible.

The basis of the EGS theorem is the Liouville equation for photons, which tells
us that if a radiation field (ie,~a solution of the Liouville equation) exists
such that for every observer on some timelike congruence the radiation field is
isotropic, then that congruence is (parallel to) a conformal Killing vector
(CKV). This may be expressed more formally as follows (Ehlers \etal~1968;
Ferrando \etal~1992):
\begin{thm}
\label{isoradthm}
A spacetime will admit an isotropic radiation field if and only if it is
conformal to a stationary spacetime, which happens if and only if there is a
velocity field~$u^a$ satisfying
\ba
\sigma_{ab}=0,\label{shear=0}\\
\del_{[a}\left(\udot_{b]}-\sfrac{1}{3}\theta u_{b]}\right) = 0,
\label{Qcondition}
\ea
where $\sigma_{ab}$, $\udot^a$ and~$\theta$ are the shear, acceleration and
expansion of~$u^a$, respectively.
\end{thm}
(Then~$u^a$ is the velocity field relative to which the radiation is isotropic,
and is parallel to the CKV.)

The proof of Theorem~\ref{isoradthm} is very straightforward in the 1+3
formalism, and follows from the results of \S\ref{CMB anisotropies section} and
\S\ref{conformal trans section} in Chapter~\ref{relativistic cosmol chapter-----------------}.

{\raggedleft\emph{Proof}}

With reference to \S\ref{CMB anisotropies section} a radiation field is
isotropic (with respect to $u^a$) iff
$q_{\!_R}^a\equiv\pi_{\!_R}^{ab}\equiv\Pi_{A_\ell}=0~\forall~\ell>2$, and
from~(\ref{monopole evolution-CMB})-(\ref{quadrupole evolution-CMB}) we find
that this condition implies
\ba
\dot\mu_{\!_R}+\;43\theta\mu_{\!_R}\li=\li0\nonumber\\
4\mu_{\!_R}\udot_a+\sdel_a\mu_{\!_R}\li=\li0\nonumber\\
\sigma_{ab}\li=\li0.\label{EGS-proof-1}
\ea
Now, if we write $Q\eqdef-\:14\ln\mu_{\!_R}$ in the first two conditions
of~(\ref{EGS-proof-1}) we find that
\be
\theta=3\dot Q,~~~\udot_a=\sdel_aQ:~~~\Longleftrightarrow~~~
\del_{[a}\left(\udot_{b]}-\sfrac{1}{3}\theta u_{b]}\right) = 0;
\ee
which, together with~(\ref{kinematic_trans}) and~$\sigma_{ab}=0$, tells us that
the spacetime is conformally
stationary~$\hat\theta=\hat{\udot}^a=\hat{\sigma}_{ab}=0$.{$\hfill\Box$}

In the absence of some statement about the matter content of a spacetime, or
further assumptions about the congruence~$u^a$, Theorem~\ref{isoradthm} is all
that can be said. In a cosmological context the simplest, and most common,
assumption is that the matter is dust (implying that~$u^a$ is geodesic), which
leads to:
\begin{thm}[Ehlers, Geren, and Sachs]
\label{EGSthm}
If the fundamental observers in a dust spacetime see an isotropic radiation
field, then the spacetime is locally FLRW.
\end{thm}

{\raggedleft\emph{Proof}:}

This is a straight forward extension of the proof of Theorem~\ref{isoradthm}:
we have for a dust congruence~$\udot^a\equiv0$ which implies $\sdel_aQ=0$. Now,
by~(\ref{sdel comute}) we find for $\theta\neq0$ that $\dot Q\omega_{ab}=0$:
hence the spacetime is a perfect fluid with geodesic, irrotational, shear-free
flow, and is therefore an FLRW model (see Krasi\'nski~1997).{$\hfill\Box$}

Alternatively, we can simply assume that~$u^a$ is geodesic. The existence of an
isotropic radiation field then ensures (for non-zero expansion) that the energy
flux relative to~$u^a$ is zero. If the anisotropic stress tensor is zero at any
instant (so that the energy-momentum tensor has perfect fluid form) then it
will remain zero and the spacetime will be FLRW (Ferrando \etal~1992, Corollary
1; but note that their statement that the anisotropic stress is invariant along
$u^a$ in general is misleading~-- from Eqs. (31) and (40) of Ellis~1998 we have
$\dot\pi_{\langle ab\rangle}\propto\theta\pi_{ab}$).

It is worth emphasising that in applications of the above results to cosmology
the motion of the fundamental observers must be identified with the
congruence~$u^a$. For example, in~\S\ref{CFsolns} \emph{all} Stephani models
are conformally flat, and therefore conformally stationary, but for most of
these spacetimes the fluid congruence is \emph{not} aligned with the timelike
CKV.

The matter content of the universe is not precisely known. Certainly, there is
a large number of possible contributors, including hot and cold dark matter (in
their various manifestations), electromagnetic fields~etc., as well as the more
obvious radiation and baryonic matter. In particular, the type~Ia supernova
results of Perlmutter~\etal\ (1999)\nocite{perl-99} suggest that an important
component may be a `quintessential' scalar field. However, the forms of matter
that are thought to contribute significantly to the energy-momentum tensor may
be treated in general as perfect fluids. That is, their energy-momentum tensor
may be written in the form
\be
                      T_{ab}=\mu u_au_b+ph_{ab},
\label{perfectfluid}
\ee
where $\mu$ and~$p$ are the energy density and pressure, $u^a$~is the timelike
velocity congruence of the fluid, and $h_{ab}$ is the spatial projection tensor
associated with~$u^a$. Scalar fields may also be written in this form,
with~$u^a$ parallel to the gradient of the scalar field, provided that this is
timelike (see~\S\ref{QCDM}). Note that if several such components are present
there is no reason why their fundamental congruences (the~$u^a$'s) should be
parallel. If they are, then they may be treated as effectively a single perfect
fluid with the energy densities and pressures added together. If not, the
decomposition of the energy-momentum with respect to~$u^a$ for one such fluid
(as in equation~(14) of Ellis~1998) will contain energy flux and anisotropic
stress terms from the other fluids (again, see~\S\ref{QCDM}). The fundamental
observers will be associated with one such congruence. Usually the fundamental
observers in standard models of the universe are associated with a dust-like
($p=0$) component, with the result that the acceleration~$\dot{u}^a$ of the
fundamental congruence is zero. However, we wish to study the consequences of
relaxing this assumption and consider models with acceleration. This
acceleration must be caused by some non-gravitational force (typically pressure
gradients for perfect fluid spacetimes, but in principle it could be the result
of a coupling between the fluid and some other component such as the
electromagnetic field).

With this in mind, in this paper we consider perhaps the two simplest
generalisations of the dust hypothesis. Firstly we imagine that the dominant
form of matter is a single irrotational perfect fluid. We do not specify what
form of matter this corresponds to, but we allow pressure gradients that give
rise to acceleration. Secondly we consider cosmological models in which more
than one matter component makes a significant contribution to the energy
density and dynamics of the universe. Specifically we consider `QCDM' models,
containing a non-interacting mixture of radiation, dust (CDM) and a scalar
field (or a cosmological constant). The observers are associated with the CDM
component, and are therefore geodesic. The difference between this and other
theorems assuming geodesic observers is that the the scalar field component can
introduce effective energy flux and anisotropic stresses relative to the dust
congruence, and so the matter need not behave as a perfect fluid.

In the following section we find all irrotational perfect fluid solutions
admitting an isotropic radiation field for the fundamental observers, showing
that they form a subclass of the Stephani spacetimes and are FLRW if and only
if the acceleration vanishes. Then in~\S\ref{QCDM} we examine QCDM models and
prove that such models must be homogeneous and isotropic if they admit an
isotropic radiation field. Finally, in~\S\ref{concs} we emphasise the
importance of acceleration for these results and show that the acceleration of
the fundamental congruence is, in principle, detectable, and measureable, in
galaxy surveys. Two appendices contain results relating to~\S\ref{CFsolns}.

\section{The Irrotational Perfect Fluid Solutions.}
\label{IPFSolns}

We wish to consider the constraints imposed by the existence of an isotropic
radiation field for the fundamental observers on perfect fluid spacetimes in
which the rotation of the fundamental congruence is zero. Since it follows from
Theorem~\ref{isoradthm} that the shear of the fundamental congruence must also
be zero we immediately know that all the acceptable solutions are members of
the Stephani-Barnes family, which is the family of \emph{all} shear-free,
irrotational, expanding (or contracting) perfect fluids (see
Krasi\'{n}ski~1989,~1997\nocite{kras97}\nocite{kras89}). It only remains, then,
to impose condition~(\ref{Qcondition}) of Theorem~\ref{isoradthm} and thus find
the sub-class of Stephani-Barnes models which admit an isotropic radiation
field for all fundamental observers.

The Stephani-Barnes family contains the Barnes solutions, which are of Petrov
type~D, and the Stephani models, which are conformally flat, although these two
classes overlap where the Barnes solutions degenerate to type O (these
solutions then become Stephani models with symmetry). The FLRW models are a
subcase of these solutions. The Barnes spacetimes all possess symmetry (see
below), whereas the Stephani spacetimes, in general, do not. In all cases the
metric in comoving coordinates can be written in the form (we use the same
notation as Krasi\'{n}ski 1997):
\be
ds^2=V^{-2}\left\{-(FV_{,t})^2dt^2+dx^2+dy^2+dz^2\right\}\label{metric}
\ee
where $F=F(t)$ and $V=V(t,x,y,z)$. $F(t)$ is arbitrary, but there are some
restrictions on the form of $V$ depending on the symmetries of the solution,
and these will be discussed in due course (but the impatient reader may wish to
note Eqs.~\ref{SS-plane-V},~\ref{hyperbolic-V},~\ref{hyp-V-de},
and~\ref{V_stephani}). Expressions for the energy density and pressure can be
found in Krasi\'{n}ski (1997).

The fluid velocity is given by (without loss of generality we can assume
that~$V>0$)
\be
          u^a=\frac{V}{|FV_{,t}|}\delta^a_{\phantom{a}0},
\label{velocity}
\ee
with expansion
\be
\theta=-\hbox{sign}(V_{,t})\frac{3}{|F|},
\label{expansion}
\ee
and acceleration
\be
          \udot_0=0, \hskip 1.0cm
          \udot_i=\frac{\partial}{\partial x^i}\ln\frac{V_{,t}}{V},
\label{acceln}
\ee
where $i=1,2,3$, $x^i=\{x,y,z\}$. Note that~(\ref{expansion}) differs from the
expression usually given for the expansion (in Krasi\'{n}ski 1997, equations
(4.1.4) and~(4.9.6), for example) by the inclusion of the
$-\hbox{sign}(V_{,t})$ factor. Neglect of this factor is inconsistent since~$F$
enters the metric only quadratically, so the sign of the expansion cannot
depend on the sign of~$F$. The sign of~$\theta$ \emph{does} depend on that
of~$V_{,t}$, though: for the Friedmann subcase of the Stephani-Barnes models,
for example, $V$ is related to the scale factor~$R(t)$ by~$V=1/R$,
and~$|FV_{,t}/V|=1$ (see~\S\ref{CFsolns}), so
that~$\theta=3\dot{R}/R=-3V_{,t}/V=-3\,\hbox{sign}(V_{,t})/|F|$. This is
important here because the constraint~(\ref{Qcondition}) contains the
expansion. Note, too, that~$F$ is not a true degree of freedom parameterising
distinct spacetimes, but rather represents a coordinate freedom, corresponding
to different choices of the time coordinate.

From (\ref{velocity}), (\ref{expansion}) and~(\ref{acceln}), the
condition~(\ref{Qcondition}) leads to the constraint:
\be
\frac{\partial^2}{\partial x^i\partial t}\ln V_{,t}=0,
\label{deVcond}
\ee
which is satisfied if and only if the function~$V$ has the form
\be
V(t,x,y,z)=T(t)X(x,y,z)+Y(x,y,z),\label{Vcondition}
\ee
where $T$, $X$, and~$Y$ are arbitrary functions. This equation is the key
additional constraint on the Stephani-Barnes solutions.

It is worth noting that it follows from~(\ref{deVcond}) that the acceleration
scalar is constant along the fluid flow for every observer, ie,~$\udot_{,t}=0$
(where $\udot^2=\udot_a\udot^a$), as can be seen by
calculating~$(\udot^2)_{,t}$ from~(\ref{acceln}). In fact, it can be verified
more generally that for any conformally stationary spacetime (ie,~a spacetime
satisfying (\ref{shear=0}) and~(\ref{Qcondition})), even with rotation, the
acceleration scalar evolves according to
\[
            u^a\del_a \udot^2 = \sfrac23 \udot^b\sdel_b\theta,
\]
where $\sdel$ denotes the spatially projected gradient (see Ellis~1998). It
follows from equation~(32) of Ellis~(1998) that~$\sdel_b\theta=0$ whenever the
rotation vanishes (for a perfect fluid), which is the case for the
Stephani-Barnes models.

Note that if~$V$ has the form~(\ref{Vcondition}) we can immediately write the
metric in a manifestly conformally static form
\[
ds^2=\frac{X^2}{V^2}\left\{-d\tau^2+X^{-2}(dx^2+dy^2+dz^2)\right\},
\]
where $d\tau=T_{,t}F \,dt$, which shows that these models will indeed be
conformally stationary, as required by Theorem~\ref{isoradthm}. We now discuss
each subcase in turn.

\subsection{The Barnes Solutions.}

The Barnes solutions all have spherical, plane or hyperbolic symmetry (ie,~they
possess three-dimensional isometry groups acting on two-dimensional
orbits,~\emph{cf}.~\S\ref{CFsolns}). The restrictions on the metric
function~$V$ depend on which of these symmetries the spacetime possesses (see
Krasi\'nski~1997). For the solutions with spherical symmetry (the
Kustaanheimo-Qvist solutions) or planar symmetry we introduce a new independent
variable~$u(x,y,z)$ defined by $u=r^2=x^2+y^2+z^2$ in the spherical case
and~$u=z$ in the planar case. Then~$V$ is defined by~$V=V(t,u)$, subject to the
condition:
\be
\frac{\partial^2 V}{\partial u^2}=f(u)V^2.\label{SS-plane-V}
\ee
where~$f$ is an arbitrary function. Since~$V=V(t,u)$ we know
from~(\ref{Vcondition}) that in order to admit an isotropic radiation field for
all observers~$V$ must have the form
\be
V(t,u)=T(t)X(u)+Y(u).\label{Vcondition2}
\ee
Equation~(\ref{SS-plane-V}) then imposes a constraint on the functions $X$
and~$Y$, which will be outlined below.

For the solutions with hyperbolic symmetry the constraint on~$V$ is very
similar. This time we introduce the variable~$u=x/y$. $V$~can then be written
\be
V(t,x,y)=yW(t,u),
\label{hyperbolic-V}
\ee
where~$W$ satisfies
\be
\frac{\partial^2 W}{\partial u^2}=f(u)W^2,
\label{hyp-V-de}
\ee
with~$f$ once again a free function. The condition~(\ref{Vcondition}) now gives
\[
V(t,x,y)=T(t)X(x,y)+Y(x,y)=yW(t,u).
\]
Dividing by~$y$ and redefining $X$ and~$Y$ in an obvious fashion we obtain
\be
W(t,u)=T(t)\tilde{X}(u)+\tilde{Y}(u),\label{Wcondition}
\ee
in which $\tilde{X}$ and~$\tilde{Y}$ are again constrained by~(\ref{hyp-V-de}).

For all three symmetries of the Barnes solutions, then, the constraints on the
metric function~$V$ essentially reduce to the second-order differential
equations~(\ref{SS-plane-V}) or~(\ref{hyp-V-de}). Imposing the
condition~(\ref{Vcondition}) introduces the additional constraint
(\ref{Vcondition2}) or~(\ref{Wcondition}). Substitution of (\ref{Vcondition2})
or~(\ref{Wcondition}) into (\ref{SS-plane-V}) or~(\ref{hyp-V-de}) respectively
and differentiating twice with respect to time, dividing by~$T_{,t}$ each time
(and recognising that $T_{,t}\ne 0$, $X\ne 0$, so that $V_{,t}\ne 0$
in~(\ref{metric})), leads directly to the condition
\[
                            f(u)=0.
\]
Barnes solutions with~$f(u)=0$ are conformally flat (Krasi\'{n}ski~1997,~p.142)
and are therefore actually a subcase of the Stephani models. That is, proper
Barnes spacetimes can be ruled out: they do not admit an isotropic radiation
field. It only remains to apply the condition~(\ref{Vcondition}) to the
Stephani models, which we do in the next section.

\subsection{The Conformally Flat Solutions}
\label{CFsolns}

The conformally flat sub-case is the entire class of conformally flat,
expanding, perfect fluid solutions, and is the Stephani solution (Stephani
1967a,b).\nocite{step67a,step67b} The function~$V$ is most often written in the
form (see Krasi\'{n}ski~1997; Barrett and Clarkson~1999a,b):
\be
                                  V(t,x,y,z) =
   \frac{1}{R(t)}\left(1+\sfrac14 k(t)|{\mathbf{x}}-{\mathbf{x}}_0(t)|^2\right)
\label{V_stephani}
\ee
In general the five functions of time in~$V$ are free. For our purposes,
however, it turns out to be more convenient to use
\be
V=a(t)+b(t)r^2-2{\mathbf{c}}(t)\cdot{\mathbf{r}},\label{Vstephani2}
\ee
as in Barnes~(1998)\nocite{barn98}, again with five free functions (we adopt
three-dimensional vector notation, so that~${\mathbf{c}}=(c_1,c_2,c_3)$, for
example). In fact, this form is slightly more general
than~(\ref{V_stephani})~-- see Barnes~(1998).

We must be able to write~$V$ in the form~(\ref{Vcondition}) for the spacetime
to admit an isotropic radiation field for all fundamental observers.
From~(\ref{V_stephani}) or~(\ref{Vstephani2}) it is clear that the functions
$X(x^i)$, and~$Y(x^i)$ in~(\ref{Vcondition}) can be at most quadratic in
the~$x^i$. Writing $X$, and~$Y$ as quadratics in~(\ref{Vcondition}) and
equating in~(\ref{Vstephani2}) all powers of~$x^i$, we obtain the following
constraint equations:
\begin{eqnarray}
                a(t) \li = \li a_1 T(t) + a_2, \nonumber \\
                b(t) \li = \li b_1 T(t) + b_2, \label{Steph_constr}\\
     {\mathbf{c}}(t) \li = \li {\mathbf{c}}_1 T(t) + {\mathbf{c}}_2, \nonumber
\end{eqnarray}
where~$T(t)$ is a free function of time and the $a_{1,2}$, $b_{1,2}$
and~${\mathbf{c}}_{1,2}$ are ten independent constants. Not all of~$a_1$, $b_1$
and~${\mathbf{c}}_1$ can be zero (in order that~$V,_t\ne 0$ in~(\ref{metric})).

Equations~(\ref{Steph_constr}), along with (\ref{metric})
and~(\ref{Vstephani2}), provide the complete set of irrotational perfect fluid
spacetimes admitting an isotropic radiation field. Not all of the possible
choices of parameters give rise to distinct spacetimes, though, and we outline
in appendix~\ref{app-trans} how coordinate transformations may be used to
eliminate many of the parameters in~(\ref{Steph_constr}), and determine when
the models can be reduced to manifestly spherically symmetric
form~(${\mathbf{c}}={\mathbf{0}}$).

\subsubsection{The Constraints on $R$, $k$ and~${\mathbf{x}}_0$.}
\label{app-Rkx}


To find the constraints on $R$, $k$ and~${\mathbf{x}}_0$ in~(\ref{V_stephani})
corresponding to~(\ref{Steph_constr}) first equate powers of~$x^i$ in
(\ref{V_stephani}) and~(\ref{Vstephani2}) to obtain
\begin{eqnarray}
   a(t)\li=\li \frac{1}{R}+\frac{k}{4R}|{\mathbf{x}}_0|^2, \label{A}\\
   b(t)\li=\li \frac{k}{4R}, \label{B}\\
   {\mathbf{c}}(t)\li=\li \frac{k}{4R}{\mathbf{x}}_0. \label{C}
\end{eqnarray}
Solving these equations for $R$, $k$ and~${\mathbf{x}}_0$ gives
\begin{eqnarray}
   R(t) \li=\li \frac{b}{ab-|{\mathbf{c}}|^2}, \label{R}\\
   \frac14 k(t) \li=\li \frac{b^2}{ab-|{\mathbf{c}}|^2}, \label{k}\\
   {\mathbf{x}}_0 \li=\li \frac{\mathbf{c}}{b}, \label{x}
\end{eqnarray}
which are valid whenever ${ab-|{\mathbf{c}}|^2}\ne 0$ (otherwise $V$ cannot be
written in the form~(\ref{V_stephani})).

To impose the constraints~(\ref{Steph_constr}) perform the transformations of
appendix~\ref{app-trans} so that ${\mathbf{c}}=c{\mathbf{\hat{z}}}$ as
in~\ref{constr-app}) and~$a_1\ne 0$. Then $|{\mathbf{c}}|^2 = c^2$, and $b$
and~$a$ are related by
\be
          b = \frac{b_1}{a_1}a + \left(b_2-\frac{b_1a_2}{a_1}\right)
            \equiv \gamma a+\delta,
\label{bofa}
\ee
where $\gamma=b_1/a_1$ and~$\delta=b_2-b_1a_2/a_1$ are constants. Using this in
(\ref{R}) and~(\ref{k}) leads, after some rearrangement, to a quadratic
relationship between $k$ and~$R$:
\be
    \left(\frac{k}{4}\right)^2 - (\gamma + \delta R)\left(\frac{k}{4}\right)
              - \gamma c^2 R^2 = 0.
\label{kReqn}
\ee
In addition to this constraint relating $k$ and~$R$ we can trivially rewrite
(\ref{C}) or~(\ref{x}) as
\be
          {\mathbf{x}}_0 = \frac{4R}{k} c{\mathbf{\hat{z}}}.
\label{ckR}
\ee

Equations (\ref{kReqn}) and~(\ref{ckR}) are the constraint equations on $R$,
$k$ and~${\mathbf{x}}_0$ corresponding to the equations~(\ref{Steph_constr}),
or rather~(\ref{constr-app}). If desired (\ref{kReqn}) can be solved to obtain
\[
         \frac{k}{4} = \frac12 \left[\gamma+\delta R
                       \pm \sqrt{(\gamma+\delta R)^2 + 4\gamma c^2R^2}\right].
\]

From~(\ref{kReqn}) it is clear that the spherically symmetric Stephani
spacetimes admitting an isotropic radiation field, for which~$c=0$, satisfy
\[
   \frac{k}{4}\left(\frac{k}{4} - (\gamma + \delta R)\right) = 0,
\]
which has the solutions $k=0$ and $R(t)$~free (flat Friedmann), or $k$ linearly
related to~$R$,
\[
         \frac14 k(t)=\gamma +\delta R(t),
\]
with $R(t)$ again free (when~$\delta=0$ these become Friedmann models with
curvature~$k=4\gamma$). This fixes notation for the rest of this thesis.

At this point we can say that perfect fluid spacetimes admitting an isotropic
radiation field for all fundamental observers are FLRW if and only if the
acceleration of the fundamental observers is zero. This follows because the
Stephani models with zero acceleration are FLRW (see Krasi\'{n}ski~1997,
although it can easily be seen from~(\ref{acceln}): if $\dot{u}=0$ then
$V=T(t)X(x^i)$ for some functions~$T$ and~$X$, showing that $a$, $b$
and~${\mathbf{c}}$ depend on the single free function~$T$ and so must be FLRW
by the results of the next section). Thus we have proved:
\begin{thm}
\label{IPFthm}
The irrotational perfect fluid spacetimes admitting an isotropic radiation
field for the fundamental observers are Stephani models with the free functions
restricted by~(\ref{Steph_constr}) (or~(\ref{constr-app})). These spacetimes
are FLRW if and only if the acceleration of the fundamental observers is zero.
\end{thm}

Spacetimes satisfying Theorem~\ref{IPFthm} will be referred to as IRF models.
It is worth noting that all of the models admitting an isotropic radiation
field are manifestly conformal to (part of) the Einstein static spacetime
(\emph{cf}.~\S\ref{horizon}) once they have been transformed so
that~${\mathbf{c}}=c{\mathbf{\hat{z}}}$ as outlined in
appendix~\ref{app-trans}. From (\ref{Vstephani2}) and~(\ref{constr-app}) we
obtain
\[
        V_{,t} = a_1T_{,t}\left( 1 + \frac{b_1}{a_1} r^2 \right),
\]
since~${\mathbf{c}}_{,t}={\mathbf{0}}$ when~${\mathbf{c}}=c{\mathbf{\hat{z}}}$
(as noted in appendix~\ref{app-trans} we can assume~$a_1\ne 0$). Changing the
time coordinate via $dt\mapsto a_1 T_{,t}F\,dt$, the metric~(\ref{metric})
becomes
\be
    ds^2  = \frac{(1+\sfrac14\Delta r^2)^2}{V(z,r,t)^2}
    \left\{ -dt^2 + \frac{1}{(1+\sfrac14\Delta r^2)^2}(dx^2+dy^2+dz^2)\right\},
    \label{step metric-IRF}
\ee
where~$\Delta=4b_1/a_1$. The factor in braces is the Einstein static metric. In
the following chapter we use this conformal relationship to simplify the study
of the observational characteristics of these models.

\subsection{Symmetry and Thermodynamic Schemes.}
\label{Symmetry and Thermodynamic Schemes}

Having obtained the conditions~(\ref{Steph_constr}) for a Stephani model to
admit an isotropic radiation field it is possible to say immediately that all
such spacetimes possess (at least) a three-dimensional symmetry group acting on
two-dimensional orbits (just as for the general Barnes models). This follows
from the work of Barnes~(1998), who showed that the dimension of the isometry
group of any Stephani spacetime is determined by the dimension~$d$ of the
linear space spanned by the five free functions $a$, $b$, and~${\mathbf{c}}$:
\begin{enumerate}
   \item if $d=4$ or~$5$ (ie,~at least four of the free functions are linearly
independent), then the spacetime has no Killing vectors;
   \item if $d=3$ there is a one-dimensional isometry group;
   \item if $d=2$ there is a three-dimensional isometry group acting on
two-dimensional orbits;
   \item if $d=1$ there are six Killing vectors and the spacetime is
Robertson-Walker.
\end{enumerate}
It is clear from~(\ref{Steph_constr}) that $a$, $b$, and~$\mathbf{c}$ depend on
(at most) only two functions of time: $f_1(t)=T(t)$ and the constant
function~$f_2(t)\equiv 1$ (since~$V_{,t}\ne 0$ we must have~$T_{,t}\ne 0$, so
that these are necessarily linearly independent). Thus,~$d=2$ and the solutions
have three-dimensional isometry groups as claimed. If
$a_2=b_2={\mathbf{c}}_2=0$ in~(\ref{Steph_constr}) then~$d=1$ and the spacetime
is~FLRW (see appendix~\ref{app-trans}).

It follows further from this and the work of Bona and Coll~(1988) (see also
Krasi\'nski, Quevedo and Sussman~1997\nocite{kras-97}) that all Stephani models
that admit an isotropic radiation field for all fundamental observers also
admit a strict thermodynamic scheme (that is, entropy and temperature functions
can be found that depend on the energy density and pressure and satisfy the
second law of thermodynamics). The converse is not true, however, since there
are Stephani spacetimes with~$d=2$ (which must admit a thermodynamic scheme)
that cannot have an isotropic radiation field (these are models for which the
second independent function~$f_2(t)$ is not restricted to be~$1$). So, we have
the following corollary to Theorem~\ref{IPFthm}:
\begin{cor}
An irrotational perfect fluid spacetime that admits an isotropic radiation
field has spherical, planar or hyperbolic symmetry and admits a strict
thermodynamic scheme.
\end{cor}

On the subject of thermodynamics, let us mention for completeness that the
thermodynamic scheme occurring most often in the literature is that of a
barotropic equation of state. It is known that the only Stephani models with a
barotropic EOS are precisely the FLRW models (Bona and Coll~1988; Krasi\'{n}ski
1997). Thus, the only spacetimes with a barotropic EOS admitting an isotropic
radiation field are FLRW models. This also follows (when~$\mu+p\ne 0$) from a
theorem of Coley~(1991)\nocite{coley91}. See also Collins and
Wainwright~(1983)\nocite{col-wai83}.

\section{QCDM Models.}
\label{QCDM}

The type~Ia supernova results of Riess \etal~(1998)\nocite{riess-98}, Schmidt
\etal~(1998)\nocite{schmidt-98} and Perlmutter \emph{et al.}~(1999), which
suggest that the expansion of the universe is accelerating, have led to an
increased interest in cosmological models in which a significant contribution
to the energy density comes from either a cosmological constant or a scalar
field (quintessence component), which is capable of driving the expansion
(Peebles and Ratra~1988\nocite{peebl-88}; Ratra and
Peebles~1988\nocite{ratra-88}; Caldwell \etal~1998\nocite{cald-98}; Zlatev
\etal~1999\nocite{zlat-98}). In QCDM models the matter is an admixture of
non-interacting cold dark matter (CDM), ie,~dust, and a scalar field. The
quintessence component may be thought of as fairies pushing the galaxies
apart\footnote{G. F. R. Ellis, private communication}.  The fundamental
observers (galaxies) are implicitly identified with the geodesic congruence of
the CDM. Note that there is no reason \emph{a priori} why the scalar field
gradient (which defines a natural `velocity' field) should be aligned with the
CDM congruence (although it will turn out that they are aligned if the
fundamental observers see an isotropic radiation field).

It is interesting to ask whether the EGS theorem can be extended to this case.
We demonstrate that it can by proving the following theorem:
\begin{thm}
\label{QCDMthm}
Any solution to Einstein's equations in which the matter consists of
non-interacting radiation, expanding dust (CDM), and a scalar field (or
cosmological constant), and for which the dust sees an isotropic radiation
field, must either be an FLRW model, or have the gradient of the scalar field
orthogonal to the dust congruence.
\end{thm}
(Note that the latter possibility means that gradient of the scalar field is
spacelike, and is usually rejected as unphysical~-- although see below.)

{\raggedleft \emph{Proof:}}

We may divide this proof into two parts: first we demonstrate from Einstein's
equations in the 1+3 formalism that any energy flux component with respect to
the CDM frame must be zero if the CDM observers see isotropic radiation, then
we show that the contribution to the energy flux (with respect to the CDM
frame) from the scalar field is zero if and only if the gradient of the scalar
field is parallel (or orthogonal) to the CDM velocity~$u^a_{_{CDM}}$, so we
deduce that the velocity fields are parallel (or orthogonal). The case where
the field gradient is orthogonal to~$u^a_{_{CDM}}$ is probably unphysical, and
will be rejected. Thus, the mixture of radiation, CDM and scalar field can be
written as a single perfect fluid with geodesic fundamental
congruence~$u^a=u^a_{_{CDM}}$, and it follows from the results
of~\S\ref{IPFSolns} that the model is necessarily FLRW. \nocite{ell98}

Since the radiation is isotropic for the dust observers~$u^a$ the
energy-momentum tensor for radiation may be written in the perfect fluid
form~(\ref{perfectfluid}) with~$p=\sfrac13\mu$ and the total energy-momentum
tensor is:
\be
T_{ab}=\underbrace{\mu u_au_b+\sfrac{1}{3}\mu
h_{ab}}_{\textrm{Radiation}}+\underbrace{\vphantom{\sfrac13}\rho
u_au_b}_{\textrm{CDM}} +\underbrace{\phi_{,a}\phi_{,b}
-g_{ab}\left(\sfrac{1}{2}\phi_{,c}\phi^{,c}+\Phi(\phi)\right)}_{\textrm{Scalar
Field}},
\label{T-QCDM}
\ee
where $\Phi(\phi)$ is the scalar field potential (often assumed to be zero, in
which case the scalar field can be interpreted as a stiff perfect fluid). Note
that the cosmological constant case can be included by setting
$\phi=\Lambda=$constant,~$\Phi(\phi)=\phi$.

\begin{enumerate}

\item
The fundamental congruence  $u^a$ is geodesic ($\udot^a=0$) because the CDM
component does not interact with the other matter. This, in fact, implies that
the rotation of~$u^a$ must also vanish: from the momentum conservation equation
for the radiation, we can write
\[
\udot_a=-\frac14\sdel_a\ln\mu=0
\]
so that (using (\ref{sdel comute}))
\[
0=\sdel_{[a}\udot_{b]}=\frac14\sdel_{[b}\sdel_{a]}\ln\mu=
\frac14\omega_{ab}\frac{\dot\mu}{\mu}=\frac13\omega_{ba}\theta,
\]
and we see that $\omega_{ab}=0$ when~$\theta\ne 0$.

When $\dot{u}_a$ and~$\omega_{ab}$ are zero, (\ref{Qcondition}) becomes
\[
\del_{[a}(\theta u_{b]})= u_{[b}\del_{a]}\theta = 0.
\]
This implies (since $\del_a\theta=\sdel_a\theta-\dot{\theta}u_a$) that
\be
\sdel_a\theta=0.
\label{gradexp}
\ee
(ie,~the expansion is homogeneous). From the constraint equation relating the
divergence of the shear to other kinematical quantities (\ref{div shear}) we
see that any energy flux component with respect to the CDM velocity field must
vanish:
\be
q_a=\frac23\sdel_a\theta=0.
\label{qa}
\ee
This is the key step in the proof.

\item
Decomposing ~(\ref{T-QCDM}) with respect to $u^a$ we find that the relative
energy flux component is
\be
0=q_a=-h_a^{~b}u^cT_{bc}=-\dot\phi\sdel_a\phi.
\ee
So $q_a=0$ if $\dot{\phi}=0$ (the scalar field gradient is orthogonal to~$u^a$,
and therefore spacelike), or if~$\sdel_a\phi=0$ (the scalar field gradient is
parallel to~$u^a$). We take the latter case to be most important since the
gradient of a scalar field is usually assumed to be timelike.

Since $\del_a\phi=\sdel_a\phi-u_a\dot\phi=-u_a\dot\phi$ it is possible to
write~(\ref{T-QCDM}) as a perfect fluid with geodesic, shear-free,
rotation-free velocity field; it is thus an FLRW model by~\S\ref{IPFSolns}.
{\hfill $\Box$}

\end{enumerate}

It is easy to see from the above proof that the fact that the fundamental
observers correspond to dust-like matter was not used, only that they followed
geodesics. Hence the above result applies for more general perfect fluids in
place of the CDM component, as long as the fundamental congruence is geodesic.

The idea of a spacelike scalar field gradient seems physically unappealing.
However, such a field can (depending on the potential~$\Phi$) satisfy the weak,
strong and dominant energy conditions. The strong energy condition will be
satisfied if and only if~$\Phi(\phi)\le 0$ everywhere, whereas as the weak and
dominant energy conditions will be satisfied if~$\Phi(\phi)\ge 0$, although not
only so. Thus, a massless scalar field (stiff perfect fluid) with spacelike
gradient satisfies all energy conditions. It should be borne in mind, though,
that scalar fields arising in cosmological contexts often fail to satisfy the
energy conditions. This case may deserve further consideration. As can easily
be seen, the scalar field component gives rise to anisotropic stresses in the
energy-momentum tensor, so such spacetimes are not FLRW. Note that this theorem
also applies to any number of non-interacting perfect fluids~-- that is, a
spacetime consisting of dust seeing isotropic radiation and some perfect fluids
will be FLRW.

\section{Conclusions.}
\label{concs}

We have proved that the irrotational perfect fluid spacetimes admitting an
isotropic radiation field are Stephani models restricted
by~(\ref{Steph_constr}) (see also equations~(\ref{constr-app})), and are FLRW
if and only if the acceleration~$\dot{u}^a$ of the fundamental congruence is
zero (Theorem~\ref{IPFthm}). It follows from the fact that the
constraints~(\ref{Steph_constr}) depend on only two independent functions of
time that all of the acceptable models possess three-dimensional symmetry
groups acting on two-dimensional orbits (ie,~have spherical, planar, or
hyperbolic symmetry) and therefore possess a thermodynamic interpretation. We
have also shown that spacetimes containing a mixture of radiation, dust and
scalar field (QCDM models) for which the dust observers see the radiation as
isotropic must always be homogeneous and isotropic (Theorem~\ref{QCDMthm})
unless the scalar field gradient is spacelike and orthogonal to the CDM
congruence~-- a possibility we reject as unphysical. This result also relies on
the geodesic nature of the fundamental congruence.

Crucial, therefore, to the proof of homogeneity and the verification of the
cosmological principle is the non-acceleration of the fundamental observers.
Despite the intuitive appeal of cosmological models in which the fundamental
observers are associated with a dust-like matter component, it is unacceptable
to simply assume that we are geodesic observers, especially when such an
assumption is, in principle, testable. Acceleration leaves a characteristic
dipole signature in the redshifts of nearby galaxies that may be detectable
using galaxy surveys. The physical principle underlying this effect is easy to
see. For a set of uniformly accelerated observers (`galaxies') in Minkowski
space each observer will see other galaxies redshifted or blueshifted in a
dipole pattern, with the blueshifted galaxies lying in the direction of the
acceleration, because during the light-travel time between galaxy and observer
the observer's velocity has increased relative the velocity at emission, so
that the galaxies the observer is travelling towards are blueshifted, and those
it is travelling away from redshifted. It also follows from this that the
magnitude of the dipole increases with distance, simply because the
light-travel time from more distant galaxies is larger. In a cosmological
context this acceleration dipole must be added to other terms contributing to
the redshift of nearby objects, in particular the expansion. The method of
Kristian and Sachs (1966)\nocite{kris-sac66} and MacCallum and Ellis
(1970)\nocite{Mac-Ellis-II} gives, for any cosmological model with fundamental
congruence~$u^a$, the lowest-order term in the redshift~$z$ as a function of
distance~$r$ and direction~$e^a$ (a spacelike unit vector orthogonal to~$u^a$,
denoting the direction of observation):
\be
z=
\left. H_0r\left(1-\frac{\udot_ae^a}{H_0}+\frac{\sigma_{ab}e^ae^b}{H_0}\right)
\right|_0+{\cal{O}}(r^2),
\label{H0}
\ee
where $H_0=\sfrac13\theta_0$ is Hubble's constant and the last term in brackets
indicates the quadrupole introduced by the presence of shear. In~(\ref{H0}) $r$
can be any cosmological distance measure (area distance, for example) because
for small~$r$ all such measures agree to first order. Note that just as the
monopole (expansion) term increases linearly with distance according to the
Hubble law, so does the acceleration dipole. This is important, because it
allows the acceleration dipole to be distinguished from any dipole resulting
from the peculiar velocity of our galaxy with respect to the cosmological
average rest frame (usually identified with the CMB frame). Equation~(\ref{H0})
applies in this rest frame, and any peculiar motion results in a doppler shift
for each galaxy, which introduces an additional dipole component into the
galaxy redshifts. This dipole is just a constant depending only on the peculiar
velocity of our galaxy. A boost to the `correct' rest frame can eliminate this
constant component, but cannot remove the acceleration dipole because it is
distance dependent. It is important to note in this context that the
acceleration referred to here is not the same as the `acceleration dipole'
resulting from the gravitational attraction by the Great Attractor overdensity,
which is often calculated using galaxy surveys (see Schmoldt
\etal~1999\nocite{schmoldt-99}).

Galaxy surveys are often used to measure a possible bulk flow of the local
universe, that is, the difference, if any, between the rest frame of the local
universe and the CMB frame, which in standard cosmological models should be the
same (see Willick 1998\nocite{willick}). A simple extension of these techniques
(Clarkson, Rauzy and Barrett, in preparation) permits the acceleration to be
constrained by observations. However, preliminary results suggest that the
constraints on~$\dot{u}^a$ are quite weak: it appears not to be possible to
conclude definitively that we are geodesic observers (\ref{acceleration
constraint}). The accuracy of~$\dot{u}^a$ determinations is limited both by
uncertainties in the distance estimates to galaxies as well as the peculiar
velocities of galaxies.

Even if the acceleration was measured to be zero, it is still necessary to show
that there are no anisotropic stresses (Ferrando \etal~1992) before the
cosmological principle can be verified. It follows from equation~(31) of
Ellis~(1998) that this is equivalent to determining that the electric part of
the Weyl tensor is zero. It is not clear how this may be achieved using
observations.

Of course, the Copernican principle (which is a vital element of EGS-type
theorems, allowing the high isotropy of the CMB here to be assumed for other
points in the universe) remains a purely philosophical assumption. It has been
suggested by a number of authors that the Sunyaev-Zeldovich effect might be
used to place constraints on the anisotropy of the CMB at distant positions,
but it seems unlikely that such observations will provide a definitive
verification of the Copernican principle. Nevertheless, the arguments in favour
of the Copernican principle are quite powerful, and it is a much weaker
assumption than the cosmological principle. Note that if the acceleration
\emph{is} measured to be zero here, the Copernican principle must also be
applied to give geodesic observers everywhere for the results of this chapter
to hold.

Finally, one might expect that the `almost' version of Theorem~\ref{IPFthm}
would lead to spacetimes that are almost the Stephani models
of~(\ref{Steph_constr}). However, when the assumption of geodesic observers is
relaxed it is no longer possible to constrain the rotation to be small, and the
class of perfect fluid spacetimes with an almost isotropic CMB may well include
examples with distinctly non-zero rotation, unless other constraints are
brought to bear. It would be interesting to determine the class of \emph{all}
perfect fluid spacetimes (including those with rotation) admitting an isotropic
radiation field, and if it turns out that they are indeed all irrotational then
the Stephani spacetimes defined by~(\ref{Steph_constr}), (\ref{metric})
and~(\ref{Vstephani2}) are indeed the complete set.

The rest of this thesis concerns the IRF models. We derive all the relevant
observational relations, for all observer locations, and demonstrate that the
models are acceptable on observational grounds, while retaining the Copernican
principle.

\forget{
\section*{Acknowledgements}

We would like to thank Roy Maartens for a great deal of input and
encouragement, Bruce Bassett for being very helpful, and for looking out for
the `little fish,' and Stephane Rauzy for useful discussions. CAC was funded by
a PPARC Studentship.


\newpage
}

}
\forgetmenot{
\chapter{Derivations of the Observational Relations}
\label{step chapter obs derivations----------------}
This chapter presents a derivation of all the important classical observational
tests of cosmology for the IRF models. These were presented in
chapter~\ref{relativistic cosmol chapter-----------------} for the FLRW models,
but the method here is slightly different. This is principally because the
models have less symmetry (although the conformal symmetry allows us to bypass
this). In contrast to the FLRW relations, none of the relevant quantities in
this chapter have an integral in them (eg, (\ref{pogson-FLRW})); all the
relations here, though, are given in parametric form, which makes them more
difficult computationally.

All the results in this chapter are new; \dab~(1995)\nocite{dab95} has
presented observational relations for the Stephani models before, but only as a
series expansion to first order, using the method of Kristian and Sachs~(1966).

\section{The Stephani Models Which Admit an Isotropic Radiation Field.}
\label{Stephani}

Stephani models are the most general conformally flat (expanding) perfect fluid
spacetimes. They have vanishing shear and rotation, but non-zero acceleration
and expansion. Although the general Stephani model has no symmetry at all, we
only consider here the class possessing a three dimensional isometry group
acting on two dimensional orbits, derived in chapter~\ref{EGS
chapter---------------------------------}. The metric in comoving (and
isotropic) coordinates can then be written (see~\S\ref{CFsolns}):
\be
ds^2  = \frac{(1+\sfrac14\Delta r^2)^2}{R(t)^2V(r,\vartheta,t)^2}
\left\{ -c^2dt^2 + \frac{R(t)^2}{(1+\sfrac14\Delta r^2)^2}(dr^2+r^2d\Omega^2)\right\},
\label{metric_gen}
\ee
where $c$ is the speed of light,\footnote{Factors of the speed of light will be
kept in this chapter, and throughout the rest of this thesis, to facilitate
comparison with observations.} $d\Omega^2 = d\vartheta^2 +\sin^2\vartheta
d\varphi^2$ is the usual angular part of the metric and the function $V$ is
given by (using~\ref{x} and $\mathbf{x_0}=\zeta\mathbf{\hat{z}}$ to avoid
confusion)
\be
 V=V(r,\vartheta,t)=\;1R\left(1+\frac{1}{4}\kappa (t)r^2\right)-
 2\zeta r\cos\vartheta+4\zeta^2\;{R}{\kappa}, \label{Vdef}
\ee
$R(t)$ is the Stephani version of the \FRW\ scale factor (note that we are
using a different time coordinate to~(\ref{step metric-IRF}) which uses
\emp{conformal time}), and $V(r,\vartheta,t)$ is a generalisation of the
\FRW\ spatial curvature factor (which is $1+\frac{1}{4}kr^2$ in isotropic
coordinates, with $k=0,\pm 1$, but is more familiar as $1-kr^2$ in Friedmann
coordinates). Since $\kappa$ is a function of~$t$ the spatial curvature can
vary from one spatial section to the next. In fact, it is possible for a closed
universe to evolve into an open universe, or vice versa, in stark contrast to
\FRW\ models (see Krasi\'{n}ski~1983). The models which admit an isotropic
radiation field have $\kappa$ restricted by~(\ref{kReqn}); viz,
\be
\kappa(t)\eqdef\:12\Delta+2\delta R\pm2\sqrt{\left(\:14\Delta+\delta R\right)^2
 +\Delta\zeta^2 R^2}.\label{kappa_IRF_general}
\ee
It is worth noting that this is the most general form for $\kappa$, for the
metric~(\ref{metric_gen}) with conformal factor~(\ref{Vdef}) to be a perfect
fluid; any change to $\kappa$ will introduce energy flux into the
energy-momentum tensor -- and, incidentally, make the expansion inhomogeneous.
Note that the FLRW limit is given by $\delta=\zeta=0$.

The four-velocity of the fluid is given directly from~(\ref{metric_gen}) since
$u^a\propto\delta^a_0$ in comoving coordinates and $u^a u_a=-c^2$, and is
\be
u^0=\;{c}{\sqrt{|g_{tt}|}}=\;{R V}{(1+\:14\Delta r^2)}.
\ee
The expansion is homogeneous, depending only on time,
\ba
\theta=\theta(t)\li=\li\del_a u^a=-\;{3R V_{,t}}{1+\:14\Delta r^2}
=-\;{3R}{\Delta}\left(\;{\kappa}{R}\right)_{,t}\nonumber\\
 \li=\li\;{3R_{,t}}{2 R}\left\{1\pm\;{\:14\Delta+\delta R}{
 \sqrt{\left(\:14\Delta+\delta R\right)^2
 +\Delta\zeta^2 R^2}}\right\},~~~~~~~\Delta\neq0,~\zeta\neq0\nonumber\\
 \li=\li\;{3R_{,t}}{R}~~~~~~~\zeta=0,~\hbox{or if}~\Delta=0;\label{expansion-IRF-gen}
\ea
(note that $\kappa$ must have the form~\ref{kappa_IRF_general}).

 The acceleration, $\dot{u}_a=u^b\del_b u_a$ has radial and angular
components:
\ba
\dot{u}_r\li=\li c^2\left[\ln\;{1+\:14\Delta r^2}{V(r,\vartheta,t)}\right]_{,r}
= c^2\frac{\Gamma,_r}{\Gamma}\nonumber\\
\li=\li c^2\left\{\;{1+\:12\Delta r}{1+\:14\Delta r^2}-\;2V\left[\;{\kappa
r}{4R}-\zeta\cos\vartheta\right]\right\},\nonumber\\
\udot_\vartheta\li=\li-\;{c^2}{r^2}\left[\ln V(r,\vartheta,t)\right]_{,\vartheta}
=-\;{2c^2\zeta\sin\vartheta}{rV}
\label{accel_gen}
\ea
where $\Gamma=|g_{tt}|^{1/2}$ is the square root of the time component of the
metric~(\ref{metric_gen}).

The density and pressure for these models become (using equations (5) and~(6)
of Krasi\'{n}ski~(1983, 1997); Eq. (3) in Barnes~(1998), and the results
of~\S\ref{app-Rkx})
\ba
\;{8\pi G}{c^2}\mu\li=\li\;{3\kappa}{R^2}+\;{\theta^2}{3c^2}=\;{3\kappa}{R^2}+
\;{3R^2}{c^2\Delta^2}
\left[\left(\;{\kappa}{R}\right)_{,t}\right]^2\nonumber\\
p\li=\li-\mu c^2+\;13\mu_{,t}c^2\;{V}{V_{,t}}.
\ea
We see that the density is homogeneous on surfaces of constant time but the
pressure is not~-- see also~(\ref{pressV}), below. So, for any fixed~$t$, the
spatial curvature, $\kappa(t)$, the expansion and the density are homogeneous,
but there are pressure gradients that lead to acceleration of the fluid. It is
interesting to note once again that the spacetime is locally \FRW\ if and only
if $\dot{u}^a=0$, which happens if and only if $\dot{\kappa}=0$, which in turn
is equivalent to the existence of a barotropic equation of state, $p=p(\mu)$
Krasi\'{n}ski~(1983\nocite{Kras83}). \forget{Note that in the limit
$\Delta\rightarrow0$, we have $\kappa/R=4\delta$, and the expansion becomes
zero, while the density and pressure behave as $mu\propto1/R,~p\propto\mu r^2$,
which may be plausible as a stellar model.}

\subsection{Geometry.}

The metric~(\ref{metric_gen}) is manifestly conformal to a Robertson-Walker
metric with curvature~$\Delta$. However, if we multiply through by the
conformal factor we see that the \emph{actual} spatial curvature is time- and
position-dependent and is given by the curvature factor $\kappa(\ti)$
in~$V(r,\ti)$ in the spherically symmetric case, but is quite complex in
general, with variations depending on angle and distance. When $\zeta\neq0$ the
spatial sections have a hyperbolic geometry~-- and thus a center of symmetry.
However the spatial sections foliate in such a way as to make this `center'
wander around. Indeed there is no reason to assume that this central worldline
will be causal, and an observer who happens upon this worldline may be able to
see it in their past. It is possible in principle to write the metric in
coordinates to make this apparent, but we do not do this here.

\forget{
In the case of the
\dab\ models, for any time~$\ti$ the spatial sections are homogeneous and
isotropic and their geometry depends on the sign of~$\kappa(\ti)$. If, at some
point during the evolution of the universe,~$\ti=-(b\pm 1)/2a$, the curvature
changes sign, as can easily be seen from (\ref{kappa}) and~(\ref{Roft}). For
example, if $|b|<1$ and~$a>0$ then the spatial sections are closed ($\kappa>0$)
for~$\ti<(1-b)/2a$. The spheres increase in size until $\ti$~passes through
$\ti=(1-b)/2a$, when the universe `opens up' and acquires a hyperbolic
geometry. This does not happen in \FRW\ models, where the spatial
curvature,~$k$, is fixed. The distinction between the time-dependent true
geometry ($\kappa$) and the fixed conformal geometry ($\Delta$) should be borne
in mind throughout what follows. As the universe opens up and the sections
become hyperbolic the coordinate~$\chi$ represents a conformal mapping from a
hyperbolic surface onto a sphere: spatial infinity will then correspond to some
finite value of~$\chi<\pi$. For a more detailed explanation of this see
theorems~4.1~--~4.4 of Krasi\'{n}ski~(1983)\nocite{Kras83}. }

(\ref{metric_gen}) and~(\ref{metric_spec}) are the usual forms in which the
Stephani metric is presented (see D\c{a}browski 1993,~1995; Krasi\'{n}ski
1983,~1997). However, for our purposes they are not the most advantageous
forms. The conformal geometry of the models is most easily studied by changing
from the stereographic coordinate,~$r$, to the `angle' coordinate,~$\chi$ (see
figure~\ref{stereo},
\begin{figure}[t!]
\centerline{\psfig{file={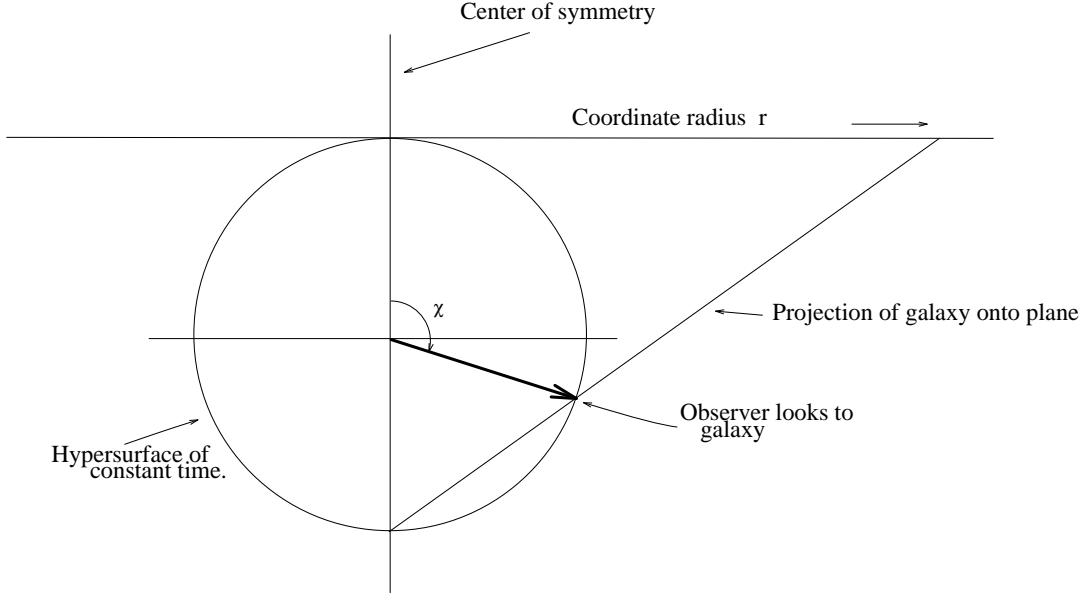},width=8cm,angle=90}}
\vspace{0cm}
\caption{\small Stereographic projection of a sphere showing the relationship
between the coordinates $r$ and~$\chi$. Note that one can envisage the angle
$\chi$ increasing steadily round the sphere \emph{ad infinitum}.
\label{stereo}}
\end{figure}
and equation~(5.15) of Hawking and Ellis, 1973), appropriate to the value
of~$\Delta$. Furthermore, for models with closed spatial sections (which will
be our principal concern here) it is more convenient to choose a radial
coordinate that is better able to reflect the fact that light rays can circle
the universe many times. In such models the spatial surfaces have two centers
or symmetry, $r=0$ and~$r=\infty$. Physically, there is nothing extraordinary
about the point $r=\infty$: it is not infinitely far away from the centre, and
it is quite possible for light rays to pass through it. This last point is
particularly important for subsequent discussions, so we make the coordinate
change
\be
r=\frac{2}{\sqrt{\Delta}}\left|\tan \frac{\chi}{2}\right|.\label{r_chi}
\ee
Then $r\rightarrow\infty$ as $\chi\rightarrow\pi$. As a coordinate $\chi$ is
restricted to the range $0\le\chi <\pi$. However, it will prove convenient to
use $\chi$ not just as a coordinate but as a parameter along light rays. In the
latter role its value can increase without bound. Strictly speaking, we should
distinguish these two uses, but it should not lead to confusion. The absolute
value is taken in~(\ref{r_chi}) so that, when $\chi$ increases beyond~$\pi$,
$r$ remains positive. Using~(\ref{r_chi}) in the metric~(\ref{metric_spec})
gives
\be
ds^2 = \frac{1}{W(\chi,\vartheta,\ti)^2} \left\{-c^2d\ti^2+
\frac{R(\ti)^2}{\Delta}(d\chi^2+\sin^2\chi d\Omega^2) \right\}\label{metric_W}
\ee
where
\ba
W(\chi,\vartheta,\ti)\li=\li \cos^2\frac{\chi}{2} V(r(\chi),\vartheta,\ti)
R(\ti)\nonumber\\
\li
=\li\left(1+4\;{\zeta^2R(\ti)^2}{\kappa(\ti)}\right)\cos^2\;{\chi}{2}+
\;{\kappa(\ti)}{\Delta}\sin^2\;{\chi}{2}-
2\;{\zeta R(\ti)}{\sqrt{\Delta}}\sin\chi\cos\vartheta\label{conf-W-general}
\ea
in general; and
\[
W(\chi,\ti)= \cos^2\frac{\chi}{2} V(r(\chi),\ti) R(\ti) = \cos^2\frac{\chi}{2}
           + \frac{\kappa(\ti)}{\Delta} \sin^2\frac{\chi}{2},
\]
in the spherically symmetric case. Singularities in the conformal factor
($1/W$) correspond to spatial and temporal infinity.

We can easily calculate the acceleration in these coordinates. Again it has
only radial ($\chi$) and angle~($\vartheta$) components:
\ba
 \dot{u}_\chi \li=\li -c^2 \frac{W_{,\chi}}{W}\nonumber\\ 
 \udot_\vartheta \li=\li-c^2\;{W_{,\vartheta}}{W}
\label{acceln2}
\ea
A simple calculation shows that the acceleration scalar, which we will need
below, is just
\be
  \dot{u} \equiv \left( \dot{u}_a \dot{u}^a \right)^{1/2}
  =\;{c^2}{R}\sqrt{\Delta[(W_{,\chi})^2+\csc^2\chi(W_{,\vartheta})^2]}.
\label{accscalar}
\ee

For completeness we mention that when the IRF models are conformal to an
\FRW\ spacetime with hyperbolic geometry ($\Delta<0$) the coordinate
transformation is obtained from that just given by replacing trigonometric
functions with their hyperbolic equivalent.

\subsection{Non-Central Observers.}
\label{noncent}

Since the purpose of this thesis is to provide insight into the observational
characteristics of the IRF models for all observer positions, we must find
expressions for the distance-redshift relations and other observable properties
of the models from any point in the spacetime. For a general inhomogeneous
metric this is far from trivial, but the IRF models have features that make
this problem tractable. In particular, they are conformally flat. As has
already been noted, in equation~(\ref{metric_W}) the part of the metric in
braces is exactly the form of the \FRW\ metric in `angle' coordinates, so that
the IRF models are manifestly conformal to the (homogeneous)
\FRW\ spacetimes. This means that there is a group of transformations acting
transitively on surfaces of constant time that \emph{preserve the form of the
\FRW\ part of the metric} (but not the conformal factor). If the conformal spatial
sections are closed (3-spheres), flat or hyperbolic (according to the value
of~$\Delta$), the transformations are rotations, translations or `Lorentz
transformations', respectively. After such a transformation the metric will
have the form of an \FRW\ metric \emph{centred on the new point}, multiplied by
a modified conformal factor.

As will be shown in~\S\ref{energy_cond}, we will be dealing exclusively with
closed models ($\Delta>0$), and so will concentrate on this case.

\subsubsection{$\zeta=0$ -- The Spherically Symmetric Case}

In the spherically symmetric case, to find the coordinate transformation to a
non-central position, we perform a rotation of the spatial part of the metric,
moving the origin ($\chi=0$) to the point $\chi=\psi$ ($\psi$~is the observer's
position in what follows). In appendix~\ref{chitrans} we derive this
transformation. The old~$\chi$ is given in terms of the new (primed)
coordinates by~(\ref{newchi}), which we reproduce here:
\ba
\cos \chi=\cos\psi\cos\chi' - \sin\psi\sin\chi'\cos\vartheta',
\ea
 where $\vartheta'$ denotes the direction
of the new location. The conformal factor now becomes, dropping the primes on
the coordinates
\be
W\rightarrow W(\chi,\vartheta,t)=
\;12\left(1+\;\kappa\Delta\right)+\;12\left(1 -\;\kappa\Delta\right)
(\cos\psi\cos\chi - \sin\psi\sin\chi\cos\vartheta),
\ee
while the rest of the metric~(\ref{metric_W}) stays the same, but now in terms
of the new coordinates.

This transformation makes the study of our inhomogeneous models significantly
easier, and allows us to find \emph{exact} observational relations valid for
any observer.

\subsubsection{$\zeta\neq0$ -- The General Case}

When $\zeta\neq0$, things are not so simple, and the form of the conformal
factor~(\ref{conf-W-general}) is not general enough. This is because
in~(\ref{Vdef}) we took the direction of the center of symmetry to be aligned
with the $z$ axis. We desire the generality of the case above: we wish to be
able to move the observers to any location in the spacetime, regardless of the
direction to the center. In the spherically symmetric case we only had to worry
about the radial coordinate; in the axially symmetric case we need to worry
about radial and an angle coordinate. Instead of performing a general
coordinate transformation, we instead define our direction to the center to lie
somewhere in the $y=0$ plane, in the direction given by $\xi$;
\be
{\mathbf x_0}\eqdef\zeta\left(
      \begin{array}{c}
            \sin\xi\\
            0\\
            \cos\xi
      \end{array}
      \right).
\ee
Now we may write
\be
 V=V(r,\vartheta,t)=\;1R\left(1+\frac{1}{4}\kappa (t)r^2\right)
  +4\zeta^2\;{R}{\kappa}-2\zeta(x\sin\xi+z\cos\xi).
\ee
In order to find $W$, we must first note that in terms of the mixed coordinates
$\chi,~r,~x,~z$ we have
\be
W=\;12\left(1+\;\kappa\Delta+4\zeta^2\;{R^2}{\kappa}\right)+\;12\left(1 -
\;\kappa\Delta+4\zeta^2\;{R^2}{\kappa}\right)\cos\chi
-2\;{\zeta R}{\sqrt{\Delta}}\;{x\sin\xi+z\cos\xi}{1+\:14\Delta r^2}.
\ee
In order to use the transformations in appendix~\ref{chitrans} we note that the
cartesian coordinates used there are not the same as the ones used here, but
are related by
\be
\tilde x^i\propto\;{\sqrt\Delta x^i}{1+\:14\Delta r^2}
\ee
(this amounts to the transformation between the FLRW and stereographic
coordinates) and the rotation is in terms of the $\tilde x^i$, so that we have
in the new primed coordinates
\ba
\;{x}{1+\:14\Delta r^2}\li=\li\;{1}{\sqrt\Delta} \sin\chi'\sin\vartheta'
\cos\varphi
\nonumber\\
\;{z}{1+\:14\Delta r^2}\li=\li\;{1}{\sqrt\Delta}
(\sin\psi\cos\chi'+\cos\psi\sin\chi'\cos\vartheta').
\ea
We now have
\ba
W\li=\li\;12\left\{1+\;\kappa\Delta+4\zeta^2\;{R^2}{\kappa}\right\}+\;12\left\{1
-
\;\kappa\Delta+4\zeta^2\;{R^2}{\kappa}\right\}\left(\cos\psi\cos\chi -
\sin\psi\sin\chi\cos\vartheta\right)\nonumber\\
\li-\li2\;{\zeta R}{\sqrt{\Delta}}\left(\sin\chi[\sin\xi\sin\vartheta\cos\varphi+
\cos\xi\cos\psi\cos\vartheta] +\cos\xi\sin\psi\cos\chi\right).
\ea
Note that $W$ is of the form
\be
W=A+B\sin\vartheta\cos\varphi+C\cos\vartheta=A+Bx+Cz,
\ee
where $A, B,$ and~$C$ are functions of time only, then we can rotate in the
$x-z$ plane (by $\tan^{-1}(B/C)$) so that $W$ depends only on $z$, and we have
in this new coordinate system
\be
W=A+\sqrt{B^2+C^2}\cos\vartheta.
\ee
This shows that, as in the spherically symmetric case, we have only a dipole
moment in~$W$.

To summarise, we have changed the coordinate system in order to center our
coordinates on any point in the spacetime. In these new coordinates the metric
is of the form
\be
ds^2 = \frac{1}{W(\vartheta,\chi,\ti)^2} \left\{-c^2d\ti^2+
\frac{R(\ti)^2}{\Delta}(d\chi^2+\sin^2\chi d\Omega^2) \right\}\label{metric_W2}
\ee
where
\ba
\li W\li=\Phi_++\Phi_-\cos\psi\cos\chi-
\;{2\zeta R}{\sqrt\Delta}\cos\xi\sin\psi\cos\chi
\nonumber\\ \li+\li\sin\chi\cos\vartheta\sqrt{\Phi_-^2\sin^2\psi
+4\;{\zeta^2 R^2}{\Delta}(1-
\cos^2\xi\sin^2\psi)+2\;{\zeta R\Phi_-}{\sqrt\Delta}\cos\xi\sin2\psi},\label{W_complete}
\ea
and
\be
\Phi_\pm\eqdef\;12\left[1\pm\;\kappa\Delta+4\zeta^2\;{R^2}{\kappa}\right].
\ee

\section{Observational Relations}
\label{obs-constr}

So far we have considered only some of the `global' physical properties of IRF
models, but to really assess their potential viability as cosmological models
it is necessary to confront them with observations. In this section we derive
the distance-redshift relations that form the basis of the classical
cosmological tests which are used later to compare them with available
observational constraints to see whether any regions of parameter space are
capable of providing a fit.

Deriving the observational relations (redshift, angular size or area distance,
luminosity distance and number counts) means relating the coordinates and
metric functions to observable quantities. This requires knowledge of the
observer's motion (4-velocity), which can, strictly speaking, be specified
independently of the background geometry. However, the IRF models contain
perfect fluid, so we will identify the observer's motion with the fluid
velocity. We are not obliged to do this, and, given the strange form of the
matter, it might be thought advantageous to instead assume that observers
(ie,~galaxies) constitute a dust-like test fluid moving freely through the
spacetime whose geometry is determined by the exotic matter. It should be clear
from chapter~\ref{EGS chapter---------------------------------} that if we were
to make this assumption a large dipole anisotropy in the CMB would result
(although the dipole in~$H_0$ would be eliminated~-- see~(\ref{specific H_0})
with~$\dot{u}=0$) because such a flow will, in general, have a significant
velocity relative to the IRF fluid flow, which, it will turn out, is very
nearly in the rest frame of the CMB everywhere. This should not be a surprise
because chapter~\ref{EGS chapter---------------------------------} was devoted
to proving just this.

\subsection{Redshift}
\label{redshift-IRF-derivation}

Probably the most important distance measure in cosmology is redshift. In
general, it is no simple task to find analytic expressions for the redshift in
any cosmological model; derivations usually rely on symmetries of the spacetime
or other simplifying factors to solve the equations of null geodesics. Here we
can take advantage of the conformal flatness of Stephani models (strictly
speaking, of the fact that the IRF models are manifestly conformal to
\FRW\ cosmologies) using the results of \S\ref{conformal redshift section},
 although we can also derive the redshift formula as a time
dilation effect which we do here. When the true spacetime is conformal to a
spherically symmetric spacetime the radial null geodesics connecting any point
with an observer at the centre (of the conformal part) are obviously purely
radial (since their paths are not affected by the conformal factor). They are
therefore given (in terms of coordinates $r$ and~$t$ with respect to which the
spherical symmetry is manifest) by some
function~$t_{\scriptscriptstyle{O}}(r_{\scriptscriptstyle{E}},
t_{\scriptscriptstyle{E}})$ relating the time,~$t_{\scriptscriptstyle{O}}$ ,
that the light ray is received by the observer, to the time of
emission,~$t_{\scriptscriptstyle{E}}$, for an object at
radius~$r_{\scriptscriptstyle{E}}$. This is just the lookback-time relation.
Redshift, as the ratio of proper time intervals at the observer to proper time
intervals at the emitter, is then given by
\ba
    1+z \equiv \frac{d\tau_{\scriptscriptstyle{O}}}{d\tau_{\scriptscriptstyle{E}}}
          \li = \li \frac{d\tau_{\scriptscriptstyle{O}}}{dt_{\scriptscriptstyle{O}}}
             \frac{dt_{\scriptscriptstyle{O}}}{d\tau_{\scriptscriptstyle{E}}}
            = \frac{d\tau_{\scriptscriptstyle{O}}}{dt_{\scriptscriptstyle{O}}}
              \left(\frac{\partial t_{\scriptscriptstyle{O}}}{\partial r_{\scriptscriptstyle{E}}}
                    \frac{dr_{\scriptscriptstyle{E}}}{d\tau_{\scriptscriptstyle{E}}}
               +    \frac{\partial t_{\scriptscriptstyle{O}}}{\partial t_{\scriptscriptstyle{E}}}
                    \frac{dt_{\scriptscriptstyle{E}}}{d\tau_{\scriptscriptstyle{E}}}\right) \nonumber\\
           \li = \li \frac{1}{u^t_{\scriptscriptstyle{O}}}
                 \left(\frac{\partial t_{\scriptscriptstyle{O}}}{\partial r_{\scriptscriptstyle{E}}}
                    u^r_{\scriptscriptstyle{E}}
               +    \frac{\partial t_{\scriptscriptstyle{O}}}{\partial t_{\scriptscriptstyle{E}}}
                    u^t_{\scriptscriptstyle{E}}\right).
\label{zDtdef}
\ea
(When the coordinates $r$ and~$t$ are comoving~-- $u^r=0$~-- the $r$-derivative
term disappears.) This will provide an analytic expression for the redshift
whenever the lookback-time equation can be integrated. For the IRF models:
\be
          u^\chi=0,  \hskip 1.0cm   u^\ti = \frac{c}{|g_{00}|^{1/2}} = W
\label{ut}
\ee
and the lookback time can be derived directly from the metric: on the past null
cone of the observer $ds=0=d\vartheta=d\varphi$, leading to an expression
for~$d\chi/d\ti$, which, when integrated, gives
\be
              \chi(t) = c\sqrt{\Delta}\int_\ti^\T \frac{d\ti'}{R(\ti')},
\label{lookback_gen}
\ee
where $T$ is the coordinate age. Differentiating this with respect to~$\ti$ at
fixed~$\chi$ then gives
\[
     \frac{\partial t_{\scriptscriptstyle{O}}}{\partial t_{\scriptscriptstyle{E}}} \equiv
     \frac{\partial \T}{\partial \ti} = \frac{R_{\scriptscriptstyle{O}}}{R_{\scriptscriptstyle{E}}},
\]
which, together with (\ref{zDtdef}), (\ref{ut}), and~(\ref{metric_W2}) we find
\be
1+z(\psi,\chi,\vartheta)=\frac{R_0}{W_0}\frac{W(\psi,\chi,\vartheta;\ti)}{R(\ti)},
\label{redshift_new}
\ee
where $R_0=R(\T)$, $W_0=W(\psi,\T)$ and $\ti$ and~$\chi$ are related by
equation~(\ref{lookback_gen}).

A more elegant approach is to use the result of
\S\ref{conformal redshift section}.  For the IRF models
$\tilde{g}_{ab}$ is an \FRW\ metric, the conformal factor is $e^Q=1/W$
(see~(\ref{metric_W2})) and $\tilde{u}^a$ is the usual
\FRW\ comoving velocity field. The well-known expression for redshift in
\FRW\ spacetimes, $1+\bar{z} =
R_{\scriptscriptstyle{O}}/R{\scriptscriptstyle{E}}$, then gives
\be
          1+z = \frac{R_{\scriptscriptstyle{O}}}{W_{\scriptscriptstyle{O}}}
                \frac{W_{\scriptscriptstyle{E}}}{R_{\scriptscriptstyle{E}}}.
\label{redshift}
\ee
(Alternatively, we could change to conformal time~(\ref{conftime}), making the
metric manifestly conformal to the Einstein static spacetime,~(\ref{ESS}), for
which the redshift is zero. The conformal factor is
then~$e^Q=R/(\Delta^{1/2}W)$, which also gives~(\ref{redshift}).)

Now, if we take our large expression for $W$~(\ref{W_complete}) and write it as
\be
W=\Psi_m+\Psi_d\cos\vartheta
\ee
where
\ba
\Psi_m\li\eqdef\li\Phi_++\Phi_-\cos\psi\cos\chi-
\;{2\zeta R}{\sqrt\Delta}\cos\xi\sin\psi\cos\chi,\\
\Psi_d\li\eqdef\li\sin\chi\sqrt{\Phi_-^2\sin^2\psi
+4\;{\zeta^2 R^2}{\Delta}(1-
\cos^2\xi\sin^2\psi)+2\;{\zeta R\Phi_-}{\sqrt\Delta}\cos\xi\sin2\psi}
\ea
then we see that
\be
1+z=\;{R_0}{R(T)W_0}\left(\Psi_m+\Psi_d\cos\vartheta\right):
\ee
all inhomogeneity simply results in a dipole variation in $z$ around the sky.

\subsection{Distance-Redshift Relations}

For metrics with spherical symmetry about the observer the angular size (and
area) distance is given directly from the coefficient in front of the angular
part of the metric, because symmetry ensures that for radial rays $\vartheta$
and~$\varphi$ are constant along the trajectory. For our models we do not have
spherical symmetry about every observer, but the metric is everywhere conformal
to a spherically symmetric metric, as can be seen from~(\ref{metric_W2}). Since
null rays are not affected by the conformal factor they also remain at fixed
$\vartheta$ and~$\varphi$, so we can again obtain the angular size
distance,~$r_A$, from the coefficient of the angular part of the metric:
\be
r_A(\psi,\chi,\theta)=\frac{R(\ti)}{\sqrt{\Delta}W(\psi,\chi,\theta;\ti)
}|\sin\chi|
\label{area dist-new}
\ee
(again, $\chi$ and~$\ti$ are related by~(\ref{lookback_gen})). The absolute
value of $\sin\chi$ is taken to keep $r_A$ positive~-- which we are perfectly
entitled to do since only $\sin^2\chi$ appears in the metric~(\ref{metric_W2}).
We can find now, for the first time, the \emph{exact} angular size distance
relation parametrically by combining equations~(\ref{area dist-new})
and~(\ref{redshift_new}), which bypasses the rather cumbersome method of
Kristian and Sachs~(1966)\nocite{kris-sac66}, used in \dab~(1995) to find a
series relation. This is valid for \emph{any} null-connected points in the
spacetime. As far as I am aware, nobody has done this for an inhomogeneous
model before.

The other classical tests can now be written down. Luminosity distance,~$r_L$,
is related to~$r_A$ by the reciprocity theorem:
\be
r_L=(1+z)^2r_A,
\label{recip}
\ee
see MacCallum and Ellis~(1970)\nocite{Mac-Ellis-I}\nocite{Mac-Ellis-II} and
Ellis~(1998)\nocite{ell98}, or \S\ref{area dist section}. This then allows the
magnitude-redshift relation to be determined in the usual way: the apparent
magnitude~$m$ of an object of absolute magnitude~$M$ is given in terms of the
luminosity distance by
\be
          m-M-25 = 5\log_{10} r_L.
\label{pogson2}
\ee

\subsection{Number-Count-Redshift Relation}

Finally, for completeness, we discuss the number count-redshift relation,
although we do not use it here. The only important consideration is the
identification of some sensible comoving number density
distribution,~$n_c(\chi)$, for the observers (fluid particles). This is not
necessary for homogeneous spacetimes because it is natural to simply take~$n_c$
also to be homogeneous (and independent of time, assuming no evolution).
Furthermore, when there is fluid pressure $n_c$ cannot be directly related to
the energy density of the matter, as is possible with a dust, and, in general,
there need be no obvious choice for~$n_c$~-- the only constraint it must
satisfy (again assuming no number evolution, such as would be caused by
mergers, for example) is particle number conservation:
\[
                       \del_a (n_c u^a )= 0.
\]
Since the metric~(\ref{metric_W}) is in comoving coordinates we can satisfy
this by simply assuming that $n_c$~is independent of time~$\ti$, being given by
whatever distribution we specify at~$\ti=0$. Moreover, the fact that the IRF
models become homogeneous ($W\rightarrow 1$) as~$\ti\rightarrow 0$ means that
it makes sense to define $n_c$ to be independent of spatial position too. We
are then back to the same situation as with \FRW\ models, except that now
observations of galaxy numbers constrain comoving volume, but not \emph{proper}
volume (note that the proper density of particles is \emph{not} constant,
because the conformal factor modifies the proper volume by different amounts at
different positions and times), and so are less directly a constraint on the
radial component of the metric (which determines proper distance). The number
of particles within some radius~$\chi$ of any observer is simply proportional
to the volume of the \FRW\ 3-sphere of curvature~$\Delta$ out to that radius:
\[
       N(\chi) \propto \frac{4\pi}{\Delta^{3/2}}\int_0^\chi \sin^2\chi' d\chi'
                    = \frac{2\pi}{\Delta^{3/2}}\left(\chi-\sin\chi \cos\chi\right)
\]
exactly as in \FRW\ models. The difference from the \FRW\ $N-z$ relation comes
when we use~(\ref{redshift_new}) to relate number counts to redshift: $z(\chi)$
will be different to that for \FRW\ spacetimes. Nevertheless, it is clear from
this discussion that the fluid particles (observers) in the IRF spacetime are
distributed \emph{uniformly with respect to the \FRW\ volume on the spatial
sections}. Since these hypersurfaces of constant time are just spheres in the
\FRW\ metric we know that the Copernican principle can easily be applied: the
probability that an observer lies in some region of space is exactly the volume
of that region \emph{with respect to the \FRW\ volume} (divided by the total
volume of the sphere).

\section{Discussion}

In this chapter we have, for the first time, derived all the relevant classical
cosmological distance-redshift relations \emph{exactly}: that is, we have given
relations connecting observable quantities, which can now be used to pit the
models against real data. Moreover we have derived these relations in such a
way as to be valid \emph{for any observer in the spacetime}. We have not
discussed, though, observations connected with a blackbody (ie, the CMB); this
will be done in the following chapters (in fact, the temperature evolves simply
as $1/(1+z)$).

In the following chapters we use these observational relations (with a specific
form for the scale factor in order to integrate the lookback time relation)  to
demonstrate that these models are quite suitable as a cosmological model from
an observational point of view -- regardless of observer position. This is
important because previous studies of non-standard inhomogeneous cosmologies
have always assumed a privileged observer position (at a center of symmetry),
or in a small neighborhood of such a location.

}
\forgetmenot{
\chapter{Observational Characteristics With $\zeta=0$}
\label{step chapter 1------------------------------}

In chapter~\ref{step chapter obs derivations----------------} we derived
expressions for all the important observational relations used to test a
cosmological model, and we took the time to derive them for any location in the
spacetime. Throughout the rest of this thesis we wish to demonstrate that these
observational aspects of the models are compatible with present observational
constraints. In this chapter we restrict ourselves to the spherically symmetric
class, $\zeta\equiv0$.

These models are far more general than we will need because there are no
restrictions on the function $R(t)$. In fact we need a specific form for $R$ in
order to find the lookback time relation~(\ref{lookback_gen}). We want a family
of models with a small number of free parameters to simplify the analysis and
produce graphs etc., so we restrict attention to the spherically symmetric
two-parameter family derived in~\S{IV} of D\c{a}browski (1993). These are just
an extension of Model~I of D\c{a}browski~(1995) (in these papers the
parameter~$b$ in equation~(\ref{Roft}), below, was set to~$1$). They were
chosen largely because of their simplicity, and because they proved to be the
most useful models in D\c{a}browski and Hendry~(1998). As described in
\dab~(1993), we choose the metric functions in terms of this new~$\ti$ to
be of the form:
\begin{eqnarray}
       R(\ti) \li \eqdef \li c\ti(a\ti+b) \label{Roft}\\
  \kappa(\ti) \li = \li \Delta-\frac{4a}{c}R(\ti)~(=1-{R}(\ti)^2_{,t}/c^2),\label{kappa}
\end{eqnarray}
where $a\eqdef-c\delta$ and~$b$ are the free parameters and
\be
         \Delta \eqdef 1-b^2,
\label{Delta}
\ee
and $\zeta\equiv0$. The metric is then,
\be
ds^2= \frac{(1+\frac{1}{4}\Delta r^2)^2}{R(t)^2V(r,\ti)^2}\left\{ -c^2 d\ti ^2
+\frac{R(\ti)^2}{(1+\frac{1}{4}\Delta r^2)^2}\left( dr^2+r^2d\Omega^2\right)
\right\}. \label{metric_spec}
\ee
We will henceforth refer to these models as D\c{a}browski models. Note that in
D\c{a}browski (1993,~1995) the quadratic dependence of $R(\ti)$ also includes a
constant term,~$d$. We restrict attention to models which have a big bang (a
density singularity in the language of D\c{a}browski~1993), which means that
$R(\ti)$ must have a root. We can therefore choose the origin of time so
that~$d=0$. Furthermore, we require that after the big bang (and before any big
crunch) $R>0$, which forces
\be
                   b \ge 0
\label{bpos}
\ee

We make the coordinate changes used in chapter~\ref{step chapter obs
derivations----------------}; viz.
\be
r=\;{1}{\sqrt\Delta}\left|\tan\;\chi2\right|,
\ee
so that the metric becomes
\be
ds^2 = \frac{1}{W(\chi,\ti)^2} \left\{-c^2d\ti^2+
\frac{R(\ti)^2}{\Delta}(d\chi^2+\sin^2\chi d\Omega^2) \right\};
\ee
 and we can use~(\ref{kappa}) to simplify the
expression for~$W$, resulting in
\be
     W(\chi,\ti) = 1-\frac{4aR}{c\Delta}\sin^2\frac{\chi}{2}.
\label{confW}
\ee
This has no dependence on $\vartheta$. The conformal factor ($1/W^2$) is
non-singular for all $\chi$ if~$a\le 0$. When $a>0$ singularities~$W=0$
correspond to spatial and temporal infinity, and indicate that the universe has
`opened up'.

We transform to a non-central location, as discussed in \S\ref{noncent}, so
that $W$ transforms as
\be
W\rightarrow W(\chi,\psi,\vartheta;\ti)=1-\frac{2aR(\ti)}{c\Delta} (1-
               \cos\psi \cos\chi + \sin\psi \sin\chi \cos\vartheta),
\label{new W}
\ee
while the rest of the metric retains its original form (but now in terms of the
new coordinates).

In (\ref{Roft})-(\ref{Delta}) we have retained factors of the speed of
light,~$c$, to facilitate comparison with the references given above and with
observations. The units we will use are as follows: $[c]=$~km~s$^{-1}$, $r$~is
dimensionless, $R$~is in Mpc and $[\ti]=$~Mpc~s~km$^{-1}$$=[1/H_0]$, so that
$[a]=$~km~s$^{-1}$~Mpc$^{-1}$$=[H_0]$ and $b$~is dimensionless. Note that these
units are slightly different to those used in D\c{a}browski~(1995) because the
parameters $a$ and~$b$ in that paper contain a factor of~$c$ (so
$[b]=[c]$,~etc.). This explains the appearance of~$c$ in~(\ref{Roft}).

We will use~$\T$ to denote the coordinate time of a specific epoch of
observation along some observer's worldline (ie,~the coordinate age of the
universe), again in Mpc~s~km$^{-1}$, and $\tau$ to denote \emph{proper} time
along a particular flow line. When we state ages they will generally be given
in~Gyr: $\T_{\mathrm{Gyr}}\approx 978\T$.

We will impose constraints on the value of~$H_0$, age, size (the meaning of
which will be explained below) and the anisotropy of the microwave
background\footnote{This may come as a surprise, given that the models were
derived specifically to have an isotropic radiation field. The CMB has physics
behind it though, and consideration of this produces anisotropy.}, leaving the
wealth of data available from galaxy surveys and high-redshift supernovae for
consideration in a later paper; the complexities involved in interpreting such
data and applying it to inhomogeneous cosmological models require separate
treatment.

The task now is to limit $a$, $b$, and~$\T$ using present observational
constraints. A full discussion of each constraint is made in the following
sections, and they are followed by exclusion diagrams showing the regions of
parameter space for which $a$ and~$b$ give a plausible cosmological model for
\emph{all} observer locations in these models. First off though we demonstrate
that the choice of scale factor~(\ref{Roft}) gives an immediate solution to the
horizon problem~(cf. Rindler, 1956\nocite{rind56}).

In the case of the \dab\ models, for any time~$\ti$ the spatial sections are
homogeneous and isotropic and their geometry depends on the sign
of~$\kappa(\ti)$. If, at some point during the evolution of the
universe,~$\ti=-(b\pm 1)/2a$, the curvature changes sign, as can easily be seen
from (\ref{kappa}) and~(\ref{Roft}). For example, if $|b|<1$ and~$a>0$ then the
spatial sections are closed ($\kappa>0$) for~$\ti<(1-b)/2a$. The spheres
increase in size until $\ti$~passes through $\ti=(1-b)/2a$, when the universe
`opens up' and acquires a hyperbolic geometry. This does not happen in \FRW\
models, where the spatial curvature,~$k$, is fixed. The distinction between the
time-dependent true geometry ($\kappa$) and the fixed conformal geometry
($\Delta$) should be borne in mind throughout what follows. As the universe
opens up and the sections become hyperbolic the coordinate~$\chi$ represents a
conformal mapping from a hyperbolic surface onto a sphere: spatial infinity
will then correspond to some finite value of~$\chi<\pi$. For a more detailed
explanation of this see theorems~4.1~--~4.4 of
Krasi\'{n}ski~(1983)\nocite{Kras83}.

When $a>0$, there is a difficulty in applying the Copernican principle, which
is central to this thesis. This is because once the universe has opened up it
becomes infinite in extent; applying the Copernican principle becomes extremely
difficult, because there is no way to consider `all' observer locations. Even
worse, `most' observers cannot be defined in any obvious way. This is not a
problem in the homogeneous open FLRW models simply because it \emph{is}
homogeneous; \emph{a priori} every location is equivalent to every other. In
the Stephani case this is not true: we are attempting to show, through what one
can observe, that all locations are equivalent. Obviously this cannot be done
for $a>0$. This isn't a problem though, because in \S\ref{energy_cond} we
reject the case $a>0$ because it fails the energy conditions. However see the
discussions in chapter~\ref{conclusions_thesis}.

\forget{
 Using~(\ref{new W}) this is
\be
  1+z=\frac{R_0}{W_0}\left\{\frac{1}{R(\ti)}-
\frac{2a}{c\Delta}\left(1-\cos\psi\cos\chi\right) -
\frac{2a}{c\Delta}\sin\psi\sin\chi\cos\theta\right\};
\label{zdipole}
\ee
showing that, for objects at any fixed~$\chi$, the inhomogeneity of universe
manifests itself in the redshift as a pure dipole in angle around the sky
($\cos\theta$~term). This will be important in~\S\ref{CMB}. }

\section{Lookback Time and the Horizon.}
\label{horizon}

If, in equation~(\ref{metric_W}), we absorb a factor of $R^2(\ti)/\Delta$ into
the conformal factor and change to conformal time,~$\eta$, defined by
$d\eta=c\sqrt{\Delta}d\ti/R(\ti)$, so that
\be
                      \ti(\eta)=b/(e^{-b\eta/\sqrt{\Delta}}-a)
\label{conftime}
\ee
then the resulting metric is clearly conformal to the completely homogeneous
metric
\be
ds_E^2=-d\eta^2+d\chi^2+\sin^2\chi d\Omega^2,
\label{ESS}
\ee
which is precisely the Einstein static spacetime (ESS~-- Hawking and
Ellis~1973,~\S5.1--5.3). Equation~(\ref{conftime}) implies that the big bang
($\ti=0$) happens at $\eta=-\infty$. When~$a\le 0$ the big crunch ($\ti=-b/a$)
occurs at~$\eta=+\infty$, so the D\c{a}browski models with~$\Delta>0$ are
conformal to \emph{all} of the ESS when~$a\le 0$. If~$a>0$ the models are not
conformal to all of the ESS because then~$\eta$ will be bounded above and, what
is more, the coordinate~$\chi$ will not take on all values once the universe
has opened up (the no-horizon argument outlined below applies for~$a>0$ too,
though, because it only relies on the D\c{a}browski models being conformal to a
region of the ESS that is unbounded below for all~$\chi$). Since the null
structures of conformally related spacetimes are identical we can find the null
geodesics of the D\c{a}browski models directly from those of the ESS, which can
be derived trivially from the metric~(\ref{ESS}): the past null rays from
$\chi=0$,~$\eta=\eta_0$ satisfy~$\eta=\eta_0 -\chi$. That is, they are straight
lines at $45^o$ in the $(\chi,\eta)$-plane. If we represent the ESS (and the
D\c{a}browski models) on the Einstein cylinder (see
figure~\ref{lookback_cylinder}),
\begin{figure}[p!]
\centerline{\psfig{file={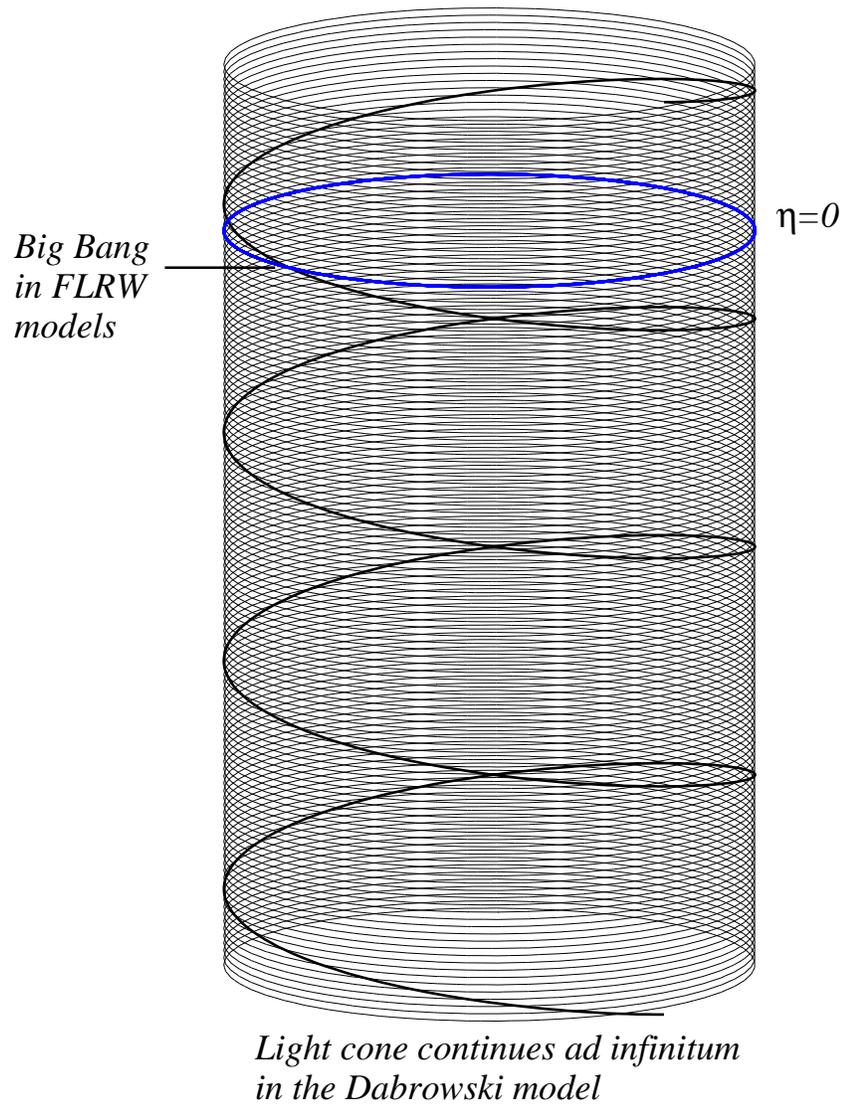},width=12cm,angle=0}}
\caption{\small Lookback time and the past null cone on the Einstein cylinder. This
shows the null structure for any model conformal to the Einstein static
spacetime, including the D\c{a}browski and \FRW\ models. In \FRW\ models, the
big bang occurs at $\eta=0$, whereas in \dab\ models, the spacetime extends to
$\eta=-\infty$.
\label{lookback_cylinder}}
\end{figure}%
where $\chi$ is shown as a circle ($0\le\chi\le 2\pi\equiv 0$), the null rays
just circle around the cylinder indefinitely into the past, going round
infinitely many times regardless of the initial~$\eta_0$ (or~$t_0$). Thus, the
whole of the big-bang surface is contained within the causal past of
\emph{every} point of both spacetimes, and there is no horizon problem for the
D\c{a}browski models.

We can see this in another way if we calculate the lookback time in our models
(ie,~the time,~$\ti$, at which a galaxy at some position~$\chi$ emits the light
that the observer sees now at time $\T$), which we can do
using~(\ref{conftime}) (or directly from the metric~(\ref{metric_W})~--
see~\ref{lookback_gen}). From~(\ref{conftime}) and the equations for null
geodesics in conformal time,~$\eta=\eta_0(\T)-\chi$ (with~$\eta_0(\T)$ given by
the inverse of~(\ref{conftime})) we get
\be
t(\chi)=\frac{bT}{(aT+b)\exp{(b\chi /\sqrt{\Delta}) }-aT}.
\label{lookback}
\ee
Note that this function is continuous through $\chi=\pi$. Again, $t\rightarrow
0$ if and only if~$\chi\rightarrow\infty$. We can visualise this using the
`D\c{a}browski cylinder' of figure~\ref{lookback_cylinder2}
\begin{figure}[here!]
\centerline{\psfig{file={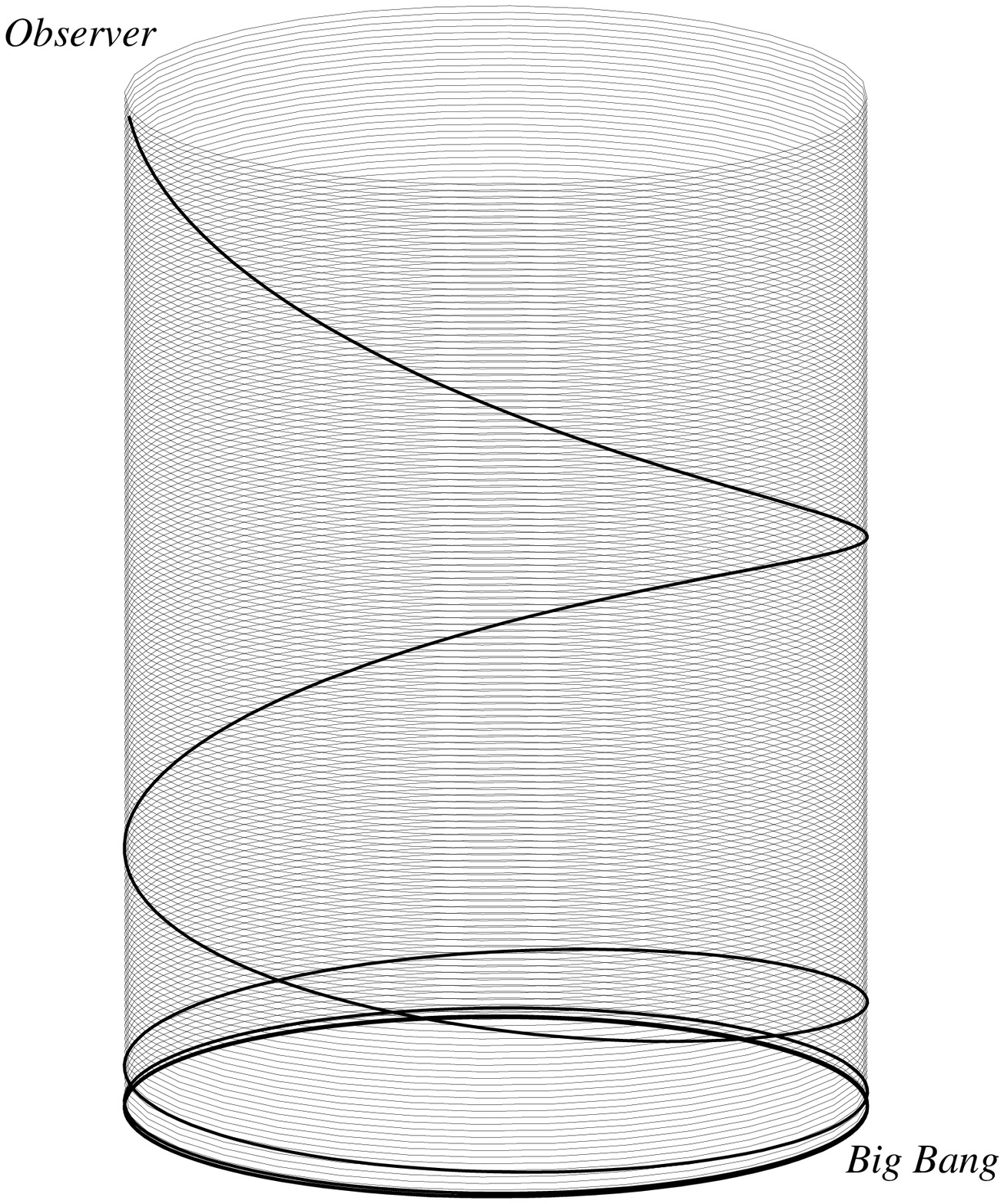},width=11cm,angle=0}}
\caption{\small As in figure~\ref{lookback_cylinder}, but now using the actual
D\c{a}browski model time coordinate, $\ti$.
\label{lookback_cylinder2}}
\end{figure}%
(like the Einstein cylinder but with~$t$ instead of~$\eta$ as the time
coordinate). The null rays still asymptote to the big-bang surface, but this
diagram reflects the way that in reality the spatial sections shrink as time
decreases to zero: as the sections shrink (going back in time) light rays can
travel all the way round the universe in a shorter and shorter time, and if the
spatial sections shrink sufficiently slowly as we move back towards the big
bang this travel time can approach zero without the light ray hitting the
big-bang surface.

This is all in sharp contrast to \FRW\ universes and de Sitter space, which are
conformal only to parts of the Einstein cylinder (they are bounded below
at~$\eta=0$): at early times the particle horizon is finite and contains only a
small part of the big-bang surface, so that widely separated points can share
no common influences. See figures 17 and~21 in Hawking and
Ellis~(1973)\nocite{hawk-ell}.

The existence of a horizon is directly related to the rate at which the (\FRW\
or Stephani) scale factor grows for small~$t$. If $R\sim t^\alpha$ at early
times, then conformal time goes as
\[
 \eta \sim\int\frac{dt}{R(t)}\sim \frac{1}{1-\alpha}t^{1-\alpha}\hskip 0.5cm
 \hbox{ (or $\log t$ for $\alpha=1$).}
\]
So, when $\alpha<1$ (as it is for dust \FRW\ models) $\eta=0$ at~$t=0$, the
spacetime is conformal to only part of the Einstein cylinder and a horizon
exists. Otherwise, as $t\rightarrow 0$ $\eta\rightarrow -\infty$ and the
horizon is absent.

Other methods which lead to the solution of the horizon problem include using
the LTB models; C\'el\'erier and Schneider~(1998)\nocite{cel-sch98}. See also
Rindler~(1956) who first reported the horizon problem.

\subsection{Matter Content and Energy Conditions.}
\label{energy_cond}

The Stephani models do not have an equation of state in the strict sense, with
the relationship between pressure and energy density being position dependent.
Along each flow line, however, there is a relation of the form $p=p(\mu)$ --
see Krasinski~(1983)\nocite{Kras83}. The particular models we are using have an
equation of state at the centre of symmetry (or everywhere in the homogeneous
limit~$a\rightarrow 0$) of the (exotic) form $p=-\frac{1}{3}\mu$. The matter
content of these models with regard to the natural (comoving) velocity field is
a perfect fluid with energy density
\be
                       \frac{8\pi G}{c^2}\mu = \frac{3}{R(t)^2},
\label{density}
\ee
and pressure given by
\ba
     p & = & \mu c^2 \left( \frac{2}{3}
             \frac{V(r,t)}{1+\frac{1}{4}\Delta r^2}-1 \right) \label{pressV}\\
       & = & -\frac{1}{3}\mu c^2
              \left(1+\frac{8aR}{c\Delta}\sin^2\frac{\chi}{2}\right),\label{pressW}\\
       & = & -\frac{1}{3}\mu c^2
      \left(1+\sqrt{\frac{24}{\pi
G\mu}}\frac{a}{\Delta}\sin^2\frac{\chi}{2}\right),
\label{pressure}
\ea
where~(\ref{density}) has been used to express the $\ti$-dependence of pressure
(which arises through the appearance of~$R(\ti)$ in~$V(r,\ti)$~-- see
(\ref{Vdef}) and~(\ref{kappa})) in terms of the density. In other words, we
have an `equation of state' of the form
\be
p=-\frac{1}{3}\mu c^2 + \epsilon(\chi)\mu^{1/2}.
\ee
The appearance of $-\frac{1}{3}\mu$ as the dominant contribution to the
equation of state immediately suggests a quintessential or scalar field model
(see Frampton~1999\nocite{framp99}; Liddle~1999\nocite{lid99}; Coble, Dodelson
and Frieman~1997\nocite{coble-97}; Liddle and Scherrer~1998\nocite{lid-sch98}),
although it has been shown (Vilenkin~1981\nocite{Vilenkin}; D\c{a}browski and
Stelmach~1989\nocite{dab-stel89}) that cosmic strings also give rise to this
EOS. The interpretation of the Stephani matter content as a perfect fluid is by
no means required; other interpretations, such as a scalar field, are equally
valid, and the fact that there is no true EOS might even suggest a
two-component interpretation. It is certain from chapter~\ref{EGS
chapter---------------------------------} that they will admit a thermodynamic
scheme. We will discuss this in detail in a future paper. For the moment,
however, this is as far as we will go to provide a physical motivation for the
matter in the D\c{a}browski models: in this paper we are only interested in the
observational consequences of the geometry; although when we impose the energy
conditions below we show that the matter is certainly not obviously unphysical.

We note that the Bianchi identities lead to the following conservation
equations (Ellis~1998; Wainwright and Ellis~1997\nocite{wain-ell97})
\begin{eqnarray*}
\sdel_a\epsilon=-\dot{u}_a(\epsilon+\sfrac{2}{3}) \\
\dot{\mu}=-3H(\epsilon+\sfrac{2}{3})\mu
\end{eqnarray*}
the second of which is exactly the form of the energy conservation equation for
\FRW\ models with a `$\gamma$'~equation of state~-- note also that it shows that
$\dot\mu$ is not homogeneous.

We can see that there are singularities of density and pressure as
$R(\ti)\rightarrow 0$ (ie,~at $\ti=0,-b/a$), which correspond to the big bang
and crunch for these models (the metric becomes singular at these points
too)~-- see D\c{a}browski~(1993)\nocite{dab93}. We can also have a
\emph{finite-density} singularity, where only the pressure becomes singular.
This happens when $r\rightarrow 2/\sqrt{-\Delta}$. Such infinite pressure is
clearly not physical, so we can reject models with~$\Delta<0$. For models
with~$\Delta=0$, it is difficult to compare them directly with $\Delta>0$
models due to the different geometries of the spatial sections, so we will not
consider them in this thesis, and we are left with $\Delta>0$, ie,~$b<1$~--
see~(\ref{Delta}). As we explained in~\S\ref{Stephani} the natural assumption
that $R(\ti)>0$ after the big bang ensures that $b\ge 0$ ($\Delta\le 1$).

Having calculated the pressure and density of the fluid we can now investigate
its physical viability through the energy conditions. It is more convenient to
use the original stereographic coordinates for this (ie,~expression
(\ref{pressV}) for the pressure), since we wish to consider all values
of~$\Delta$, until we find reasons to the contrary. The weak energy condition
states that $\mu\ge 0$ and $p+\mu c^2\ge 0$, whereas the strong energy
condition is equivalent to $3p+\mu c^2\ge 0$ (see, for example,
Wald~1984\nocite{wald} for a discussion). The weak energy condition does not
constrain our models at all: it implies that $V\ge 0$, but this is always true
since~$V\rightarrow 0$ only at spatial (and temporal) infinity (even though the
coordinates themselves may be finite). The strong energy condition, however,
implies that $\kappa(\ti)\ge\Delta$ for all~$\ti$. From~(\ref{kappa}) we can
see that this is equivalent to~$a\le 0$ (since~$R>0$), so the models must have
a big crunch ($R(\ti)$~is an `upside-down' quadratic).

The dominant energy condition is more interesting. It states that $|p|\le\mu
c^2$, from which (\ref{density}) and~(\ref{pressure}) immediately give
\[
0\le\frac{1}{3}\frac{V(r,\ti)}{1+\frac{1}{4}\Delta r^2}\le 1.
\]
The left inequality requires only that $\Delta\ge 0$, which rules out
infinities in the pressure (finite-density singularities). The inequality on
the right says that for all $\ti$ and~$r$
\be
(\kappa(\ti)-3\Delta)r^2\le 8
\label{domc}
\ee
must hold. This condition is always true for~$a\ge 0$ (see~(\ref{kappa})), as
long as~$\Delta\ge 0$. When~$a<0$, $r$~is unbounded (for~$\Delta\ge 0$, because
$\kappa$~is then positive~-- see~(\ref{Vdef})), so the left hand side
of~(\ref{domc}) must always be negative, ie,~$\kappa(\ti)\le 3\Delta$. It is
easy to show that $R(\ti)/c\le -b^2/4a$, so, from~(\ref{kappa}),
\[
        \kappa(\ti) \le 1.
\]
Then $\kappa\le 3\Delta$ for all~$\ti$ provided
\be
          b\le \sqrt{\frac{2}{3}}\approx 0.82.
\label{bdom}
\ee
For~$b$ larger than this the dominant energy condition will be broken at some
time, in regions of the universe at large~$r$. We will not consider further the
intricacies of this. A glance at the exclusion diagrams, figures
\ref{exclusion_H=50} and~\ref{exclusion_H=70}, shows that~(\ref{bdom}) does not
eliminate a significant area of the allowed region. In light of this we will,
for the moment, overlook~(\ref{bdom}) and investigate the properties of all
models with~$0<b<1$.

To summarise: we have used basic physical requirements, such as the occurrence
of a big bang and the avoidance of pressure singularities, and energy
conditions to restrict the ranges that the two parameters $a$ and~$b$
(or~$\Delta$) can take. The results are:
\be
a\le 0,\pha 0<b<1 \pha \hbox{(ie, $0<\Delta<1$)}\label{paramlims}
\ee
(we reject~$b=1$ for simplicity, as explained above, and we refrain from
invoking~(\ref{bdom}) until~\S\ref{conclusions}). Of course, we are not forced
to accept the energy conditions. Although they seem physically very reasonable
conditions to impose on any form of matter there are many examples of
cosmological models based on matter that does not satisfy them (quantum fields,
for example, can exhibit negative energy density). However, we will assume
their validity and in most of what follows we only investigate the properties
of the models satisfying~(\ref{paramlims}).

\section{Constraining the Model Parameters Using Observations.}
\label{obs-constr-sss}

So far we have considered only the `global' physical properties of
D\c{a}browski models, but to really assess their potential viability as
cosmological models it is necessary to confront them with observations. In this
section we derive the distance-redshift relations that form the basis of the
classical cosmological tests and compare them with available observational
constraints to see whether any regions of parameter space are capable of
providing a fit. We will impose constraints on the value of~$H_0$, age, size
(the meaning of which will be explained below) and the anisotropy of the
microwave background, leaving the wealth of data available from galaxy surveys
and high-redshift supernovae for consideration in a future paper; the
complexities involved in interpreting such data and applying it to idealised
cosmological models require separate treatment.

\forget{
Deriving the observational relations (redshift, angular size or area distance,
luminosity distance and number counts) means relating the coordinates and
metric functions to observable quantities. This requires knowledge of the
observer's motion (4-velocity), which can, strictly speaking, be specified
independently of the background geometry. However, the D\c{a}browski models
contain perfect fluid, so we will identify the observer's motion with the fluid
velocity. We are not obliged to do this, and, given the strange form of the
matter, it might be thought advantageous to instead assume that observers
(ie,~galaxies) constitute a dust-like test fluid moving freely through the
spacetime whose geometry is determined by the exotic matter. It will become
clear in~\S\ref{CMB} that if we were to make this assumption a large dipole
anisotropy in the CMB would result (although the dipole in~$H_0$ would be
eliminated~-- see~(\ref{specific H_0}) with~$\dot{u}=0$) because such a flow
will, in general, have a significant velocity relative to the D\c{a}browski
fluid flow, which, it will turn out, is very nearly in the rest frame of the
CMB everywhere. }

\forget{
Probably the most important distance measure in cosmology is redshift. In
general, it is no simple task to find analytic expressions for the redshift in
any cosmological model; derivations usually rely on symmetries of the spacetime
or other simplifying factors to solve the equations of null geodesics. Here we
take advantage of the conformal flatness of Stephani models (strictly speaking,
of the fact that the D\c{a}browski models are manifestly conformal to
\FRW\ cosmologies), although we can also derive the redshift formula as a time
dilation effect. These procedures are outlined in appendix~\ref{appen}. Using
(\ref{redshift}) and~(\ref{metric_W}) we find }

To recap from chapter~\ref{step chapter obs derivations----------------}, when
the change of coordinates is made to an arbitrary location, we have the
metric~(\ref{metric_W2}) with $W$ restricted to
\be
W\rightarrow W(\chi,\psi,\vartheta;\ti)=1-\frac{2aR(\ti)}{c\Delta} (1-
               \cos\psi \cos\chi + \sin\psi \sin\chi \cos\vartheta).
\label{new W2}
\ee

Using the expression for the lookback time~(\ref{lookback}) and the expression
for the redshift~(\ref{redshift} and our simplified expression for
$W$~(\ref{new W}) we find
\ba
1+z(\psi,\chi,\theta)\li=\li\frac{R_0}{W_0}\frac{W(\psi,\chi,\theta;\ti)}
{R(\ti)}\nonumber\\
\li=\li\frac{R_0}{W_0}\left\{\frac{1}{R(\ti)}-
\frac{2a}{c\Delta}\left(1-\cos\psi\cos\chi\right) -
\frac{2a}{c\Delta}\sin\psi\sin\chi\cos\vartheta\right\};
\label{zdipole}
\ea
where $R_0=R(\T)$, $W_0=W(\psi,\T)$, showing that, for objects at any
fixed~$\chi$, the inhomogeneity of universe manifests itself in the redshift as
a pure dipole in angle around the sky ($\cos\vartheta$~term). This will be
important in~\S\ref{CMB}.

For metrics with spherical symmetry about the observer the angular size (and
area) distance is given directly from the coefficient in front of the angular
part of the metric, because symmetry ensures that for radial rays $\vartheta$
and~$\varphi$ are constant along the trajectory. For our models we do not have
spherical symmetry about every observer, but the metric is everywhere conformal
to a spherically symmetric metric, as can be seen from~(\ref{metric_W}). Since
null rays are not affected by the conformal factor they also remain at fixed
$\vartheta$ and~$\varphi$, so we can again obtain the angular size
distance,~$r_A$, from the coefficient of the angular part of the metric:
\be
r_A(\psi,\chi,\vartheta)=\frac{R(\ti)}{\sqrt{\Delta}W(\psi,\chi,\vartheta;\ti)
}|\sin\chi|
\label{area dist-new2}
\ee
(again, $\chi$ and~$\ti$ are related by~(\ref{lookback})). We can find, for the
first time, the \emph{exact} angular size distance relation parametrically by
combining equations~(\ref{area dist-new}) and~(\ref{redshift_new}), which
bypasses the rather cumbersome method of Kristian and
Sachs~(1966)\nocite{kris-sac66}. This is valid for \emph{any} null-connected
points in the spacetime.

Luminosity distance,~$r_L$, is related to~$r_A$ by the reciprocity theorem:
\be
r_L=(1+z)^2r_A,
\label{recip2}
\ee
see MacCallum and Ellis~(1970)\nocite{Mac-Ellis-I}\nocite{Mac-Ellis-II} and
Ellis~(1998)\nocite{ell98}. This then allows the magnitude-redshift relation to
be determined in the usual way: the apparent magnitude~$m$ of an object of
absolute magnitude~$M$ is given in terms of the luminosity distance by
\be
          m-M-25 = 5\log_{10} r_L.
\label{pogson3}
\ee

\forget{
Finally, for completeness, we discuss the number count-redshift relation,
although we do not use it here. The only important consideration is the
identification of some sensible comoving number density
distribution,~$n_c(\chi)$, for the observers (fluid particles). This is not
necessary for homogeneous spacetimes because it is natural to simply take~$n_c$
also to be homogeneous (and independent of time, assuming no evolution).
Furthermore, when there is fluid pressure $n_c$ cannot be directly related to
the energy density of the matter, as is possible with a dust, and, in general,
there need be no obvious choice for~$n_c$~-- the only constraint it must
satisfy (again assuming no number evolution, such as would be caused by
mergers, for example) is particle number conservation:
\[
                       \del_a (n_c u^a )= 0.
\]
Since the metric~(\ref{metric_W}) is in comoving coordinates we can satisfy
this by simply assuming that $n_c$~is independent of time~$\ti$, being given by
whatever distribution we specify at~$\ti=0$. Moreover, the fact that the
D\c{a}browski models become homogeneous ($W\rightarrow 1$) as~$\ti\rightarrow
0$ means that it makes sense to define $n_c$ to be independent of spatial
position too. We are then back to the same situation as with \FRW\ models,
except that now observations of galaxy numbers constrain comoving volume, but
not \emph{proper} volume (note that the proper density of particles is
\emph{not} constant, because the conformal factor modifies the proper volume by
different amounts at different positions and times), and so are less directly a
constraint on the radial component of the metric (which determines proper
distance). The number of particles within some radius~$\chi$ of any observer is
simply proportional to the volume of the \FRW\ 3-sphere of curvature~$\Delta$
out to that radius:
\[
       N(\chi) \propto \frac{4\pi}{\Delta^{3/2}}\int_0^\chi \sin^2\chi' d\chi'
                    = \frac{2\pi}{\Delta^{3/2}}\left(\chi-\sin\chi \cos\chi\right)
\]
exactly as in \FRW\ models. The difference from the \FRW\ $N-z$ relation comes
when we use~(\ref{redshift_new}) to relate number counts to redshift: $z(\chi)$
will be different to that for \FRW\ spacetimes. Nevertheless, it is clear from
this discussion that the fluid particles (observers) in the D\c{a}browski
spacetime are distributed \emph{uniformly with respect to the \FRW\ volume on
the spatial sections}. Since these hypersurfaces of constant time are just
spheres in the \FRW\ metric we know that the Copernican principle can easily be
applied: the probability that an observer lies in some region of space is
exactly the volume of that region \emph{with respect to the \FRW\ volume}
(divided by the total volume of the sphere).}

The task now is to limit $a$, $b$, and~$\T$ using present observational
constraints. A full discussion of each constraint is made in the following
sections, and they are followed by exclusion diagrams showing the regions of
parameter space for which $a$ and~$b$ give a plausible cosmological model for
\emph{all} observer locations in these models.

\subsection{Hubble's Constant.}

The expansion rate of the universe has been measured with reasonable accuracy.
Hubble's constant is believed to lie in the range $50\lapp H_0\lapp 80$~\hu,
and we will use these limits to constrain the D\c{a}browski models. The Hubble
\emph{parameter} for these models is independent of
position~(\ref{expansion-IRF-gen}):
\be
H\eqdef\frac{\theta}{3}=\frac{{R}(\ti)_{,t}}{R(\ti)}\phan\ \Longrightarrow
\phan\
      H_0=\frac{{R}_{,t}(\T)}{R(\T)}
\label{Hubble parameter}.
\ee
We can use this to place constraints on the time at which observations can be
made at any position: for our models~$H$ decreases monotonically, so it will
only lie in the observed range of~$H_0$ for some range of~$T$. For any observer
with coordinate age~$\T$, we require
\be
               50\lapp\frac{2a\T+b}{a\T^2+b\T}\lapp 80.
\label{minimum H0}
\ee
Usually, for simplicity, we will choose a specific value for~$H_0$ (almost
invariably that which produces the `worst case'). Then we can
solve~(\ref{Hubble parameter}) for~$T$.

However, when $H_0$ is actually \emph{measured}, it is not necessarily equal to
the expansion rate. What is measured in practice is the lowest order term in
the
\mofz\ relation, which gives the measured Hubble's constant, $H^m_0$:
\be
H^m_0 = \left.\frac{k^a k^b \del_a u_b}{(u_ck^c)^2}\right|_{0}
\label{measured H_0}
\ee
(where~$k^a$ denotes the wave-vector of the incoming photons); see MacCallum
and Ellis~(1970). Equivalently, we can consider the gradient of the
redshift-area distance curve at the observer (\emph{cf}.~Humphreys
\etal~1997\nocite{humph-97}). The covariant derivative of the velocity field
that appears in~(\ref{measured H_0}) can be decomposed in terms of the
expansion, acceleration, shear and rotation of the flow. In the case of the
Stephani models these last two are zero, so
\be
\del_b u_a = H h_{ab}-\dot{u}_a u_b.
\ee
Therefore the measurement of $H^m_0$ depends upon the \emph{acceleration} of
the observer. If we measure the magnitude-redshift relation for objects in some
direction, then, comparing equation~(\ref{measured H_0}) with the expansion
given by~(\ref{Hubble parameter}), we find
\be
H^m_0(\theta) = H_0-\frac{\dot{u}}{c}\cos\theta,
\label{specific H_0}
\ee
where $\dot{u}$ is the acceleration scalar and~$\theta$ is the angle between
the acceleration vector and the direction of observation (which is opposite to
the direction in which the photons are travelling). Since the acceleration is
non-zero in the Stephani models there will be a dipole moment in~$H^m_0$. The
size of this in the D\c{a}browski models is given directly
from~(\ref{accscalar}). If this is large in any model we can probably reject
that model because a large dipole moment in~$H_0$ is not observed. However,
nearby it is difficult to measure~$H_0$ accurately due to peculiar motions and
the discreteness of galaxies. There \emph{is} a dipole moment in observations
of more distant objects, which is assumed to be due to the fact that the Local
Group is falling into the potential well produced by Virgo and the Great
Attractor. The question is: what upper bound can be placed on the acceleration
by observations? This issue will be discussed in~\S\ref{local-dipole}. For now
we simply use~(\ref{minimum H0}) to constrain the epoch of observation,~$\T$.

\subsubsection{Hubble Normalised Scalars}
\label{Hubble Normalised Scalars}

If we wish to define, in analogy with FLRW models, \emp{Hubble normalised
scalars}, then we can do so as follows. We define
\ba
\Omega\li=\li\;{24\pi G\mu }{\theta^2}=(2at+b)^{-2},\\
\Omega_p\li=\li\;{24\pi G p}{c^2\theta^2}=-\;13\Omega\left(1+\;{8aR}{c\Delta}\sin^2\;\chi2\right).
\ea
We can get a fairly good idea how these models behave, in comparison with what
is know from local observations. We recall that
\be
0.3\lapp\Omega_0\lapp1,~~~0\lapp\Omega_{\Lambda_0}\lapp1
\ee
for the present-day values of the density parameter and cosmological constant.
For the \dab\ models, we note that in the limit $t\rightarrow0$ we have
\be
\Omega=\;{1}{b^2},~~~\Omega_{,t}=-\;{4a}{b^3}.
\ee
We know that $b$ is positive in order that the scale factor is positive, which
implies that $\Omega_{,t}$ is negative at the big bang only if $a>0$. However
we also see that as $t\rightarrow0$, then $\Omega$ approaches some value larger
than $1$ if $b<1$, as required by neglecting models with a finite density
singularity. This means that if we desire $\Omega\lapp1$ today, then we force
$a>0$, in contradiction with the strong and dominant energy conditions (but not
the weak one).

As far as the quantity `$\Omega_p$' is concerned, we may `equate' it with the
cosmological constant term of the standard model. If we consider the expression
for $q$, the deceleration parameter, given by equation~(\ref{q-from
raychaudhuri}) (which is essentially the Raychaudhuri equation) then the
standard model has
\be
q=\:12\Omega-\Omega_\Lambda
\ee
while for the \dab\ models we may write
\be
q\approx \:12\Omega+\:12\Omega_p
\ee
neglecting the acceleration terms (which are zero at the center). We may
therefore identify
\be
\Omega_p\simeq-2\Omega_\Lambda.
\ee
Thus we may equate a positive cosmological constant with a negative pressure
parameter. If we require this for all locations today (which means that $R$ is
quite small) then the sign of $a$ doesn't matter: however, there will be a time
(at $t=-b/2a+\sqrt{3b^2-1}/2\sqrt2a$) when the sign of $\Omega_p$ changes if
$a$ is negative. Of course we should take into account acceleration in the
`definition' of $\Omega_\Lambda$, but this is only a qualitative discussion.

What is interesting is that while the strong and dominant energy conditions
favour $a<0$, this seems to be ruled out by current measurements of the matter
density of the universe.

This may give cause for concern for these models, but it does not really
concern us here. We will carry on with the analysis of models with $a<0$
throught this thesis, because we are more concerned with the Copernican
principle in a non-standard cosmology, rather than finding the `best-fit' IRF
model (although, it will be shown in
\S\ref{resolution} that SNIa data favour $a>0$ too).
This will be discussed later.

\subsection{The Age of the Universe.}
\label{Age}

The original inspiration for D\c{a}browski and Hendry~(1998) to study Stephani
models was the potential resolution of the age problem that they provided,
which at that time seemed to be virtually insurmountable within the framework
of
\FRW\ models (even when a non-zero cosmological constant was invoked): the high
measured value of~$H_0$ suggested an age,~$\tau_0$, of at most about $11$~Gyr
for a
\FRW\ cosmology, whereas globular cluster ages were thought to be at least
 $12$-$13$~Gyr
(Hendry and Tayler,~1996; Chaboyer \etal~1998\nocite{chab-97}). D\c{a}browski
and Hendry~(1998) and Paper~I showed that, for the particular Stephani models
they considered, this apparent paradox disappears: the D\c{a}browski models
have ages that are consistently 1-4~Gyr older than their \FRW\ counterparts
(for an observer at the centre of symmetry, at least). However, the age problem
has recently been alleviated by a recalibration of the RR Lyrae distance scale
and globular cluster ages in the light of Hipparcos (Chaboyer
\etal~1997\nocite{chab-97}), which has reduced the globular cluster ages
considerably, to~$\sim 10$~Gyr. The fit is still marginal, but the new ages are
generally accepted as they allow a flat \FRW\ model to fit the observations
provided that $H_0\lapp 67$~\hu.

We certainly require, then, that at the epoch of observation our models are
older than~$10$~Gyr. However, we will also consider the stronger
constraint~$\tau_0>12$~Gyr, partly to be conservative, but also because the
diagrams for the $12$~Gyr constraint are often clearer. Since it was shown in
Paper~I that best fit D\c{a}browski models are significantly older than their
\FRW\ counterparts, we do not expect problems from this constraint. There is a
surprise in store, though, when we consider non-central observers, and age will
turn out to be the dominant constraint on these models.

The age of the universe according to an observer at position~$\psi$ and at
coordinate time~$\T$ is simply the proper time elapsed from the big bang
($\ti=0$):
\forget{
\ba
\tau_0 \li = \li \int_0^{\T}\frac{d\ti}{W(\psi,\ti)} \nonumber\\
       \li = \li \frac{\Delta}{4|a|\sin\frac{\psi}{2}\sqrt{1-b^2\cos^2\frac{\psi}{2}}}
       \ln\left\{
\frac{\Delta-\left(b\sin\frac{\psi}{2}+\sqrt{1-b^2\cos^2\frac{\psi}{2}}\right)
      2a\T\sin\frac{\psi}{2}}
     {\Delta-\left(b\sin\frac{\psi}{2}-\sqrt{1-b^2\cos^2\frac{\psi}{2}}\right)
      2a\T\sin\frac{\psi}{2}}
          \right\}
\label{proper age}
\ea}
\ba
\tau_0 \li = \li \int_0^{\T}\frac{d\ti}{W(\psi,\ti)} \nonumber\\
       \li = \li \frac{\Delta}{4|a|\sin\frac{\psi}{2}\sqrt{1-b^2\cos^2\frac{\psi}{2}}}
       \ln\left\{
\frac{\Delta-\Sigma_+}
     {\Delta-\Sigma_-}
          \right\}
\label{proper age}
\ea
where
\[
\Sigma_\pm=\left(b\sin\frac{\psi}{2}\pm\sqrt{1-b^2\cos^2\frac{\psi}{2}}\right)
      2a\T\sin\frac{\psi}{2}
\]
(for~$a\ne 0$ and~$\psi\ne 0$: otherwise $W\equiv 1$ and~$\tau_0=\T$), where we
take the value of~$\T$ given by the solution of equation~(\ref{Hubble
parameter}) as our constraint on the coordinate time for any specific~$H_0$. In
figure~\ref{3D proper age a-b 12Gyr cont}
\begin{figure}[here!]
\centerline{\psfig{file={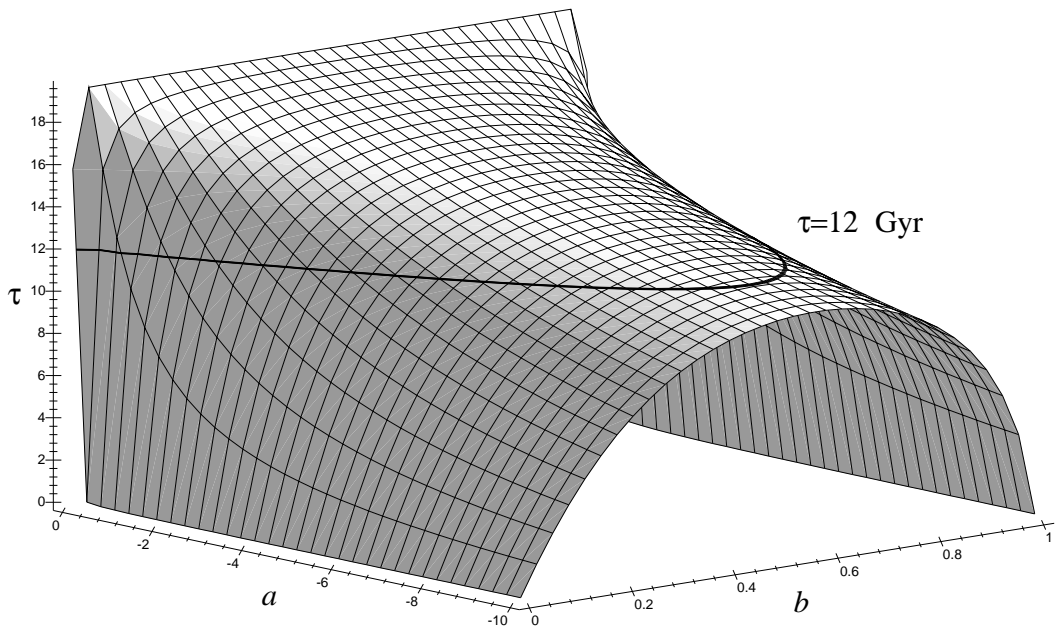},width=10cm,angle=90}}
\caption{\small  Surface plot of the proper age of the universe  for $H_0=50$~\hu\
with the observer at $\psi=\pi$. \label{3D proper age a-b 12Gyr cont} }
\end{figure}%
we show proper age for observers at $\psi=\pi$ with $H_0=50$~\hu\ (which shows
the behaviour of the age function most effectively) as $a$ and~$b$ vary. The
contour at~$\tau_0=12$~Gyr is also drawn. As can be seen, below this curve the
values of $a$ and~$b$ can be rejected~-- the proper age is too low.

In figure~\ref{exclusion-age-noncentral}
\begin{figure}[here!]
\centerline{\psfig{file={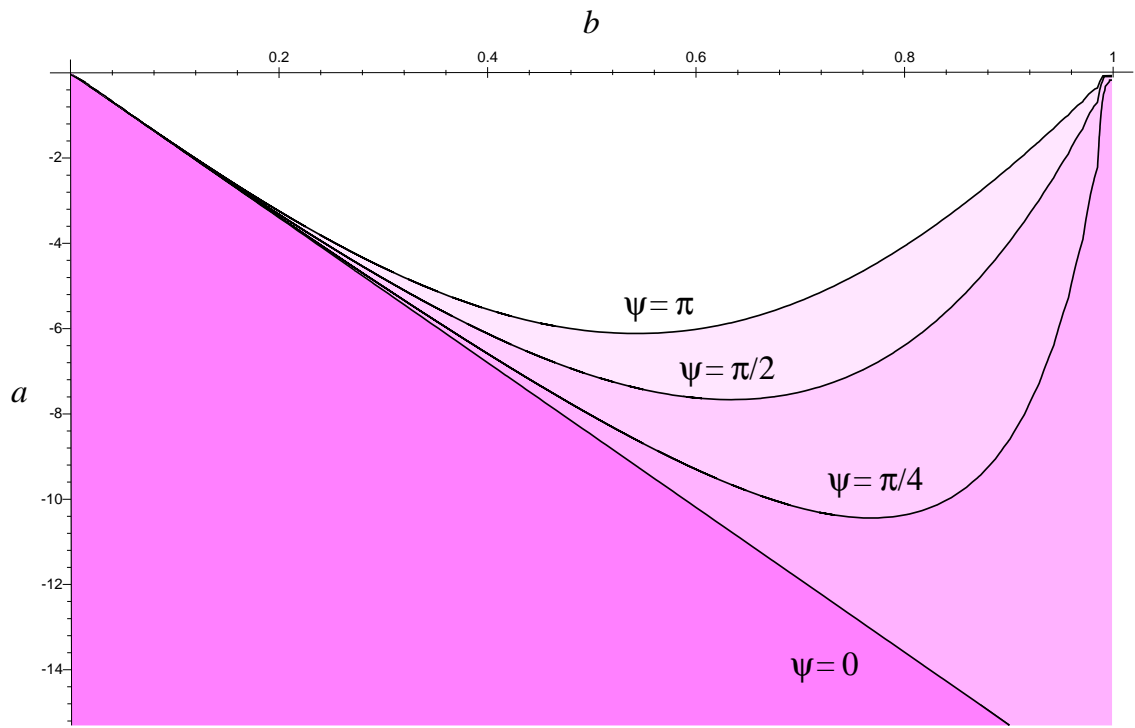},width=10cm,angle=90}}
\caption{\small $\tau_0=12$~Gyr exclusion plot for different observer positions, for
$H_0=60$~\hu. In the shaded areas the models are not old enough. $\psi=\pi$ is
clearly the most restrictive case.
\label{exclusion-age-noncentral} }
\end{figure}%
the $\tau_0=12$~Gyr contours of the proper age function are plotted
for~$H_0=60$~\hu\ and for several observer positions,~$\psi$, showing how the
age of an observer varies with~$\psi$ for different parameters $a$ and~$b$. The
shaded regions contain models that are less than $12$~Gyr old for at least one
of the observer positions. This demonstrates that the proper age of an observer
is smallest at the antipodal centre of symmetry,~$\psi=\pi$, so the contour at
$\tau_0=12$~Gyr in figure~\ref{3D proper age a-b 12Gyr cont} marks the limit of
viable models for this age constraint. Consequently, we will always use proper
age at~$\psi=\pi$ to constrain the model parameters. We could weaken this
constraint by requiring only that \emph{most} observers are old enough, which
would allow us to consider instead the age of observers at~$\psi\le\pi/2$ while
still satisfying the Copernican principle (half of the observers would lie in
this region). For simplicity, though, we will not do this here.

Finally, in figure~\ref{exclusion_age_T=10}
\begin{figure}[here!]
\centerline{\psfig{file={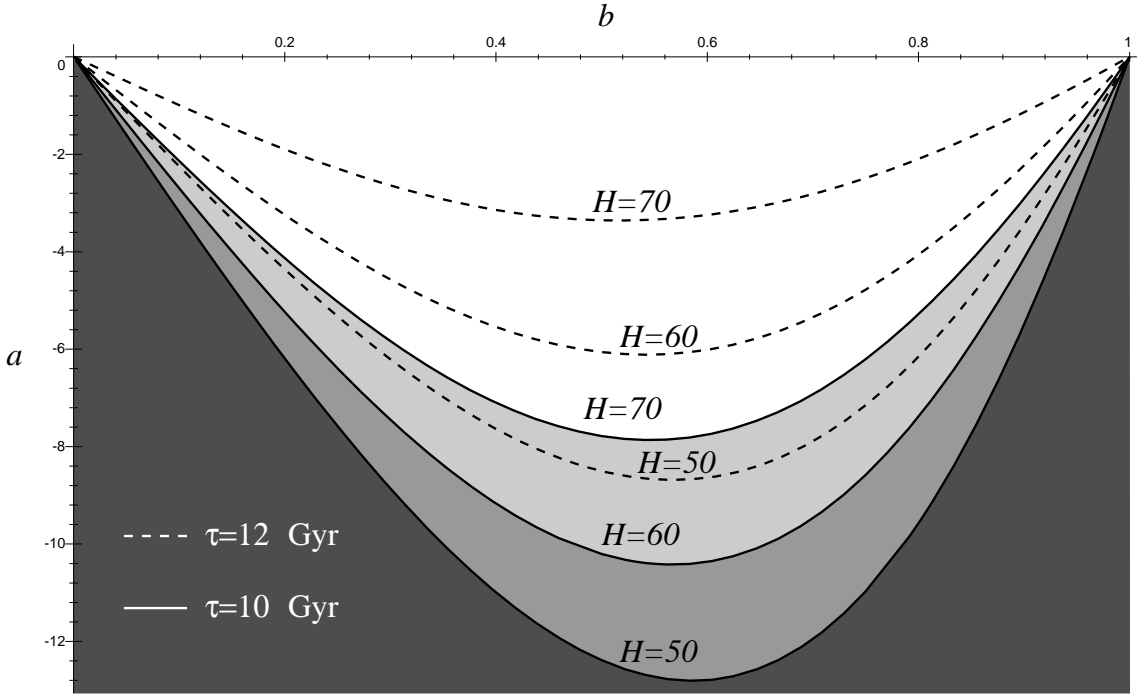},width=10cm,angle=90}}
\caption{\small The age exclusion diagram for various~$H_0$ and proper
age~$\tau=10$~Gyr. The shaded region represents the prohibited area. Also shown
as dashed lines are the age limits for~$\tau=12$~Gyr. (Note that the region
excluded for~$H_0=70$~\hu\ contains the excluded regions for lower~$H_0$~-- the
progressively darker shading indicates this.)
\label{exclusion_age_T=10} }
\end{figure}%
we show the age exclusion plot for the models (based on proper age
at~$\psi=\pi$). We use three values of $H_0$: $50$, $60$, and $70$~\hu, and a
proper age of~$10$~Gyr (although the limits for~$12$~Gyr are also indicated).
The shaded regions are excluded. It can be seen that unless we require the
universe to be particularly old or the expansion rate high there is still a
significant region of parameter space that cannot be excluded on the basis of
age.

It should be noted here that these plots are meaningless for~$b=1$, because
then~$\Delta=0$ and the model is conformal to an \FRW\ model with \emph{flat}
spatial sections, for which~$\chi$ is not a good coordinate~-- as can easily be
seen from~(\ref{r_chi}).

\subsection{Size and the Distance-Redshift Relation.}
\label{size}

When the spatial sections of a cosmological model are 3-spheres it may be
possible for light rays to circle the entire universe, perhaps a large number
of times. This is indeed the case for the models we are considering here. What
will the signature of this be in the various distance-redshift relations? For
the D\c{a}browski models this is a particularly easy question to answer because
the fact that they are manifestly conformal to \FRW\ models allows us to
determine the paths of light rays directly from the null geodesics of the
underlying \FRW\ space. From this it is clear that light rays from a point
directly opposite the observer (ie,~from the antipode,~$\chi=\pi$) will spread
out around the universe isotropically from the antipode until they pass the
`equator' ($\chi=\pi/2$), where they will begin to converge and be focused onto
the observer. As a result, a point source positioned exactly at the antipode
will fill the entire sky when seen by the observer, so that its angular size
distance,~$r_A =$ (physical length/apparent diameter), is zero. Similarly, the
refocusing of light onto a point produces an infinite \emph{flux} at the
observer, and therefore the luminosity distance,~$r_L$, is also zero
($m\sim\log_{10}r_L=-\infty$). Precisely the same argument applies to light
that leaves the observer's position, travels through the antipode (where it
will be focused to a point), and returns to the observer (being focused a
second time). In fact, it is obvious that whenever the light rays travel
through a parameter distance~$\chi$ that is an exact multiple of~$\pi$,
$r_A=r_L=0$: this is reflected by the factor of $\sin\chi$ in~(\ref{area
dist-new}).

This effect can be seen clearly in figures~\ref{r_A_two_b}-\ref{r_L_z_two_b},
where we show the two principle measures of distance as they vary both with
coordinate distance $\chi$ and redshift. Viewed as a function of~$\chi$, in
figure~\ref{r_A_two_b},
\begin{figure}[t!]
\centerline{\psfig{file={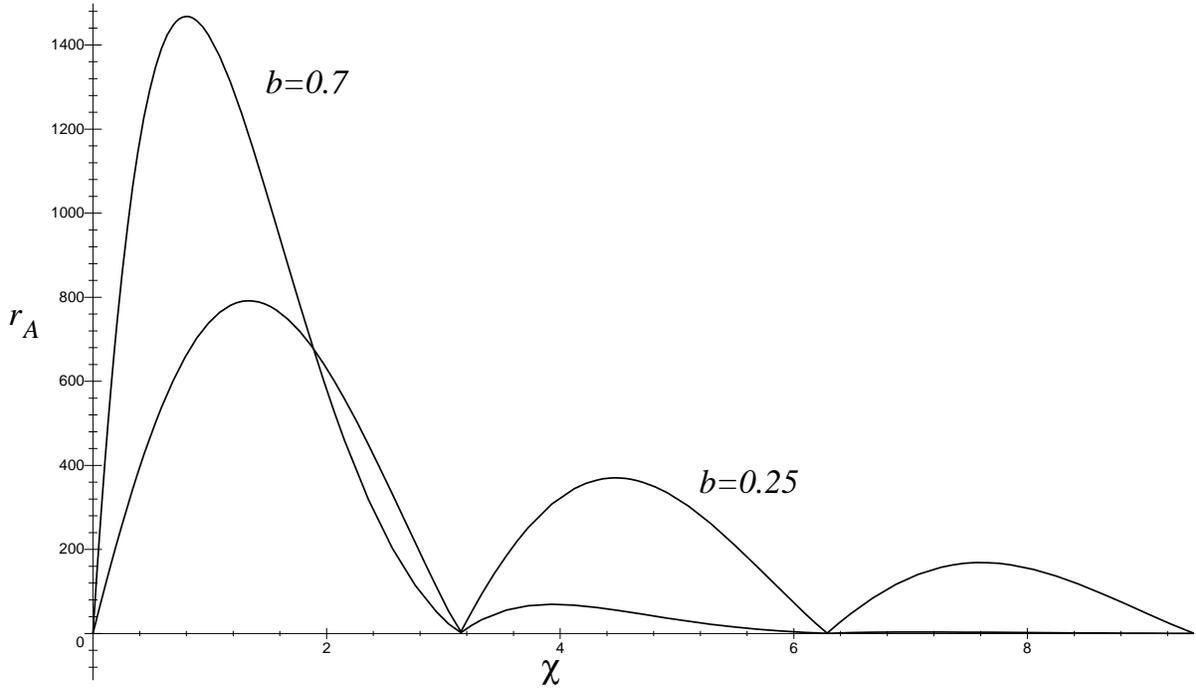},width=10cm,angle=90}}
\caption{\small
 Area distance from the centre as a function of~$\chi$ for two values of~$b$.
$\T=15$~Gyr,~$a=-1$. \label{r_A_two_b}}
\end{figure}%
the zeros of the angular diameter distance clearly occur at multiples of~$\pi$
for all model parameters. Looked at in terms of redshift, though
(figure~\ref{r_A_z_two_b}),
\begin{figure}[here!]
\centerline{\psfig{file={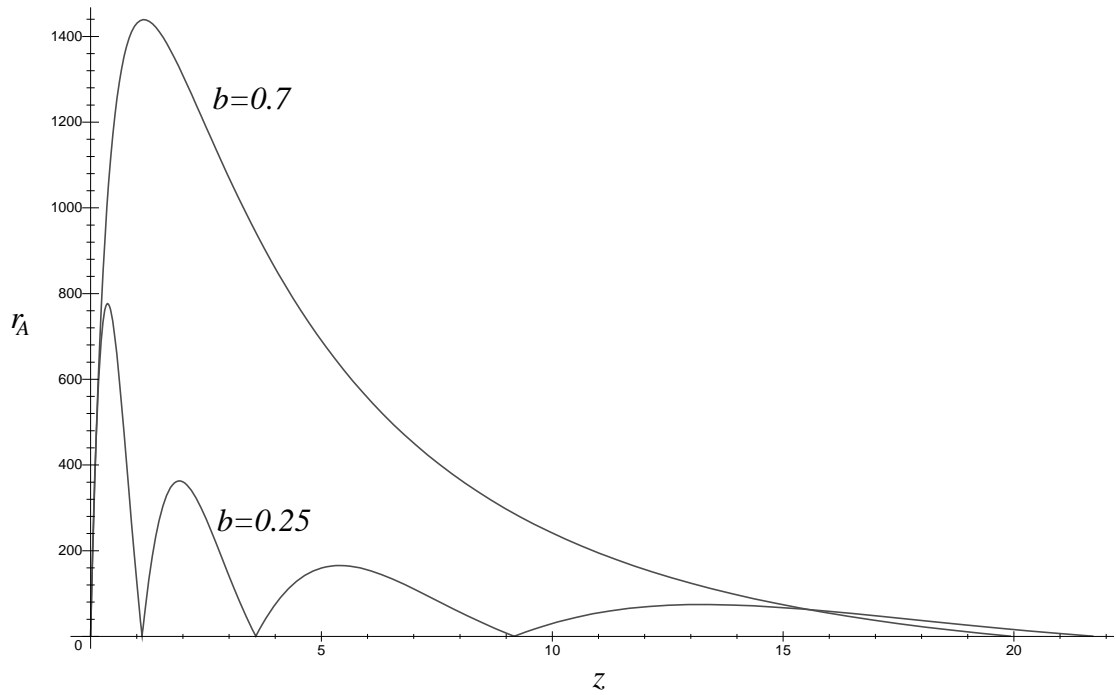},width=10cm,angle=90}}
\caption{\small
Area distance from the centre as a function of redshift for the same parameters
as figure~\ref{r_A_two_b}. For small~$b$ the angular size distance oscillates
far too rapidly.
\label{r_A_z_two_b}}
\end{figure}%
\forget{
\begin{figure}[here!]
\centerline{\psfig{file={E:/data/tex/thesis/area_dist_red_zeta_0pm70_cent.ps},width=\textwidth,angle=0}}
\caption{\small
Area distance from the `centre' as a function of redshift for $a=-10,~b=0.5$,
and $T=15$Gyr, with the value of $\zeta$ to be read as $\zeta/c$. This plot
explicitly demonstrates that a large value of $\zeta$ is quite unacceptable due
to the large blueshifts nearby because we note that in~(\ref{conf-W-general})
replacing $\zeta\rightarrow-\zeta$ is exactly the same as the identification
$\vartheta=0\rightarrow\vartheta=Pi$; ie, an observer would see large
bleushifts in one direction, and redshifts in the other.
\label{r_A_z_two_zeta}}
\end{figure}%
\begin{figure}[here!]
\centerline{\psfig{file={E:/data/tex/thesis/m-z-zeta=70_opposite_dirction.ps},width=\textwidth,angle=0}}
\caption{\small
Apparent magnitude from the `centre' as a function of redshift for
$a=-10,~b=0.5$, and $T=15$Gyr, with the value of $\zeta$ to be read as
$\zeta/c$. The two lines represent the \mofz\ relation in opposite directions
in the sky. We see clearly from this that the situation
in~Fig.~\ref{r_A_z_two_zeta} is completely ridiculous.
\label{r_A_z_two_zeta}}
\end{figure}%
}
it is clear that for small~$b$ the zeros are much closer together than for
larger~$b$, with the first zero occurring at~$z\approx 1$ for~$b=0.25$.
Figure~\ref{r_L_z_two_b}
\begin{figure}[here!]
\centerline{\psfig{file={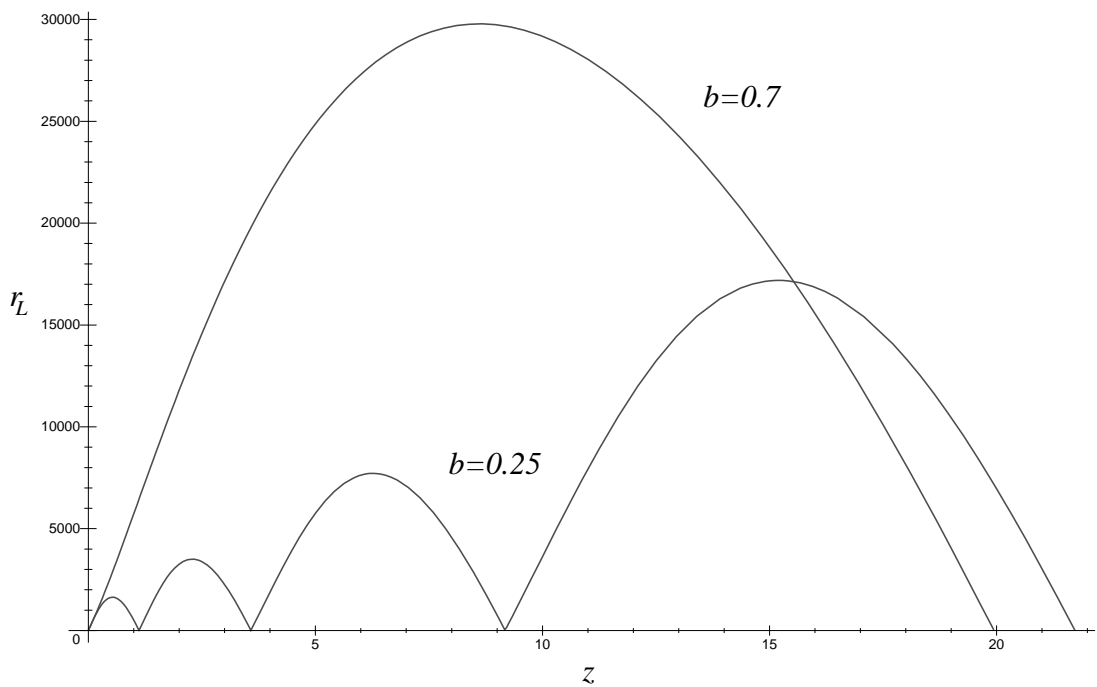},width=10cm,angle=90}}
\caption{\small
Luminosity distance from the centre as a function of redshift for the same
parameters as figures \ref{r_A_two_b} and~\ref{r_A_z_two_b}.
\label{r_L_z_two_b}}
\end{figure}
shows the luminosity distance-redshift relation, for comparison. These effects
are not as unusual as they look, and can be found also in \FRW\ geometries for
models with positive~$\Lambda$~-- see~\S4.6.1 in Ellis~(1998)\nocite{ell98} and
references therein. It is noteworthy that a blackbody situated at the opposite
pole to the observer would look exactly like the CMB if it was at the right
redshift. For example, a star of surface temperature 3000K at a redshift
$z=1000$ would have an apparent temperature of 3K, and would cover the whole
sky.

Can we rule out such apparently aberrant behaviour? Theories of structure
formation are fairly well developed (see Bertschinger
\etal~1997\nocite{bert-97} for a thorough discussion and references), and the
evolution of galaxies and the star formation rate~(SFR), while not accurately
known, are at least qualitatively understood. In particular, the SFR, which is
very important for determining the luminosity of distant, young galaxies, is
believed to fall off beyond~$z\sim 3$ (see, for example,
Loeb,~1999\nocite{loe99}). As a result, one could argue that there will be
relatively few bright objects beyond some redshift~$z_{SF}$ that corresponds to
the epoch at which galaxies `turned on' and the SFR began to increase
significantly. This would mean that the zeros in the distance-redshift
relations would be essentially unobservable if they occurred at redshifts
larger than~$z_{SF}$ because there would be no luminous objects to be seen
magnified in the sky, whereas if the zeros occurred at lower redshifts
than~$z_{SF}$ one could reasonably argue that there ought to be some signature
of this in the observations. Since galaxies have only been observed (in the
Hubble Deep Field, for example) with redshifts up to~$z\sim 5$ we take
$z_{SF}=5$ (although the constraints imposed by larger~$z_{SF}$ can easily be
inferred from the exclusion diagram, figure~\ref{rA_exclusion}). If these
arguments are not completely convincing, then, at a simpler level, the fact
that the \emph{observed} magnitude-redshift relation is known accurately out
to~$z\sim 1$ from type~Ia supernovae (Perlmutter~1999), and is certainly not
dipping down, allows us to say that there is no zero of luminosity distance
below~$z\sim 2$, say. We therefore also consider the constraint that results
from requiring that there are no zeros below~$z=2$.

We wish, then, to constrain the parameters of our models by rejecting any
models for which the first zero in the distance-redshift relations occurs
at~$z\le z_\pi$, where $z_\pi=2$ or~$z_\pi = z_{SF}=5$.
Using~(\ref{redshift_new}), this means
\be
  1+z(\chi=\pi)=\frac{R_0}{W_0}\frac{W(\psi,\pi,\ti_\pi)}{R(\ti_\pi)}>1+z_\pi,
\label{redshift-rA=0}
\ee
where $R_0=R(\T)$, $W_0=W(\psi,\T)$ (giving the conformal factor at the
observer) and $\ti_\pi$ denotes the lookback time~(\ref{lookback})
at~$\chi=\pi$. Again we determine the epoch of observation (ie,~the observer's
coordinate time~$\T$) using~(\ref{Hubble parameter}). The solution
of~(\ref{redshift-rA=0}) for $a$ and~$b$ is shown in figure~\ref{rA_exclusion}
as an exclusion diagram.
\begin{figure}[here!]
\centerline{\psfig{file={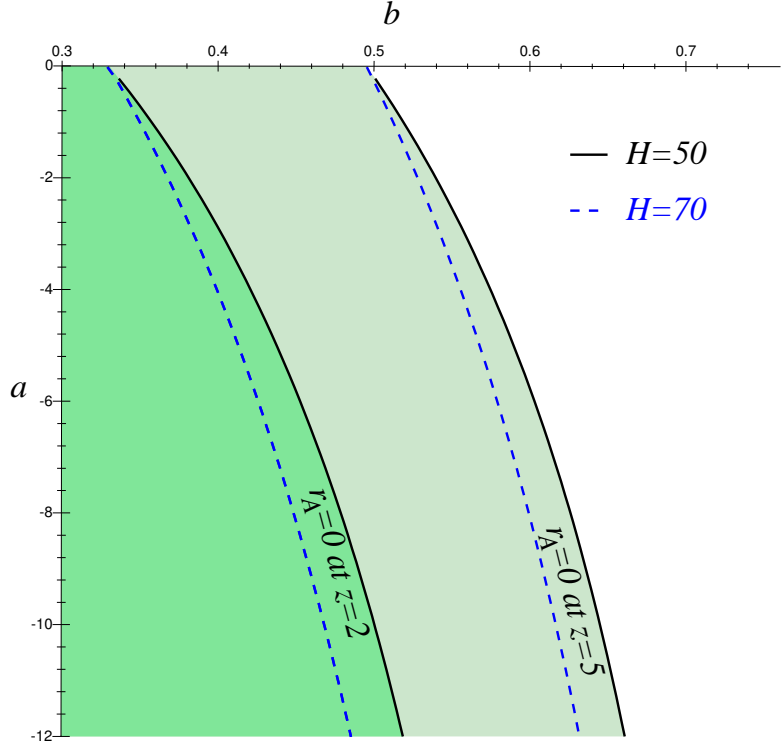},width=10cm,angle=90}}
\caption{\small Exclusion plot obtained by requiring that the first zeros of~$r_A(z)$
occur at~$z>z_\pi$, for $z_\pi=2$ and~$z_\pi=5$. The shaded regions are
excluded. Curves are given for $H_0=50$ and $H_0=70$~\hu.
\label{rA_exclusion}}
\end{figure}%
The effect of this constraint is to rule out small values of~$b$, for any~$a$.
This is a reflection of the fact that, loosely speaking, $b$~measures the
`size' of the universe: for small times the scale factor goes as~$b\ti$, so
that when $b$~is small the spatial sections are small, light rays don't take
long to travel from antipode to observer, the scale factor changes relatively
little during this time and the redshift of the antipode (which is dominated by
$R_0/R(t_\pi)$ as in \FRW\ models) is small.

As a coda to this section we consider the effect of demanding that the first
zero of~$r_A$ is effectively unobservable as a result of being `hidden' behind
the CMB. Figure~\ref{log redshift_b}
\begin{figure}[here!]
\centerline{\psfig{file={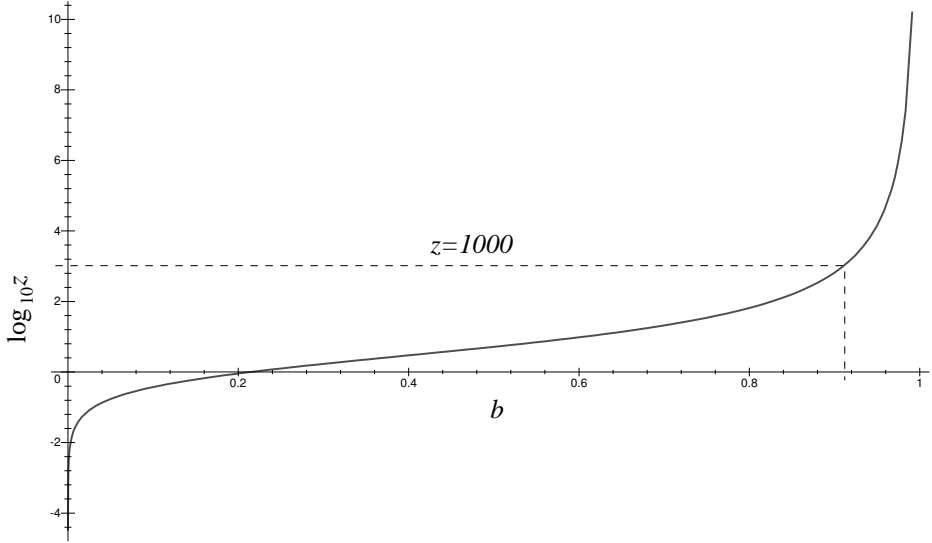},width=8cm,angle=90}}
\caption{\small
Logarithmic plot of the redshift at which the first zero of~$r_A$ occurs as a
function of~$b$ for $H_0=50$~\hu\ and~$a=-1$. For the first zero to occur
at~$z>1000$, so that recombination occurs `nearer to us' than the antipode~--
ie,~at~$\chi<\pi$~-- requires $b\approx 1$.
\label{log redshift_b}}
\end{figure}%
shows how the redshift of the first zero of $r_A(z)$ varies with~$b$. If,
instead of choosing~$z_\pi=z_{SF}$ as our primary constraint, we want the first
zero of $r_A(z)$ to happen at a redshift large enough for the universe to be
opaque (ie,~before decoupling), then figure~\ref{log redshift_b} shows that $b$
must be quite close to unity. This figure also allows the extent to which
values of~$b$ are excluded for any~$z_\pi$ to be estimated.

\subsection{The Microwave Background Anisotropy.}
\label{CMB}

The CMB is observed today to be a blackbody at a temperature of
$T_0=2.734\pm0.01$K, with a dipole moment of $T_1=3.343\pm0.016\times10^{-3}$K
and quadrupole moment as big as $T_2=2.8\times10^{-5}$K (see
Partridge~1997\nocite{part97} for details and references). It was emitted at a
time when the radiation was no longer hot enough to keep hydrogen ionised,
causing it to decouple from matter, which happens at $T_{\mathrm{dec}}\sim
3000$K. Idealised cosmological models do not have realistic thermodynamics
(that is, they do not, in general, describe the thermodynamic evolution of the
gas and radiation mixture that fills the real universe). In \FRW\ models the
epoch at which decoupling occurs is simply defined to be that corresponding to
the redshift necessary to shift the temperature at decoupling to the observed
mean temperature of the CMB,~$T_0$. From the redshift relation applied to the
temperature of a blackbody ($T$~will be used to denote temperature in this
section),
\be
                        T_{\mathrm{obs}}=\frac{T_{\mathrm{dec}}}{1+z},
\label{CMB temperature}
\ee
we infer that the CMB is formed at a redshift~$z\approx 1000$. This definition
is fine for homogeneous models, leading to a consistent definition of the time
of decoupling for every observer at the same cosmic time, but raises an
interesting point for the inhomogeneous D\c{a}browski models, because the
redshift depends on both the observer's position,~$\psi$, and the angle around
the sky,~$\theta$. If we simply define the redshift of the CMB at any point to
satisfy~(\ref{CMB temperature}) with $T_{\mathrm{obs}}=T_0$ then, by
definition, we obtain a perfectly isotropic CMB for that observer, but we must
choose a different emitting surface for each different observer. Such an
observer-based definition of the CMB surface is clearly unsatisfactory. Instead
we propose several alternative ways to define what is meant by the CMB surface
(it is important in these definitions to distinguish between the dominant
strange matter that is responsible for the geometry of the D\c{a}browski
models~-- see~\S\ref{energy_cond}~-- and the putative `real' gas that
decouples):
\begin{enumerate}
\item \label{costdef} On a surface of constant cosmic time,~$\ti=\ti\cmb$~--
since the D\c{a}browski models possess a cosmic time coordinate with respect to
which only the pressure is inhomogeneous this is a natural extension of the
\FRW\ definition;
\item In general inhomogeneous models we could choose hypersurfaces of constant
density, which at least has some physical basis~-- for the D\c{a}browski models
this is equivalent to~\ref{costdef}. because the density is homogeneous on
cosmic time surfaces;
\item \label{rhodef} On surfaces such that $p/\rho=$constant (which gives constant
temperature for an ideal gas)~-- for our models this would mean that decoupling
occurs at different times at different places in the universe (probably a good
thing in an inhomogeneous universe); however, the D\c{a}browski matter is not
an ideal gas so the validity of this definition for these models is debatable;
\item \label{ptdef} On a surface of constant \emph{proper} time (based on the
assumption of some common evolution for the ideal gas component at different
positions)~-- see equation~(\ref{proper age});
\item \label{EGSdef} Finally, we could avoid all consideration of the physics of
decoupling and simply assume that it happens at such an early time that we can
effectively define the CMB to be free-streaming radiation `emitted at the big
bang', as is implicitly assumed in the EGS theorem (Ehlers, Geren and
Sachs~1968\nocite{ehlers-68}). This only really makes sense if there is some
natural definition of the radiation field at early times. For example, if the
model is homogeneous and isotropic at early times, we can define a homogeneous
and isotropic radiation field. Our models have exactly this property of
homogeneity at early times (as can be seen from~(\ref{confW}),
if~$\ti\rightarrow 0$ then~$W\rightarrow 1$).
\end{enumerate}
Although~\ref{EGSdef}) is interesting we will not consider it here because it
does not reflect the physics of the CMB, and, on a more practical level, it
prevents us from constraining the CMB anisotropy, because there isn't one.
Also, the homogeneity of D\c{a}browski models at early times means that for
small~$\ti$ the proper age is virtually identical to the coordinate time~$\ti$
(equation~(\ref{proper age}) with~$W\approx 1$). It turns out that for times of
observation that reproduce the observed~$H_0$ any reasonable definition of the
CMB surface puts it at an early time, which means definition~\ref{ptdef}) is
virtually identical to~\ref{costdef}) We will therefore define the CMB
according to~\ref{costdef}) in this section.

It still remains, though, to decide exactly \emph{which} surface of constant
cosmic time the CMB originates from. Consider, for an observer at
position~$\psi$, the temperature distribution on the sky that the CMB would
have if it were emitted from the surface~$\ti=\ti\cmb$ (related by the lookback
time formula~(\ref{lookback}) to some distance~$\chi\cmb$):
\be
T_{\mathrm{obs}}(\psi,\chi\cmb,\theta)=\frac{T_{\mathrm{dec}}}{1+z(\psi,\chi\cmb,\theta)}.
\ee
Equation~(\ref{zdipole}) shows that we can write
\[
  1+z(\psi,\chi\cmb,\theta) = 1+z_0(\psi,\chi\cmb) +
                      z_1(\psi,\chi\cmb)\cos\theta
\]
where
\ba
   1+z_0(\psi,\chi\cmb) & = & \frac{R_0}{W_0}\left[\frac{1}{R\cmb}-
\frac{2a}{c\Delta}\left(1-\cos\psi\cos\chi\cmb\right)\right],\nonumber \\
   z_1(\psi,\chi\cmb) & = & -\frac{2a}{c\Delta}\frac{R_0}{W_0}\sin\psi\sin\chi\cmb
                        = \frac{\dot{u}(\psi)}{c^2}\frac{R_0}{\sqrt{\Delta}W_0}\sin\chi\cmb
\label{z_1}
\ea
(using~(\ref{accscalar}) in the last equality and assuming~$a\le 0$). The mean
redshift of the CMB surface is~$z_0$; $z_1$~gives rise to an anisotropy in the
CMB. We can therefore define the location of the CMB surface to be
the~$\ti\cmb$ (or~$\chi\cmb$) that gives a mean redshift of~1000. That is,
$\chi\cmb$ is the solution of
\be
         z_0(\psi,\chi\cmb) = 1000
\label{zcmbdef}
\ee
for any observer position~$\psi$.

Having found~$\chi\cmb$ we can evaluate the anisotropy in the temperature of
the CMB. Since $T_{\mathrm{obs}}$ depends on the reciprocal of~$1+z$ the dipole
moment in~$z$ will give rise to higher multipoles when expanded as a binomial
series:
\be
T_{\mathrm{obs}}(\theta)=\frac{T_{\mathrm{dec}}}{1+z_0}\left[1-\frac{z_1}{1+z_0}\cos\theta
    +\left(\frac{z_1}{1+z_0}\right)^2 \cos^2\theta
    + O(\cos^3\theta)\right],
\ee
that is,
\be
\frac{\delta T(\theta)}{T}=-\frac{z_1}{1+z_0}\cos\theta
    +\left(\frac{z_1}{1+z_0}\right)^2 \cos^2\theta
    + O(\cos^3\theta).
\label{variation of CMB}
\ee
The dipole moment of the CMB temperature is then (using~(\ref{zcmbdef}))
\be
                \delta_1 = \frac{z_1}{1+z_0} \approx 10^{-3}z_1.
\label{Tdipole}
\ee

Measurements of the CMB can now be used to constrain the model parameters. We
at least require that the dipole moment should be no larger than the observed
dipole anisotropy,~$|\delta_1|<T_1/T_0\approx 10^{-3}$ (ie,~$|z_1|<1$). If this
is satisfied for any model then it is clear that the quadrupole and higher
multipole moments will all be~$\lapp 10^{-6}$~-- certainly no larger than their
observed values. In fact, such a constraint on~$z_1$ is very weak, leaving vast
tracts of parameter space entirely untouched. Moreover, there are very good
reasons for believing that there is a significant contribution to the observed
dipole moment from the peculiar velocity of the Local Group as a result of
infall towards the Great Attractor (Lynden-Bell
\etal~1988\nocite{lynd-bell88}), which can be measured with moderate accuracy
using galaxy surveys, and seems to be consistent with the motion of the Local
Group with respect to the CMB (Schmoldt \etal~1999\nocite{schmoldt-99}). In an
inhomogeneous cosmological model, though, one expects local anisotropies that
could be interpreted as bulk flows, and it is not beyond the bounds of
possibility that all of the CMB dipole \emph{and} the local anisotropies could
be explained purely as cosmological effects in, say, a Stephani universe
without any need to invoke peculiar motions and local inhomogeneities
(ie,~perturbations of the perfect background cosmology). Such perturbations
must exist, of course, and will make some contribution to the observed
anisotropies, but it has not been shown that they are the only, or even the
dominant, contributions. However, we choose here to reject iconoclasm in favour
of the more conservative viewpoint that most of the observed dipole \emph{is}
due to the peculiar motions induced by local inhomogeneities, but that there
remains some leeway~-- up to 10\% of the observed dipole~-- due to
observational uncertainties, for there to be a purely cosmological contribution
to the CMB dipole. Then the largest dipole moment that we can accept from our
models is~$|\delta_1|<10^{-4}$, or
\be
                           |z_1|<0.1.
\label{maxzdipole}
\ee

Given any model parameters and some observer position we adopt the following
procedure. First we use~(\ref{Hubble parameter}) to determine the epoch of
observation for some~$H_0$, as usual, then we solve for~$\chi\cmb$
using~(\ref{zcmbdef}). Having found all the parameters we need to
determine~$z_1$ we simply check~(\ref{maxzdipole}) to see whether the model, or
at least that observer position, must be rejected. In practice we can simply
solve~(\ref{maxzdipole}) as an equality to obtain $a$ as a function of~$b$ at
the boundary of the allowed region, and this is what is shown in
figure~\ref{cmb_exclusion}
\begin{figure}[ht!]
\centerline{\psfig{file={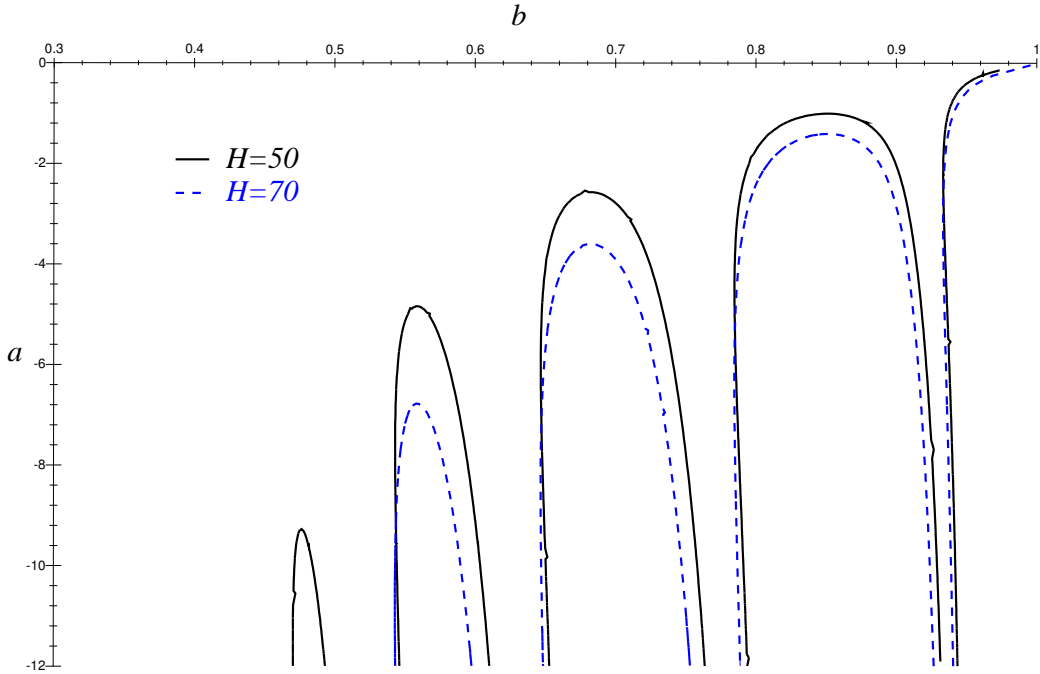},width=9cm,angle=90}}
\caption{\small The exclusion diagram from CMB anisotropies for an observer
at $\psi=\pi/2$. The curves for $H_0=50$, $70$~\hu\ are shown. As $H_0$
increases, the `fingers' move down to more negative~$a$. The excluded region
lies within the fingers.
\label{cmb_exclusion}}
\end{figure}%
(for $\psi=\pi/2$), where it can be seen that a low value of $H_0$ constrains
our models most -- in contrast to the age constraint. This is because~$H_0$
decreases monotonically with time, so small~$H_0$ corresponds to a later time
of observation and therefore a later time for the CMB surface, which means
that~$W$ has evolved to become more inhomogeneous. We choose $\psi=\pi/2$
because, as is clear from (\ref{acceln}) and~(\ref{z_1}), the anisotropy is
generally worst there, so if a model is rejected at~$\psi=\pi/2$ it will be
unacceptable everywhere, in order to satisfy the Copernican principle. Again,
therefore, for models not excluded in figure~\ref{cmb_exclusion} detection of a
CMB anisotropy of the magnitude that we observe would be typical, and the
Copernican principle need not be abandoned for these models to be viable.

The finger-like excluded regions in figure~\ref{cmb_exclusion} appear because
for different model parameters~$\chi\cmb$ takes on different values, and for
some parameters this value is very close to~$\pi$, so that the CMB surface is
almost at the antipode of the observer. This means that the entire CMB is
emitted from virtually a single point. Since redshift depends only on the
relative conformal factors at emitter and observer for our conformally flat
models, the CMB must be almost exactly uniform however inhomogeneous the model
(ie,~whatever the value of~$a$).

\subsection{The Local Dipole Anisotropy}
\label{local-dipole}

Although we are not in a position to use real observations to constrain the
dipoles that would be detected in observations of the `local' universe (in
galaxy surveys, for example, where~$z\lapp 0.01$, or with type~Ia supernova
data, for which~$z\lapp 1$), we can at least consider these effects
qualitatively.

From (\ref{zdipole}) and~(\ref{accscalar}) the redshift dipole for objects at
any radius~$\chi$ from an observer at~$\psi$ is seen to be
\[
   z_1(\psi,\chi) = \frac{\dot{u}(\psi)}{c^2}\frac{R_0}{\sqrt{\Delta}W_0}\sin\chi
\]
(\emph{cf}.~equation~(\ref{z_1})). It is clear that any constraint on the
anisotropy in the properties of objects at some distance from the observer will
amount to a constraint on the acceleration,~$\dot{u}$, of the fundamental
observers, and vice versa. In the same way, constraints on the acceleration
place limits on the possible dipole moment in the measured~$H_0^m$
in~(\ref{specific H_0}). Defining~$\delta H= \dot{u}/c$ (according
to~(\ref{specific H_0})), the dipole in redshift becomes
\be
        z_1(\psi,\chi) = \frac{R_0\sin\chi}{\sqrt{\Delta}W_0} \frac{\delta H}{c}.
\label{z_1gen2}
\ee

If we assume that the time of observation,~$\T$, is fairly close to~$\ti=0$ (as
is generally the case for the values of~$H_0$ we allow), then $W_0\approx 1$
and~$a\T\ll b$, which means that $R_0/c\approx b\T$ and~$\dot{R}_0/c\approx b$.
Equation~(\ref{z_1gen2}) then becomes, with the help of~(\ref{Hubble
parameter}),
\be
     z_1(\psi,\chi) \approx \frac{1}{\sqrt{\Delta}}\frac{\dot{R}_0\sin\chi}{cH_0}
                                      \delta H
                    \approx \frac{b}{\sqrt{\Delta}}\sin\chi\frac{\delta H}{H_0}.
\label{z_1gen3}
\ee
Note that dependence on distance from the observer only arises through
the~$\sin\chi$ factor, so for observations of objects at any distance from the
observer
\be
        |z_1| \lapp \frac{b}{\sqrt{\Delta}}\frac{\delta H}{H_0}.
\label{z_1gen4}
\ee
Local measurements in principle, therefore, constrain the acceleration and the
anisotropy at all redshifts. For example, applying this to the CMB dipole,
(\ref{z_1gen4}) shows that the constraint~(\ref{z_1}) imposed in~\S\ref{CMB}
will always be satisfied provided that
\be
             \frac{\delta H}{H_0}\le 0.1 \frac{\sqrt{\Delta}}{b}.
\label{dHCMB}
\ee
To ensure that the variation in~$H_0$ is less than~20\% requires
$\sqrt{\Delta}/b\le 2$, or~$b\ge 1/\sqrt{5}\approx 0.45$. For a variation of
less than~10\%, we must have~$b\ge 1/\sqrt{2}\approx 0.71$. Since we are
interested in all values of~$b\gapp 0.5$ (see figure~\ref{exclusion_H=50}) we
must be prepared to countenance variations in~$H_0$ around the sky of up
to~20\% if the local anisotropy is not to impose tighter constraints on the
model parameters than those already derived from the CMB. Note, though,
that~(\ref{dHCMB}) is considerably stronger than is really required for most
model parameters, owing to the fact that the~$\sin\chi$ factor
in~(\ref{z_1gen3}) was neglected. This amounts to adopting as the CMB
constraint the envelope of the fingers in figures~\ref{cmb_exclusion},
\ref{exclusion_H=50} and~\ref{exclusion_H=70}, which would obviously overlook
large areas of parameter space that should really be allowed.

Although it would seem that the easiest way to constrain the acceleration in
the D\c{a}browski models would be to measure the local $H_0$~dipole, it is not
really possible to measure~$\delta H/H_0$ accurately, because that would
require accurate measurements of galaxies in different directions at very small
redshifts; there are relatively few such galaxies, and those that there are
possess significant random peculiar motions that make a large contribution to
the errors on any estimate of~$\delta H/H_0$. Moreover, there is known to be a
dipole in observations of galaxies at somewhat larger redshifts ($z\sim 0.01$),
which is usually interpreted as the effect of infall of the Local Group towards
the Great Attractor (Lynden-Bell \etal~1988\nocite{lynd-bell88}; Schmoldt
\etal~1999\nocite{schmoldt-99}). This infall manifests itself as a systematic
relative motion of the Local Group with respect to distant galaxies, at a
velocity~$v\approx 600$~km~s$^{-1}$, corresponding exactly to the Local Group
motion relative to the CMB frame. In order not to conflict with these
observations we at least require that at redshifts~$z_0\approx 0.01$ the dipole
moment due to the D\c{a}browski acceleration is no larger than the observed
dipole:
\[
                   c z_1 \lapp v = 600\hbox{~km~s$^{-1}$}.
\]
Using (\ref{redshift}) and~(\ref{area dist-new}) it is possible to
rewrite~(\ref{z_1gen2}) in the form
\[
       z_1(\psi,\chi) = (1+z_0)r_A(\psi,\chi) \frac{\delta H}{c},
\label{z_1gen5}
\]
where we implicitly identify $z_0$ and~$r_A$ as the mean redshift and angular
size distance of objects at coordinate distance~$\chi$. Then, since at low
redshift the Hubble law is valid, $cz_0=v=H_0r_{\mathrm{prop}}$
($r_{\mathrm{prop}}$~denotes proper distance), and $r_A=r_L=r_{\mathrm{prop}}$,
we have
\[
        z_1 = \frac{cz_0}{H_0}\frac{\delta H}{c}
\]
That is,
\be
\frac{\delta H}{H_0} = \frac{cz_1}{cz_0} \lapp \frac{600\hbox{~km~s$^{-1}$}}{0.01c} = 0.2;
\label{GAdipole}
\ee
at most a 20\% variation in~$H_0$ around the sky, consistent with the CMB
result~(\ref{dHCMB}).

Finally, we consider the magnitude-redshift relation in the low-redshift limit
and apply it to type~Ia supernovae (for example, to the relatively low-redshift
supernovae of Hamuy \etal~1996\nocite{hamuy-96}~-- although it is not
unreasonable to assume that this gives at least an order-of-magnitude estimate
of the size of the anisotropy for observations at higher~$z$ characteristic of
the supernova data of Perlmutter \etal~1999). If~$z\ll 1$, then $r_L=cz_0/H_0$.
Equations (\ref{pogson2}) and~(\ref{specific H_0}) give
\[
         m-M-25 = 5\log_{10} \frac{cz_0}{H_0^m}
 = 5\log_{10} \frac{cz_0}{H_0} - 5\log_{10}\left(1-\frac{\delta H}{H_0}\cos\theta\right)
         \approx 5\log_{10} \frac{cz_0}{H_0} + \;{5}{\ln 10}\frac{\delta H}{H_0}\cos\theta
\]
(assuming that~$\delta H/H_0$ is small). That is,
\be
           \delta m = \;{5}{\ln 10}\frac{\delta H}{H_0}.
\label{deltam}
\ee
The dispersion on the magnitudes of type~Ia supernovae (due to both
observational errors and intrinsic dispersion) is estimated to be~$\delta m
\sim 0.3$ (Hamuy \etal~1996\nocite{hamuy-96}; Perlmutter
\etal~1999). For the dipole resulting from the acceleration to be undetectable
via type~Ia supernovae we must have
\be
            \frac{\delta H}{H_0} \lapp 0.14.
\label{SNlim}
\ee
This is somewhat stronger than the other local constraints derived above, and
would suggest that the supernova data will indeed restrict the model parameters
more than the CMB anisotropy. Comparison with (\ref{dHCMB})
and~(\ref{GAdipole}) suggest that $b$ may only be half or a quarter as large as
would be permitted by the CMB constraint (see equations (\ref{specific H_0})
and~(\ref{accscalar})). However, these are only very approximate results, and
require more detailed analysis. In particular, there is \emph{observed} to be a
dipole in the supernova data (explained in exactly the same way as the other
dipoles in the conventional interpretation), and this means that the
constraint~(\ref{SNlim}) must be re-evaluated: the real constraint is likely to
be somewhat weaker than~(\ref{SNlim}).

It should be borne in mind that throughout the preceding discussion we have
only considered the modulus of the dipole, not its direction. It is clear
from~(\ref{z_1}) that the local dipoles may be in the same direction as the CMB
dipole or in the opposite direction, depending on the sign of~$\sin\chi\cmb$.
What is more, the variation of the dipole with distance is controlled entirely
by~$\sin\chi$, so the dipole will change sign whenever~$\chi$ is a multiple
of~$\pi$.

In this section we have examined the local anisotropies resulting from
acceleration in a rough and qualitative way. Although certainly not conclusive,
the results indicate that it is important to give full consideration to the
constraints imposed by local observations on these anisotropies.

\subsection{The Combined Exclusion Diagrams.}

When we combine all of the constraints derived in this section (figures
\ref{exclusion_H=50} and~\ref{exclusion_H=70})
\begin{figure}[ht!]
\centerline{\psfig{file={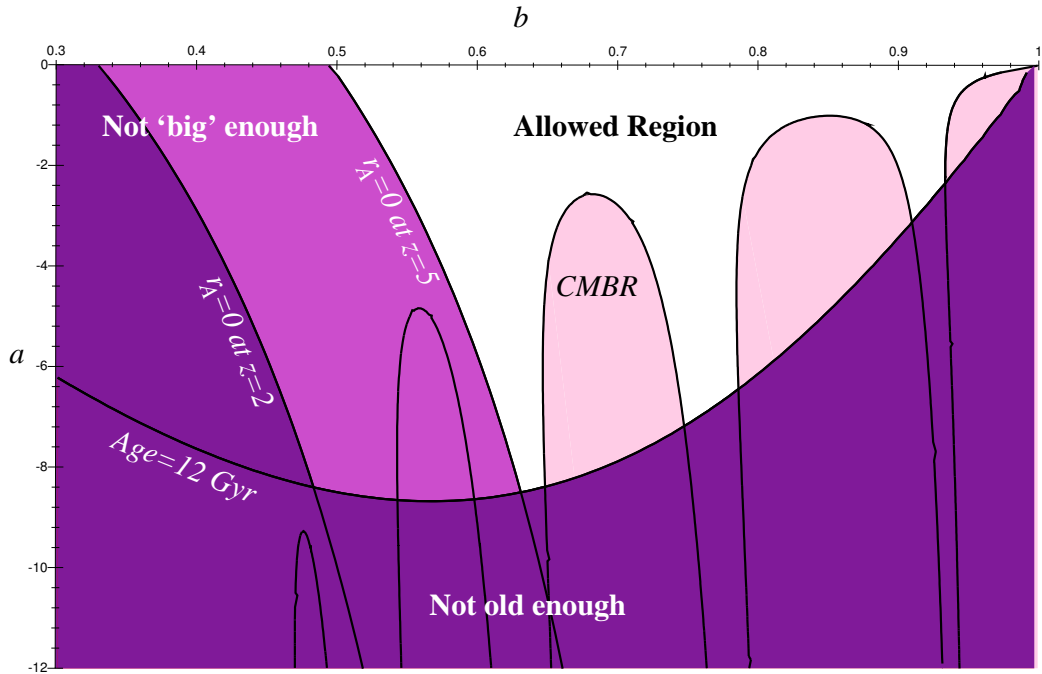},width=9cm,angle=90}}
\caption{\small The complete exclusion diagram for all the observational constraints
studied (age, size and the CMB anisotropy), for~$H_0=50$~\hu. We have taken the
$12$~Gyr age constraint, to be conservative. The dominant energy condition
should be added to these constraints: it eliminates models with~$b>0.82$
(equation~(\ref{bdom})).
\label{exclusion_H=50}}
\end{figure}%
we can see that for~$H_0=50$~\hu\ the strongest constraint comes from the CMB,
with age placing somewhat weaker limits on the allowable degree of
inhomogeneity (which is measured largely by the size of~$a$~--
see~\S\ref{inhomog}). The `size' restriction of~\S\ref{size} eliminates quite a
large region of parameter space for small~$b$, but this is not really a
constraint on the inhomogeneity, which is our principle concern.

Perhaps rather surprisingly, given the results of D\c{a}browski and
Hendry~(1998) and Paper~I, the strongest constraint for larger~$H_0$ really
comes from the age. As can be seen in figure~\ref{exclusion_H=70}, the
exclusion plot for~$H_0=70$~\hu,
\begin{figure}[ht!]
\centerline{\psfig{file={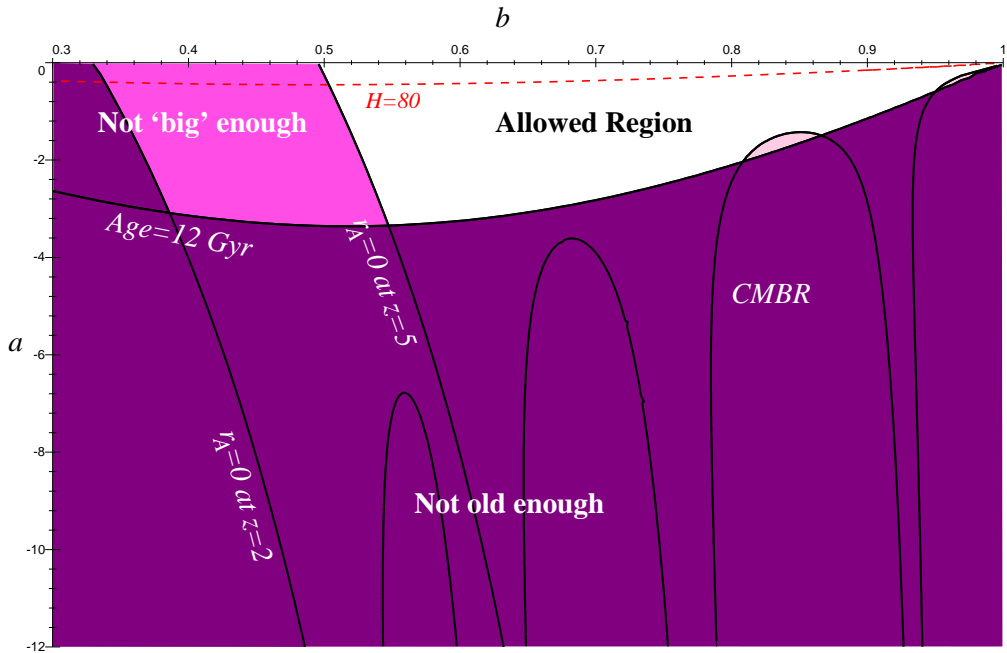},width=9cm,angle=90}}
\caption{\small As in figure~\ref{exclusion_H=50}, but for~$H_0=70$~\hu. The age
constraint for~$H_0=80$~\hu\ is also shown as a dashed line (close to the
$b$-axis): high~$H_0$ means a very low age, just as for the \FRW\ models.
\label{exclusion_H=70}}
\end{figure}%
the CMB constraint pokes out in places to eliminate certain regions, and the
size constraint cuts off low values of~$b$, but age does most of the dirty
work. It can also be seen that for $H_0=80$~\hu\ age imposes a very strong
constraint on the models (dashed line in figure~\ref{exclusion_H=70}): the
models must be very nearly homogeneous. However, if we relaxed the age
constraint to (a probably quite acceptable) $10$~Gyr the CMB anisotropy would
be the dominant limitation for most values of~$H_0$ in the currently
fashionable range ($50\lapp H_0\lapp 80$~\hu).

We should not forget, at this point, to reintroduce the
restriction~(\ref{bdom}) from the dominant energy condition, which rules out
high~$b$. This is not shown on the diagrams, in order to avoid clutter. Most of
the models eliminated by this constraint have already been ruled out by the age
or CMB constraints, and models that are rejected solely by the dominant energy
condition are not hugely inhomogeneous (see the next section).

\section{The Size of the Inhomogeneity.}
\label{inhomog}

While we have considered many different aspects of the D\c{a}browski models,
what we have not done is to assess the extent to which the models that are not
excluded are inhomogeneous. It is obvious from the exclusion plots,
figures~\ref{exclusion_H=50} and~\ref{exclusion_H=70}, that the homogeneous
D\c{a}browski models (those with~$a=0$) are the 'most acceptable', in that all
the constraints favour small~$a$. This should not be surprising, as far as
anisotropy constraints are concerned, at least. What is not clear is whether
the allowed region only contains models that are very nearly homogeneous. We
will show that it does not.

The most natural way to assess the degree of inhomogeneity of the models is to
examine the variation of the pressure over surfaces of constant cosmic time. It
can be seen from~(\ref{pressW}) that the extremes of pressure occur at $\chi=0$
and~$\chi=\pi$, so we define the inhomogeneity factor~$\Pi$ to be the relative
pressure difference between the two poles:
\be
          \Pi = \left|\frac{p(\pi)-p(0)}{p(0)}\right| = \frac{8|a|R}{c\Delta}.
\label{Pidef}
\ee
If~$\Pi\gapp 1$ then it is reasonable to say that the models are truly
inhomogeneous, whereas if~$\Pi\ll 1$ they are obviously nearly \FRW. Note,
though, that~$\Pi$ depends on the cosmic time surface under consideration: for
small~$\ti$, $\Pi\approx 0$, and~$\Pi$ reaches its maximum at~$\ti=-b/2a$
(see~(\ref{Roft})), at which time
\[
               \Pi = \frac{2b^2}{1-b^2} = 2\frac{1-\Delta}{\Delta}
\]
so that the models are significantly inhomogeneous ($\Pi\gapp 1$) when
\be
          b\gapp \frac{1}{\sqrt{3}}\approx 0.58.
\label{binhom}
\ee
Most of the allowed models in figures~\ref{exclusion_H=50}
and~\ref{exclusion_H=70} are unaffected by~(\ref{binhom})~-- models with
smaller~$b$ have already been eliminated by the size constraint
in~\S\ref{size}.

This is not really a fair reflection of the inhomogeneity of the models at the
times of observation that are relevant here, though, because the~$H_0$
constraint~(\ref{Hubble parameter}) generally ensures that the epoch of
observation is quite early on in the evolution of the universe when the scale
factor is somewhat smaller than its maximum size. To evaluate the impact of
this, consider two specific examples. From the allowed region
of~(\ref{exclusion_H=50}) choose the model at $a=-7$,~$b=0.75$.
For~$H_0=50$~\hu\ the solution of~(\ref{Hubble parameter}) is $\T\approx
0.016$~Mpc~s~km$^{-1}$, which gives~$\Pi=1.33$. For the model at $a=-8$,
$b=0.64$ we get $\T = (3-\sqrt{5})/50 \approx 0.015$~Mpc~s~km$^{-1}$
and~$\Pi=0.86$, which is close enough to~$1$. These models are certainly not
`close to \FRW'.

Nevertheless, it could be said that the models are not massively inhomogeneous,
and that their degree of inhomogeneity only reflects the looseness of the
constraints applied. This is not so. Firstly, we have at every stage chosen
stronger limits than were strictly necessary, especially as far as the age is
concerned, where we could have adopted a $10$~Gyr limit, rather than $12$~Gyr,
which would have increased the allowed region considerably. Most importantly of
all, though, is the fact that even for models that are only inhomogeneous at
the 10\% level ($\Pi=0.1$), the CMB anisotropy,~$\delta_1<10^{-4}$, is at least
\emph{three orders of magnitude smaller} than the inhomogeneity it permits.
This certainly conflicts with the spirit of the almost EGS theorem of Stoeger,
Maartens, and Ellis~(1995)\nocite{stoeg-95} (although not the actuality, since
our models have acceleration, whereas the theorem deals with geodesic
observers), which says that small CMB anisotropies give rise to small
perturbations from homogeneity.

\section{Fitting to FLRW Models}
\label{resolution}

So far we have demonstrated that the \dab\ models do not have any problem with
regard to the `global' constraints we have considered. As far as `local'
constraints are concerned we have only considered the size of the dipole, and
ensured it was not too large. We still have to show that the \dab\ models are
capable of providing an acceptable fit to the observed magnitude-redshift
relation. A glance at
figure~\ref{m-z-opp-directions-a=-10_b=0.5_zeta=100_psi=0} (with reference to
figure~\ref{magnitude vs redshift FLRW} for the FLRW case) shows that these IRF
models are quite capable of producing very incompatable $m-z$ relations. In
order to demonstrate that the \dab\ models are acceptable in this regard we
`match up' the \dab\ $m-z$ relation to the FLRW one. Because of the local
dipole considered above we will simply fit the two curves for an observer
located at the center. (Obviously a best fit from any location will pick out
the center as the location most able to fit the FLRW models.)  This should
provide confidence, if not conclusive evidence, that the \dab\ models are
acceptable in terms of this test.

The complete solution of fitting the two $m-z$ relations is quite complicated.
Given an actual data set it is very easy to decide whether either model is
consistent with it: form a $\chi^2$ statistic and check whether it is small
enough for the model to be consistent at the desired confidence level. To
distinguish two \emph{theoretical models}, however, it is, strictly speaking,
necessary to adopt a sort of `two-level' approach and ask: assuming that the
Stephani model is the true model of the universe, what is the
probability,~$\rho_1$, that we would, for any realistic data set, reject the
\FRW\ model at some confidence level,~$\rho_2$? To answer this rigorously requires
consideration of all possible data sets drawn from both the \FRW\ \emph{and}
the Stephani distributions. In fact, we will use a much simpler and more
intuitive resolution criterion based on the mean-square difference between the
theoretical models. This will be more than adequate for our purposes. For
brevity we give a detailed derivation for the $m-z$~relations only, any other
case being similar.

The mean-square difference between $m_S(z)$ and~$m_F(z)$ over some $z$-range
$0\le z\le z_\mathrm{max}$, for which we expect to have observations, is
\begin{equation}
  D_2 = \frac{1}{z_\mathrm{max}}\int_0^{z_\mathrm{max}} [m_S(z)-m_F(z)]^2\,dz
\label{meansquare}
\end{equation}
We assume (as is justifiable for the case we consider here) that the difference
$m_S(z)-m_F(z)$ is smooth and slowly varying, and also that the integral does
not need weighted~-- ie, that the number density is uniform in redshift. Then
we imagine that the difference is just a constant; $m_S(z)-m_F(z)=\sqrt{D_2}$.
We can accept or reject this difference on the basis of the usual $\chi^2$
analysis.

We adopt the null hypothesis that the FLRW model is correct so that the data
set consists of $n$ data points $\{m_i, z_i\}$ such that the errors on the
$m_i$ are normally distributed with variance $\sigma^2$ say; ie,
$m_i=m_F(z_i)+\epsilon_i$ where $\epsilon_i\sim N(0,\sigma^2)$. Then, clearly
\be
m_S(z)-m_F(z)=\sqrt{D_2}\leq\;{\sigma}{\sqrt n}
\ee
(since $m_S(z)-m_F(z)=$const. implies the best fit value of $m_S(z)-m_F(z)$ is
$\<m_i-m_F(z)\>=\<\epsilon_i\>$ and the variance on this estimate goes as
$\sigma^2/n$).

Now, for the SNIa of Perlmutter \etal~(1999) $n=42$ and their error on the
intrinsic magnitude dispersion is $\sigma\approx0.15$. This implies
\[
D_2\leq\;{\sigma^2}{n}\approx5\times10^{-4}
\]
for these data. We take $z_{\mathrm{max}}=1$. When we fit the Stephani to the
FLRW model we require $D_2\lapp10^{-4}$ to demonstrate that the Stephani models
are capable of providing at least as good a fit to the SNIa data as an FLRW
model.

The procedure we use is to fix $H_0$ and $\Omega_0$ and find the `best fit'
parameters $a, b$ and $T$ for a range of $q_0$ (effectively $\Omega_\Lambda$).
For simplicity we consider two values of $\Omega_0=0.3, 1.0$. In
figures~\ref{bestfit-a-q} to~\ref{bestfit-T-q} we show the best fit values of
$a, b$ and $T$ against $q_0$. We can see that a negative $q_0$ requires $a>0$
as expected from the strong energy condition. In figure~\ref{bestfit-D-q} we
show the goodness of fit: provided $D_2<10^{-4}$ then these models will fit the
SNIa data acceptably. Finally, in figure~\ref{bestfit-a-T} we demonstrate that
a positive value of $a$ will enable an old enough universe.

\begin{figure}[ht!]
\centerline{\psfig{file={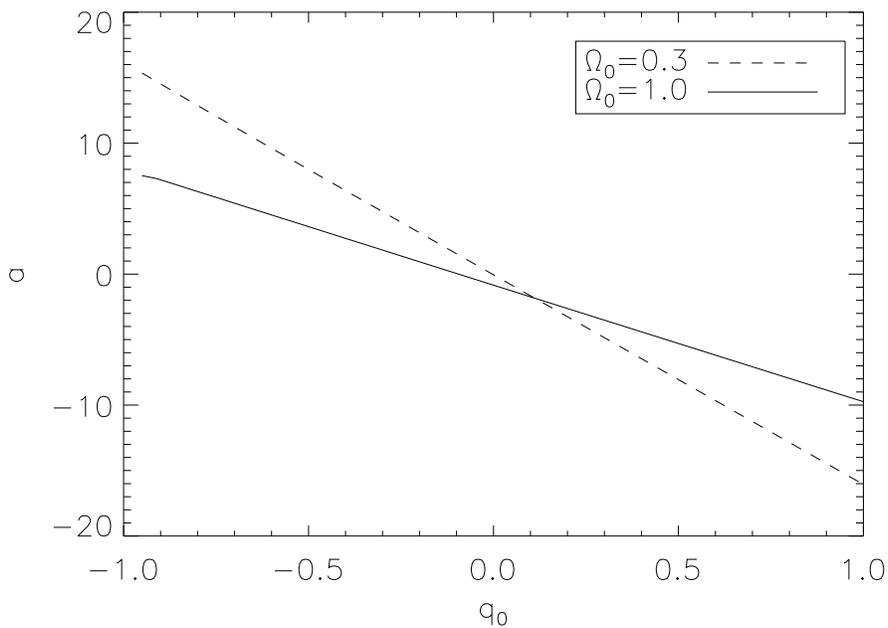},width=13cm,angle=0}}
\caption{\small The best fit value of $a$ over a range of $q_0$ for two values
of $\Omega_0$.
\label{bestfit-a-q}}
\end{figure}%
\begin{figure}[ht!]
\centerline{\psfig{file={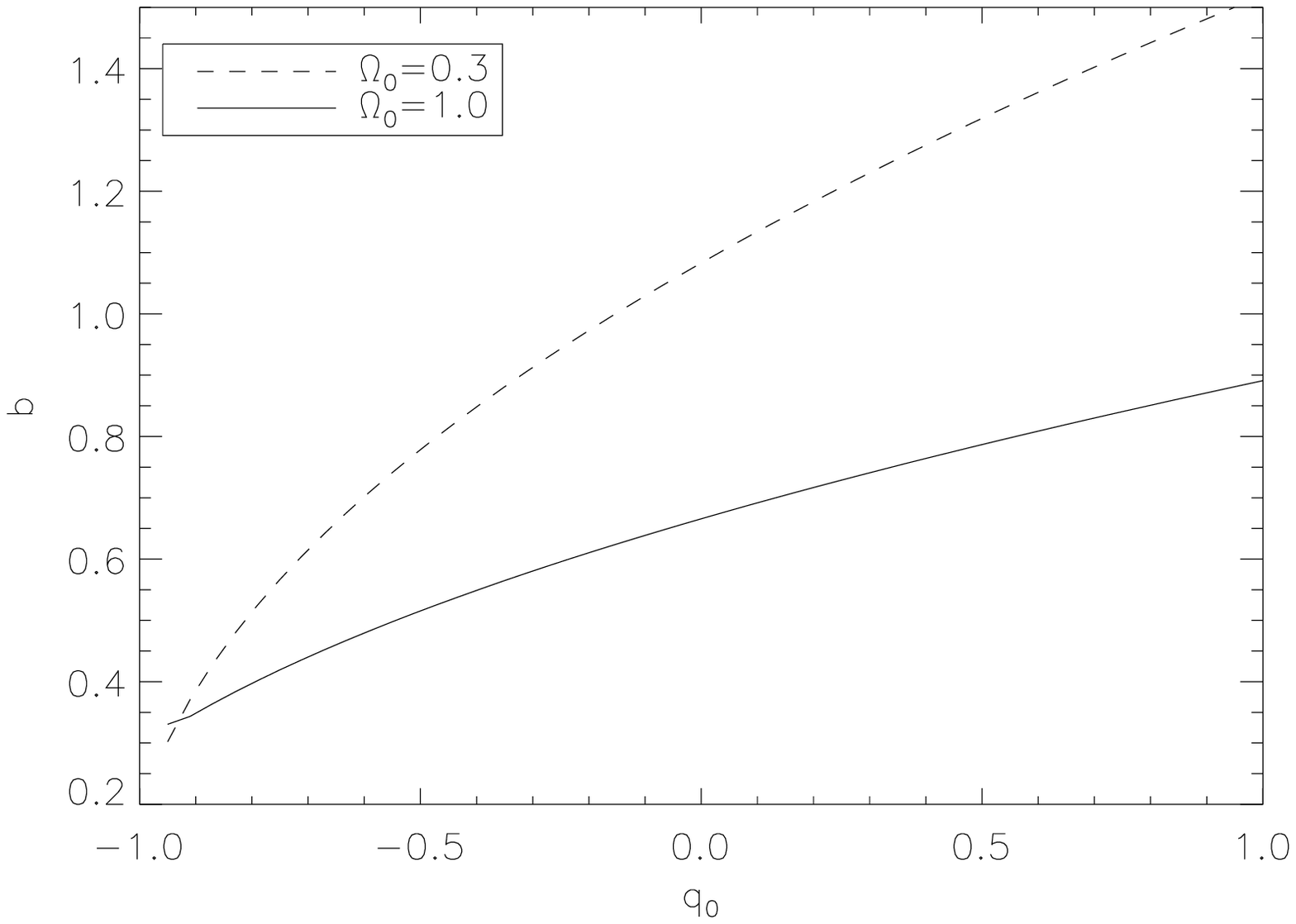},width=13cm,angle=0}}
\caption{\small The best fit value of $b$ over a range of $q_0$ for two values
of $\Omega_0$.
\label{bestfit-b-q}}
\end{figure}%
\begin{figure}[ht!]
\centerline{\psfig{file={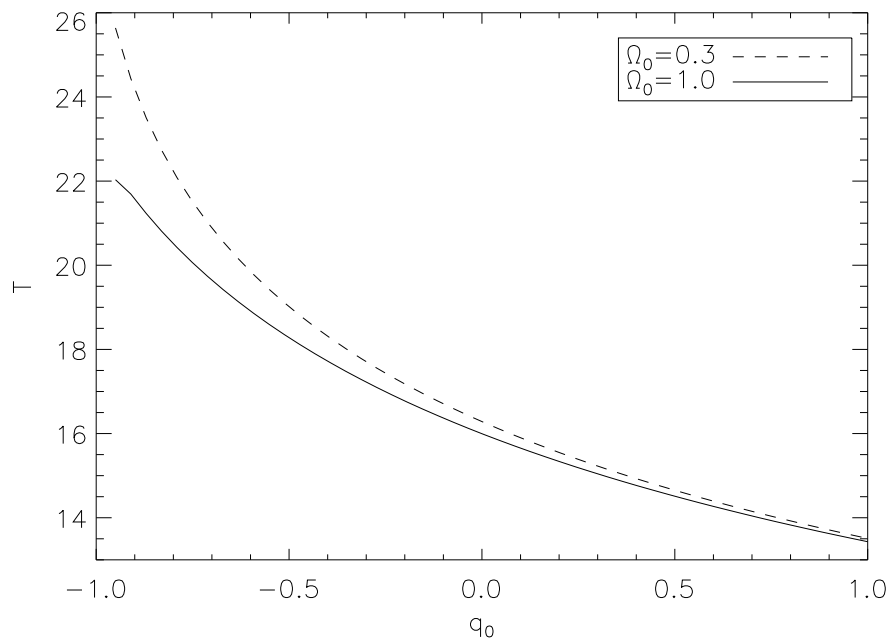},width=13cm,angle=0}}
\caption{\small The best fit value of $T$ (in Gyr) over a range of $q_0$ for two values
of $\Omega_0$.
\label{bestfit-T-q}}
\end{figure}%
\begin{figure}[ht!]
\centerline{\psfig{file={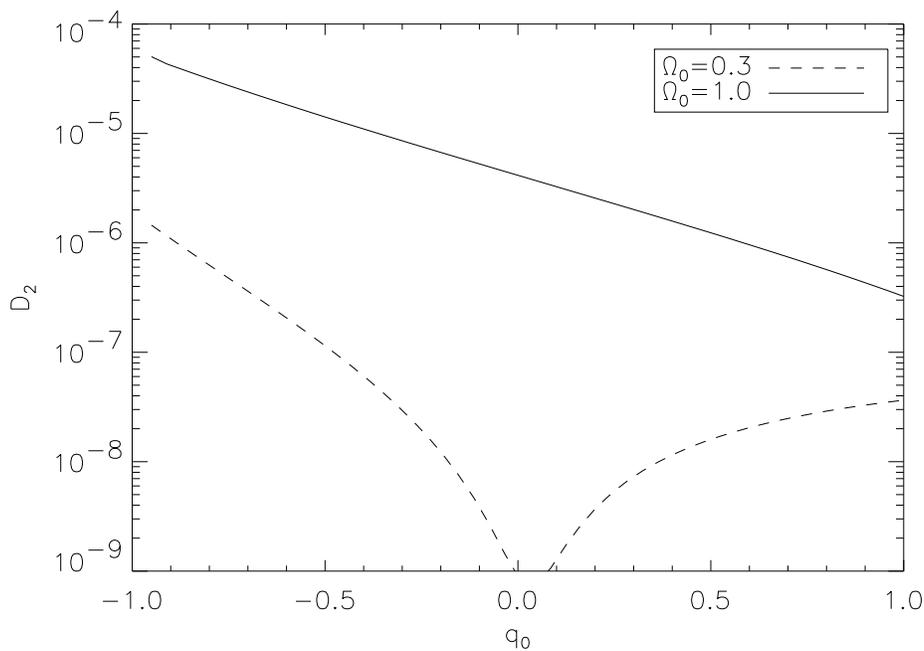},width=13cm,angle=0}}
\caption{\small The mean-squared difference as a function of $q_0$. Provided
this function satisfies $D_2<10^{-4}$ then the SNIa data will not be able to
distinguish between the FLRW and Stephani models. This is clearly satisfied in
these cases.
\label{bestfit-D-q}}
\end{figure}%
\clearpage
\begin{figure}[ht!]
\centerline{\psfig{file={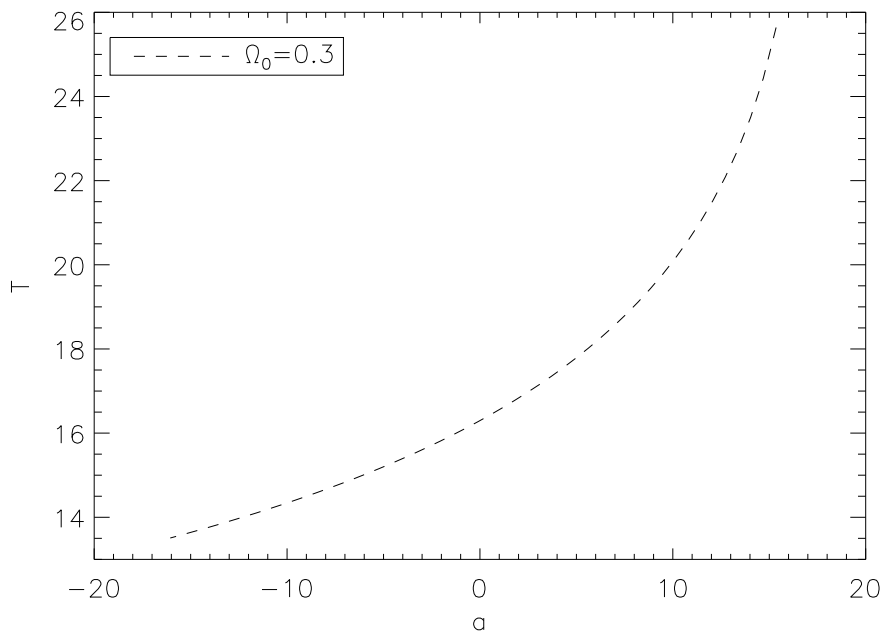},width=13cm,angle=0}}
\caption{\small The best fit values of $a$ vs. $T$. This clearly shows that a
positive value of $a$ will provide a sufficiently old universe.
\label{bestfit-a-T}}
\end{figure}%

\forget{
Any real data set will contain the measured magnitudes (and redshifts) of some
number,~$n$, of objects (hopefully evenly distributed between $z=0$
and~$z=z_\mathrm{max}$), each with some measurement error,~$\sigma$ (which we
will take to be uniform). With this in mind we approximate the integrand
in~(\ref{meansquare}) by assuming it is constant over~$n$ bins of
width~$z_\mathrm{max}/n$ to get
\[
   D_2 \approx \frac{1}{z_\mathrm{max}}\sum_{i=0}^n
                   ({m_S}_i-{m_F}_i)^2\,\frac{z_\mathrm{max}}{n}
       = \frac{\sigma^2}{n}\sum_{i=0}^n
                   \frac{({m_S}_i-{m_F}_i)^2}{\sigma^2}
       \equiv \frac{\sigma^2}{n}\, \chi^2_\mathrm{eff}
\]
where ${m_S}_i$ denotes the value of~$m_S$ in the $i$th bin, etc., and
$\chi^2_\mathrm{eff}$ is really defined by the last equality, but its close
relationship to the usual statistical~$\chi^2$ is obvious. Then, if we can give
values for $n$ and~$\sigma$ that represent the best data currently available we
can say that the Stephani and \FRW\ $m-z$~relations are indistinguishable
provided that the difference~$D_2$ in~(\ref{meansquare}) satisfies
\begin{equation}
        D_2 \le \frac{\sigma^2}{n}\, \chi^2_\mathrm{max}(\nu |p),
\label{resol}
\end{equation}
where $\chi^2_\mathrm{max}(\nu |p)$ is defined as follows. Imagine that we have
a real data set to which we fit one of the models, calculating the
usual~$\chi^2$. If $n_\mathrm{par}$ denotes the number of free parameters of
the model ($n_\mathrm{par}=2$ for the Stephani model, say), so that $\nu =
n-n_\mathrm{par}$ is the number of degrees of freedom in the fit, then the
$p$th confidence limit on $\chi^2$ is~$\chi^2_\mathrm{max}(\nu |p)$. That is,
we can accept the hypothesis that the model is a good fit to the data with
confidence~$p$ provided $\chi^2\le \chi^2_\mathrm{max}(\nu |p)$ (see
??\emph{Numerical Recipes},~\S15). Applying this confidence limit to the case
of two theoretical distributions results in~(\ref{resol}).

Probably the best distance estimator available at high redshift is that derived
from type Ia supernovae (SNIa). For the SNIa data of Perlmutter
\emph{et~al.}~(1999) $n=42$ and the errors consist of
\begin{enumerate}
\item an intrinsic dispersion in the absolute magnitudes of
$\sigma_M=0.26$, or $\sigma_M=0.17$ using the correction for light curve shape
(Hamuy \emph{et al.} 1995, 1996),
\item an error on the observed bolometric magnitude (which varies from
supernova to supernova) of between $\sigma_m = 0.06$ and $\sigma_m = 0.18$ for
the uncorrected magnitudes, or between $\sigma_m = 0.11$ and $\sigma_m = 0.35$
after applying the light-curve correction.
\end{enumerate}
In addition there is an uncertainty in the absolute magnitude calibration, but
this is the same for all SNIa and just leads to a small offset in the observed
$m-z$~relation, so we will ignore it here. For simplicity we will,
conservatively, take the smallest $\sigma_m$s quoted for the corrected and
uncorrected magnitudes and add them in quadrature to the intrinsic dispersions,
giving
\begin{equation}
       \sigma_\mathrm{uncorr} = 0.27,\,\,\,\, \sigma_\mathrm{corr} = 0.20
\label{sigmas}
\end{equation}
Naturally, we take the smaller of these to define the highest resolution that
can be obtained with present data (via~(\ref{resol})). Since the upper limit
on~$D_2$ will depend on $n$, $n_\mathrm{par}$ and the confidence level~$p$,
which will vary in the following sections, we will not calculate it here.

If we expect only to have observations of a single object (or of several
objects clustered at some redshift~-- as is largely true of the
Perlmutter~(1999) supernovae, which all have~$z\sim 0.4-1$), or if, rather than
being uniform across the redshift range, the difference between the Friedmann
and Stephani relations is large only in a narrow range of~$z$ (which is not the
case) it may be more appropriate to define resolution via the $\infty$-norm
(ie,~the largest absolute difference):
\begin{equation}
           D_\infty = \max_{0\le z\le z_\mathrm{max}} |m_S(z)-m_F(z)|.
\label{infnorm}
\end{equation}
Since $|m_S(z)-m_F(z)|\le D_\infty$, $D_2$ and $D_\infty$ are related by
\begin{equation}
        D_2 \le D_\infty^2
\label{D2Dinf}
\end{equation}
Intuitively, if the $n$ observations are all at the same redshift the two
models cannot be distinguished (at the $1-\sigma$ level) provided
\begin{equation}
       D_\infty \le \frac{\sigma}{\sqrt{n}},
\label{resolinf}
\end{equation}
which, through~(\ref{D2Dinf}), is consistent with~(\ref{resol}).

\forget{
For the angular-size distance relation there is really no good data at higher
redshifts, what data there is being ??mainly appropriate for testing models of
galaxy evolution. In any event, as a result of the reciprocity theorem
(see~(\ref{recipro})) angular-size distance does not provide a constraint on
the cosmological model independent of that imposed by the observed
$m-z$~relation. Since the difficulty of measuring diameters at high redshift,
not to mention the effect of evolution, severely limits the power of angular
diameter measurements it is reasonable to assume that any models that have
consistent $m-z$~relations (according to the criterion~(\ref{resol})) also have
consistent $r_A-z$~relations. In the sequel, therefore, we will only briefly
mention  $r_A-z$~relations. }

Given this then we are now able to fit the two $m-z$ relations. We do this in
two ways; first is simply a free fit over the parameters $a, b, T$; while the
second is to fix $H_0^{\mathrm{steph}}=H_0$ and then fit over the parameters~--
this will ensure that the two $m-z$ relations match up at very low redshift. We
choose a fixed value of $H_0, \Omega_0$, and we vary $q_0$ over the range $-1$
to $1$. For simplicity, we fix $H_0=60$~\hu, and pick two values of
$\Omega_0=0.3, 1$.}

\section{Conclusions.}
\label{conclusions}

We have discussed the observational relations and other properties of a family
of inhomogeneous perfect fluid cosmological models which, in contrast to \FRW\
models, have pressure gradients and consequently acceleration of the
fundamental observers. It was demonstrated that these models do not suffer from
particle horizons (although they also do not contain a radiation field, which,
were it to be present, would dominate at early times, leading to a more rapid
evolution of the scale factor at early times and the reintroduction of
horizons). More importantly, our studies have shown that there is a significant
subset of this family that are markedly inhomogeneous but cannot be excluded on
the basis of the tests considered here. It is possible, \emph{for every
observer} in each of the models in the allowed regions of figures
\ref{exclusion_H=50} and~\ref{exclusion_H=70}, to choose the epoch of
observation so that the observed value of~$H_0$ is reproduced, the age is
greater than the measured age of the universe and there are no obviously
unacceptable features at low-redshift ($z\lapp 5$) in the observational
relations. It has also been shown that it is possible for these models to
reproduce the FLRW \mofz\ relation to high accuracy. Most importantly, though,
the dipole in the cosmic background radiation would be considerably smaller
than the observed CMB dipole: despite the inhomogeneity of the models the
anisotropy they produce is very small. The fact that this is true for every
observer means that it is not possible to reject these models by appealing to
the Copernican principle. As a result, the standard assumption that the
observed high degree of isotropy about us combined with the Copernican
principle necessarily forces the universe to be homogeneous (the cosmological
principle) is seriously undermined.

It is clear from \ref{exclusion_H=50} and~\ref{exclusion_H=70} that the
constraints on the inhomogeneity are much more severe for values of~$H_0$ at
the upper end of the currently accepted range ($H_0\gapp 70$~\hu), and from the
results of~\S\ref{Age} it can be see that a higher age also results in stricter
limits. It should be noted, though, that the constraints we have adopted are
invariably stronger than is strictly required by observations, so that the
results are conservative. Moreover, the principal interest in this thesis is
the relationship between anisotropy and inhomogeneity, particularly with regard
to the CMB. The fact that age places a strong constraint on the models for
large~$H_0$ is less interesting than the fact that the anisotropy induced in
the CMB is smaller for larger~$H_0$: it seems quite reasonable to believe that
it should be possible to find models that are similarly isotropic, though
inhomogeneous, but that have much more efficacious age characteristics. In fact
when $a>0$, which we require for $q_0$ according to the best fits, are much
older than their $a<0$ counterparts.

In the following chapter we investigate the properties of the more general
$\zeta\neq0$ models. We will demonstrate that a substantial deviation from
spherical symmetry does not cause major problems for the `global' tests we have
considered so far. However, the local dipole is large enough to possibly
preclude some of these models.

\forget{
Before we can accept the conclusion that the cosmological principle must be
rejected it is obviously necessary to extend the analysis to the low-redshift
relations for which the largest quantity of observational data is available.
This we will do in a subsequent paper. The most important point to be
considered is the extent to which the dipole (and other multipoles) introduced
in the observational relations can be rejected on the basis of presently
available constraints. This is a complicated issue, because the nearby universe
is not isotropically distributed about us (the standard interpretation of this
is that the local universe contains a significant inhomogeneity~-- the Great
Attractor~-- towards which the Local Group is falling), so there \emph{are}
anisotropies in the data, and it is necessary to assess not just their
magnitude but also their distribution on the sky (the Great Attractor
interpretation, for example, predicts a small dipole nearby but a large
quadrupole due to the geodesic shear flow into the overdensity). Furthermore,
for the supernova data (Perlmutter \etal~1998b) analysis of any anisotropy
requires positions as well as magnitudes, making the fitting procedure much
more difficult. Preliminary results suggest, though, that low-redshift
observations (galaxy surveys and supernovae) may also be unable to rule out the
D\c{a}browski models.

Even if, as seems quite possible, these models do force us to reject the
standard justification for the cosmological principle~-- based in no small
measure on the EGS and almost-EGS theorems of Ehlers, Geren and
Sachs~(1968)\nocite{ehlers-68} and Stoeger, Maartens, and
Ellis~(1995)\nocite{stoeg-95}~-- it is necessary to look more closely at the
D\c{a}browski models, particularly at their matter content, to see whether they
possess obviously undesirable properties at some more subtle level than has
been considered here that would lead us to reject them on basic physical
grounds. For the moment, though, we must take seriously the possibility that
the D\c{a}browski spacetimes may be viable inhomogeneous models of the universe
and, perhaps more importantly, we must recognise that they represent an
`existence proof': if the D\c{a}browski models are capable of providing a
satisfactory fit to presently available observational constraints there must be
an enormous space of other inhomogeneous models that would prove similarly
satisfactory. }

}
\forgetmenot{
\chapter{Observational Characteristics When $\zeta\neq0$}
\label{step chapter hyperbolic----------------------}
In chapter~\ref{step chapter 1------------------------------} we looked into
the general observational characteristics of the spherically symmetric Stephani
solutions ($\zeta=0$) which admit an isotropic radiation field, with the scale
factor $R$ restricted to be a quadratic, and we demonstrated that the models
have no significant problems with fitting the broader aspects of our universe.
It was also shown that some of the allowed models are significantly
inhomogeneous. There is no particular reason for choosing a scale factor of
this form, and it is highly likely that the freedom in $R(t)$ could mean that
there are even better forms for the scale factor to fit observations. This is a
degree of freedom which we do not exploit here.

This chapter is devoted to giving the results for the more general case of
$\zeta\neq0$ (the `isotropic radiation field' or IRF models). The idea is
exactly the same as before, but the algebra is a bit more messy. In fact none
of the observational relations can be given in analytic form, so graphical
methods are used to show that there is a distinctly non-zero `volume' of
parameter space that cannot be rejected by current observations.

\section{Hyperbolic Stephani Models: Overview}
\label{Hyperbolic Stephani Models: Overview section}

We have commented before on the conformal nature of the IRF models, which have
allowed us to perform rotations to arbitrary locations in the spacetime. The
same conformal nature has allowed us to find all the relevant observational
relations for these models. The inhomogeneity of the models is `contained' in
the conformal factor $W$. In fact, because the light rays are unaffected by the
conformal factor, we see inhomogeneities as they were in the past, which allows
the CMB to be so isotropic (the universe is homogeneous at early times.

We have the conformal factor of the metric~(\ref{metric_W2})
\ba
W\li=\li\Phi_++\Phi_-\cos\psi\cos\chi-
\;{2\zeta R}{c\sqrt\Delta}\cos\xi\sin\psi\cos\chi
\nonumber\\ \li+\li\sin\chi\cos\vartheta\sqrt{\Phi_-^2\sin^2\psi
+4\;{\zeta^2 R^2}{c^2\Delta}(1-
\cos^2\xi\sin^2\psi)+2\;{\zeta R\Phi_-}{c\sqrt\Delta}\cos\xi\sin2\psi},
\label{W_complete2}
\ea
where
\be
\Phi_\pm\eqdef\;12\left[1\pm\;\kappa\Delta+4\zeta^2\;{R^2}{c^2\kappa}\right].
\ee
We have $\kappa$ restricted by~(\ref{kappa_IRF_general});
\be
\kappa(t)\eqdef\;12\Delta-2\;ac R+ 2\Sigma_\pm
\sqrt{\left(\:14\Delta-\;ac R\right)^2
 +\;{\Delta\zeta^2}{c^2} R^2},
\ee
where
\be
\Sigma_\pm\eqdef\hbox{sign}\left(\;12\Delta-\;{2a}{c} R\right).
\ee
We take the sign in front of the square root term to be the sign of
$\:12\Delta-2aR/c$ in~(\ref{kappa_IRF_general}) to give the \dab\ models in the
limit $\zeta\rightarrow0$ whatever the sign of $\:12\Delta-2a R/c$. We take $R$
to be
\be
R(t)=ct(at+b).
\ee
Note that $\Delta\eqdef 1-b^2$ and we replaced $\delta\eqdef-a/c$.

The units we will use are as follows: $[c]=$~km~s$^{-1}$, $\chi, \psi, \xi$~are
dimensionless, $R$~is in Mpc and $[\ti]=$~Mpc~s~km$^{-1}$$=[1/H_0]$, so that
$[a]=$~km~s$^{-1}$~Mpc$^{-1}$$=[H_0]$ and $b$~is dimensionless. Note that we
have redefined $\zeta\mapsto\zeta/c$ to give the units of $[\zeta]=[H_0]$.

We will now use all the constraints of chapter~\ref{step chapter
1------------------------------}.

\section{Energy Conditions}
\label{energy_cond_hyp}

As before, we don't have a specific form of thermodynamic scheme for the IRF
models, although chapter~\ref{EGS chapter---------------------------------}
proved the existence of one. Knowing the specific form of such a scheme would
be desirable, but rather difficult (Sussman,~1999).  Rather we wish to limit
the models just using the energy conditions:
\ba
\mu\li\geq\li0\\
\mu c^2+p\li\geq\li0\\
\mu c^2+3p\li\geq\li0\\
\mu c^2-|p|\li\geq\li0.
\ea
The first must always be satisfied; the second is the weak energy condition,
while the other two are the strong and dominant energy conditions respectively.
They say that for all reasonable forms of matter the energy density is positive
and the pressure cannot be very large and negative; the dominant conditions
goes even further and says that positive pressures can't be large either.  For
the IRF models we have
\ba
\;{8\pi G}{c^2}\mu\li=\li\;{3\kappa}{R^2}+\;{\theta^2}{3c^2}=\;{3\kappa}{R^2}+
\;{3R^2}{c^2\Delta^2}
\left[\left(\;{\kappa}{R}\right)_{,t}\right]^2\nonumber\\
p\li=\li-\mu
c^2+\;13\mu_{,t}c^2\left.\;{W}{R}\right/\left(\;{W}{R}\right)_{,t}.
\ea
The conditions are pretty complicated and will not submit to an analytic
treatment. We can turn to graphical methods to give a rough guide. It turns out
that, as in the \dab\ case the strong condition requires $a<0$, and the
dominant energy condition fails at certain times and places. As in
\S\ref{energy_cond}, things are worst when the universe is half way through its evolution,
$t=-b/2a$; similarly on one timelike slice the condition will hold in some
places (near $\psi=0$) and fail at others ($\psi=\pi$). We test the condition
at this spacetime point ($t=-b/2a,~\psi=\pi$), on the assumption that if the
models are unacceptable there then they will be unacceptable everywhere.

In figure~\ref{dominant_EC_b-vs-zeta-3a_psi=pi},
\begin{figure}[here!]
\centerline{\psfig{file={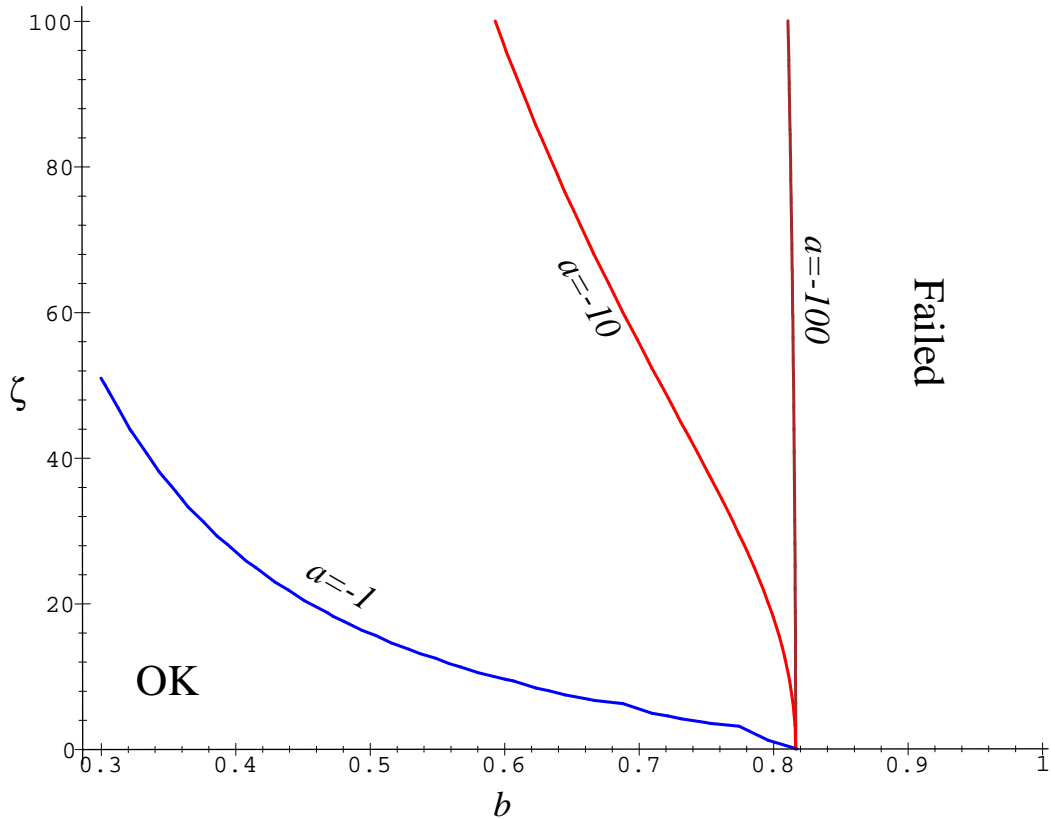},width=14cm,angle=0}}
\caption{\small Regions of the $b-\zeta$ plane rejected by the dominant energy
condition, at $t=-b/2a,~\psi=\pi$ for 3 values of $a$. What is surprising is
that a small value of $a$ (when $\zeta\neq0$) rules out more of the $b-\zeta$
parameter space.
\label{dominant_EC_b-vs-zeta-3a_psi=pi}}
\end{figure}%
we show that a high value of $\zeta$ combined with a \emph{small} value for $a$
fails the dominant energy condition. As $a$ becomes large and negative, $\zeta$
becomes less and less important, with $b$ being most important for determining
the acceptability of the models. As in the case $\zeta=0$, any value of
$b\geq\sqrt{2/3}$ is always disallowed.

To try to get a feeling for this, consider
figure~\ref{dominant_EC_a-vs-b-3zeta_psi=pi},
\begin{figure}[here!]
\centerline{\psfig{file={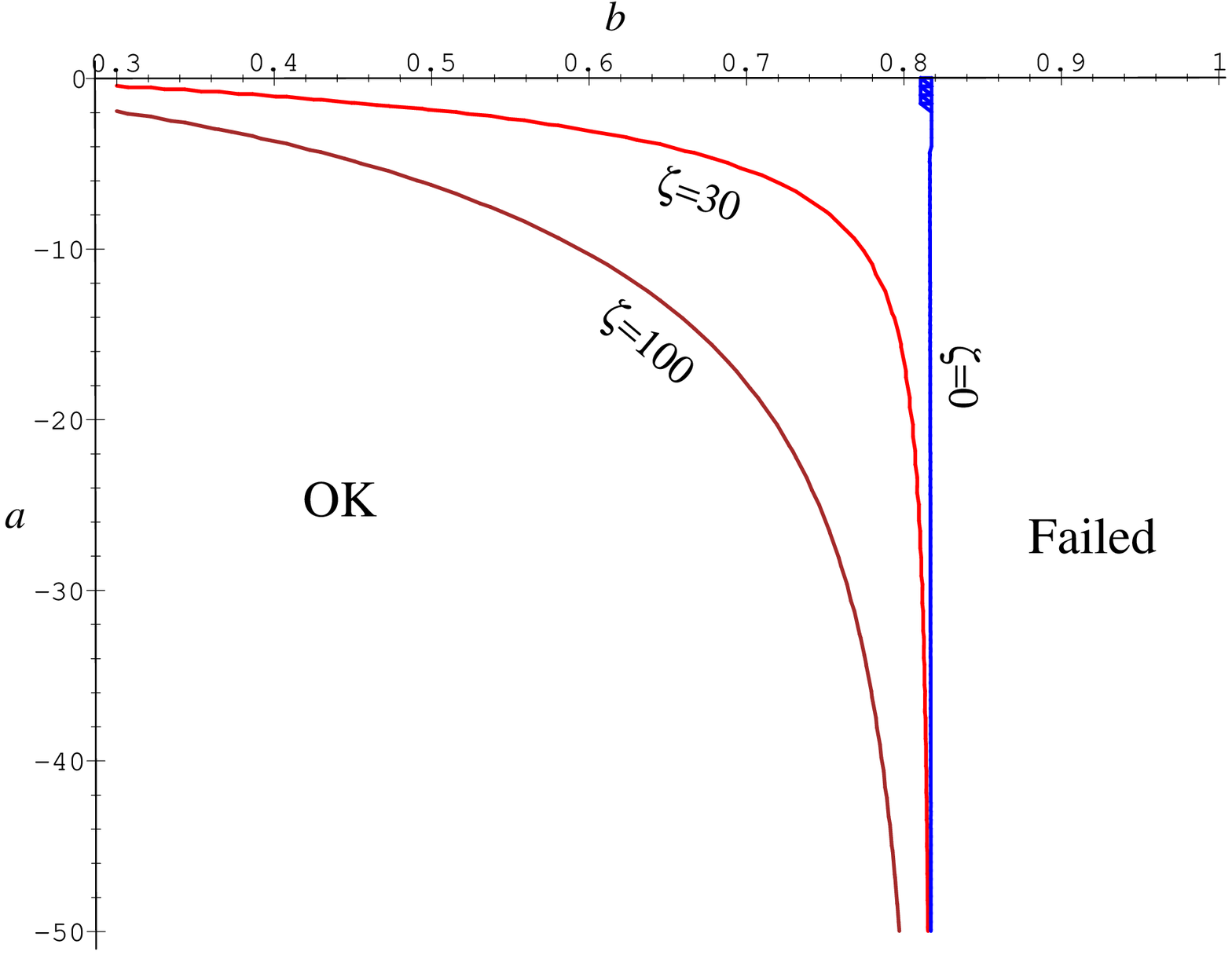},width=14cm,angle=0}}
\caption{\small Regions of the $a-b$ plane rejected by the dominant energy
condition, at $t=-b/2a,~\psi=\pi$ for 3 values of $\zeta$. Once again a small
value of $a$ (when $\zeta\neq0$) rules out more of the parameter space.
\label{dominant_EC_a-vs-b-3zeta_psi=pi}}
\end{figure}%
where we show the same as figure~\ref{dominant_EC_b-vs-zeta-3a_psi=pi}, but now
in the $a-b$ plane. As $\zeta$ becomes large, $a$ must be large and negative or
the dominant energy condition will be broken.

The Stephani models have enough problems with their matter content without
chastising them for failing this condition. We will banish this problem from
our minds, in the knowledge that it actually won't really matter because, as we
will see below, the dominant energy condition doesn't rule out anything that
isn't more or less ruled out by other conditions. The matter is `reasonable
enough' to pass both the weak and strong conditions.

\section{$H_0$}

We must have our observers living at a coordinate time which is representative
of us. We do this by ensuring that the observers measure a (mean) Hubble's
constant (ie, $\theta/3$) the same as we measure. It will not change over a
surface of constant time.

As before, we require the expansion rate to lie somewhere in the range $50\leq
H_0\leq80$~\hu. This means that for any given choice of parameters the
age,~$T$, of the universe is given by the solution of
\be
H_0=\left.-\;{R}{\Delta}\left(\;{\kappa}{R}\right)_{,t}\right|_{t=T}.
\ee
For simplicity, in this chapter we take a fixed value of $H_0=60$~\hu, because
the added complication of considering a range of $H_0$ does not warrant the
effort: in chapter~\ref{step chapter 1------------------------------} the
differences were only small. So, for any given values of the parameters $\zeta,
a, b$ we have a value of the coordinate age $T$ given by the solution of
\be
\left.-\;{R}{\Delta}\left(\;{\kappa}{R}\right)_{,t}\right|_{t=T}=60,
\label{coord age H_0 soln}
\ee
which is pretty complicated (it's a sixth order polynomial in $T$).

\section{Age}
\label{age hyperbolic section}

It is known that the universe is older than 10 Gyr old, and probably older than
12 Gyr (cf, \S\ref{Age}). The standard model has gone through various crises
over this problem, but at the moment things seem to be acceptable, provided
that $H_0\lapp67$~\hu. We require that any observer in the spacetime, with a
local expansion rate $H_0=60$~\hu\ be older than 10 or 12~Gyr. The proper age
of any observer is given by
\be
\tau_0\eqdef\int_0^T\;{dt}{W(0,t)}
\ee
where $T$ is the solution of (\ref{coord age H_0 soln}). This integral may have
an analytical solution, but we just use a simple numerical analysis, as for all
the other tests. To gauge what is going on, in
figure~\ref{age_exclusion_3d_b-vs-zeta_a=-5},
\begin{figure}[here!]
\centerline{\psfig{file={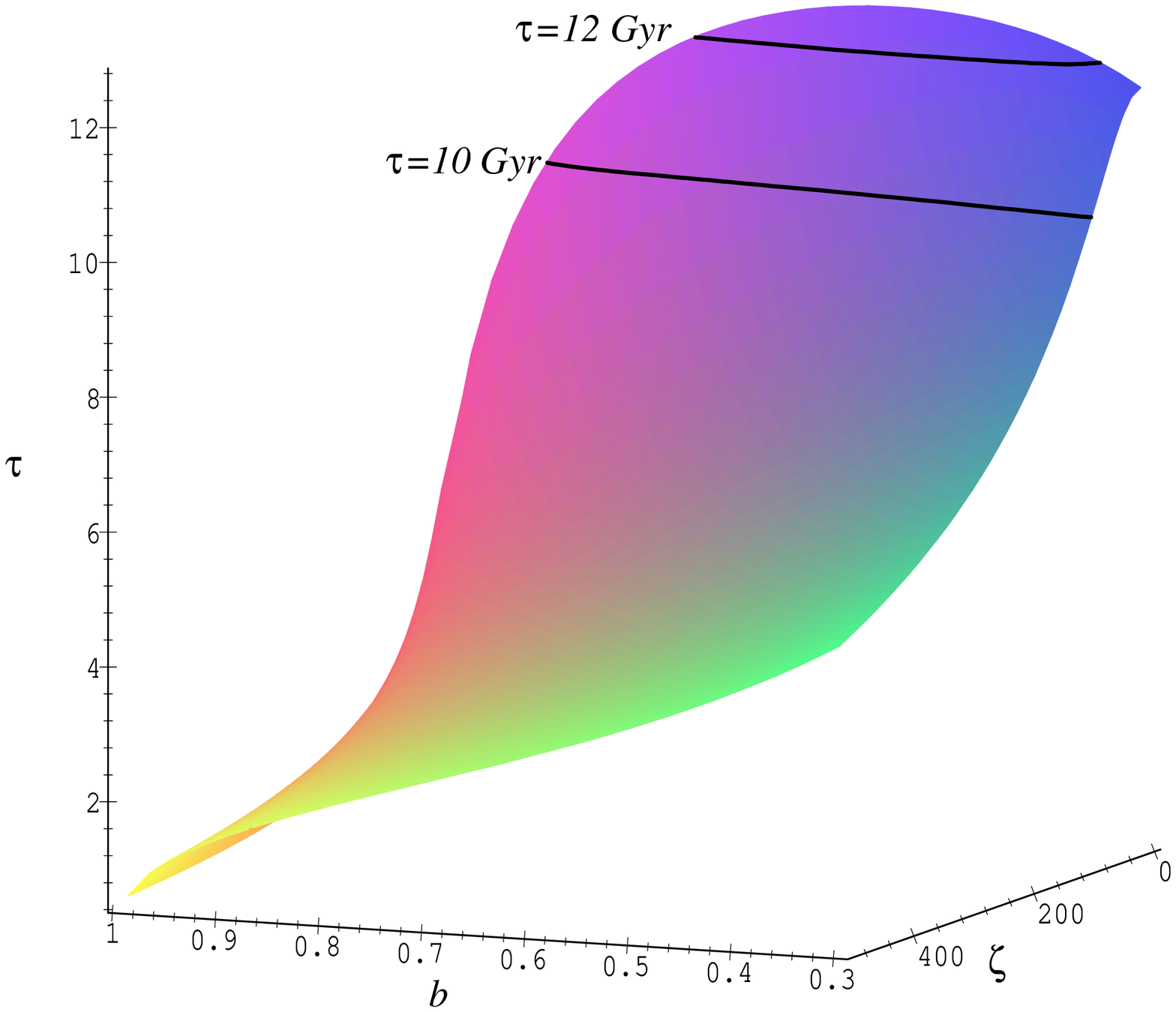},width=14cm,angle=0}}
\caption{\small Age exclusion plot of $\zeta$ vs $b$ for $a=-5$ and
$H_0=60$~\hu\ at $\psi=\pi$. The contours mark the limits of the allowed ages.
\label{age_exclusion_3d_b-vs-zeta_a=-5}}
\end{figure}%
we show a surface plot of proper age verses $b$ and $\zeta$ for $a=-5$.
contours at $10$ and $12$ Gyr are marked on, demonstrating that for high
$\zeta$ the age is far too low, for the expansion rate we measure. This means
that if we introduce a large $\zeta$ the expansion rate will be too low when
the universe is old enough to accommodate globular clusters. This can also be
seen in figures~\ref{age_exclusion_a-vs-zeta_b=0.5,0.7}
and~\ref{age_exclusion_a-vs-b_3zeta}.

\begin{figure}[here!]
\centerline{\psfig{file={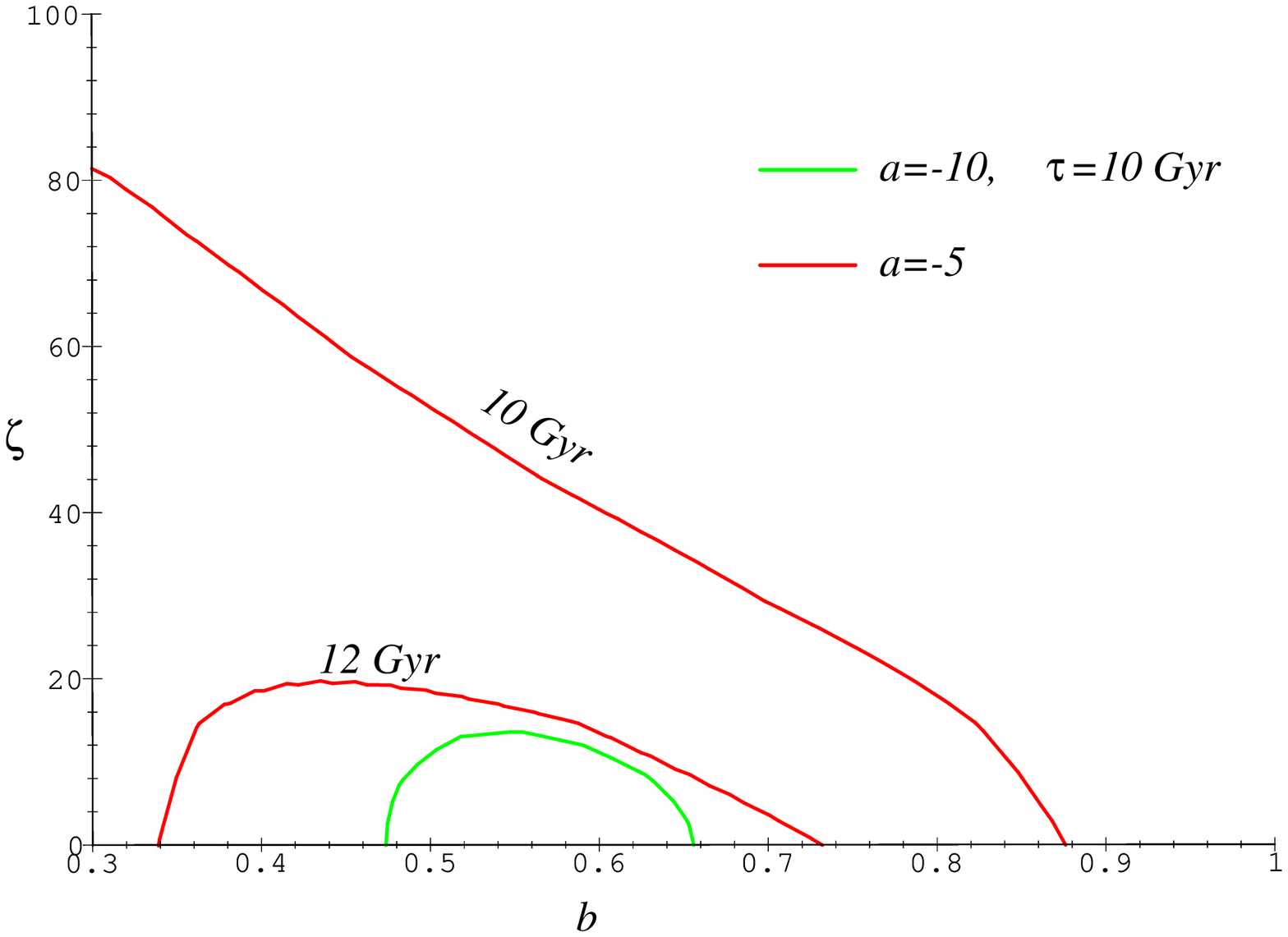},width=14cm,angle=0}}
\caption{\small Age exclusion plot of $\zeta$ vs $b$ for various $a$ with
$H_0=60$~\hu\ at $\psi=\pi$. The contours mark the limits of the allowed ages.
\label{age_exclusion_b-vs-zeta_a=-5}}
\end{figure}%

\begin{figure}[here!]
\centerline{\psfig{file={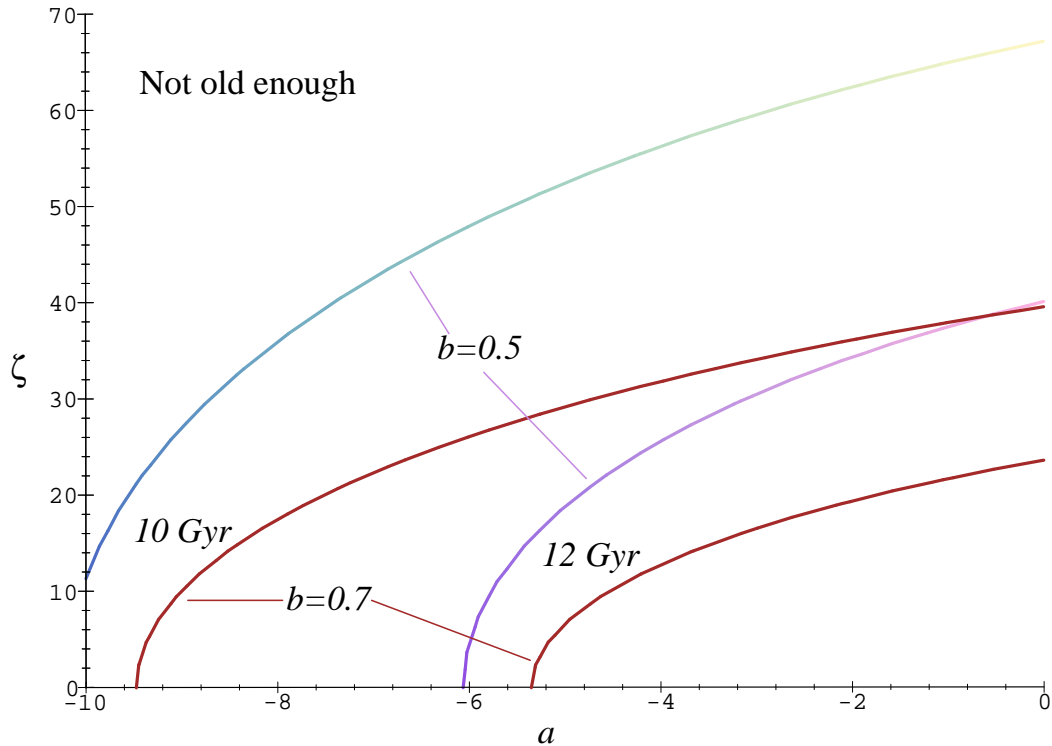},width=14cm,angle=0}}
\caption{\small Age exclusion plot of $\zeta$ vs $a$ for $a=-5$ and
$H_0=60$~\hu\ at $\psi=\pi$. The contours mark the limits of the allowed ages.
As in the case $\zeta=0$ a large negative $a$ is not allowed by this test.
\label{age_exclusion_a-vs-zeta_b=0.5,0.7}}
\end{figure}%

\begin{figure}[here!]
\centerline{\psfig{file={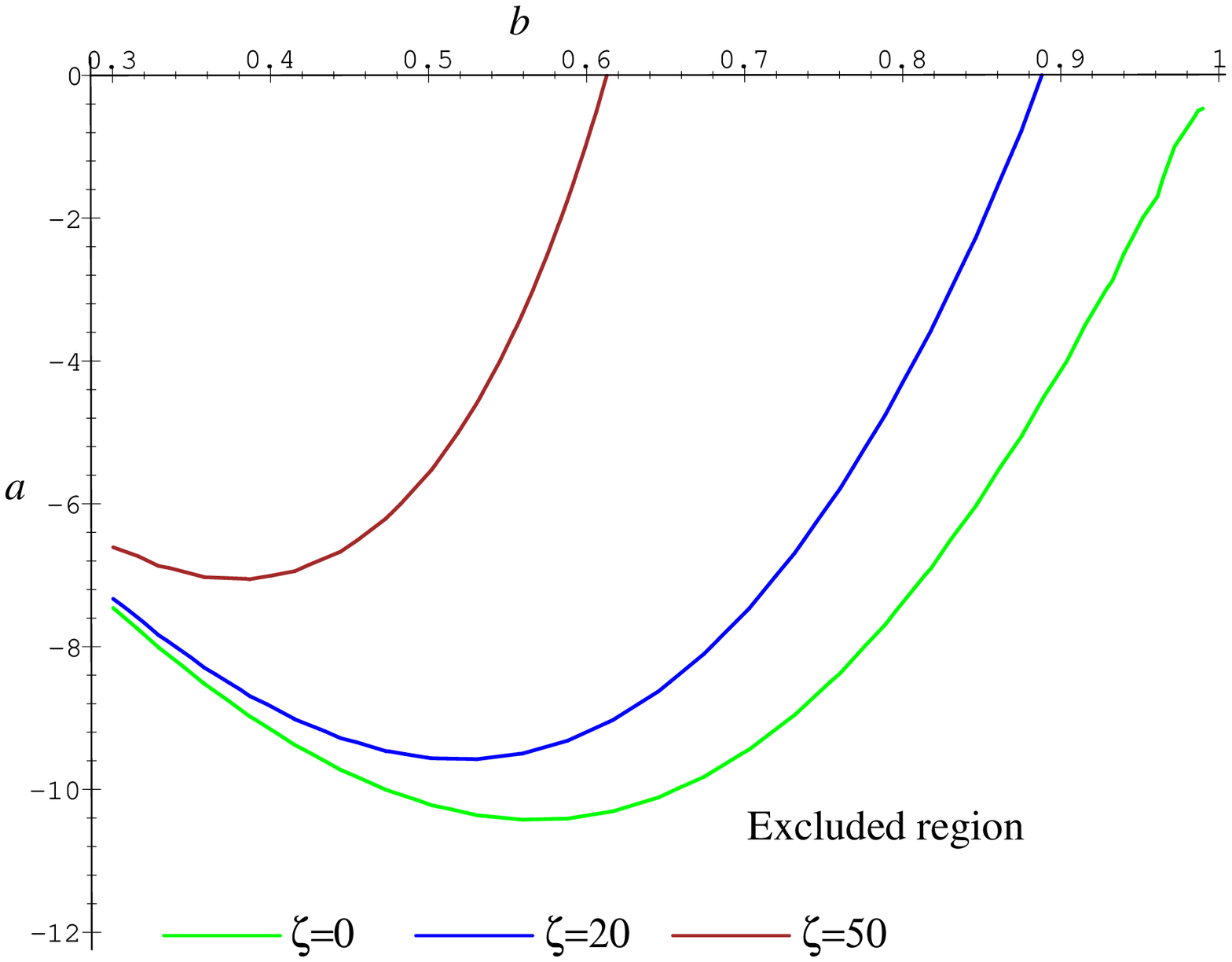},width=14cm,angle=0}}
\caption{\small Age exclusion plot of $b$ vs $a$ for various values of $\zeta$ and
$H_0=60$~\hu\ at $\psi=\pi$. These curves are all for an age of $10$ Gyr.
Increasing $\zeta$ restricts high values of $b$.
\label{age_exclusion_a-vs-b_3zeta}}
\end{figure}%

\section{Area Distance}
\label{area distance hyperbolic -- section}

In \S\ref{size} we considered the area distance-redshift relation, with respect
to the `size' of the observed spatial sections. This is because the zero in the
distance-redshift relation at small redshift, for certain model parameters
(namely small $b$), is probably ruled out by present observations. Since the
spacetime is conformal to a spherically symmetric spacetime, a zero in $r_A(z)$
would correspond to a star situated opposite to an observer (ie, at
$\chi=\pi$~for all~$\psi$) would become spread all over the observers sky and
have an infinite apparent magnitude -- it would look as bright as if the star
were right next to the observer. Similarly, an object close to a pole would
become distorted in a similar manner to normal gravitational lensing~-- see
MacCallum and Ellis (1970) for the distortion formula. As in \S\ref{size}, we
make the assumption that such an effect would have been detected\footnote{One
might argue that if such a phenomenon were to be detected it would require a
systematic comparison of close-by galaxies with ones much further away, and at
much earlier stages of their evolution. This would involve deciding what a
particular galaxy would look like at say 1/4 or 1/2 its age.}. As before we
consider the cases $z_\pi\geq2$ and $z_\pi\geq5$.

To proceed, then, we simply need look at the $z(\chi)$ function at $\chi=\pi$
and at a coordinate age given by~(\ref{coord age H_0 soln}); ie, we must find
the range of parameter space for which
\be
z(\chi=\pi)\geq2,~\hbox{or}~5.
\ee
We are dealing with a 3-dimensional parameter space so we must proceed with
care. In figure~\ref{ra_z=0_3d_exclusion_a-vs-zeta_b=0-7}
\begin{figure}[t!]
\centerline{\psfig{file={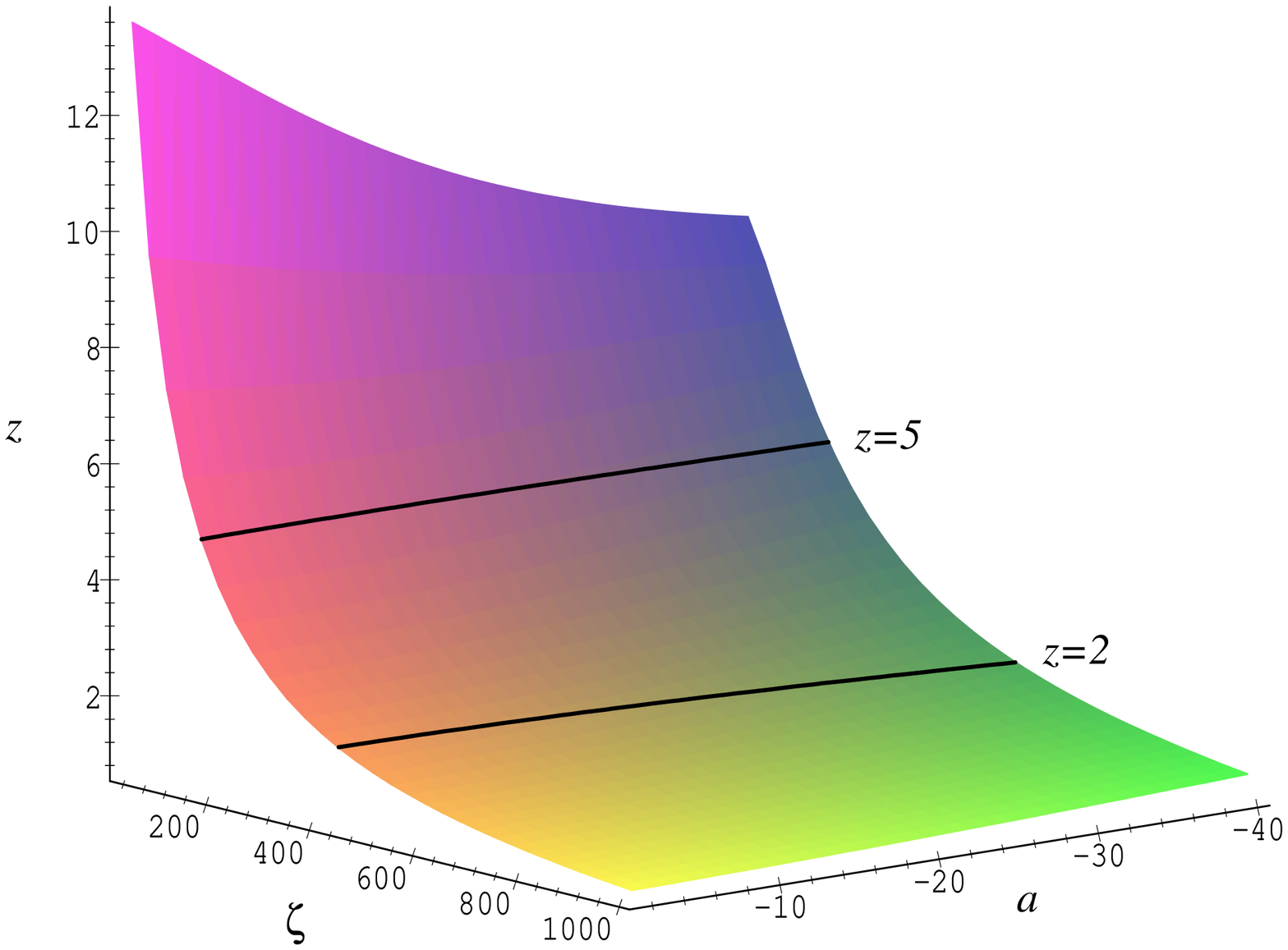},width=14cm,angle=0}}
\caption{\small Exclusion plot of $\zeta$ vs. $a$ for $b=0.7$ and $H_0=60$~\hu\
for the first zero of the $r_A(z)$ function.
\label{ra_z=0_3d_exclusion_a-vs-zeta_b=0-7}}
\end{figure}%
we show a 3-dimensional plot of $z$ for a variation of the parameters $a$ and
$\zeta$, for $b=0.7$. We see from this that for large $\zeta$ the redshift at
which the first zero in the $r_A(z)$ function occurs at a very low redshift.
The second thing to notice is that, even though $a$ varies between $-40$ and
$0$ the curves don't change much. In fact, if we look at
figure~\ref{ra_z=0_exclusion_zeta-vs-a},
\begin{figure}[t!]
\centerline{\psfig{file={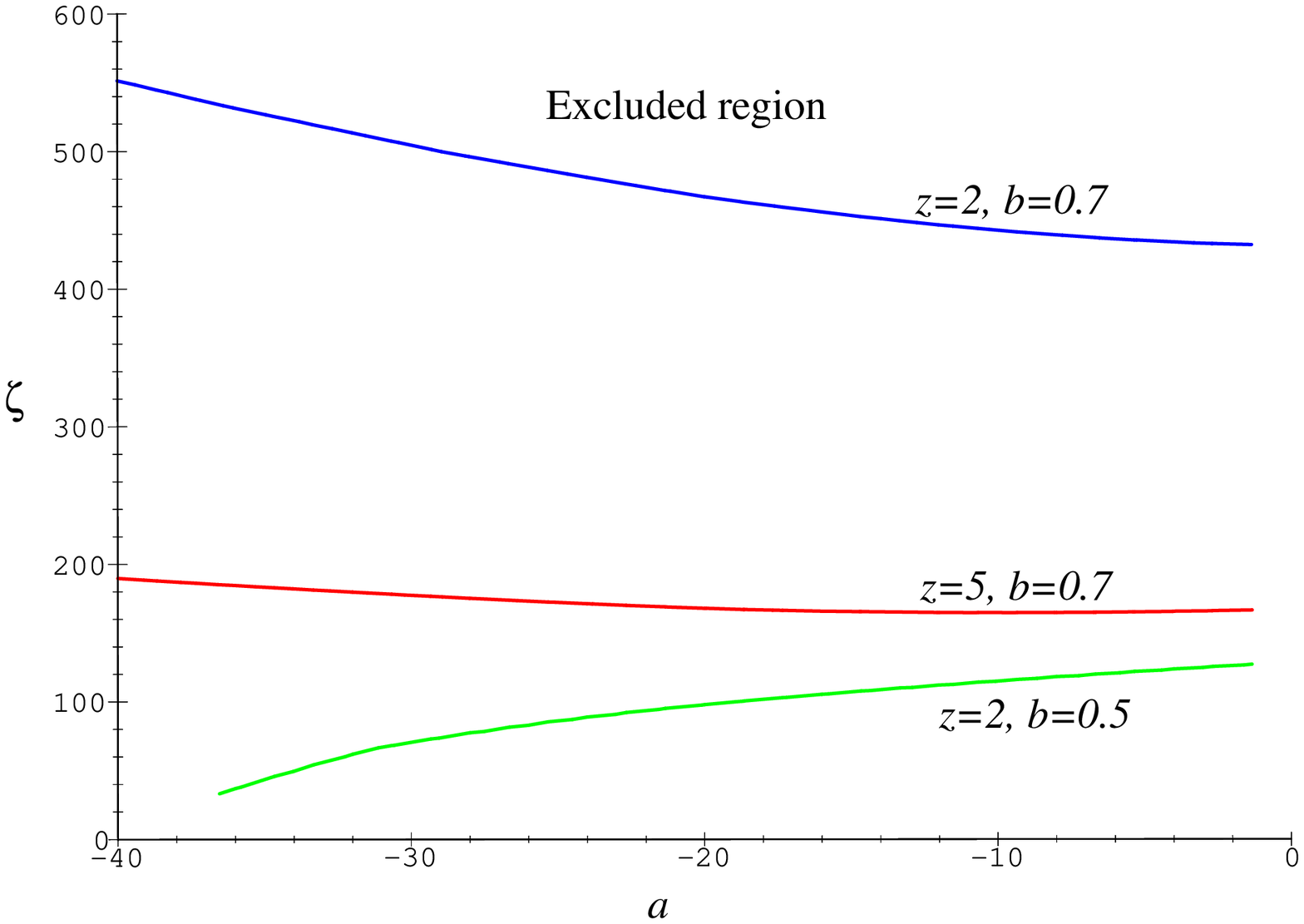},width=14cm,angle=0}}
\caption{\small Exclusion plot of $\zeta$ vs. $a$ for $b=0.7, 0.5$ and $H_0=60$~\hu\
for the first zero of the $r_A(z)$ function. This clearly shows that only $b$
affects this function significantly.
\label{ra_z=0_exclusion_zeta-vs-a}}
\end{figure}%
we see that $a$ doesn't do much damage at all: $b$ and $\zeta$ seem to be most
important. In chapter~\ref{step chapter 1------------------------------} this
was also the case; most of the problems came from small $b$ (cf,
fig~\ref{rA_exclusion}). In figure~\ref{ra_z=0_exclusion_a=-10_b-zeta},
\begin{figure}[th!]
\centerline{\psfig{file={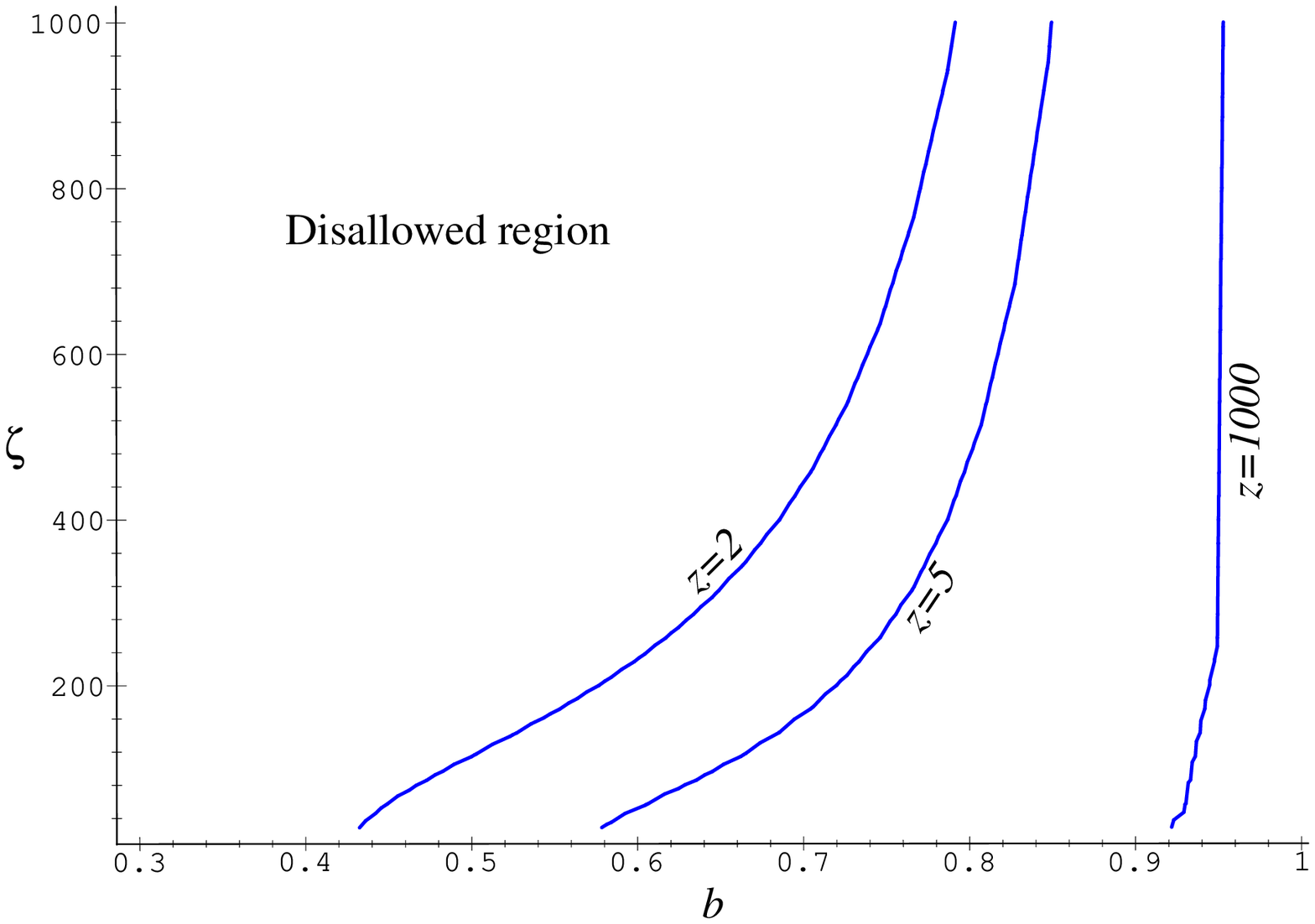},width=14cm,angle=0}}
\caption{\small Exclusion plot of $\zeta$ vs. $b$ for $a=-10$ and $H_0=60$~\hu\
for the first zero of the $r_A(z)$ function.
\label{ra_z=0_exclusion_a=-10_b-zeta}}
\end{figure}%
we demonstrate that this is the case: small $b$ is ruled out, while large
$\zeta$ is only permitted if $b$ is sufficiently close to 1.

Also in figure~\ref{ra_z=0_exclusion_a=-10_b-zeta}, we show that for the first
zero to occur `behind the CMB' requires $b\sim1$.

In order to understand the effects that $\zeta$ has on the spacetime, we need
to study the $a-b$ plane for various values of $\zeta$.
\begin{figure}[here!]
\centerline{\psfig{file={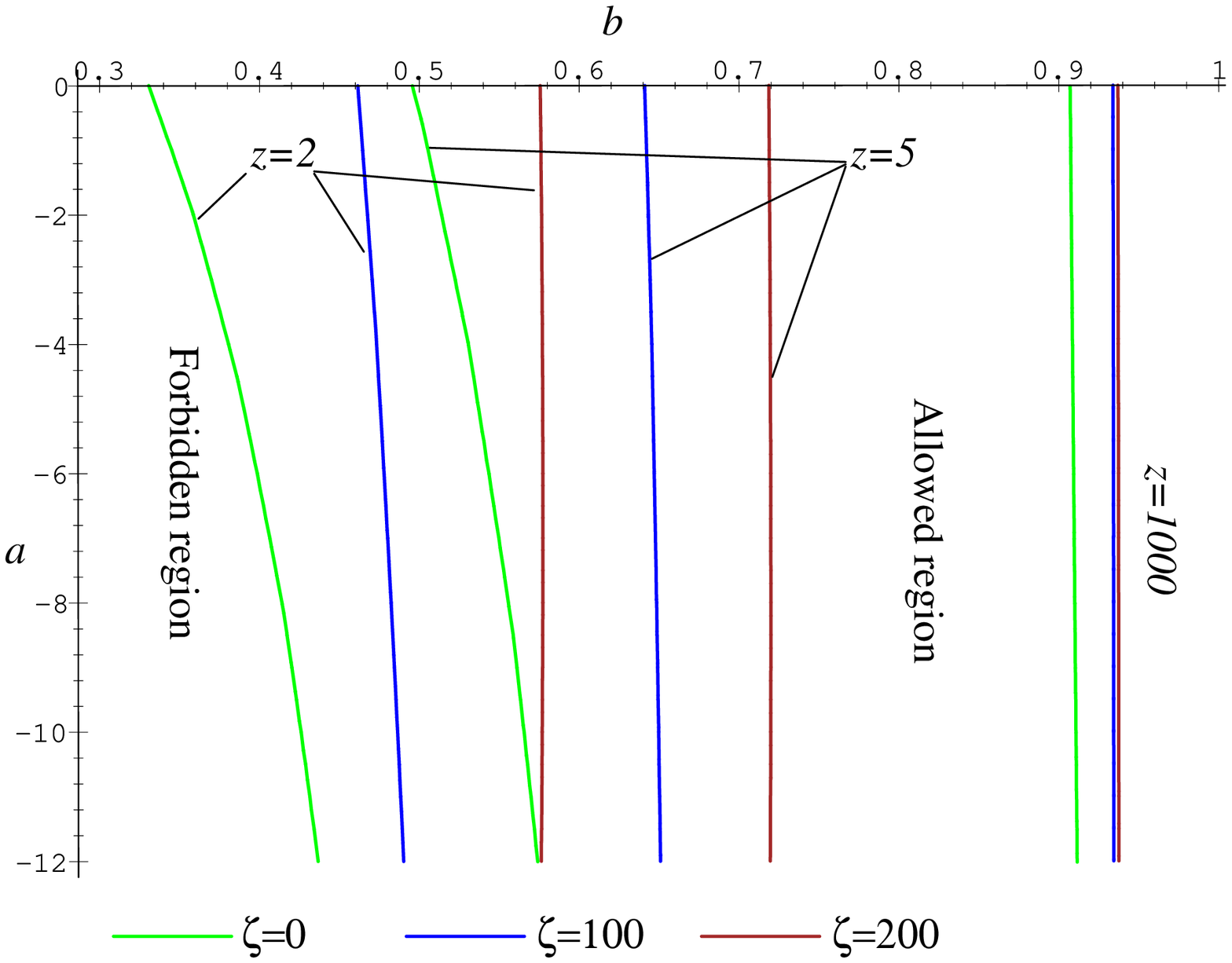},width=14cm,angle=0}}
\caption{\small Exclusion plot of $a$ vs. $b$ for various values of $\zeta$
for the first zero of the $r_A(z)$ function.
\label{ra_z=0_a-vs-b_3zeta}}
\end{figure}%
In figure~\ref{ra_z=0_a-vs-b_3zeta} we show this exclusion plot, from which we
can immediately see that a large $\zeta$ has the effect of moving the excluded
region towards larger $b$, while at the same time making $a$ more and more
redundant as a parameter for this test. Physically the effect of a large
$\zeta$ is to decrease the `size' of the spatial sections, as one looks into
the past; the first zero in the $r_A(z)$ function will occur closer to the
observer.

\section{The CMB Anisotropy}
\label{CMB anisotropy hyperbolic section}

\forget{Although it has been shown in chapter~\ref{EGS
chapter---------------------------------} that the IRF models admit an IRF,
this does not actually imply that the CMB will necessarily be isotropic,
although we will demonstrate that this is nearly true. In fact it is a
radiation field emitted `at the big bang' which is isotropic, and not a
radiation field emitted afterwards, such as the CMB. It is because we use a
more physical definition of the CMB here that it becomes `almost isotropic.'}

We assume for simplicity that the CMB is a blackbody which occurs at $T=3000$K,
which is roughly the temperature of decoupling. As it is a blackbody, the
temperature evolves according to (see MacCallum and Ellis~1970)
\be
\left.T\right|_{_\mathsf{today}}=\;{T\cmb}{1+z}.
\ee
\forget{As far as we're concerned, the mean redshift of the CMB is $z_0\sim1000$, and
has a dipole moment of $\sim10^{-3}$, which is attributed to our peculiar
motion, and higher-order moments $\sim10^{-5}$, which are accounted for by
various physical means, such as perturbations of one kind or another before and
after decoupling. As in \S\ref{CMB} we must ensure that any anisotropy
occurring from the Stephani spacetime inhomogeneity must be smaller that those
observed and accounted for.}

Our observer at some location given by $\psi$ will measure the temperature of
the CMB to be given by
\be
T_o(\psi,\chi\cmb,\vartheta)=\;{T\cmb}{1+z(\psi,\chi\cmb,\vartheta)}.
\ee
Now from~(\ref{redshift_new}) and~(\ref{W_complete}) we may write
\be
1+z(\psi,\chi\cmb,\vartheta)=1+z_0(\psi,\chi\cmb)+z_1(\psi,\chi\cmb)\cos\vartheta,
\label{zcmbdef2}
\ee
where
\ba
1+z_0\li\eqdef\li\;{R_0}{W_0 R\cmb}\left\{\Phi_{+\cmb}+\Phi_{-\cmb}
\cos\psi\cos\chi\cmb\right.\nonumber\\&&\hfill\left.
-
\;{2\zeta R\cmb}{c\sqrt\Delta}\cos\xi\sin\psi\cos\chi\cmb\right\}\label{z_0 gen},\\
z_1\li\eqdef\li\;{R_0\sin\chi\cmb}{W_0 R\cmb}\left\{\Phi_{-\cmb}^2\sin^2\psi
+4\;{\zeta^2 R\cmb^2}{c^2\Delta}(1-
\cos^2\xi\sin^2\psi)\right.\nonumber\\
&&\hfill\left.+2\;{\zeta
R\cmb\Phi_{-\cmb}}{c\sqrt\Delta}\cos\xi\sin2\psi\right\}^{\;12},
\ea
where $`X\cmb$'$\equiv `X(t(\chi\cmb))$'. The multipole moments in the CMB
temperature fluctuations will then become
\be
T_{\mathrm{obs}}(\vartheta)=\frac{T_{\mathrm{dec}}}{1+z_0}\left[1-
\frac{z_1}{1+z_0}\cos\vartheta
    +\left(\frac{z_1}{1+z_0}\right)^2 \cos^2\vartheta
    + {\cal O}(\cos^3\vartheta)\right],
\ee
that is,
\be
\frac{\delta T(\vartheta)}{T}=-\frac{z_1}{1+z_0}\cos\vartheta
    +\left(\frac{z_1}{1+z_0}\right)^2 \cos^2\vartheta
    + O(\cos^3\vartheta).
\label{variation of CMB2}
\ee

The strategy is then as follows:
\begin{enumerate}
\item Given a set of parameters, $a, b, \zeta, \xi, \psi$ we solve~(\ref{coord age H_0
soln}) for the coordinate age~$T$.
\item Given our complete parameter set $a, b, \zeta, \xi, \psi, T$ we then find
the mean distance to the CMB surface $\chi\cmb$ by solving
\be
z_0(\chi\cmb)=1000
\ee
for $\chi\cmb$. This must be done numerically because the equations are such a
mess.
\item We can then calculate the temperature anisotropy by calculating $z_1$
using the value of $\chi\cmb$ found above. The dipole moment of the CMB
temperature is then (using~(\ref{zcmbdef2}))
\be
                \delta_1 = \frac{z_1}{1+z_0} \approx 10^{-3}z_1.
\label{Tdipole2}
\ee
\end{enumerate}
We do not wish that the IRF models account for the entire dipole moment of the
CMB, because there is good evidence that this is due to our peculiar velocity
of the local group. However, there is no reason to say that a small part of the
dipole cannot be of a cosmological origin: a variation of $10\%$ of the dipole
could in principle be of a cosmological nature. We take this constraint here;
that is $\delta_1\sim10^{-4}$, or $z_1\sim0.1$.

So, that's the plan. We are principally interested in the effects a non-zero
$\zeta$ will have. In
figure~\ref{cmb_3dexclusion_b-vs-zeta_a=-5_psi=0.25pi_xi=0.5pi}
\begin{figure}[here!]
\centerline{\psfig{file={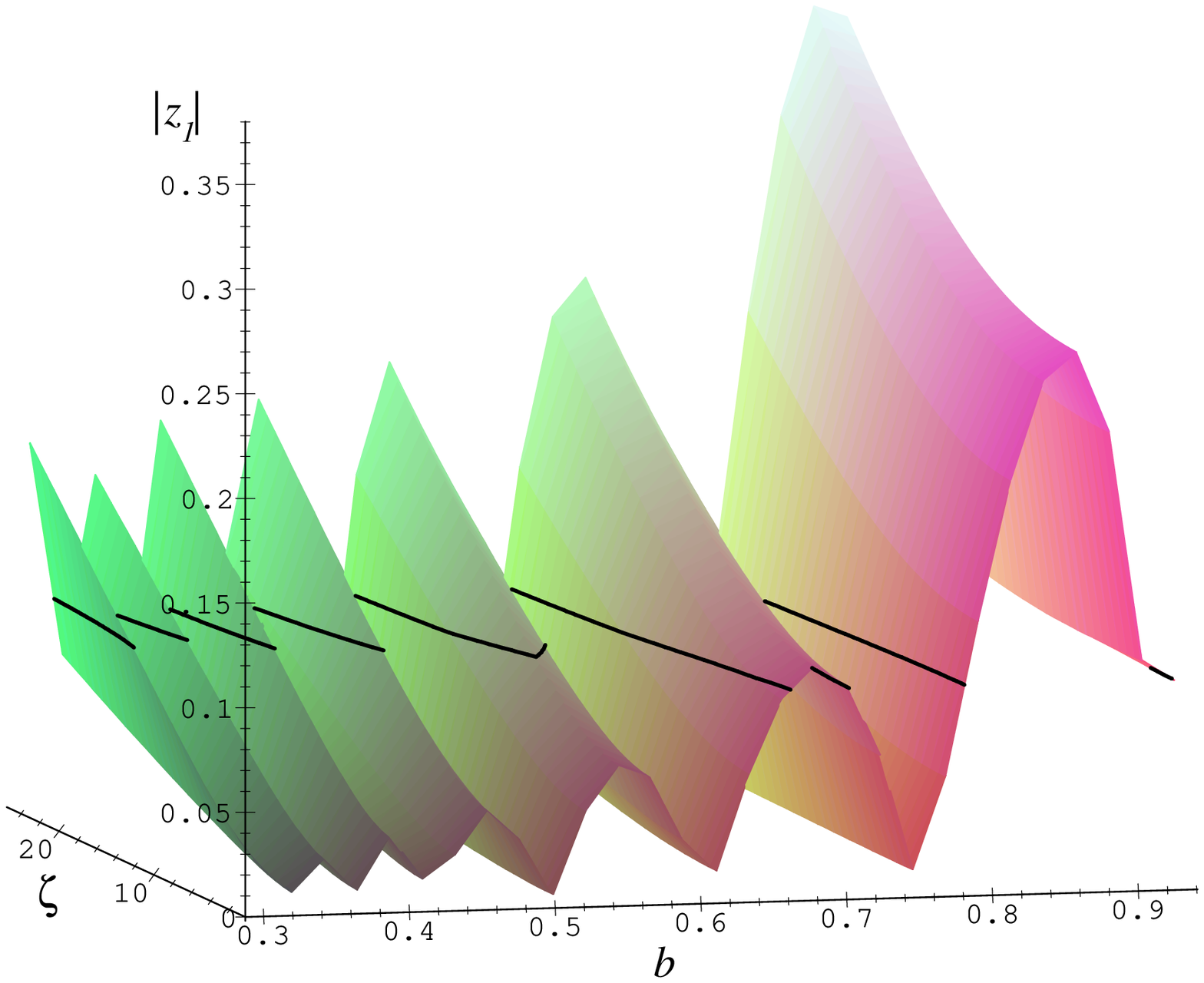},width=14cm,angle=0}}
\caption{\small Exclusion plot of $\zeta$ vs. $b$ for $\psi=1/4\pi$
for the CMB anisotropy. $a=-5$. Regions below the black contour is acceptable.
\label{cmb_3dexclusion_b-vs-zeta_a=-5_psi=0.25pi_xi=0.5pi}}
\end{figure}%
we show a surface plot of $|z_1|$ against $b$ and $\zeta$. The contour
$z_1=0.1$ is shown, with the surface below this line representing the
acceptable parameter range. This is for an observer at $\psi=\pi/4$. We wish to
constrain the parameters from the most restrictive position; it is not obvious,
when $\zeta\neq0$ where the `worst' position is.
Figure~\ref{cmb_exclusion_b-vs-zeta_a=-5_psi=0.25-0.5pi_xi=0.5pi}
\begin{figure}[here!]
\centerline{\psfig{file={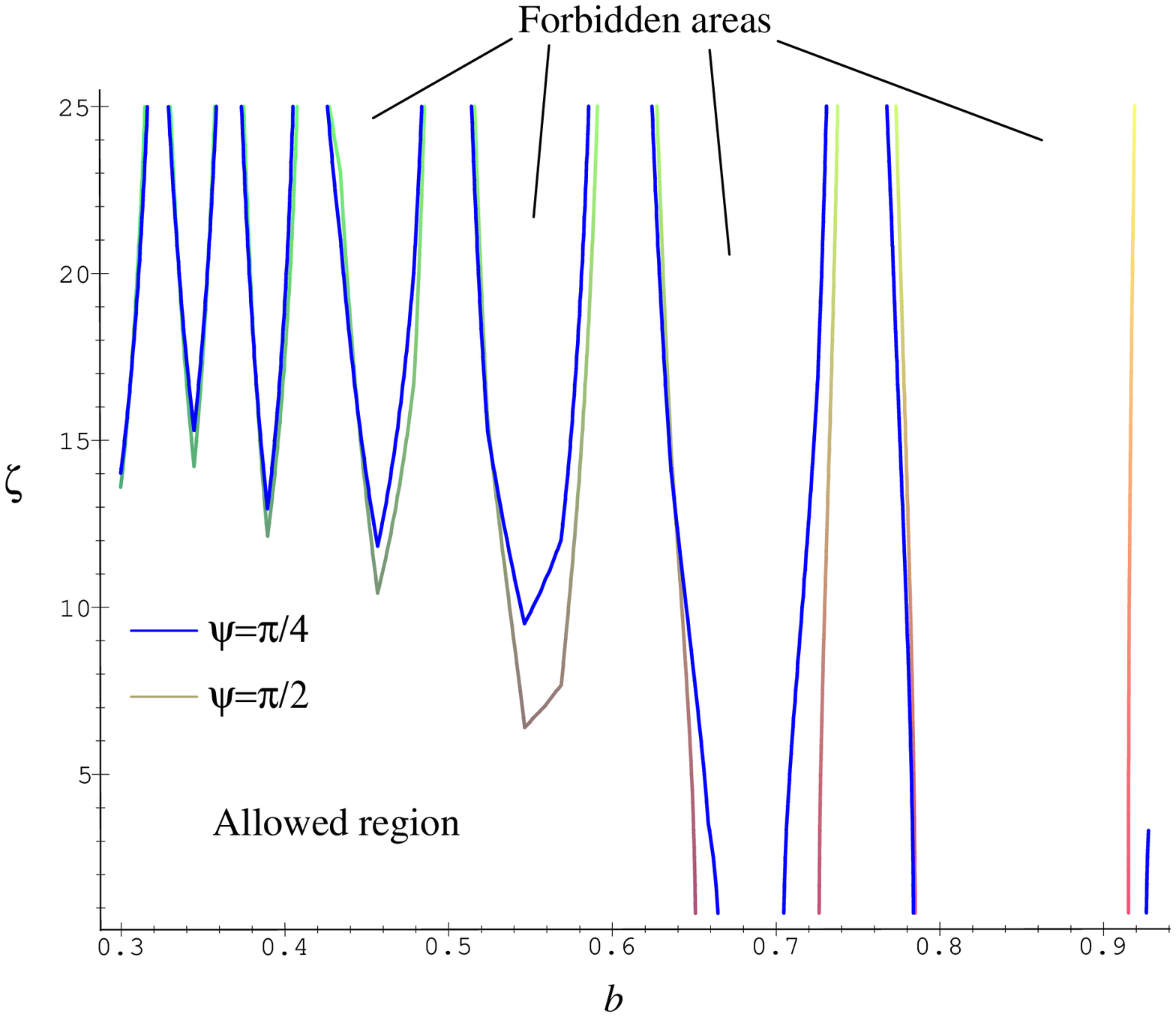},width=14cm,angle=0}}
\caption{\small Exclusion plot of $\zeta$ vs $b$ for two values of $\psi$
for the CMB anisotropy. $a=-5$.
\label{cmb_exclusion_b-vs-zeta_a=-5_psi=0.25-0.5pi_xi=0.5pi}}
\end{figure}%
demonstrates that the most restrictive position for this test is an observer at
$\psi=\pi/2$, and we can use that in what follows.

The other thing to notice in
figure~\ref{cmb_exclusion_b-vs-zeta_a=-5_psi=0.25-0.5pi_xi=0.5pi} is that
$\zeta$ behaves very much as $a$ does in \S\ref{CMB} (cf.
Fig.~\ref{cmb_exclusion}), that is the excluded regions are `fingers', with the
position depending on $b$, and the size depending on $\zeta$.

In figure~\ref{cmb_exclusion_a-vs-b_3zeta_psi=0.5pi_xi=0.5pi},
\begin{figure}[here!]
\centerline{\psfig{file={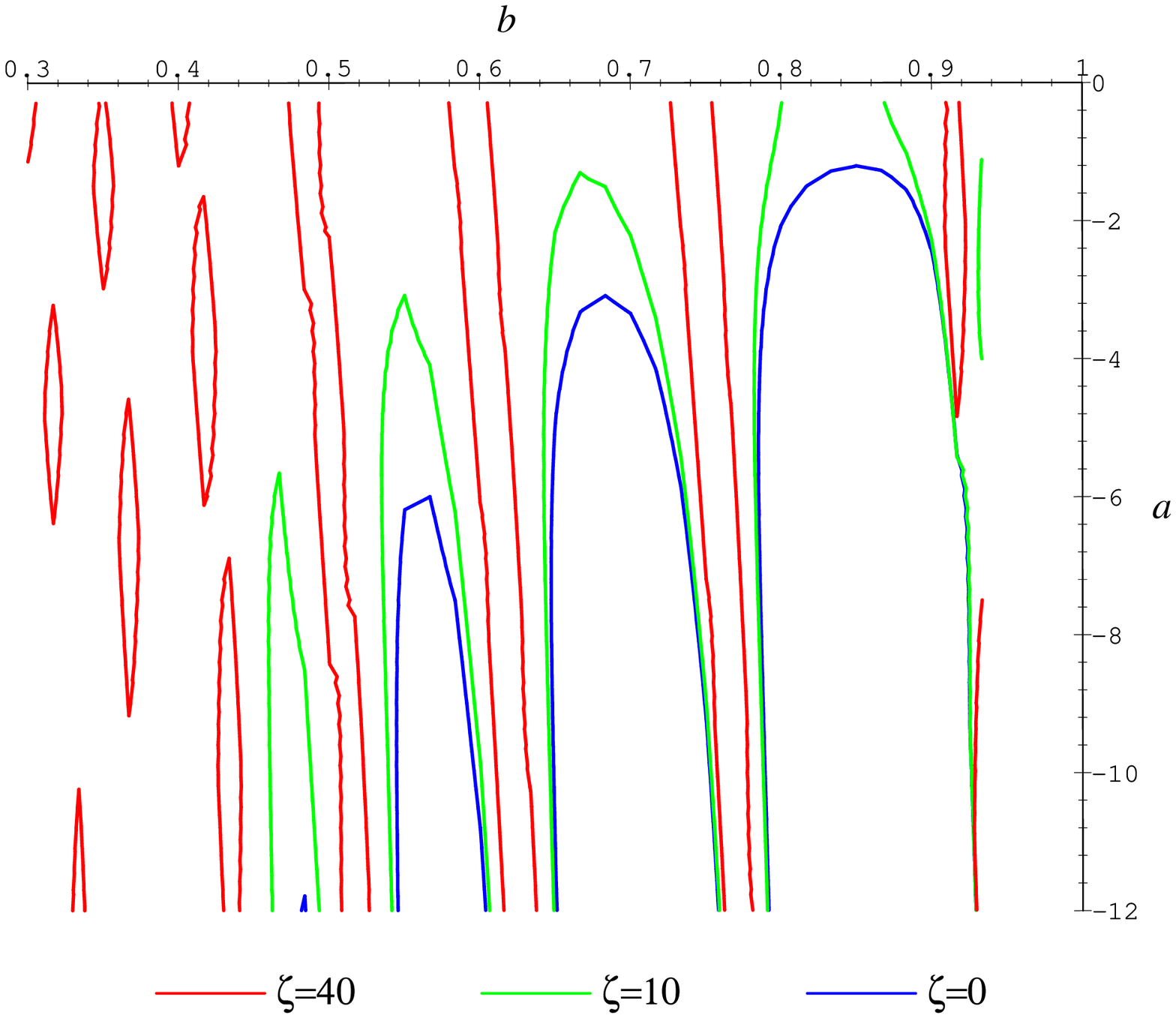},width=14cm,angle=0}}
\caption{\small Exclusion plot of $a$ vs $b$ for three values of $\zeta$
for the CMB anisotropy. As $\zeta$ increases, the `fingers' move in to exclude
more and more of the $a-b$ plane, leaving a few narrow strips and isolated
pockets.
\label{cmb_exclusion_a-vs-b_3zeta_psi=0.5pi_xi=0.5pi}}
\end{figure}%
we show the exclusion plot in the $a-b$ plane, as in
figure~\ref{cmb_exclusion}, but for different values of $\zeta$. As can be
seen, a high value of $\zeta$ is not exactly `ruled out,' but makes things more
`unlikely' to fit. For any value of $a$, the range of allowed $b$ decreases its
total size, but never (except for very high $\zeta$) rules out the models
completely. This is in contrast to the other constraints, such as the size
constraint, which completely rules out small values of $b$; or the age
constraint, which rules out large negative values of $a$. Note that all the
plots don't quite get close to $b=1$, as the calculation becomes very
(computationally) difficult there; it should be clear from the previous chapter
that this region is ruled out.

\section{The Magnitude-Redshift Relation}

Due to the complexity of the equations -- basically the conformal factor -- a
proper analysis of the local inhomogeneity is really beyond the scope of  this
thesis, although not impossible in principle. Instead we will consider some
examples to get an idea of the effect a large value of $\zeta$ will have. It is
perhaps appropriate to use the \mofz\ relation for this purpose, because it is
easier to compare with observations.

To recap, we have
\be
m(z)-M-25=5\log_{10}r_L=5\log_{10}\;{R_0}{\sqrt\Delta W_0}+5\log_{10}\left\{
(1+z)|\sin\chi(z)|\right\},
\ee
where $\chi(z)$ is given by the solution of
\ba
1\li+\li z(\chi)=\;{R_0}{W_0 R(\chi)}\left\{\Phi_{+}(\chi)+\Phi_{-}(\chi)
\cos\psi\cos\chi-
\;{2\zeta R(\chi)}{c\sqrt\Delta}\cos\xi\sin\psi\cos\chi\right.\nonumber\\
\li+\li\left.\sin\chi\cos\vartheta\left[\Phi_{-}(\chi)^2\sin^2\psi
+4\;{\zeta^2 R(\chi)^2}{c^2\Delta}(1-
\cos^2\xi\sin^2\psi)\right.\right.\nonumber\\
&&\hfill\left.\left.+2\;{\zeta
R(\chi)\Phi_{-}(\chi)}{c\sqrt\Delta}\cos\xi\sin2\psi\right]^{\;12}\right\},
\ea
with $R(\chi)\equiv X(t(\chi))$, etc.

We note first of all in
figure~\ref{m-z-opp-directions-a=-10_b=0.5_zeta=100_psi=0},
\begin{figure}[here!]
\centerline{\psfig{file={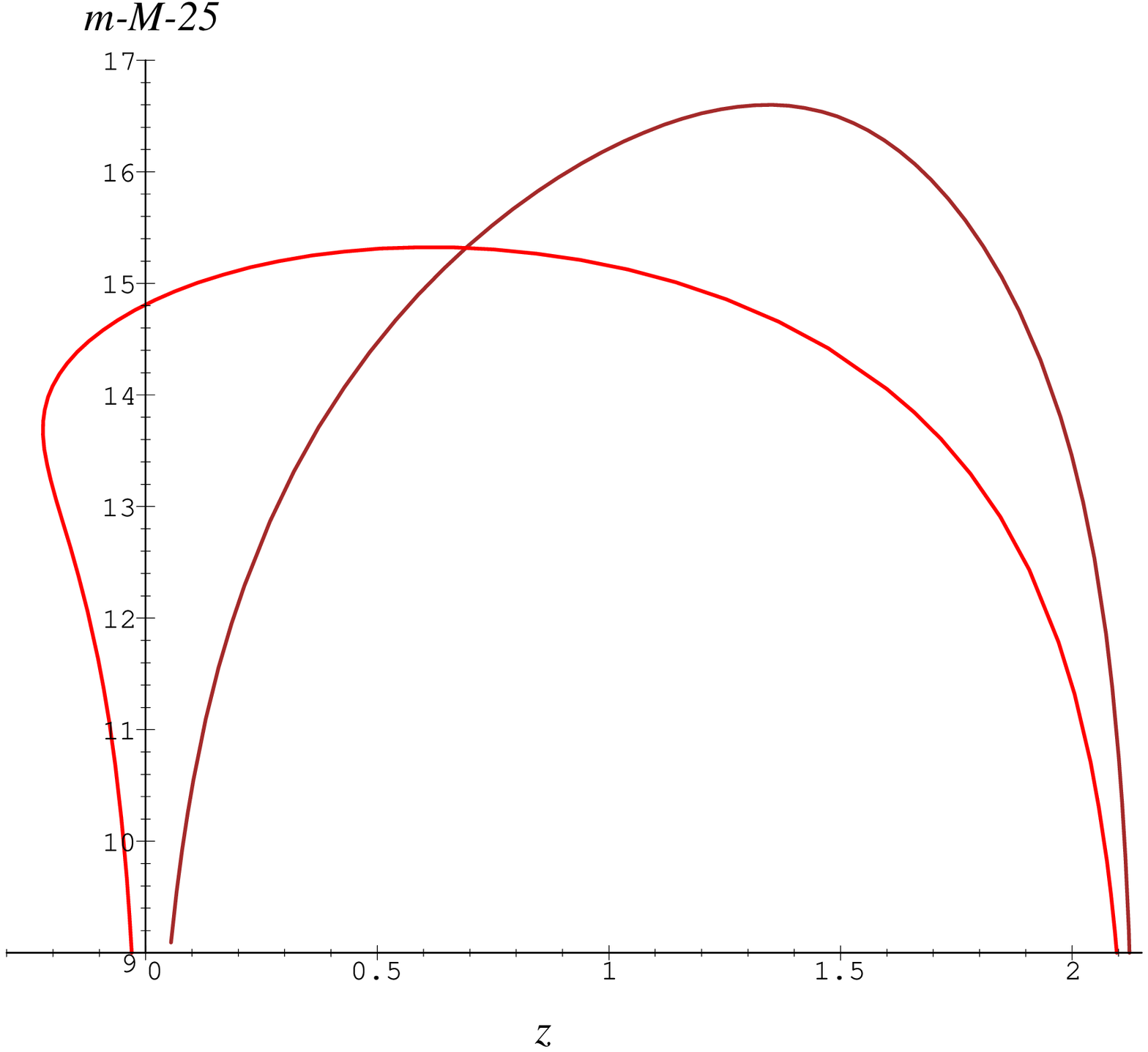},width=14cm,angle=0}}
\caption{\small The \mofz\ relation in opposite directions in the sky for an
observer at $\psi=0$. We have $b=0.5, a=-10$, and $\zeta=100$.
\label{m-z-opp-directions-a=-10_b=0.5_zeta=100_psi=0}}
\end{figure}%
we can see that a large $\zeta$ can cause all sorts of problems, with the
situation illustrated in
figure~\ref{m-z-opp-directions-a=-10_b=0.5_zeta=100_psi=0} being completely
ridiculous. It is not all the fault of $\zeta$ though, as
figure~\ref{m-z-opp-directions-a=-10_b=0.95_zeta=100_psi=0},
\begin{figure}[here!]
\centerline{\psfig{file={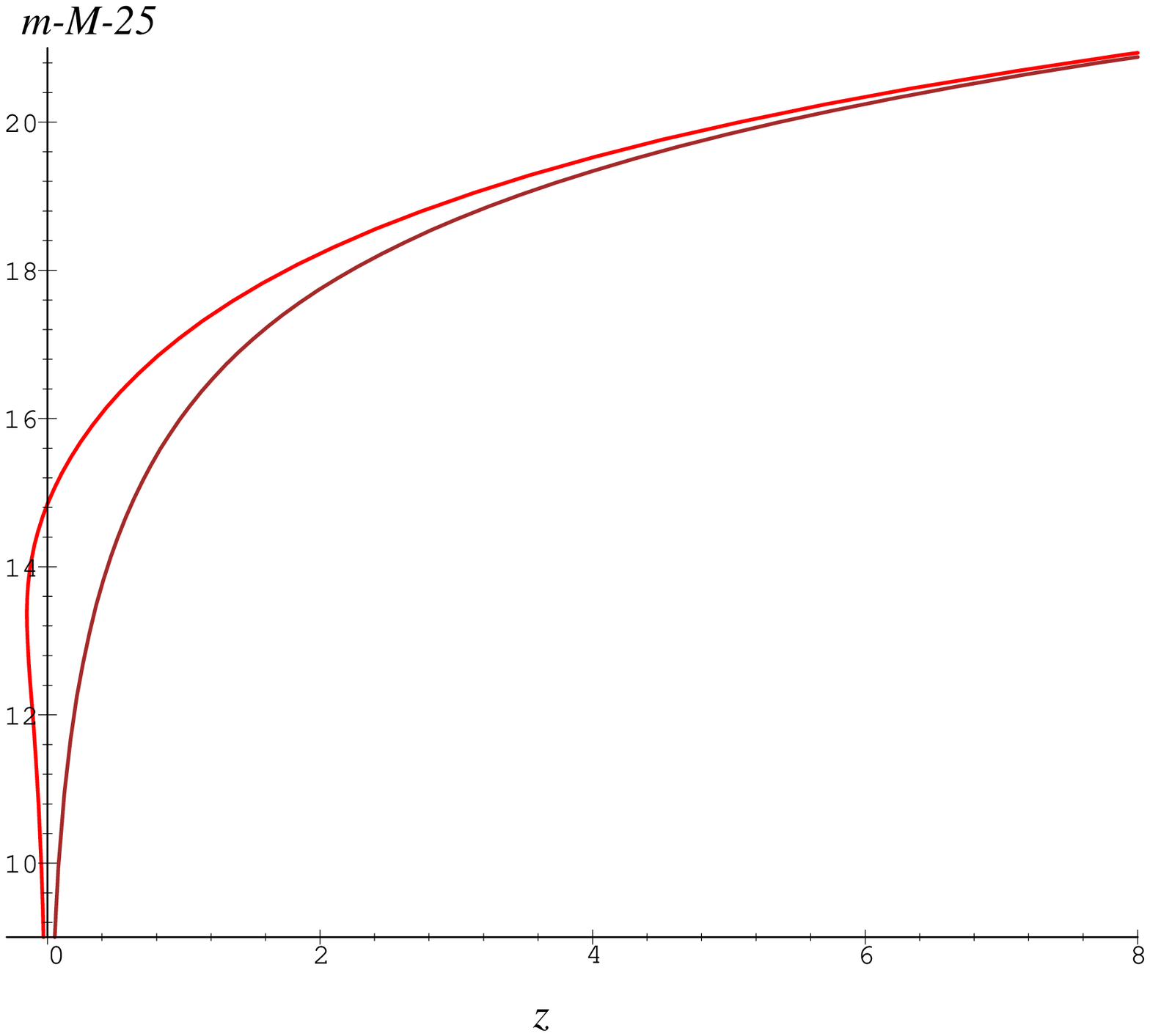},width=14cm,angle=0}}
\caption{\small The \mofz\ relation in opposite directions in the sky for an
observer at $\psi=0$. We have $b=0.95, a=-10$, and $\zeta=100$.
\label{m-z-opp-directions-a=-10_b=0.95_zeta=100_psi=0}}
\end{figure}%
demonstrates. A much higher value of $b$ is used, and the problem is slightly
relieved. However, the situation is still unacceptable as a cosmological model.
It is clear that a blueshift in one direction in the sky is not reasonable --
the results are similar whatever the observers location in the spacetime,
$\psi$. However the results do depend on $\xi$; with $\xi=\pi/2$ (as all the
graphs here have) we get a large dipole moment, which virtually disappears when
$\xi=0$. This is because when $\zeta$ is large, the dominant part of the dipole
in $1+z$ comes from the second term in the square root.

A far more reasonable situation is shown in
figure~\ref{m-z-opp-directions-a=-5_b=0.9_zeta=5_psi=0.5pi},
\begin{figure}[here!]
\centerline{\psfig{file={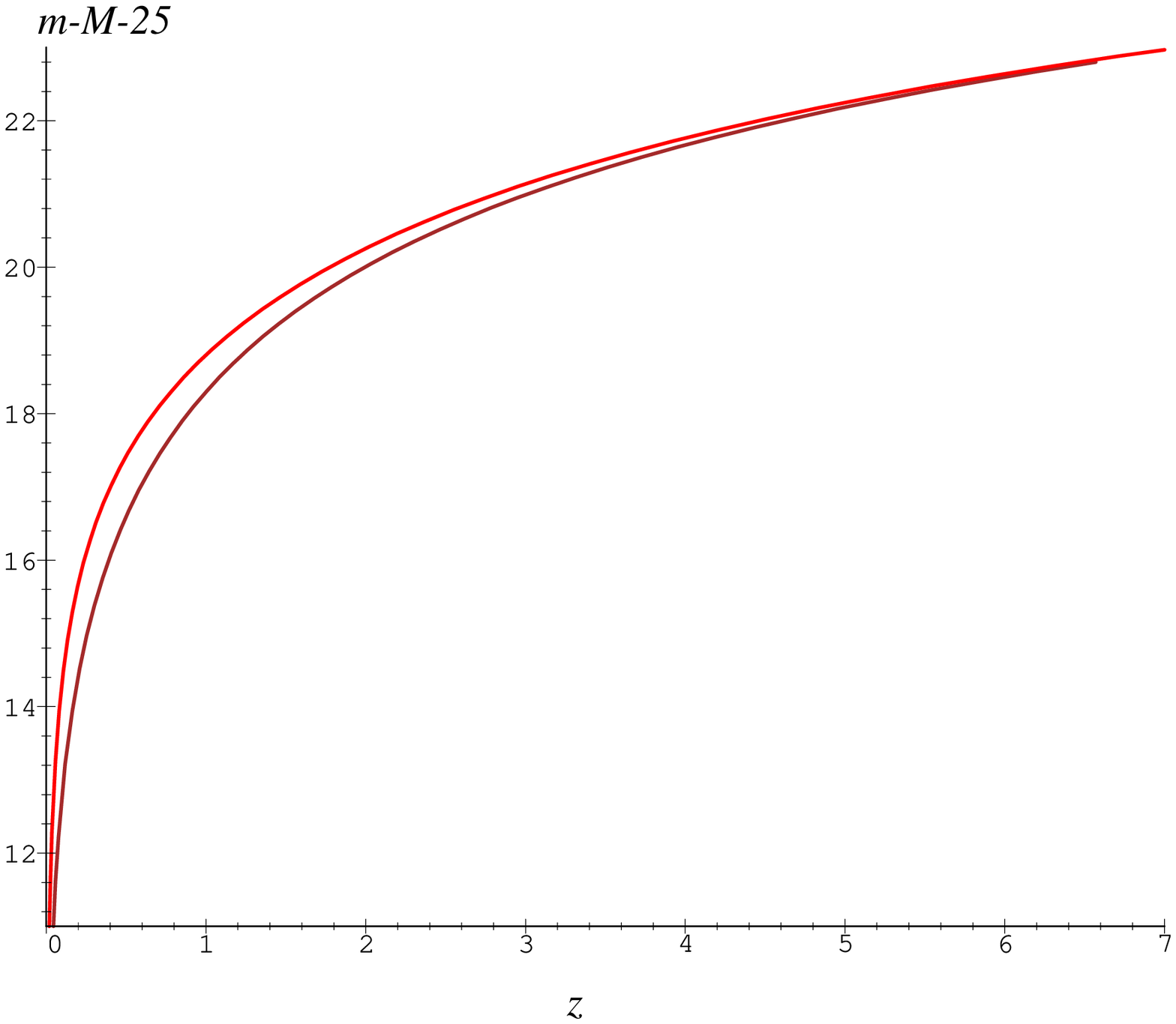},width=14cm,angle=0}}
\caption{\small The \mofz\ relation in opposite directions in the sky for an
observer at $\psi=\pi/2$. We have $b=0.9, a=-5$, and $\zeta=5$.
\label{m-z-opp-directions-a=-5_b=0.9_zeta=5_psi=0.5pi}}
\end{figure}%
with a much smaller $\zeta$. The observers location is at $\psi=\pi/2$, with
the direction to the center given by $\xi=\pi/2$, which seems to be the `worst
case' scenario.

The moral seems to be to have $\zeta$ and $|a|$ quite small, and at the same
time $b$ should be `close to' 1. It is interesting though, that when $\zeta$ is
distinctly non-zero, the dipole in the \mofz\ relation is more or less the same
regardless of the observers spatial location, but does depend heavily on their
direction to the center.

\section{Summary}

It this chapter we have considered the effects of a non-zero $\zeta$. We have
shown that the age may be high enough, the first zero of $r_A(z)$ does not
occur too close to the observer, and that the CMB anisotropies are not too
large, for non-zero values of $\zeta$. However, we have shown that a large
value of $\zeta$ will introduce a significant dipole into the \mofz\ relation.
This is probably enough to rule out these models, unless $\xi$ is close to 0:
this would violate the Copernican principle. This deserves further
consideration.




}
\forgetmenot{
\chapter{Conclusions and Future Work}
\label{conclusions_thesis}

This thesis has examined the application of the Copernican principle in
inhomogeneous spacetimes. The class of models we considered were derived in
chapter~\ref{EGS chapter---------------------------------} under the condition
that they admit an isotropic radiation field, and were shown to be a subclass
of the Stephani solutions with symmetry. A cosmological model which satisfies
the Copernican principle \emph{must} satisfy Theorem~\ref{isoradthm}, if one
wants a spacetime with observers who see an (exactly) isotropic CMB. The IRF
models we derive are generalisations of the FLRW models, that satisfy
Theorem~\ref{EGSthm} of Ehlers, Geren and Sachs~(1966). It was also
demonstrated that the models we find all admit a thermodynamic scheme. The rest
of the thesis was spent developing the relevant observational relations, and
`testing' various aspects of them.

In chapter~\ref{step chapter obs derivations----------------}, we derived all
the necessary cosmological tests. After the conformal symmetries were
recognised, and the relevant coordinate transformations were made, this became
a tractable problem. Indeed this transformation enabled us to `rotate' the
origin of coordinates to an arbitrary location, which gave the opportunity to
derive all the observational relations valid for all observers in the
spacetime. This is clearly crucial for the subsequent observational
consideration of the Copernican principle.

It chapters~\ref{step chapter 1------------------------------}, and~\ref{step
chapter hyperbolic----------------------}, we discussed observational
constraints. We limited the \emph{coordinate} time of observation using the
present constraints on $H_0$ by constraining the expansion rate
with~(\ref{minimum H0}) and~(\ref{coord age H_0 soln}). For any choice of
parameters this gave a value of the coordinate~$T$. We then proceeded to
examine the models on a variety of tests; the most important being age and the
anisotropy of the CMB. Obviously, any plausible cosmological model must have
reached the required age by the time the expansion has slowed to the rate we
measure today. This is position dependent; on a given surface of constant
\emph{coordinate} time (determined by the expansion rate) the observers at the
center will generally be older than their counterparts away from the center
(cf, figure~\ref{exclusion-age-noncentral}). As with all the tests, we consider
only the `worst case' observer position; ie, that which excludes most of the
parameter space.

The CMB is also a crucial test. Despite the fact that the IRF models admit an
isotropic radiation field by definition, they do not exhibit an exactly
isotropic CMB. This is because the models are homogeneous at the big
bang~$t=0$, or~$W=1$ (and big crunch, incidentally), which is when the
isotropic radiation field in question is `emitted'. The CMB is not emitted
then, but a short time after when the universe was inhomogeneous, but not
drastically so. The anisotropies in the observed temperature are
$\propto1/(1+z)\sim1/W$: that is all anisotropies arise from the
inhomogeneities in $W$ \emph{at the time of emission}, and have nothing to do
with the subsequent evolution and motion of observers. This means that, since
$t$ is small, $W\sim1$, and the inhomogeneities are very small too; the
conformal flatness of the spacetime allows the light from the CMB surface to
travel unmolested by any subsequent deviation from homogeneity, and we can see
the CMB as a reflection of the universe's inhomogeneity as it was at that time.
The dipole moment of the CMB is the largest moment, which we restrict to be
smaller than 10\% of the observed dipole, on the assumption that we are unable
to say, at present, that the dipole is entirely due to the motion of the local
group.

The other tests we considered were that the energy conditions are satisfied
(see below) and that the first zero of the $m-z$ relation does not occur too
close to the observer. Both these tests are independent of observer
location\footnote{The dominant energy condition actually depends on observer
position at certain times.}. As we have discussed, these do not impose strong
constraints on the available parameter space.

The IRF models we chose to look at had the scale factor (arbitrarily)
restricted to be a quadratic function of time. We must conclude, therefore that
the allowed degree of freedom in the IRF models, which manifests itself in the
freedom in choosing the scale factor, $R(t)$, gives us an enormous range of
models which would satisfy the constraints chosen here to a perfectly
acceptable level. In fact chapters~\ref{step chapter
1------------------------------}, and~\ref{step chapter
hyperbolic----------------------} may be considered as an existence `proof'
that the IRF models are acceptable models of the universe on
\emph{observational} grounds; that their matter content is unappealing does not
enter into it.

\section{A Note On `Best Fits' and $a>0$}

The discussion in chapters~\ref{step chapter 1------------------------------}
and~\ref{step chapter hyperbolic----------------------} is obviously not
complete. We should really provide a fit of the $m-z$ relation to real data.
However, this would not be simple in general, because any result would be
position dependent. Moreover, because of the dipole mentioned above, the data
would have to include the directional coordinates of the SNIa. As the complete
SNIa data is yet to be made public, we are not able to do this as yet.

However, in the spherically symmetric case, an observer at the center would see
no dipole variation of the $m-z$ relation, so it is of value to fit the models
from this special location, if only to show a proper fit (ie, as good a fit as
may be obtained with an FLRW model) is possible. In \S\ref{resolution} we
demonstrated that this was possible. In fact it is possible to show that any
FLRW $m-z$ relation may be `matched' by a suitable choice of parameters in
these IRF models (Barrett and Clarkson, in preparation), for an observer at the
center.

It should be clear from the discussion in the introduction on the active
gravitational mass, that a fit to an FLRW model with $q_0<0$ will violate the
strong energy condition\footnote{This may not be quite true in general (the
results are only preliminary); from the discussion in \S\ref{Observable
Quantities in a General Spacetime}, it should be clear that things are not that
simple when comparing observationally derived values of $q_0$ to the `length
scale' derived version. It is true at the center, however, where the
acceleration is zero and the negative pressure generates a negative $q_0$,
regardless of definition.}. This means that local observations at the center
will force $a>0$ (cf, \S\ref{energy_cond}). Moreover, we have seen from
\S\ref{Hubble Normalised Scalars}, that requiring the normalised energy density
to be less than unity today also requires $a>0$. This implies that this thesis
is incomplete. As we have noted before, $a>0$ implies that the universe will
open up at $t=(1-b)/2a$ (in the spherically symmetric case), after which the
universe will become infinite in spatial extent. Spatial infinity, in terms of
$\psi$ this will happen at $W=0$ (cf, \ref{confW});
\be
\sin^2\;{\psi_{_\mathsf{max}}}{2}=\;{c\Delta}{4aR(t)}<1~~~\hbox{for}~~~t>\;{1-b}{2a}.
\ee
Applying the Copernican principle in this case becomes difficult~-- at least in
terms of producing exclusion diagrams. It may not be possible to find a `worst
case' observer position, as we have done up to now. If we choose any
$\psi<\psi_{_\mathsf{max}}$, then the number of observers between the center
and $\psi$ will be finite; whereas, between $\psi$ and $\psi_{_\mathsf{max}}$
the number will be infinite. Thus it is impossible to say that for any $\psi$
we have considered all or most observers.

However, it may be possible to take a constraint, say the CMB dipole, and
consider the limit $\psi\rightarrow\psi_{_\mathsf{max}}$. That is we may say
that if
\be
\lim_{\psi\rightarrow\psi_{_\mathsf{max}}}\left|z_1\right|<0.1
\ee
then the model will pass the test, \emph{and} satisfy the Copernican principle.

This however, is a computational nightmare~-- as we take the limit, at any
given $\psi$, we must solve (\ref{z_0 gen}) for $\chi\cmb$. This must also be
done over the whole 3-dimensional parameter space. This is a chore for the
future.

\section{The Local Dipole}

The most interesting question this thesis has raised is the possibility that we
may be living in a distinctly inhomogeneous universe (that is, not a slightly
inhomogeneous perturbed FLRW universe).

Acceleration makes the IRF models inhomogeneous, but will leave its mark in the
local universe. This is testable (\S\ref{concs}) by considering the local
dipole moment of $H_0$~-- ie, the local dipole in the linear Hubble law. The
\emph{measured} dipole is usually interpreted as our peculiar velocity with
respect to the CMB frame; that the directions of our peculiar motion from local
studies and that of the CMB frame line up to within about $20^\circ$
(Saunders~\etal, 1999; Scaramella~\etal, 1991; Riess,
1999)\nocite{scar-91}\nocite{sau-99}\nocite{rie99} lends support to this but it
cannot be conclusive. It should be noted that in the IRF models, we also
require the local and CMB dipoles to line up.

If we wish to prove homogeneity then we must test this local dipole to
determine if it grows linearly with distance. If it does then we may use the
slope of this linear relationship to \emph{either} (cf,~\ref{specific H_0}): 1.
determine the maximum inhomogeneity of the universe, assuming we are located at
or near the `equator' ($\psi=\pi/2$); or, 2. determine our distance from the
center, in an arbitrarily inhomogeneous model. Both would require violation of
the Copernican principle to some degree. The first would require us to be
situated at the point of maximal inhomogeneity (maximum acceleration)~-- which
is a `non-typical' location; while the second \emph{may} violate it depending
on how inhomogeneous we may decide the universe to be. That is, the more
inhomogeneous the model (the greater the acceleration at the equator) the
closer we must be to the center, and we would no longer be representative of
`most' observers.

Preliminary investigations reveal that constraints on the dipole in $H_0$ are
weak~-- around 20\% may be attributed to an acceleration term (Clarkson, Rauzy,
and Barrett, in preparation). This has been investigated using the IRAS galaxy
catalogue; however, the recently released PSCz catalogue with around 15000
galaxies may provide stronger constraints.

\section{Almost EGS Considerations}

One may expect that the IRF models derived in \S\ref{IPFSolns} would extend to
an `almost' theorem, in the same manner as the almost EGS theorem of Stoeger,
Maartens and Ellis~(1995). If we take an almost isotropic radiation field
characterised by
\ba
\sigma_{ab}={\cal O}(\epsilon),\\
\del_{[a}\left(\udot_{b]}-\sfrac{1}{3}\theta u_{b]}\right) = {\cal O}(\epsilon),
\ea
where epsilon is small in some sense then we may write, from
equations~(\ref{EGS-proof-1})
\ba
\dot\mu_{\!_R}+\;43\theta\mu_{\!_R}\li=\li{\cal O}(\epsilon)\nonumber\\
4\mu_{\!_R}\udot_a+\sdel_a\mu_{\!_R}\li=\li{\cal O}(\epsilon),\\
\ea
or
\be
\udot_a=-\:14\sdel_a\ln\mu_{\!_R}+{\cal O}(\epsilon)
\ee
which gives
\be
\sdel_{[a}\udot_{b]}\sim\sdel_{[a}\sdel_{b]}\mu_{\!R}\sim\omega_{ba}\theta
\neq{\cal O}(\epsilon):
\ee
there doesn't seem to be any way to make the rotation small, unless one
artificially introduces additional assumptions. Therefore perfect fluid
solutions which admit an almost isotropic radiation field may not (at first
sight at least) be perturbations of the IRF models. However, there may be ways
to constrain rotation on the basis of \emph{local} observations: rotation leads
to anisotropic number-counts~-- see Fennelly~(1976)\nocite{fen76}. It would
seem unlikely that such observations would be anywhere near as accurate as
those of the CMB anisotropies; ie, we would be unable to claim that
$\omega_{ab}={\cal O}(\epsilon)$ from number-counts alone.

\section{Further Considerations}

The IRF models, by definition, admit an isotropic radiation field. As
\emph{all} observers see such a radiation field to be isotropic, the isotropy
of the CMB cannot be used to verify the cosmological principle. The usual
suggestions of testing the cosmological principle directly, necessarily involve
making measurements \emph{within} our past lightcone. The Sunyaev-Zel'dovich
(SZ) effect is often quoted as a suitable physical effect to produce such a
test (Goodman, 1995). Such a test will involve measuring the \emph{anisotropy}
of the CMB around other observers. It is normally assumed that if the
anisotropy is small around other observers then one may conclude that we are
not at a special location (eg, at a center of symmetry), and that the universe
is FLRW~-- hence the cosmological principle is verified.

This, of course, assumes that the acceleration is zero. The IRF models would
also give the same effect; measurement of the CMB anisotropies as seen by
another observer would be similarly small, but these would not imply that the
universe is homogeneous, by Theorem~\ref{IPFthm}. Hence, contrary to popular
belief, \emph{the SZ effect cannot possibly show that the cosmological
principle is a valid assumption.}

Clearly, the SZ effect cannot distinguish between different IRF models~-- thus
the acceleration may be arbitrarily large. This means that the \emph{Copernican
principle} may be violated, because for our \emph{local} dipole to be so small
(at most 20\% of $H_0$) would necessarily require us to be close to the center
of such an inhomogeneous universe. Therefore \emph{the SZ effect will not even
be able to provide any proof of the Copernican principle.}

\forget{
Depending on ones definition, it would be impossible to verify the
\emph{Copernican principle} via the Sunyaev-Zel'dovich effect, or any similar
effect\footnote{Similar in the sense of enabling us to see the CMB inside our
past lightcone.}, because, one may claim, the IRF models \emph{do not} satisfy
the Copernican principle~-- they are inhomogeneous, and `special' places do
exist. This would assume a very strict definition of the Copernican
principle~-- ie, a definition which requires homogeneity. Even as the
Copernican principle is defined in this thesis however, one may not conclude
from any type of CMB measurements alone that the universe is FLRW, as one
cannot place \emph{any} limits on the acceleration.}

It should be noted that acceleration would leave a small signature in such CMB
measurements: small dipoles around other observers would be correlated, rather
than a random distribution of dipoles from their peculiar motions. The
distribution of the dipoles would be a mixture of some pointing in one
direction, and others pointing in the opposite direction. In the spherically
symmetric case, the direction of the dipole would depend on the number of times
the scattered light had passed through the point opposite to us. This will
depend on the redshift of the galaxy. It is not clear what pattern the more
general IRF models would give.

If we used the SZ effect and found such a pattern of dipoles, then this would
imply that acceleration is present in the universe. Such accuracy for these
experiments is a long way off, as it would be next to impossible to disentangle
any dipole from peculiar velocities and that from acceleration~-- we have
enough difficulty in determining our own.

This poses some interesting questions. It would be interesting to carry on this
work along the lines of Stoeger, Maartens and Ellis~(1995), and Maartens, Ellis
and Stoeger~(1995)\nocite{maart-95}, which is to place limits on all the
components which characterise a cosmological model: $\{\udot^a, \theta,
\sigma_{ab}, \omega_{ab}, E_{ab}, \& H_{ab}\}$, together with the matter content.
In general this is not a simple task and, as this thesis makes clear, one
cannot rely on the CMB alone.

}

\nocite{dab95}\nocite{cla-bar99}\nocite{bar-clI99,bar-clII99,ehl93}

\forgetmenot{
\appendix
\label{appendices}
\chapter{Coordinate transformations and the Stephani Spacetimes.}
\label{app-trans}
\markright{APPENDIX \ref{app-trans}. COORDINATE TRANSFORMATIONS}

On the face of it the Stephani models admitting an isotropic radiation field
in~(\ref{Steph_constr}) depend on one free function and ten free parameters.
However, it is possible to use coordinate transformations on the spacetime to
eliminate many of these parameters, resulting in a considerable simplification.
As is shown in Barnes~(1998), conformal transformations of the coordinates on
the hypersurfaces of constant time preserve the form of the metric but change
the free functions $a$, $b$, and~$\mathbf{c}$. These transformations can be
thought of as acting on the five-dimensional space spanned by $a$, $b$
and~${\mathbf{c}}$ and constitute the Lorentz group in five
dimensions,~$SO(4,1)$: they leave~$-ab + |{\mathbf{c}}|^2$ invariant ($a$
and~$b$ are `null coordinates'). It will be convenient here to let
$a=\alpha+\beta$ and~$b=\alpha-\beta$ (so that the transformations preserve
$-\alpha^2 + \beta^2 + |{\mathbf{c}}|^2$), and to adopt five-vector
notation:~$q^\mu=(\alpha,\beta,{\mathbf{c}})$. We will use the terms `rotation'
and `boost' to refer to the transformations on~$q$, and will call~$q$ timelike,
spacelike or null if $-\alpha^2 + \beta^2 + |{\mathbf{c}}|^2$ is negative,
positive or zero, as usual. Then it is easy to visualise the transformations on
the free functions by imagining the `mass hyperboliods' of representations of
the Lorentz group in the usual way: a timelike vector can always be boosted so
that it has the form~$(\alpha,0,{\mathbf{0}})$, whereas a spacelike vector can
be boosted and rotated into~$(0,\beta,{\mathbf{0}})$, for example.

In addition to the Lorentz transformations we also have the freedom to change
basis in the function space spanned by the free functions. For the spacetimes
of interest here, described by~(\ref{Steph_constr}), it is desireable to
preserve~$f_2(t)=1$, so that the basis change is
\be
                  T \mapsto \gamma T + \delta.
\label{Ttrans}
\ee

In five-vector notation the equations~(\ref{Steph_constr}) become
\be
    q^\mu(t) \equiv (\alpha,\beta,{\mathbf{c}}) = q_1^\mu T(t) + q_2^\mu
\label{Steph_conq}
\ee
(with $q_1$ and~$q_2$ constant vectors), and the goal is to reduce as many of
the components of $q_1^\mu$ and~$q_2^\mu$ to zero as possible using Lorentz
transformations in the five-dimensional space containing $q_1$ and~$q_2$, and
the basis change~(\ref{Ttrans}). Note that the FLRW ($d=1$) subcase
of~(\ref{Steph_constr}) is characterised by the linear dependence of $q_1$
and~$q_2$.

It is easy to see that~${\mathbf{c}}$ (which breaks the spherically symmetry of
the metric~(\ref{metric})) may always be reduced to the
form~${\mathbf{c}}=c{\mathbf{\hat{z}}}$ (${\mathbf{\hat{z}}}=(0,0,1)$),
with~$c$ a constant: perform a spatial 4-rotation to reduce~$q_1^\mu$
to~$q_1^\mu=(\alpha_1,\beta_1,{\mathbf{0}})$, then a spatial rotation amongst
the ${\mathbf{c}}$-components (which obviously leaves~$q_1$ unaffected) to
give~$q_2^\mu=(\alpha_2,\beta_2,c{\mathbf{\hat{z}}})$.

It is possible in general to make further simplifications, but precisely how
$q_1$ and~$q_2$ are simplified depends on whether they are spacelike, timelike
or null. For example, if either $q_1$ or~$q_2$ is timelike (or may be made
timelike by a transformation~(\ref{Ttrans})) it is possible to reduce the model
to manifestly spherically symmetric form ($c=0$): boost so that the timelike
vector, say~$q_1$, becomes~$q_1=(\alpha_1,0,{\mathbf{0}})$ and rotate spatially
so that the ${\mathbf{c}}$-components of the other vector are also
zero,~$q_2=(\alpha_2,\beta_2,{\mathbf{0}})$ (we could then use~(\ref{Ttrans})
to eliminate more of these constants).

To summarise, we have demonstrated that it is always possible to reduce the
Stephani models of~(\ref{Steph_constr}) to the form
\begin{eqnarray}
                a(t) \li = \li a_1 T(t) + a_2, \nonumber \\
                b(t) \li = \li b_1 T(t) + b_2, \label{constr-app}\\
     {\mathbf{c}}(t) \li = \li c{\mathbf{\hat{z}}}, \nonumber
\end{eqnarray}
(where we have transformed back from $\alpha$ and~$\beta$ to $a$ and~$b$), and
when either of the $q_1$ or~$q_2$ is (or may be made) timelike we can
set~$c=0$.

Finally, note that we may always assume that~$a_1\ne 0$ in~(\ref{constr-app}),
because if~$a_1=0$ then~$b_1\ne 0$, otherwise~$V_{,t}=0$, and it is possible to
perform a coordinate inversion ${\mathbf{x}}\mapsto {\mathbf{x}}/r^2$ that
interchanges $a$ and~$b$). This does not exhaust the possibilities for
simplification: we could, for example, use~(\ref{Ttrans}) to set~$a_2=0$.

\chapter{Transformation to a Non-Central Position.}
\label{chitrans}

We want to transform from the $(\chi,\theta,\phi)$ coordinate system, whose
origin is at the centre, to coordinates centred instead on some observer
at~$\chi=\psi$, while preserving the form of the \FRW\ part of the
metric~(\ref{metric_W}). It is therefore necessary to identify the
transformations of the (homogeneous) \FRW\ spatial sections that leave the
\FRW\ metric invariant, ie,~the isometries of the spatial sections. This is
simple. Since the spatial sections of an \FRW\ model with positive curvature
constant ($\Delta>0$) are 3-spheres, the isometries we require are
4-dimensional rotations (ie,~elements of $SO(4)$, the isometry group of the
3-sphere).

A sphere of radius~$R$ in 4-dimensional space with cartesian coordinates
$(x,y,z,u)$ is defined by
\[
                x^2+y^2+z^2+u^2 = R^2.
\]
We have three coordinates on this sphere: $\chi$ and the two spherical polar
angles $\theta$ and~$\phi$. These are related to the cartesian coordinates by
\ba
       x & = & R\sin\chi \sin\theta \cos\phi \label{xofchi}\\
       y & = & R\sin\chi \sin\theta \sin\phi \\
       z & = & R\sin\chi \cos\theta \label{zofchi}\\
       u & = & R\cos\chi \label{uofchi}
\ea
The origin,~$\chi=0$, is then at $x=y=z=0$,~$u=R$. We are only interested in
rotations that move the origin, and, as the initial metric is spherically
symmetric (really spherically symmetric, not just conformally: even the
conformal factor is spherically symmetric about the centre), we need only
consider moving the observer in one direction, which we choose to be the
`$z$~direction' (ie,~to a position with non-zero~$z$, but~$x=y=0$). Clearly,
then, we are looking for a rotation in the $u-z$~plane. Since we have the
conformal factor as a function of~$\chi$ we want to find $\chi$ as a function
of the new coordinates. Starting with coordinates $\chi'$, $\theta'$
and~$\phi'$, centred on some position~$\chi=\psi$, along with their primed
cartesian counterparts $x'$, $y'$, $z'$ and $u'$ (which are related in the same
way as the unprimed coordinates in (\ref{xofchi})--(\ref{uofchi})), a rotation
back to the original coordinates is given, in cartesian coordinates, by $x=x'$,
$y=y'$ and
\ba
            z & = & \cos\psi z' + \sin\psi u', \nonumber\\
            u & = & -\sin\psi z' + \cos\psi u'. \label{urot}
\ea
(Note that at the origin of the primed coordinates, where $z'=0$ and~$u'=R$, we
have $u=R\cos\psi$, showing that~$\chi=\psi$ there, as required.)
Equation~(\ref{urot}), along with the primed versions of (\ref{zofchi})
and~(\ref{uofchi}), then immediately gives
\be
        \cos\chi = \cos\psi \cos\chi' - \sin\psi \sin\chi' \cos\theta',
\label{newchi}
\ee
and this is all we will need, since the only spatial coordinate that enters
into the original metric~(\ref{metric_W}) is~$\chi$, and that enters only
as~$\cos\chi$ ($2\sin^2\frac{\chi}{2}=1-\cos\chi$).

\forget{
\section{Redshift in Conformally Related Spacetimes.}
\label{appen}

We present two derivations of the redshift formula for spacetimes sharing some
of the simple properties of the D\c{a}browski models. The first can be used in
spacetimes that are conformal to simpler metrics for which the geodesics and
redshifts can be found. The second is valid for spacetimes that are conformal
to a spherically symmetric spacetime.

If we have two conformally related metrics, $g_{ab}=\Omega^2\bar{g}_{ab}$
($\Omega>0$), their associated metric connections are related by (see
appendix~D of Wald~1984\nocite{wald})
\be
    \del_b V^a = \bar{\del}_b V^a + (\bar{\del}_b \ln\Omega) \,V^a
               + (V^c\bar{\del}_c \ln\Omega) \,\delta^a_b
               - (\bar{\del}_d \ln\Omega) \,g^{ad}g_{bc} V^c,
\label{connect}
\ee
for any vector field~$V^a$. Null geodesics with respect to~$g$ (or, more
correctly, with respect to~$\del$) satisfy $g_{ab}k^a k^b=0$ and
\be
                     k^b \del_b k^a = 0,
\label{geodesic}
\ee
where~$k^a$ is the tangent vector to the geodesic. Applying~(\ref{connect})
gives
\begin{eqnarray*}
  0 = k^b\del_b k^a & = & k^b\bar{\del}_b k^a + 2 k^b\bar{\del}_b\ln\Omega\,k^a
            - g^{ad}\bar{\del}_d\ln\Omega\,\left(g_{bc}k^b k^c\right) \\
         & = & \frac{1}{\Omega^4} (\Omega^2 k^b)\bar{\del}_b (\Omega^2 k^a),
\end{eqnarray*}
and we see immediately that
\be
                  \bar{k}^a = \Omega^2 k^a
\label{kbar}
\ee
is the tangent vector to a null geodesic with respect to~$\bar{\del}$. So,
every null geodesic of~$\del$ corresponds to a null geodesic of~$\bar{\del}$
(and vice versa, since we can repeat the above steps interchanging $g$
and~$\bar{g}$ and putting~$\Omega\mapsto 1/\Omega$). If the geodesics with
respect to $\bar{\del}$ are known explicitly then we can find them easily for
$\del$.

To find the redshift, though, we also need the velocities of emitter and
observer. A four-velocity satisfies $g_{ab}u^au^b = -c^2$, and if we define
\be
                 \bar{u}^a = \Omega u^a
\label{ubar}
\ee
then $\bar{u}$ is a four-velocity with respect to~$\bar{g}$:
$\bar{g}_{ab}\bar{u}^a\bar{u}^b=-c^2$. Redshift is calculated from the ratio of
the emitted frequency~$\nu_{\scriptscriptstyle{E}} =
\left.u_ak^a\right|_{\scriptscriptstyle{E}}$ to the observed
frequency~$\nu_{\scriptscriptstyle{O}} =
\left.u_ak^a\right|_{\scriptscriptstyle{O}}$. Using (\ref{kbar}) and~(\ref{ubar})
we have
\[
 u_ak^a = g_{ab}u^ak^b = \Omega^2\bar{g}_{ab}\frac{\bar{u}^a}{\Omega}
            \frac{\bar{k}^b}{\Omega^2}
        = \frac{1}{\Omega}\bar{u}_a\bar{k}^a,
\]
which means that
\be
1+z=\frac{\left. u_{a}k^{a}\right|_{\scriptscriptstyle{E}}}
{\left.u_{b}k^{b}\right|_{\scriptscriptstyle{O}}}=
\frac{\Omega_{\scriptscriptstyle{O}}}{\Omega_{\scriptscriptstyle{E}}}
\frac{\left.\bar{u}_{a}\bar{k}^{a}\right|_{\scriptscriptstyle{E}}}
{\left.\bar{u}_{b}\bar{k}^{b}\right|_{\scriptscriptstyle{O}}}=
\frac{\Omega_{\scriptscriptstyle{O}}}{\Omega_{\scriptscriptstyle{E}}}(1+\bar{z}).
\label{redshift1}
\ee
where $\bar{z}$ is the redshift associated with $\bar{g}_{ab}$ for the
fundamental velocity~$\bar{u}$. If the paths of null rays in the
spacetime~$\bar{g}_{ab}$ and the redshift formula for the velocity~$\bar{u}$
are known then~(\ref{redshift}) gives the redshift in the true spacetime. For
the D\c{a}browski models $\bar{g}_{ab}$ is an \FRW\ metric, the conformal
factor is $\Omega=1/W$ (see~(\ref{metric_W})) and $\bar{u}$ is the usual \FRW\
comoving velocity field. The well-known expression for redshift in \FRW\
spacetimes, $1+\bar{z} = R_{\scriptscriptstyle{O}}/R{\scriptscriptstyle{E}}$,
then gives
\be
          1+z = \frac{R_{\scriptscriptstyle{O}}}{W_{\scriptscriptstyle{O}}}
                \frac{W_{\scriptscriptstyle{E}}}{R_{\scriptscriptstyle{E}}}.
\label{redshift}
\ee
(Alternatively, we could change to conformal time~(\ref{conftime}), making the
metric manifestly conformal to the Einstein static spacetime,~(\ref{ESS}), for
which the redshift is zero. The conformal factor is
then~$\Omega=R/(\Delta^{1/2}W)$, which also gives~(\ref{redshift}).)

When the true spacetime is conformal to a spherically symmetric spacetime the
radial null geodesics connecting any point with an observer at the centre are
obviously purely radial (since their paths are not affected by the conformal
factor). They are therefore given (in terms of coordinates $r$ and~$t$ with
respect to which the spherical symmetry is manifest) by some
function~$t_{\scriptscriptstyle{O}}(r_{\scriptscriptstyle{E}},
t_{\scriptscriptstyle{E}})$ relating the time,~$t_{\scriptscriptstyle{O}}$ ,
that the light ray is received by the observer, to the time of
emission,~$t_{\scriptscriptstyle{E}}$, for an object at
radius~$r_{\scriptscriptstyle{E}}$. This is just the lookback-time relation.
Redshift, as the ratio of proper time intervals at the observer to proper time
intervals at the emitter, is then given by
\ba
    1+z \equiv \frac{d\tau_{\scriptscriptstyle{O}}}{d\tau_{\scriptscriptstyle{E}}}
          & = & \frac{d\tau_{\scriptscriptstyle{O}}}{dt_{\scriptscriptstyle{O}}}
             \frac{dt_{\scriptscriptstyle{O}}}{d\tau_{\scriptscriptstyle{E}}}
            = \frac{d\tau_{\scriptscriptstyle{O}}}{dt_{\scriptscriptstyle{O}}}
              \left(\frac{\partial t_{\scriptscriptstyle{O}}}{\partial r_{\scriptscriptstyle{E}}}
                    \frac{dr_{\scriptscriptstyle{E}}}{d\tau_{\scriptscriptstyle{E}}}
               +    \frac{\partial t_{\scriptscriptstyle{O}}}{\partial t_{\scriptscriptstyle{E}}}
                    \frac{dt_{\scriptscriptstyle{E}}}{d\tau_{\scriptscriptstyle{E}}}\right) \nonumber\\
           & = & \frac{1}{u^t_{\scriptscriptstyle{O}}}
                 \left(\frac{\partial t_{\scriptscriptstyle{O}}}{\partial r_{\scriptscriptstyle{E}}}
                    u^r_{\scriptscriptstyle{E}}
               +    \frac{\partial t_{\scriptscriptstyle{O}}}{\partial t_{\scriptscriptstyle{E}}}
                    u^t_{\scriptscriptstyle{E}}\right).
\label{zDtdef}
\ea
(When the coordinates $r$ and~$t$ are comoving~-- $u^r=0$~-- the $r$-derivative
term disappears.) This will provide an analytic expression for the redshift
whenever the lookback-time equation can be integrated. For the D\c{a}browski
models:
\be
          u^\chi=0,  \hskip 1.0cm   u^\ti = \frac{c}{|g_{00}|^{1/2}} = W
\label{ut}
\ee
and the lookback time can be derived directly from the metric: on the past null
cone of the observer $ds=0=d\theta=d\phi$, leading to an expression
for~$d\chi/d\ti$, which, when integrated, gives
\[
              \chi = c\sqrt{\Delta}\int_\ti^\T \frac{d\ti'}{R(\ti')}.
\]
Differentiating this with respect to~$\ti$ at fixed~$\chi$ then gives
\[
     \frac{\partial t_{\scriptscriptstyle{O}}}{\partial t_{\scriptscriptstyle{E}}} \equiv
     \frac{\partial \T}{\partial \ti} = \frac{R_{\scriptscriptstyle{O}}}{R_{\scriptscriptstyle{E}}},
\]
which, together with (\ref{zDtdef}) and~(\ref{ut}), results in the
expression~(\ref{redshift}) for the redshift.

}


}


\bibliography{thesis}}

\begin{thebibliography}{}

\bibitem[\protect\astroncite{Barnes}{1998}]{barn98}
Barnes, A.: 1998,
\newblock {\em Class. Quantum Grav.} {\bf 15}, 3061

\bibitem[\protect\astroncite{Barrett and Clarkson}{999a}]{bar-clII99}
Barrett, R.~K. and Clarkson, C.~A.: 1999a,
\newblock {\em astro-ph/9911235; To appear in Class. Quantum Grav.}

\bibitem[\protect\astroncite{Barrett and Clarkson}{999b}]{bar-clI99}
Barrett, R.~K. and Clarkson, C.~A.: 1999b,
\newblock {\em To be submitted to Mon. Not. R. Astron. Soc.}

\bibitem[\protect\astroncite{Bertschinger et~al.}{1997}]{bert-97}
Bertschinger, E. et~al.: 1997,
\newblock in G. Borner and S. Gottlober (eds.), {\em The Evolution of
the Universe}, Chapt. 17., John Wiley \& Sons

\bibitem[\protect\astroncite{Bona and Coll}{1988}]{bon-col88}
Bona, C. and Coll, B.: 1988,
\newblock {\em Gen. Rel. Grav.} {\bf 20(3)}, 297

\bibitem[\protect\astroncite{Bondi}{1947}]{bond47}
Bondi, H.: 1947,
\newblock {\em Mon. Not. R. Astron. Soc.} {\bf 107}, 410

\bibitem[\protect\astroncite{Branchini et~al.}{1999}]{bra-99}
Branchini, E. et~al.: 1999,
\newblock {\em Mon. Not. R. Astron. Soc.} {\bf 308}, 1
({\em astro-ph/9901366v2})

\bibitem[\protect\astroncite{Bruni et~al.}{1992}]{bruni-92}
Bruni, M., Dunsby, P. K.~S. and Ellis, G. F.~R.: 1992,
\newblock {\em Astrophys. J.} {\bf 395}, 34

\bibitem[\protect\astroncite{Caldwell et~al.}{1998}]{cald-98}
Caldwell, R.~R., Dave, R. and Steinhardt, P.~J.: 1998,
\newblock {\em Phys. Rev. Letts.} {\bf 80(8)}, 1582

\bibitem[\protect\astroncite{Carr}{1994}]{car94}
Carr, B.: 1994,
\newblock {\em Ann. Rev. Astron. and Astrophys.} {\bf 32}, 531

\bibitem[\protect\astroncite{C\'el\'erier}{1999}]{cel99}
C\'el\'erier, M.: 2000
\newblock {\em Astron. Astrophys.} {\bf 353}, 63
({\em astro-ph/9907206})

\bibitem[\protect\astroncite{C\'el\'erier and Schneider}{1998}]{cel-sch98}
C\'el\'erier, M. and Schneider, J.: 1998,
\newblock {\em Phys. Lett. A} {\bf 249}, 37

\bibitem[\protect\astroncite{Chaboyer et~al.}{}]{chab-97}
Chaboyer, B., Demarque, P., Kernan, P.~J. and Krauss, L.~M.,
\newblock {\em Astrophys. J.} {\bf 494}, 96

\bibitem[\protect\astroncite{Challinor}{1999}]{chal99}
Challinor, A.: 1999,
\newblock {\em astro-ph/9906474}

\bibitem[\protect\astroncite{Challinor and Lasenby}{1998}]{cha-las98}
Challinor, A. and Lasenby, A.: 1998,
\newblock {\em Phys. Rev. D} {\bf 58}, 023001
\newblock 

\bibitem[\protect\astroncite{{Challinor} and {Lasenby}}{1999}]{cha-las99}
{Challinor}, A. and {Lasenby}, A.: 1999,
\newblock {\em Astrophys. J.} {\bf 513}, 1

\bibitem[\protect\astroncite{Clarkson and Barrett}{1999}]{cla-bar99}
Clarkson, C.~A. and Barrett, R.~K.: 1999,
\newblock {\em Class. Quantum Grav.} {\bf 16}, 3781

\bibitem[\protect\astroncite{Coble et~al.}{1997}]{coble-97}
Coble, K., Dodelson, S. and Frieman, J.~A.: 1997,
\newblock {\em Phys. Rev. D} {\bf 55(4)}, 1851

\bibitem[\protect\astroncite{Coley}{1991}]{coley91}
Coley, A.~A.: 1991,
\newblock {\em Class. Quantum Grav.} {\bf 8}, 955,
\newblock 

\bibitem[\protect\astroncite{Coley and Tupper}{1983}]{col-tup83}
Coley, A.~A. and Tupper, B. O.~J.: 1983,
\newblock {\em Gen. Rel. Grav.} {\bf 15(10)}, 977

\bibitem[\protect\astroncite{Coley and Tupper}{1984}]{col-tup84}
Coley, A.~A. and Tupper, B. O.~J.: 1984,
\newblock {\em Astrophys. J.} {\bf 280}, 26

\bibitem[\protect\astroncite{Coley and Tupper}{1985}]{col-tup85}
Coley, A.~A. and Tupper, B. O.~J.: 1985,
\newblock {\em Astrophys. J.} {\bf 288}, 418

\bibitem[\protect\astroncite{Coley et~al.}{1996}]{col-96}
Coley, A.~A., van~den Hoogen, R.~J. and Maartens, R.: 1996,
\newblock {\em Phys. Rev. D.} {\bf 54(2)}, 1393

\bibitem[\protect\astroncite{Collins and Wainwright}{1983}]{col-wai83}
Collins, C.~B. and Wainwright, J.: 1983,
\newblock {\em Phys. Rev. D} {\bf 27(6)}, 1209

\bibitem[\protect\astroncite{Cornish and Spergel}{1999}]{cor-spe99}
Cornish, N.~J. and Spergel, D.~N.: 1999,
\newblock {\em astro-ph/9906401}

\bibitem[\protect\astroncite{\dab\ and Stelmach}{1989}]{dab-stel89}
\dab\, M.~P. and Stelmach, J.: 1989,
\newblock {\em Astron. J.} {\bf 97(4)}, 978

\bibitem[\protect\astroncite{D\c{a}browski}{1993}]{dab93}
D\c{a}browski, M.~P.: 1993,
\newblock {\em J. Math. Phys.} {\bf 34(4)}, 1447,
\newblock 

\bibitem[\protect\astroncite{D\c{a}browski}{1995}]{dab95}
D\c{a}browski, M.~P.: 1995,
\newblock {\em Astrophys. J.} {\bf 447}, 43

\bibitem[\protect\astroncite{D\c{a}browski and Hendry}{1998}]{dab-hen98}
D\c{a}browski, M.~P. and Hendry, M.~A.: 1998,
\newblock {\em Astrophys. J.} {\bf 498}, 67

\bibitem[\protect\astroncite{Dicke et~al.}{1965}]{dic-65}
Dicke, R.~H., Peebles, P. J.~E., Roll, P.~G. and Wilkinson, D.~T.: 1965,
\newblock {\em Astrophys. J.} {\bf 142}, 414

\bibitem[\protect\astroncite{Dom\'{\i}nguez-Tenreiro}{981a}]{dom81a}
Dom\'{\i}nguez-Tenreiro, R.: 1981a,
\newblock {\em Astrophys. J.} {\bf 247}, 1

\bibitem[\protect\astroncite{Dom\'{\i}nguez-Tenreiro}{981b}]{dom81b}
Dom\'{\i}nguez-Tenreiro, R.: 1981b,
\newblock {\em Astron. Astrophys.} {\bf 93}, 306

\bibitem[\protect\astroncite{Drell et~al.}{1999}]{dre-99}
Drell, P.~S., Loredo, T.~J. and Wasserman, I.: 1999,
\newblock {\em astro-ph/9905027}

\bibitem[\protect\astroncite{Efstathiou et~al.}{1999}]{efs-98}
Efstathiou, G., Bridle, S. L., Lasenby, A. N., Hobson, M. P. and
Ellis, R. S.: 1999,
\newblock {\em Mon. Not. R. Astron. Soc.} {\bf 303}, L47
({\em astro-ph/9812226})

\bibitem[\protect\astroncite{Efstathiou}{1999}]{efs99}
Efstathiou, G.: 1999,
\newblock {\em Mon. Not. R. Astron. Soc.} {\bf 310}, 842
({\em astro-ph/9904356})

\bibitem[\protect\astroncite{Ehlers}{1971}]{ehl71}
Ehlers, J.: 1971,
\newblock in B.~K. Sachs (ed.), {\em General Relativity and Cosmology},
  Academic Press, New York

\bibitem[\protect\astroncite{Ehlers}{1993}]{ehl93}
Ehlers, J.: 1993,
\newblock {\em Gen. Rel. Grav.} {\bf 25(12)}, 1225,
\newblock Translated 1961, Abh.math.-Nat.,kl.,Mainz Akad. Wiss.u.Lit. Nr 11

\bibitem[\protect\astroncite{Ehlers et~al.}{1968}]{ehlers-68}
Ehlers, J., Geren, P. and Sachs, R.~K.: 1968,
\newblock {\em J. Math. Phys.} {\bf 9(9)}, 1344

\bibitem[\protect\astroncite{Ehlers and Newman}{1999}]{ehl-new99}
Ehlers, J. and Newman, E.~T.: 1999,
\newblock {\em gr-qc/9906065}

\bibitem[\protect\astroncite{{Ehlers} and {Rindler}}{1989}]{ehl-rin89}
{Ehlers}, J. and {Rindler}, W.: 1989,
\newblock {\em Mon. Not. R. Astron. Soc.} {\bf 238}, 503

\bibitem[\protect\astroncite{Einstein and de~Sitter}{1932}]{eis-des32}
Einstein, A. and de~Sitter, W.: 1932,
\newblock {\em Proc. Natl. Acad. Sci. (USA)} {\bf 18}, 213

\bibitem[\protect\astroncite{{Eisenstein} et~al.}{998a}]{eis-98a}
{Eisenstein}, D.~J., {Hu}, W. and {Tegmark}, M.: 1998a,
\newblock {\em Astrophys. J. Lett.} {\bf 504}, L57

\bibitem[\protect\astroncite{{Eisenstein} et~al.}{998b}]{eis-98b}
{Eisenstein}, D.~J., {Hu}, W. and {Tegmark}, M.: 1998b,
\newblock in {\em Evolution of Large-Scale Structure: From Recombination to
  Garching}, p.~E1

\bibitem[\protect\astroncite{Ellis}{1971}]{ellis71}
Ellis, G. F.~R.: 1971,
\newblock in B.~K. Sachs (ed.), {\em General Relativity and Cosmology}, pp
  104--182, Academic Press, New York

\bibitem[\protect\astroncite{Ellis}{1975}]{ellis75}
Ellis, G. F.~R.: 1975,
\newblock {\em The Quarterly Journal of the Royal Astronomical Society} {\bf
  16}, 245

\bibitem[\protect\astroncite{Ellis et~al.}{1978}]{ellis-78}
Ellis, G. F.~R., Maartens, R. and Nel, S.~D.: 1978,
\newblock {\em Mon. Not. R. Astron. Soc.} {\bf 184}, 439

\bibitem[\protect\astroncite{Ellis et~al.}{983a}]{ellis-83-I}
Ellis, G. F.~R., Matravers, D.~R. and Treciokas, R.: 1983a,
\newblock {\em Ann. Phys.} {\bf 150}, 455

\bibitem[\protect\astroncite{Ellis et~al.}{983b}]{ellis-83-II}
Ellis, G. F.~R., Matravers, D.~R. and Treciokas, R.: 1983b,
\newblock {\em Ann. Phys.} {\bf 150}, 487

\bibitem[\protect\astroncite{Ellis et~al.}{1985}]{ellis-85}
Ellis, G. F.~R., Nel, S.~D., Maartens, R., Stoeger, W.~R. and Whitman., A.~P.:
  1985,
\newblock {\em Physics Reports} {\bf 124(5 \& 6)}, 315

\bibitem[\protect\astroncite{Ellis and van Elst}{}]{ell98}
Ellis, G. F.~R. and van Elst, H.
: 1998, in M. Lachieze-Rey (ed.),
{\em Theoretical and Observational Cosmology}, NATO Science Series,
Kluwer Academic Publishers
({\em gr-qc/9812046v3})

\bibitem[\protect\astroncite{Etherington}{1933}]{eth33}
Etherington, I. M.~H.: 1933,
\newblock {\em Phil. Mag.} {\bf 15}, 761,
\newblock 

\bibitem[\protect\astroncite{Fennelly}{1976}]{fen76}
Fennelly, A.~J.: 1976,
\newblock {\em Astrophys. J.} {\bf 207}, 693

\bibitem[\protect\astroncite{Ferrando et~al.}{1992}]{ferran-92}
Ferrando, J.~J., Morales, J.~A. and Portilla, M.: 1992,
\newblock {\em Phys. Rev. D} {\bf 46(2)}, 578

\bibitem[\protect\astroncite{Finkbeiner and Schlegel}{1999}]{fin-sch99}
Finkbeiner, D.~P. and Schlegel, D.~J.: 1999,
\newblock in A. de Oliveira-Costa and M. Tegmark (eds.)
{\em Microwave Foregrounds}, Publications of the Astronomical
Society of the Pacific ({\em astro-ph/9907307})

\bibitem[\protect\astroncite{Frampton}{1999}]{framp99}
Frampton, P.~H.: 1999,
\newblock {\em astro-ph/9901013}

\bibitem[\protect\astroncite{Freedman}{1998}]{freed99b}
Freedman, W.: 1998,
\newblock {\em Physics Reports} {\bf 307}, 45
({\em astro-ph/9909076})

\bibitem[\protect\astroncite{Freedman}{1999}]{free99}
Freedman, W.: 1999,
\newblock {\em astro-ph/9905222}

\bibitem[\protect\astroncite{Gariel and Denmat}{1994}]{gar-led94}
Gariel, J. and Denmat, G.~L.: 1994,
\newblock {\em Phys. Rev. D} {\bf 50(4)}, 2560

\bibitem[\protect\astroncite{Gebbie and Ellis}{1998}]{geb-ell99}
Gebbie, T. and Ellis, G. F.~R.: 1998,
\newblock {\em astro-ph/9804316v3}

\bibitem[\protect\astroncite{Goicoechea and Martin-Mirones}{1987}]{goi-mm97}
Goicoechea, L.~J. and Martin-Mirones, J.~M.: 1987,
\newblock {\em Astron. Astrophys.} {\bf 186}, 22

\bibitem[\protect\astroncite{Goliath and Ellis}{1999}]{gol-ellis98}
Goliath, M. and Ellis, G. F.~R.: 1999,
\newblock {\em Phys. Rev. D} {\bf 60}, 023502

\bibitem[\protect\astroncite{Goodman}{1995}]{goo95}
Goodman, J.: 1995,
\newblock {\em Phys. Rev. D} {\bf 52(4)}, 1821

\bibitem[\protect\astroncite{Gratton et~al.}{1997}]{gra-97}
Gratton, R.~G. et~al.: 1997,
\newblock {\em Astrophys. J.} {\bf 491}, 749

\bibitem[\protect\astroncite{Gunzig et~al.}{1998}]{gunz-97}
Gunzig, E., Maartens, R. and Nesteruk, A.~V.: 1998,
\newblock {\em Class. Quantum Grav.} {\bf 15}, 923

\bibitem[\protect\astroncite{Hamuy et~al.}{1996}]{hamuy-96}
Hamuy, M., Phillips, M.~M., Maza, J., Suntzeff, N.~B., Schommer, R.~A. and
  Aviles, R.: 1996,
\newblock {\em Astrophys. J.} {\bf 112}, 2391

\bibitem[\protect\astroncite{Hawking and Ellis}{1973}]{hawk-ell}
Hawking, S.~W. and Ellis, G. F.~R.: 1973,
\newblock {\em The Large Scale Structure of Space-Time},
\newblock Cambridge Monographs on Mathematical Physics, Cambridge University
  Press

\bibitem[\protect\astroncite{Hellaby and Lake}{1984}]{hel-lak-I84}
Hellaby, C. and Lake, K.: 1984,
\newblock {\em Astrophys. J.} {\bf 282}, 1

\bibitem[\protect\astroncite{Hellaby and Lake}{1985}]{hel-lak85}
Hellaby, C. and Lake, K.: 1985,
\newblock {\em Astrophys. J.} {\bf 290}, 381

\bibitem[\protect\astroncite{Humphreys et~al.}{1997}]{humph-97}
Humphreys, N.~P., Maartens, R. and Matravers, D.~R.: 1997,
\newblock {\em Astrophys. J.} {\bf 477}, 47

\bibitem[\protect\astroncite{Israel and Stewart}{1979}]{isr-ste79}
Israel, W. and Stewart: 1979,
\newblock {\em Ann. Phys.} {\bf 118}, 341

\bibitem[\protect\astroncite{Israel and Stewart}{1980}]{isr-ste80}
Israel, W. and Stewart: 1980,
\newblock in A. Held (ed.), {\em GR and Gravitation: 100 Years after the birth
  of Einstein.}, New York London : Plenum Press

\bibitem[\protect\astroncite{Kantowski and Sachs}{1966}]{kant-sachs66}
Kantowski, R. and Sachs, R.~K.: 1966,
\newblock {\em J.~Math. Phys.} {\bf 7}, 443

\bibitem[\protect\astroncite{{Kaufman}}{1971}]{kauf71}
{Kaufman}, S.~E.: 1971,
\newblock {\em Astron. J.} {\bf 76}, 751

\bibitem[\protect\astroncite{{Kaufman} and {Schucking}}{1971}]{kauf-sch71}
{Kaufman}, S.~E. and {Schucking}, E.~L.: 1971,
\newblock {\em Astron. J.} {\bf 76}, 583

\bibitem[\protect\astroncite{Kramer et~al.}{1980}]{kram-80}
Kramer, D. et~al.: 1980,
\newblock {\em Exact Solutions of Einstein's Field Equations},
\newblock Cambridge University Press

\bibitem[\protect\astroncite{Krasi\'{n}ski}{1983}]{Kras83}
Krasi\'{n}ski, A.: 1983,
\newblock {\em Gen. Rel. Grav.} {\bf 15(7)}, 673

\bibitem[\protect\astroncite{Krasi\'nski}{1989}]{kras89}
Krasi\'nski, A.: 1989,
\newblock {\em J. Math. Phys.} {\bf 30(2)}, 433,
\newblock 

\bibitem[\protect\astroncite{Krasi\'{n}ski}{1997}]{kras97}
Krasi\'{n}ski, A.: 1997,
\newblock {\em Inhomogeneous Cosmological Models},
\newblock Cambridge University Press
\newblock 

\bibitem[\protect\astroncite{Krasi\'{n}ski}{1998}]{kras98}
Krasi\'{n}ski, A.: 1998,
\newblock {\em gr-qc/9806039}

\bibitem[\protect\astroncite{Krasi\'nski et~al.}{1997}]{kras-97}
Krasi\'nski, A., Quevedo, H. and Sussman, R.~A.: 1997,
\newblock {\em J. Math. Phys.} {\bf 38(5)}, 2602

\bibitem[\protect\astroncite{Krauss}{1998}]{kra99}
Krauss, L.~M.: 1998,
\newblock {\em Physics Reports} {\bf 307}, 235
({\em astro-ph/9907308})

\bibitem[\protect\astroncite{Kristian and Sachs}{1966}]{kris-sac66}
Kristian, J. and Sachs, R.~K.: 1966,
\newblock {\em Astrophys. J.} {\bf 143(2)}, 379,
\newblock 

\bibitem[\protect\astroncite{Liddle}{1999}]{lid99}
Liddle, A.~R.: 1999,
\newblock {\em astro-ph/9901041}

\bibitem[\protect\astroncite{Liddle and Scherrer}{1998}]{lid-sch98}
Liddle, A.~R. and Scherrer, R.~J.: 1998,
\newblock {\em Phys. Rev. D} {\bf 58}, 083508

\bibitem[\protect\astroncite{{Lilje} et~al.}{1986}]{lil-86}
{Lilje}, P.~B., {Yahil}, A. and {Jones}, B. J.~T.: 1986,
\newblock {\em Astrophys. J.} {\bf 307}, 91

\bibitem[\protect\astroncite{Loeb}{1999}]{loe99}
Loeb, A.: 1999,
\newblock {\em astro-ph/9907187}

\bibitem[\protect\astroncite{Lorenz-Petzold}{1986}]{lorpet86}
Lorenz-Petzold, D.: 1986,
\newblock {\em Astrophys. Astron.} {\bf 7(4.9)}, 155

\bibitem[\protect\astroncite{Luminet and Roukema}{1998}]{lum-rou99}
Luminet, J. and Roukema, B.~F.: 1998,
\newblock in M. Lachieze-Rey (ed.), {\em Theoretical and Observational
Cosmology}, NATO Science Series, Kluwer Academic Publishers
({\em astro-ph/9901364})

\bibitem[\protect\astroncite{et~al.}{1988}]{lynd-bell88}
Lynden-Bell, D. et~al.: 1988,
\newblock {\em Astrophys. J.} {\bf 326}, 19

\bibitem[\protect\astroncite{Maartens}{1995}]{maar95}
Maartens, R.: 1995,
\newblock {\em Class. Quantum Grav.} {\bf 12}, 1455

\bibitem[\protect\astroncite{Maartens}{1996}]{maar96}
Maartens, R.: 1996,
\newblock {\em astro-ph/9609119}

\bibitem[\protect\astroncite{Maartens}{1999}]{maar99}
Maartens, R.: 1999,
\newblock {\em astro-ph/9907284}

\bibitem[\protect\astroncite{Maartens et~al.}{1995}]{maart-95}
Maartens, R., Ellis, G. F.~R. and Stoeger, W.~R.: 1995,
\newblock {\em Phys. Rev. D} {\bf 51(4)}, 1525
({\em astro-ph/9501016})

\bibitem[\protect\astroncite{Maartens et~al.}{1999}]{maar-99}
Maartens, R., Gebbie, T. and Ellis, G. F.~R.: 1999,
\newblock {\em Phys. Rev. D} {\bf 59}, 083506,
\newblock 

\bibitem[\protect\astroncite{Maartens et~al.}{1996}]{maar-95}
Maartens, R., Humphreys, N.~P., Matravers, D.~R. and Stoeger, W.~R.: 1996,
\newblock {\em Class. Quantum Grav.} {\bf 13}, 253

\bibitem[\protect\astroncite{MacCallum and Ellis}{1970b}]{Mac-Ellis-I}
MacCallum, M. A.~H. and Ellis, G. F.~R.: 1970b,
\newblock {\em Comm. Math. Phys.} {\bf 12}, 108,
\newblock 

\bibitem[\protect\astroncite{MacCallum and Ellis}{1970a}]{Mac-Ellis-II}
MacCallum, M. A.~H. and Ellis, G. F.~R.: 1970a,
\newblock {\em Comm. Math. Phys.} {\bf 19}, 31

\bibitem[\protect\astroncite{Mather et~al.}{1994}]{mat-94}
Mather, J.~C. et~al.: 1994,
\newblock {\em Astrophys. J.} {\bf 420}, 439

\bibitem[\protect\astroncite{Mattig}{1958}]{mat58}
Mattig, W.: 1958,
\newblock {\em Astron. Nachr.} {\bf 284}, 109,
\newblock 

\bibitem[\protect\astroncite{Misner et~al.}{1971}]{mtw}
Misner, C.~W., Thorne, K.~S. and Wheeler, J.~A.: 1971,
\newblock {\em Gravitation},
\newblock Freeman

\bibitem[\protect\astroncite{Moffat and Tatarski}{1995}]{mof-tat95}
Moffat, J.~W. and Tatarski, D.~C.: 1995,
\newblock {\em Astrophys. J.} {\bf 453}, 17

\bibitem[\protect\astroncite{Mustapha et~al.}{1998}]{mus-hel98}
Mustapha, N., Hellaby, C. and Ellis., G. F.~R.: 1998,
\newblock {\em Mon. Not. R. Astron. Soc.} {\bf 292}, 817

\bibitem[\protect\astroncite{Nakao et~al.}{1995}]{nakao-95}
Nakao, K. et~al.: 1995,
\newblock {\em Astrophys. J.} {\bf 453}, 541

\bibitem[\protect\astroncite{Nilsson et~al.}{1999}]{nils-99}
Nilsson, U.~S., Uggla, C., Wainwright, J. and Lim, W.~C.: 1999,
\newblock {\em Astrophys. J. Lett.} {\bf 521}, L1
({\em astro-ph/9904252})

\bibitem[\protect\astroncite{Olive}{1998}]{oli99}
Olive, K.~A.: 1998,
\newblock in M. Lachieze-Rey (ed.), {\em Theoretical and Observational
Cosmology}, NATO Science Series, Kluwer Academic Publishers
({\em astro-ph/9901231})

\bibitem[\protect\astroncite{Paczy\'{n}ski and Piran}{1990}]{pac-pir90}
Paczy\'{n}ski, B. and Piran, T.: 1990,
\newblock {\em Astrophys. J.} {\bf 364}, 341

\bibitem[\protect\astroncite{Palle}{1999}]{pal99}
Palle, D.: 1999,
\newblock {\em astro-ph/9905252}

\bibitem[\protect\astroncite{Partridge}{1997}]{part97}
Partridge, B.: 1997,
\newblock in D. Valls-Gabaud et~al. (eds.), {\em From Quantum Fluctuations to
  Cosmological Structures}, pp 141--184, ASP Conference Series, Vol. 126

\bibitem[\protect\astroncite{Peebles}{1993}]{peebles}
Peebles, P.: 1993,
\newblock {\em Principles of Physical Cosmology},
\newblock Princeton University Press

\bibitem[\protect\astroncite{Peebles and Ratra}{1988}]{peebl-88}
Peebles, P. J.~E. and Ratra, B.: 1988,
\newblock {\em Astrophys. J. Lett.} {\bf 325}, L17

\bibitem[\protect\astroncite{Penzias and Wilson}{1965}]{pen-wil65}
Penzias, A.~A. and Wilson, R.~W.: 1965,
\newblock {\em Astrophys. J.} {\bf 124}, 419

\bibitem[\protect\astroncite{{Perlmutter} et~al.}{1999}]{perl-99}
{Perlmutter}, S. et~al.: 1999,
\newblock {\em Astrophys. J.} {\bf 517}, 565,
\newblock 

\bibitem[\protect\astroncite{Ratra and Peebles}{1988}]{ratra-88}
Ratra, B. and Peebles, P. J.~E.: 1988,
\newblock {\em Phys. Rev. D} {\bf 37}, 3406

\bibitem[\protect\astroncite{Rees and Sciama}{1968}]{ree-sci68}
Rees, M.~J. and Sciama, D.~W.: 1968,
\newblock {\em Nature} {\bf 517}, 611

\bibitem[\protect\astroncite{{Regos} and {Szalay}}{1989}]{reg-sza89}
{Regos}, E. and {Szalay}, A.~S.: 1989,
\newblock {\em Astrophys. J.} {\bf 345}, 627

\bibitem[\protect\astroncite{Reid}{1997}]{rei97}
Reid, I.~N.: 1997,
\newblock {\em Astron. J.} {\bf 114}, 161

\bibitem[\protect\astroncite{Riess}{1999}]{rie99}
Riess, A.~G.: 1999,
\newblock in S. Courteau, M. Strauss and J. Willick (eds.),
{\em Towards an Understanding of Cosmic Flows}, Publications of the
Astronomical Society of the Pacific
({\em astro-ph/9908237})

\bibitem[\protect\astroncite{Riess et~al.}{1998}]{riess-98}
Riess, A.~G. et~al.: 1998,
\newblock {\em Astrophys. J.} {\bf 116}, 1009

\bibitem[\protect\astroncite{Rindler}{1956}]{rind56}
Rindler, W.: 1956,
\newblock {\em Mon. Not. R. Astron. Soc.} {\bf 6}, 662

\bibitem[\protect\astroncite{Rindler and Suson}{1989}]{rindler-sus89}
Rindler, W. and Suson, D.: 1989,
\newblock {\em Astron. Astrophys.} {\bf 218}, 15

\bibitem[\protect\astroncite{Rocha}{1999}]{roc99}
Rocha, G.: 1999,
\newblock {\em astro-ph/9907312}

\bibitem[\protect\astroncite{{Sachs} and {Wolfe}}{1967}]{sac-wol67}
{Sachs}, R.~K. and {Wolfe}, A.~M.: 1967,
\newblock {\em Astrophys. J.} {\bf 147}, 73+

\bibitem[\protect\astroncite{Saunders et~al.}{1999}]{sau-99}
Saunders, W. et~al.: 1999,
\newblock in S. Courteau, M. Strauss and J. Willick (eds.),
{\em Towards an Understanding of Cosmic Flows}, Publications of the
Astronomical Society of the Pacific
({\em astro-ph/9909190})

\bibitem[\protect\astroncite{Scaramella et~al.}{1991}]{scar-91}
Scaramella, R., Vettolani, G. and Zamorani, G.: 1991,
\newblock {\em Astrophys. J.} {\bf 376}, L1

\bibitem[\protect\astroncite{Schmidt et~al.}{1998}]{schmidt-98}
Schmidt, B.~P. et~al.: 1998,
\newblock {\em Astrophys. J.} {\bf 507}, 46

\bibitem[\protect\astroncite{Schmoldt et~al.}{1999}]{schmoldt-99}
Schmoldt, I. et~al.: 1999,
\newblock {\em Mon. Not. R. Astron. Soc.} {\bf 304}, 893
({\em astro-ph/9901087})

\bibitem[\protect\astroncite{Schneider and C\'el\'erier}{1999}]{sch-cel99}
Schneider, J. and C\'el\'erier, M.: 1999,
\newblock {\em Astron. Astrophys.} {\bf 348}, 25

\bibitem[\protect\astroncite{Schutz}{1980}]{sch80}
Schutz, B.: 1980,
\newblock {\em Geometrical Methods of Mathematical Physics},
\newblock Cambridge University Press

\bibitem[\protect\astroncite{Schutz}{1990}]{schutz-rel}
Schutz, B.~F.: 1990,
\newblock {\em A First Course in General Relativity},
\newblock Cambridge University Press

\bibitem[\protect\astroncite{Stephani}{967a}]{step67a}
Stephani, H.: 1967a,
\newblock {\em Comm. Math. Phys.} {\bf 4}, 137

\bibitem[\protect\astroncite{Stephani}{967b}]{step67b}
Stephani, H.: 1967b,
\newblock {\em Comm. Math. Phys.} {\bf 5}, 337

\bibitem[\protect\astroncite{Stephani}{1990}]{step90}
Stephani, H.: 1990,
\newblock {\em General Relativity},
\newblock Cambridge, 2 edition

\bibitem[\protect\astroncite{Stoeger et~al.}{1995}]{stoeg-95}
Stoeger, W.~R., Maartens, R. and Ellis, G. F.~R.: 1995,
\newblock {\em Astrophys. J.} {\bf 443}, 1

\bibitem[\protect\astroncite{Straumann}{1999}]{str99}
Straumann, N.: 1999,
\newblock {Eur. J. Phys.} {\bf 20}, 419
({\em astro-ph/9908342})

\bibitem[\protect\astroncite{Sunyaev and Zel'dovich}{1969}]{sun-zel69}
Sunyaev, R.~A. and Zel'dovich, Y.~B.: 1969,
\newblock {\em Astrophys. \& Space Sci.} {\bf 4}, 301

\bibitem[\protect\astroncite{Sussman}{1999}]{sus99}
Sussman, R.~A.: 1999,
\newblock {\em gr-qc/9908019}

\bibitem[\protect\astroncite{Tadros et~al.}{1999}]{tad-99}
Tadros, H. et~al.: 1999,
\newblock {\em Mon. Not. R. Astron. Soc.} {\bf 305}, 527
({\em astro-ph/9901351})

\bibitem[\protect\astroncite{Tauber and Weinberg}{1961}]{tau-wei61}
Tauber, G.~E. and Weinberg, J.~W.: 1961,
\newblock {\em Phys.Rev.} {\bf 122}, 1342,
\newblock 

\bibitem[\protect\astroncite{{Tegmark}}{1999}]{teg99}
{Tegmark}, M.: 1999,
\newblock {\em Astrophys. J. Lett.} {\bf 514}, L69

\bibitem[\protect\astroncite{Tegmark et~al.}{998b}]{teg-98b}
Tegmark, M., Eisenstein, D.~J., Hu, W. and Kron, G.: 1998b,
\newblock {\em astro-ph/9805117}

\bibitem[\protect\astroncite{Tolman}{1934}]{tolm34}
Tolman, R.~C.: 1934,
\newblock {\em Proc. Nat. Acad. Sci. (Wash.)} {\bf 20}, 169

\bibitem[\protect\astroncite{Tomita}{1995}]{tomita95}
Tomita, K.: 1995,
\newblock {\em Astrophys. J.} {\bf 451}, 1

\bibitem[\protect\astroncite{Tomita}{1996}]{tom96}
Tomita, K.: 1996,
\newblock {\em Astrophys. J.} {\bf 461}, 507

\bibitem[\protect\astroncite{Treciokas and Ellis}{1971}]{trec-ell71}
Treciokas, R. and Ellis, G. F.~R.: 1971,
\newblock {\em Comm. Math. Phys.} {\bf 23}, 1

\bibitem[\protect\astroncite{Trimble and Aschwanden}{1999}]{tri-ash99}
Trimble, V. and Aschwanden, M.: 1999,
\newblock {\em Publications of the Astronomical Society of the Pacific} {\bf
  111}, 385

\bibitem[\protect\astroncite{Turner}{1999}]{tur99}
Turner, M.~S.: 1999,
\newblock in D.~O. Caldwell (ed.), {\em The Proceedings of
Particles Physics and the Universe (Cosmo-98)}, AIP, Woodbury, New York
({\em astro-ph/9904051})

\bibitem[\protect\astroncite{Vilenkin}{1981}]{Vilenkin}
Vilenkin, A.: 1981,
\newblock {\em Phys.~Rev.~Lett.} {\bf 46}, 1169

\bibitem[\protect\astroncite{Wainwright and Ellis}{1997}]{wain-ell97}
Wainwright, J. and Ellis, G. F.~R. (eds.): 1997,
\newblock {\em Dynamical Systems in Cosmology},
\newblock Cambridge University Press

\bibitem[\protect\astroncite{Wald}{1984}]{wald}
Wald, R.~M.: 1984,
\newblock {\em General Relativity},
\newblock The University of Chicago Press

\bibitem[\protect\astroncite{{White}}{1998}]{whi98}
{White}, M.: 1998,
\newblock {\em Astrophys. J.} {\bf 506}, 495

\bibitem[\protect\astroncite{Willick}{1998}]{willick}
Willick, J.~A.: 1998,
\newblock {\em Astrophys. J.} {\bf 522}, 647

\bibitem[\protect\astroncite{Zlatev et~al.}{1999}]{zlat-98}
Zlatev, I., Wang, L. and Steinhardt, P.~J.: 1999,
\newblock {\em Phys. Rev. Lett.} {\bf 82}, 896

\end{thebibliography}


\begin{thebibliography}{}

\bibitem[\protect\astroncite{Barnes}{1998}]{barn98}
Barnes, A. 1998,
\newblock {\em Class. Quantum Grav.} {\bf 15}, 3061

\bibitem[\protect\astroncite{Coley}{1991}]{coley91}
Coley, A.~A. 1991,
\newblock {\em Class. Quantum Grav.} {\bf 8}, 955
\newblock 

\bibitem[\protect\astroncite{Ehlers et~al.}{1968}]{ehlers-68}
Ehlers, J., Geren, P., and Sachs, R.~K. 1968,
\newblock {\em J. Math. Phys.} {\bf 9(9)}, 1344

\bibitem[\protect\astroncite{Ellis}{1998}]{ell98}
Ellis, G. F.~R. 1998,
\newblock {\em gr-qc/9812046v3}

\bibitem[\protect\astroncite{Ellis et~al.}{1978}]{ellis-78}
Ellis, G. F.~R., Maartens, R., and Nel, S.~D. 1978,
\newblock {\em Mon. Not. R. Astron. Soc.} {\bf 184},
  439

\bibitem[\protect\astroncite{Ferrando et~al.}{1992}]{ferran-92}
Ferrando, J.~J., Morales, J.~A., and Portilla, M. 1992,
\newblock {\em Phys. Rev. D} {\bf 46(2)}, 578

\bibitem[\protect\astroncite{Krasi\'nski}{1989}]{kras89}
Krasi\'nski, A. 1989,
\newblock {\em J. Math. Phys.} {\bf 30(2)}, 433
\newblock 

\bibitem[\protect\astroncite{Krasi\'{n}ski}{1997}]{kras97}
Krasi\'{n}ski, A. 1997,
\newblock {\em Inhomogeneous Cosmological Models},
\newblock CUP
\newblock 

\bibitem[\protect\astroncite{Kristian and Sachs}{1966}]{kris-sac66}
Kristian, J. and Sachs, R.~K. 1966,
\newblock {\em Ap. J.} {\bf 143(2)}, 379
\newblock 

\bibitem[\protect\astroncite{MacCallum and Ellis}{1970}]{Mac-Ellis-II}
MacCallum, M. A.~H. and Ellis, G. F.~R. 1970,
\newblock {\em Comm. Math. Phys.} {\bf 19}, 31
\newblock 

\bibitem[\protect\astroncite{Nilsson et~al.}{1999}]{nils-99}
Nilsson et~al. 1999,
\newblock {\em astro-ph/9904252}

\bibitem[\protect\astroncite{{Perlmutter} et~al.}{1999}]{perl-99}
Perlmutter, S. et~al. 1999,
\newblock {\em Ap.J.} {\bf 517}, 565
\newblock 

\bibitem[\protect\astroncite{Schmoldt et~al.}{1999}]{sch-99}
Schmoldt, I. et~al. 1999
\newblock {\em astro-ph/9901087}

\bibitem[\protect\astroncite{Stephani}{967a}]{step67a}
Stephani, H. 1967a,
\newblock {\em Comm. Math. Phys.} {\bf 4}, 137

\bibitem[\protect\astroncite{Stephani}{967b}]{step67b}
Stephani, H. 1967b,
\newblock {\em Comm. Math. Phys.} {\bf 5}, 337

\bibitem[\protect\astroncite{Stoeger et~al.}{1995}]{sto-maa95}
Stoeger, W.~R., Maartens, R., and Ellis, G. F.~R. 1995,
\newblock {\em Ap. J.} {\bf 443}, 1

\bibitem[\protect\astroncite{Tauber and Weinberg}{1961}]{tau-wei61}
Tauber, G.~E. and Weinberg, J.~W. 1961,
\newblock {\em Phys.Rev.} {\bf 122}, 1342
\newblock 

\bibitem[\protect\astroncite{Treciokas and Ellis}{1971}]{trec-ell71}
Treciokas, R. and Ellis, G. F.~R. 1971,
\newblock {\em Comm. Math. Phys.} {\bf 23}, 1

\bibitem[\protect\astroncite{Willick}{1998}]{willick}
Willick, J.~A. 1998
\newblock {\em astro-ph/9812470}

\end{thebibliography}

}

\end{document}